%% file: tesi.tex
\documentclass[a4paper,12pt]{book}
\usepackage[latin1]{inputenc}
\usepackage[english]{babel}
\usepackage{latexsym}
\usepackage{amssymb}
\usepackage{graphicx}
\usepackage{amscd}
\usepackage{amsfonts,epsfig}
\input{TXSsymb}

\usepackage{fancyhdr}

\makeatletter
\def\cleardoublepage{\clearpage\if@twoside \ifodd\c@page\else
        \hbox{}
        \thispagestyle{empty}
        \newpage
        \if@twocolumn\hbox{}\newpage\fi\fi\fi}
\makeatother

\newcommand{\di}{\mathrm{d}}
\newcommand{\fr}{\frac}
\newcommand{\wi}{\widehat}
\newcommand{\tra}{\mathrm{tr}}

\newcommand{\ol}{\overline}
\newcommand{\al}{\alpha}

\newcommand{\la}{\lambda}
\newcommand{\vp}{\mathbf{p}}
\newcommand{\vq}{\mathbf{q}}

\newcommand{\h}{m_0^2}

\newcommand{\LL}{\left}
\newcommand{\RR}{\right}

\def\bsigma{\mbox{\protect\boldmath $\sigma$}}
\newcommand{\N}{{\cal N}}

\def\reff#1{(\ref{#1})}
\newcommand{\<}{\langle}
\renewcommand{\>}{\rangle}

\begin{document}

%\small
\thispagestyle{empty}
\begin{center}

{\bf
UNIVERSIT\`A DEGLI STUDI DI MILANO\\
\vskip 0.2truecm
Facolt\`a di Scienze Matematiche Fisiche e Naturali \\
\vskip 0.2truecm
CORSO DI DOTTORATO DI RICERCA IN 
\vskip 0.2truecm
Fisica, Astrofisica e Fisica Applicata, Ciclo XIX\vspace{3.cm} }

{\bf TESI DI DOTTORATO }
\vskip 0.2truecm
\Large {\bf %\vskip 0.2truecm
 Infra-Red divergences in large-$N$ expansion}
\vskip 0.5truecm
\small{\bf SETTORE SCIENTIFICO-DISCIPLINARE FIS/02}
\end{center}

\vspace{2cm}

\begin{flushleft}
{\bf
Relatore:\hspace{.59cm}Prof.\ Sergio CARACCIOLO\\
\vskip 0.2truecm
Relatore esterno:\hspace{.5cm}Prof.\ Andrea PELISSETTO\\
\vskip 0.2truecm
Coordinatore:\hspace{.5cm}Prof.\ Gianpaolo BELLINI\\
\ }
\end{flushleft}

\vspace{2cm}

\begin{flushleft}
{\bf
\hspace{8.5cm}CANDIDATO\\
\vskip 0.1truecm
\hspace{8.5cm}Bortolo Matteo MOGNETTI \\
\vskip 0.1truecm
\hspace{8.5cm}Matr. R05424 \\
\vskip 0.1truecm \hspace{8.5cm}Codice P.A.C.S.: 11.15.Pg, 11.10.Wx, \\
\vskip 0.1truecm \hspace{10.1cm} 11.30.Rd, 12.40.Ee, 75.10.Hk. }
\end{flushleft}
\vspace{1.5cm}

\begin{center}
{\bf Anno Accademico 2005-2006}
\end{center}

\tableofcontents

%
%  INTRODUZIONE
%

\chapter*{Introduction}

The investigation of non perturbative aspects 
of quantum field theories or statistical mechanic  models
is a fundamental topic in recent theoretical
 developments. These studies would
give a more complete view 
of the structure of such theories (e.g.\ phase diagram), and
are essential in order to  elucidate non
perturbative aspects like quark confinement in QCD 
or chiral symmetry restoration at finite temperature.
Non perturbative techniques are both analytical
and numerical. The most used numerical 
methods are Monte Carlo simulations
that are very popular due to their universal
setting.\footnote{
Maybe the claim is too optimistic. There
are severe penalties related to MC simulations;
the most popular are critical slowing down  
\cite{SP} or the sign problem in fermions system
(e.g.\ \cite{S-06}).}
On the other side concerning with non perturbative
analytic approaches,
the $1/N$ expansions \cite{Coleman} have a central role,
that seems to increase in the last years.
The general scheme of this expansion  
(see chapter \ref{chapter1}) is to consider 
Hamiltonians with a large number $N$ of fields (that goes to
infinity) in such a way that the equations
of motion are simplified and can be solved analytically.
Proceeding in such a way one recovers  solutions similar to
those
found using other schemes like Mean Field, variational
methods or self-consistent approximations which reduce
the interacting system to a interaction free one, with
a self consistently determined parameter. For instance
by considering
a vectorial $\phi^4$ theory, one can write
\begin{eqnarray}
(\vec \phi^2)^2 &\Longrightarrow& \vec\phi^2
\langle \vec\phi^2\rangle
\nonumber
\end{eqnarray}
and can recover a free interaction with a mass term
that must be fixed in a self consistent way,
 by computing $\langle\phi^2 \rangle$.
However the $1/N$ expansion with respect to the 
previous methods has the advantage that can be
improved in a systematic way by including $1/N$
fluctuations. 
What does one expect from such an expansion? 
Of course one would extract  informations
about the physical case (at finite $N$), so that
one would detect for instance the presence of a phase
transition studying the $N=\infty$ solution.
This is not always possible. 
Indeed, due to the fact
that the large $N$ limit puts infinite degrees
of freedom in a finite volume, one could observe
several non analyticities  in the partition function
that are due to the taken limit instead of the
thermodynamic limit $V\to \infty$. This is the problem 
of the pathologies of the large $N$ limit
and has been investigated in detail in 
\cite{SS-00} by comparing the large $N$ expansion
with the  available exact solution in $d=1$.
In the case in which the large $N$ expansion is not
fictitious one would also try to get 
$1/N$ numerical corrections to the $N=\infty$ 
solution to compare with available data
at finite $N$ (for instance MC simulations).  
Another usually addressed advantage of the 
$1/N$ approaches is that
the theory can be solved (in certain case) in a
generic dimension,\footnote{However in the
approaches we present, as soon as one include $1/N$ 
fluctuations, there are severals distinguo 
in the scheme proposed. Of course this is not unexpected (we refer to section 
\ref{chapter2-higher-dimension} fore more details).}
while for instance conventional renormalization group equations
usually can be solved only when the fixed point 
is very near the Gaussian point (like for instance
in the $\epsilon$ expansion, $\epsilon\equiv 4-d$).

In this work we want to investigate several aspects
of the $1/N$ expansion for models with
a global $O(N)$ symmetry.\footnote{For a review
of $1/N$ expansion for this kind of models
we recommend \cite{MZ-03}.} As stressed in
\cite{MZ-03} approaching a $1/N$ expansion one
has to take care of two points (that are also 
two constraints of the theory):
\begin{itemize}
\item[{\em i})]{The theory must be both Infra Red Finite
and Renormalizable.}
\item[{\em ii})]{The large $N$ expansion is just a technique,
with its own limitation. The extrapolation of  results (also 
qualitative) to finite $N$ must be discussed carefully.
}
\end{itemize}
Point {\em ii}) deals with the question (introduced above) 
of the (possible) pathologies
introduced by the large $N$ limit \cite{SS-00}.
On the other hand, 
Renormalizability of large $N$ actions [point {\em i})] 
is quite delicate 
as the action for the auxiliary fields
is usually not local.\footnote{This is due to 
the $\tr\log(\cdots)$ terms that appear in the action for the 
auxiliary fields (see for instance 
chapter \ref{chapter5} eq.\ \ref{S-chiral-fermions}),
and it is due to the integration of the physical
fields.} 
 A scheme to circumvent this difficulty
is presented in \cite{Zinn-Justin} and \cite{MZ-03}.
In this work we mainly focus ourself on problems 
related to 
Infra Red  singularities  [point
{\em i})].  The fact that $1/N$ expansion is not
Infra Red finite  is usually addressed as a failure
of  the large $N$ technique. There are several examples 
in which this happens. In this work we have studied the following
cases: (a) Heisenberg models with generic ferromagnetic near
neighbour interaction (chap.\ \ref{chapter1},
\ref{chapter3}, \ref{chapter4}); (b) fermionic models (Yukawa 
and Gross-Neveu) at finite temperature (chap.\ 
\ref{chapter5}); (c) a tricritical model with a six-th order
interaction (chap.\ \ref{chapter6}). 
First, we will show (recover)
 that all the models introduced above 
exhibit a Mean Field phase diagram for $N=\infty$. 
The failure
of $1/N$ expansion calls for point {\em ii}): do
the $N<\infty$ models show a critical point or the
$N=\infty$ critical point is an artefact \cite{SS-00}?
Scaling arguments for model (b) \cite{KSS-98},
a Monte Carlo Simulation for model (a) \cite{BGH-02},
and exact estimates for model (a) \cite{VES-02},
suggest that the phase transition is present
also at finite $N$ with a critical behavior
that is Ising like. These results 
suggest that the phase transition near an Infra Red
singular point is present also in the physical sector
$N<\infty$ so that (at least in the present cases)
the large $N$ limit does not produce artefacts.
 However a question remains open
on the nature of the Critical Point (if present).
In this work we will confirm the results of
\cite{KSS-98}, \cite{BGH-02} and \cite{VES-02} 
that for every $N$ finite one observes an Ising phase 
transition. However in the large $N$ limit the critical zone
in which Ising behavior is observed scales 
with a proper power of $1/N$ reducing to zero in the
spherical limit ($N=\infty$). This explains why only Mean Field
is observed for $N=\infty$. This claim will be supported
by the study of the effective interaction of the critical 
mode\footnote{Infra Red divergences are related to the appearance
of a null eigenvalue in the inverse of the propagator
$P^{-1}$, which eigenvector is usually  named zero mode.}
 that, for the models presented above, 
 looks like a weakly coupled $u \varphi^4 $ interaction 
(with $u\sim 1/N$) which
exhibits an interesting crossover limit (called Critical
crossover Limit) that matches Ising and Mean Field
criticality. The mechanism is the same of what
observe in medium range models
\cite{LBB-97} \cite{PRV-98}, where now $N$ 
plays the same role of $R$ ($R$ being the range of
 spin interaction).

The conclusion of this work are that for models
(a), (b) and (c) the problem of Infra Red divergences
can be solved using a generalized $1/N$ expansion 
which explains
the universal\footnote{Indeed we will show that 
the crossover functions of all the models considered are the same
apart two obvious normalization condition. Details on how
to compare numerically models a) b) c) with field theory
are given
in app.\ \ref{App}.} crossover between 
Mean Field and Ising (or Multicritical Points in
chapter \ref{chapter4}). We believe that the scheme proposed
is quite general and can be applied to other models
characterized by the presence of a zero mode.\footnote{
This claim is supported by the general considerations
presented in sec.\ \ref{sec-stable}.}
However in principle it could  happen that 
 the weakly coupled
interaction could be unstable 
(this is related to a negative mass term 
in the zero mode interaction); until now it is 
not clear to us  if this is the sign that
the $N=\infty$ critical point considered is unstable
(we have developed the theory around the
wrong point!) or the sign that now really the $N=\infty$
critical point is an artefact of the large $N$ limit 
[point {\em ii})].  Maybe, it seems that the answer
to the previous question  is not general but model
dependent. Interesting also the fact that 
the method proposed
gives a general scheme to compute $1/N$ corrections
of the  critical parameters, like for instance the
 critical temperature $T_c$. For example, in
the critical models this can be done by studying 
the mass counterterm (and the magnetic one if the
theory is not symmetric) that  for $d<4$
can be obtained  by a computation in which
only a finite number of diagrams enter\footnote{
This is due to the fact that in a bare $\varphi^4$ theory,
if $d<4$ 
only a finite number of diagram must be 
regularized.
}
(see chap.\ \ref{chapter2}).
This is a very important point in order
to check large $N$ expansions with 
MC simulations.
In sec.\ \ref{sec9} we present some numerical predictions
of a model that has been investigated in \cite{BGH-02}
by a MC simulation for $N=3$. The  fluctuations
seem to match the $N=\infty$ with the $N=3$ results
but there are not enough precision 
for any claims.
As future plan we would compute
corrections to the critical temperature 
for the model simulated in
\cite{KSS-98}. 
In this case there is the advantage (instead of \cite{BGH-02})
 that there are simulations for several $N$
($N=4,12,24$) so that (we believe) the check will 
give more precise answers.

\section*{Plan of the work}

The aim of the first four chapters is to introduce
the generalized large $N$ formalism in the most general case
(i.e.\ without any symmetry in the effective action).
In order to do that 
we introduce Heisenberg models with general 
nearest neighbour ferromagnetic interaction
\begin{eqnarray}
{\cal H} &=& -N \beta \sum_{\langle ij \rangle}
W(1+\bsigma_i \cdot\bsigma_j).
\nonumber
\end{eqnarray}
In {\bf chapter \ref{chapter1}} we show as
for a large class of interaction $W$ a 
continuous phase transition is present at 
finite temperature with  Mean Field criticality.
By an accurate choice 
of the interaction $W$, one is able 
to detect also multicritical points.
However as soon as one tries to include $1/N$ fluctuations
infra red divergences appear near the (multi)critical point
and the $1/N$ expansion fails.
This problem has been solved in {\bf chapter \ref{chapter3}}
and {\bf chapter \ref{chapter4}} (respectively for
the critical and for the multicritical case)
by a careful study of the effective action of the
critical mode\footnote{
The appearance of a critical mode at the critical point
is strictly related
to infra red singularities. 
} that looks like a weakly coupled $\varphi^4$ 
($\varphi^{2n}$) theory. This exhibits an interesting universal 
crossover between classical to non classical criticality
that has been just studied in the past for 
instance for medium range models
\cite{BB-84a} \cite{PRV-98}. The studies of weakly coupled
interactions have been presented in 
{\bf chapter \ref{chapter2}}. The discussion presented there,
instead of the medium range models  or field theoretical
results, consider also the case in which $\mathbb{Z}_2$
 symmetry is not present in the starting interaction but
must be fixed by other renormalization conditions.
The presentation presented in the first part of 
this work follows \cite{CMP-05} and \cite{CMP-07}.\footnote{
In \cite{CMP-05} the critical points in $d=2$ have been 
explicitly computed. The results have been then 
generalized in a straightforward way  to
the $d>2$ case. However in the case in which the interaction
is not symmetric in higher than two dimension the discussion is
more involved
(see sec.\ \ref{chapter2-higher-dimension}) although the conclusions
remain unchanged.
}

The problem of IR divergences is a well known problem in 
 $1/N$ expansion \cite{MZ-03} \cite{Y-97} \cite{K-89}. 
This has suggested us to 
search to apply the same scheme used to study finite
temperature (multi)critical points in  Heisenberg models
to investigate other physical interesting models that
share the problem of $1/N$ infrared divergences.
In particular in {\bf chapter \ref{chapter5}} we have
focused our attention on  system with $N_f$ fermions
coupled to a bosonic field through  a Yukawa interaction,
\begin{eqnarray}
{\cal H} &=& {1\over 2}(\partial \phi)^2 + p(\phi)
+\sum_{f=1}^{N_f}\ol \psi_f \Big(\sc{\partial}+ g \phi\Big)\psi_f
\nonumber
\end{eqnarray}
in the large $N_f$ limit.\footnote{
Similar consideration holds for the Gross Neveu model
(see \cite{Zinn-Justin} sec.\ 31.9), that
is a model with a four fermions interaction.}
 The main result of this chapter will
be to prove the claim  that the crossover function obtained in the
study of the Heisenberg model for ${\cal N}=1$ can be 
used also for the fermionic model (they are universal).\footnote{
However in this case, instead of the Heisenberg models,
it was pointed out (using scaling arguments) time ago \cite{KSS-98}
 the mechanism of the critical zone
reduction for which the region size in which Ising fluctuations become
relevant go to zero for $N_f\to \infty$, explaining the reason
for which only Mean Field is observed for $N_f=\infty$.}
We have just pointed out that an important check for the
scheme proposed would be the 
computation of the $1/N$ correction to the critical temperature,
to compare with the available results obtained in \cite{KSS-98}
for several $N$.
In {\bf chapter \ref{chapter6}} we will consider a vectorial model 
$\vec \phi=\{\phi_1,\cdots \phi_N \}$
with a $\phi^6$ interaction
\begin{eqnarray}
{\cal H} &=& \vec H\phi+{1\over 2}(\partial \vec \phi)^2 +{r\over 2}
\vec \phi^2 + {u\over 4!} (\vec \phi^2)^2 + {v\over 6!}(\vec \phi^2)^3.
\nonumber
\end{eqnarray}
The phase diagram is rather well understood since a long time 
(e.g.\ \cite{SF-78}). For $\vec H \neq 0$ there are two line of
second order phase transition (that usually are called wing lines) that
meet together in a tricritical point for $\vec H = 0$.
For the $N=\infty$ theory a Mean Field Behavior is observed along the
wing lines while, due to symmetry arguments, one expect Ising
behaviour for every $N$ \cite{SF-78}. The apparent paradox can be
solved including $1/N$ corrections in the same way of what we
have done in the Heisenberg model.

As future application we plan to apply our multicritical scheme
to the study of Yukawa models with nonzero chemical
potential and finite temperature which is believed 
to have a phase diagram with a tricritical point.
This is an interesting topic in order to investigate
the phase diagram of QCD.

All the systems we have described are characterized by a global
$O(N)$ symmetry. In models with local symmetry
the Mean Field behavior of the large $N$ limit
is questionable and currently under investigation
\cite{CJ-06} (see also \cite{Bringoltz-05}). 
It could be interesting to use our approach 
in order to recover (or not) previous results
\cite{CJ-06}.

%%%%%%%%%%%%%%%%%%%%%%%%%%%%%%%%%%%%%%%%%%%%%%%%%%%%%%%
%%%%%%%%%%%%%%%%%%%%%%%%%%%%%%%%%%%%%%%%%%%%%%%%%%%%%%%
%%%%%%%%%%%%%%%%%%%%%%%%%%%%%%%%%%%%%%%%%%%%%%%%%%%%%%%
%%%                                                 %%%
%%%                                                 %%%
%%% CHAPTER 1 -- CHAPTER 1 -- CHAPTER 1 -- CHAPTER1 %%%
%%%                                                 %%%
%%%                                                 %%%
%%%%%%%%%%%%%%%%%%%%%%%%%%%%%%%%%%%%%%%%%%%%%%%%%%%%%%%
%%%%%%%%%%%%%%%%%%%%%%%%%%%%%%%%%%%%%%%%%%%%%%%%%%%%%%%
%%%%%%%%%%%%%%%%%%%%%%%%%%%%%%%%%%%%%%%%%%%%%%%%%%%%%%%

\chapter{Large-$N$ expansions}\label{chapter1}

In this chapter we introduce the general technique 
commonly used in the study of models with  $O(N)$ global symmetry
in the limit in which  $N$ goes to infinity. The basic
idea deals with the fact that the $O(N)$ quantities self-average
for $N\to \infty$; in fact this can be seen 
as a consequence of central limit theorem
 \cite{MP-T}. Let we consider for instance 
the standard $\mathbf{ \phi}_x^4$ theory,
where {\bf $\phi$}  is a $N$ component vectorial
field and  ${\bf \phi}_x^4\equiv
(\vec \phi_x^2)^2$ is the interacting term.
In order to compute $O(N)$ symmetric  observables
one can guess that  in the large $N$ limit 
$\vec \phi_x^2$ self averages on a given expectation
value $\alpha(x)$ with fluctuations that scale
as $1/\sqrt N$. Concerning the expectation values, one can
then write the equation reported into the introduction
\begin{eqnarray}
\langle \phi(\mathbf{y})^2 \phi(\mathbf{x})^2 \rangle
= \langle \phi(\mathbf{y})^2\rangle \langle\phi(\mathbf{x})^2 \rangle
+O\left(1\over \sqrt N\right), 
\nonumber
\end{eqnarray}
where now also fluctuations are included.
In this work we implement the large $N$ limit by
introducing several auxiliary fields that 
are related to $O(N)$ invariant quantities of the model \cite{MZ-03} 
\cite{Zinn-Justin}. This allows us to write an effective action 
in terms of the fields introduced. In literature there are 
also other methods that give the same results
like for instance Hartree-Fock approximations \cite{BM-83}. 
Also in the point of view
of stochastic quantization or in the Critical Dynamic 
\cite{JSS-89} the Langevin equation becomes linear
and self-consistent for large $N$ replacing $\phi^2$ with 
$\langle\phi^2 \rangle$. On the other hand a famous
related model is the spherical model 
\cite{BK-52}, that was shown to share the same critical
behavior  with $N=\infty$ spin systems \cite{S-68}.
 However Hartree-Fock
or self-consistent conditions give only
 the $N=\infty$ limit  while the method
of auxiliary fields gives the possibility to
develop a systematic $1/N$ expansion.
This is a fundamental necessity for our work which focuses on 
$1/N$ corrections.

As a prototype for  IR divergences appearing in the large 
$N$ limit, in sec.\ \ref{chapter1-model}
we will introduce a general class of spin models (\ref{HNLSigma})
 on a lattice, which exhibits a phase transition 
at finite temperature and spin-spin correlation length.
 In sec.\ \ref{chapter1-model} we also describe
the recent interest in studying Hamiltonians (\ref{HNLSigma}).
In sec.\ \ref{chapter1-sec2}, after having  introduced the auxiliary 
fields we study the saddle point equations 
(\ref{def-gammabar}-\ref{def-alphabar}-\ref{def-taubar}) 
and the gap-equation (\ref{puntosella2}) that give
the physics for $N=\infty$. We study both
models with critical and multicritical phase transitions.
Multicritical points can be observed choosing very
peculiar Hamiltonians (\ref{HNLSigma}) so that 
the  set of equations
(\ref{eq-punto-multicritico}) can be satisfied at the 
critical temperature $\beta_c<\infty$.
We believe that the study of multicritical points is interesting
because several physical systems undergo such a transitions. 
For example in the study of chiral symmetry in QCD several
simplified models have been considered. In chapter
\ref{chapter5} we investigate a Yukawa model\footnote{This model in the large $N$ limit \cite{Zinn-Justin} behaves exactly as the Gross-Neveu model
\cite{W-72} \cite{GN-74}.} at finite temperature $T$ that shows
the same IR singularities of (\ref{HNLSigma}) near a critical point
$T_c>0$. We plan to study the model also with
a finite chemical potential\footnote{
This is relevant for recent experimental progress in the physic
of ultrarelativistic heavy-ion collisions.} $\mu >0$ which is believed to 
exhibits a tricritical point so that our present investigation
of multicritical interactions (\ref{HNLSigma}) could be 
useful.
In sec.\ \ref{chapter1-sec3} we parametrize the gap-equation 
near the (multi)critical point obtaining the  scaling fields
and the equation of state (that are Mean Field like).
In sec.\ \ref{chapter1-sec4} in order to include
$1/N$ fluctuations we develop the theory around the
saddle point solution giving the expression for the 
propagator and the vertices of the Hamiltonian for the auxiliary fields.
We show as near the critical point the propagator is singular
at zero external momentum.
In sec.\ \ref{chapter1-sec5} we show that, because this singularity,
the standard $1/N$ expansion breaks down.

\section{The model}\label{chapter1-model} 

Motivated by condensed matter or field theory,
a lot of work has been focused on the investigation 
of the following 
Hamiltonian
\begin{equation}
H = - N \beta \sum_{\<ij\>} \bsigma_i\cdot \bsigma_j,
\label{NLSigma}
\end{equation}
where $\bsigma_i$ is an $N$-dimensional unit spin and 
the sum is extended over all lattice nearest neighbours.
Indeed (\ref{NLSigma}) is usually used as the prototype of
short-range
interacting models with global $O(N)$ symmetry. 
In two dimension the model is disordered for all finite $\beta$
\cite{MW-66} and it is described for $\beta\to\infty$ 
by the perturbative renormalization group 
\cite{Polyakov-75}, \cite{BZ-76}, \cite{BLS-76}. The square-lattice
model has been extensively studied numerically
\cite{Wolff_89_90}, \cite{EFGS-92}, \cite{Kim}, \cite{CEPS-95},
\cite{CEMPS-96}, \cite{MPS-96},
checking the perturbative predictions 
\cite{FT-86}, \cite{CP-94}, \cite{CP-95}, \cite{ABC-97}
and the non-perturbative constants  
\cite{HMN-90}, \cite{HN-90}, \cite{CPRV-97}.
Recently more general models than (\ref{NLSigma}) have received attentions
\begin{equation}
H=-N\beta\sum_{<ij>}W(1+{\bsigma}_i\cdot {\bsigma}_j).
\label{HNLSigma}
\end{equation}
Indeed using exact estimates it has been pointed out in \cite{VES-02} 
that if one consider a generic $O(N)$ interaction (\ref{HNLSigma})  
then for a large class of interactions $W$, a first-order phase 
transitions  appear for $\beta > 0$ in two dimension.\footnote{
It \cite{VES-03} similar considerations are given for a gauge
theory in $2+1$ dimensions.}$^{,}$\footnote{It is interesting to observe that 
this phase transition is not present for (\ref{NLSigma}) although that 
the two continuum limit of (\ref{NLSigma}) and (\ref{HNLSigma})
formally coincide. This is related to the fact that,
according with  \cite{MW-66}, at the critical point
the spin-spin correlation length is finite.}
 In \cite{BGH-02} a particular class of interactions
depending on a parameter $p$ is considered and studied by Monte Carlo
simulations
\begin{equation}
W(x) \sim x^{p}.  
\label{H-bloete}
\end{equation}
The results  show, according with \cite{VES-02}, the existence of a 
first order phase transition for high enough $p$  at finite 
temperature.\footnote{There are other numerical works with 
similar results for different Hamiltonians than (\ref{H-bloete}),
see e.g.\ \cite{PR-03}. However Hamiltonian (\ref{H-bloete}) 
is peculiar with respect other models --like for instance
$RP^{N-1}$ models-- because it does not have any symmetry. }
Otherwise  also a continuous phase transition for 
$p \approx 5$ is observed and numerical evidences suggest
that it belongs to the Ising 
universality class. It is important to stress that according with
\cite{MW-66} at the critical point (CP) the spin-spin correlation 
length remains finite while the critical parameter seems to be the
energy $E=\bsigma_i \cdot \bsigma_j$, where $i$ and $j$ are neighbour.
The same qualitative results of \cite{BGH-02} was obtained in \cite{CP-02}
taking the spherical limit ($N\to \infty$), but instead of the Ising
criticality a Mean Field behavior was observed.

The apparent paradox
has been solved in \cite{CMP-05} where it has been shown that in the large
$N$ limit the width of the critical Ising zone  (that is the region
near $p_c$ and $\beta_c$ where fluctuations become important and 
Mean Field
behavior is suppressed) reduces as a power of $1/N$ for $N\to\infty$.
This explains why only Mean Field is detected for $N$
 strictly infinite.
In this work we present the results obtained in 
\cite{CMP-05}
 considering also interaction $W $ depending on  $\N$
tunable parameter $p_i$. We show the existence of multicritical points
(MCP)
for every model (with $\N$ odd) that can simultaneously satisfy
$\N +1$ conditions (\ref{eq-punto-critico}) at the critical point
$p_{1c}\cdots p_{\N c},
\beta_c$. Then following \cite{CMP-05} we show how $1/N$ fluctuations
don't destroy  the $N=\infty$ critical points so that CP or MCP are present
 at finite $N$ and cannot be considered an artefact of the 
large $N$ limit \cite{SS-00}. 
Otherwise the Mean Field behaviour is destroyed
as soon as one takes $N$ finite.
We explain this in term of a generalized $1/N$ expansion using a weakly coupled
$\varphi^{\N+3}$ theory that will be introduced in chapter
\ref{chapter2} and applied to (\ref{HNLSigma}) in chapters
\ref{chapter3} and \ref{chapter4}.

\section{The large-$N$ limit} \label{chapter1-sec2}

The large-$N$ limit for the model (\ref{HNLSigma})
 has been discussed in detail in Ref.~\cite{CP-02}.
Proceeding in a standard way \cite{MZ-03}
we introduce three auxiliary fields 
$\lambda_{x\mu}$, $\rho_{x\mu}$, and $\mu_x$ in order to linearize the 
dependence of the Hamiltonian on the spins $\bsigma$ and 
to eliminate the constraint\footnote{
This constraint is usually irrelevant in the large $N$ limit
(see \cite{Zinn-Justin} sec.\  30.6). } $\bsigma_x^2 = 1$. 
The partition function  becomes
\begin{equation}
Z = \int \prod_{x\mu} [d\rho_{x\mu} d \lambda_{x\mu}]
         \prod_x [d\mu_x d\bsigma_x]\,  e^{-N {\cal H}},
\end{equation}
where 
\begin{equation}
{\cal H} = -{\beta\over2} \sum_{x\mu} 
   \left[\lambda_{x\mu} + \lambda_{x\mu}\bsigma_x\cdot\bsigma_{x+\mu} - 
         \lambda_{x\mu} \rho_{x\mu} + 2 W(\rho_{x\mu})\right] +
   {\beta\over2} \sum_{x} 
     \left(\mu_x\bsigma_x^2 - \mu_x\right).
\label{azione-ausiliari}
\end{equation}
We develop the auxiliary fields near their saddle-point values
\begin{eqnarray}
\lambda_{x\mu} = \alpha + {1\over \sqrt{N}} 
   \widehat{\lambda}_{x\mu}, &
\rho_{x\mu} = \tau + {1\over \sqrt{N}} \widehat{\rho}_{x\mu}, &
\mu_{x} = \gamma + {1\over \sqrt{N}} \widehat{\mu}_{x}. 
\label{field-exp}
\end{eqnarray}
In Ref.~\cite{CP-02} $\alpha$, $\tau$ and $\gamma$ were
explicitly given.
They can be written as
\begin{eqnarray}
\gamma &=& \alpha (4 + m_0^2)/2,
\label{def-gammabar}
 \\
\alpha &=& 2 W'(\tau),
\label{def-alphabar}
 \\
{\tau} &= &
   \overline{\tau}\left(m_0^{2}\right) \equiv 2 + {m_0^2\over 4} - {1\over 4 B_1(m_0^2)},
\label{def-taubar} \\
\beta &=& {B_1(m_0^2)\over W'(\overline{\tau},p)},
\label{puntosella2}
\end{eqnarray}
where the parameter $m_0$ is related to the spin-spin correlation  length
$\xi_\sigma = 1/m_0$ and
\begin{equation}
  B_n(m_0^2) \equiv \int_\mathbf{q}
   {1\over (\hat{q}^2 + m_0^2)^n},
\label{defBn}
\end{equation}
with the integral  extended over the first Brillouin zone.
The corresponding free energy can be written as 
\begin{equation}
F = - \beta d W(\tau) + {1\over 2} \log I(m_0^2) + {1\over2} L(m_0^2),
\label{free-energy}
\end{equation}
where 
\begin{equation}
L(m_0^2) = \int_\vp\,\log (\hat{p}^2 + m_0^2).
\end{equation}

In Ref.~\cite{CP-02} it was shown that generic models may show first-order
transitions. This happens when, for given $\beta$, there are several
values of $m_0^2$ that solve the gap equation (\ref{puntosella2}). 
Here, we will be interested at the endpoint of the first-order transition
line for which the following two relations holds \cite{CP-02}
\begin{equation}
{\partial \beta\over \partial m_{0}^2} = 0, \qquad 
{\partial^2 \beta\over \partial (m_{0}^2)^2} = 0
\qquad
{\partial^3 \beta\over \partial (m_{0}^2)^3} < 0.
\label{eq-punto-critico}
\end{equation}
In order to recover the two previous conditions (\ref{eq-punto-critico})
one can use a family of one parameter of interactions $W(x)\equiv W(p;x)$ in
(\ref{HNLSigma}) and  tune both $p$ and $\beta$ to the critical point
where (\ref{eq-punto-critico}) are satisfied. 
For instance in \cite{BGH-02} and \cite{VES-02} the Hamiltonian
\ref{H-bloete} was considered as prototype of interactions $W$ that
show phase transitions. In \cite{MR-87} similar consideration to what obtained 
in \cite{CP-02} was given for the mixed O($N$)-$RP^{N-1}$ model
 $W(x)\sim x + p x^2$. In this work 
we will consider the most generic one-parameter families of interactions
that satisfy (\ref{eq-punto-critico}) for a critical point $p_c$, 
$m_{0c}^2$. More interesting we will consider also families of
interaction that depend on $\N$ odd tunable parameters 
$W(x)\equiv W(p_1,\cdots p_\N; x)$ in oder to avoid the following
multicritical conditions
\begin{equation}
{\partial^i \beta\over \partial (m_{0}^2)^i} = 0,
\qquad {\partial^{\N+2} \beta\over \partial (m_0^2)^{\N+2}} <0 
\label{eq-punto-multicritico}
\end{equation}
for $i=1,\cdots \N +1$. This permit us to claim that for
the Hamiltonian (\ref{HNLSigma})   not only Ising
behavior can be observed but also multicritical points (described
by scalar $\phi^{2\N+2}$ theory in chap.\ \ref{chapter2}). This in principle
gives us the possibility to study large-$N$ physical system
with multicritical point.
In the next sub-section we show that solutions of 
(\ref{eq-punto-multicritico}) with $\N$ even are unstable; i.e.\ 
if (\ref{eq-punto-multicritico}) is realised for $m_0=m_{0c}$ with
$\N$ even, then  another solution $\ol m$ always exists
with $F(\ol m) < F(m_{0c})$. 
The previous conclusion remains unchanged  for the point $\ol m$ that
satisfy   (\ref{eq-punto-critico})
or  (\ref{eq-punto-multicritico}) with opposite
sign in the dis-equalities. Then 
in sec.\ \ref{sec3} we will recover the $N=\infty$  phase diagram
for the model (\ref{HNLSigma}) obtaining the scaling fields
as functions of $p_i$ and $\beta$ and explaining the mean field
behavior.

\subsection{Saddle-point solutions with $\N$ even} 

\begin{figure}
\begin{center}   
%%\vspace{-7cm}
\includegraphics[angle=0,scale=0.8]{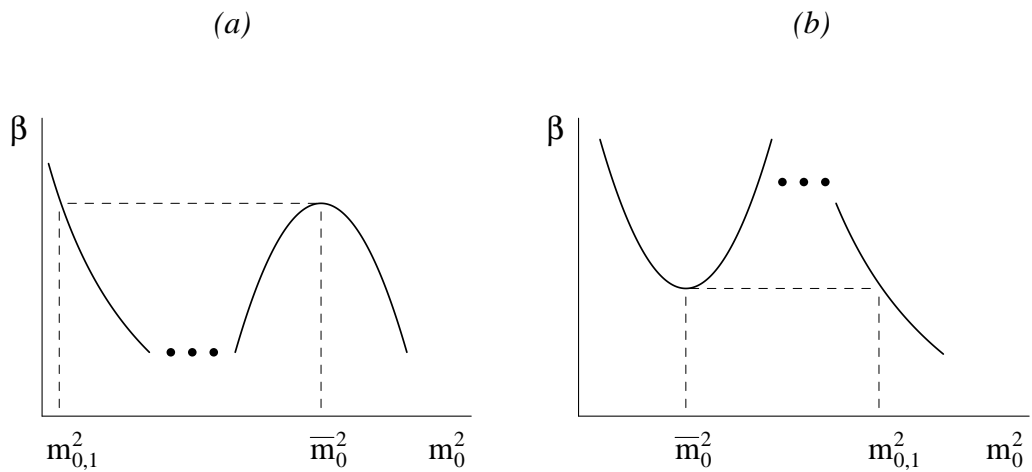}
%%\vspace{-7cm}
\caption{
The two possibilities for $\beta(m_0^2)$ in the presence of a saddle-point
with ${\cal N}$ even.
}
\label{figura-app}
\end{center}
\end{figure}

In this section we show that solutions of eq.~(\ref{eq-punto-critico})
with $\cal N$ even are of no interest. Indeed, given 
$\overline{m}_0^2$ satisfying eq.~(\ref{eq-punto-critico})
with $\cal N$ even there exists $m_{0,1}^2$ such that
\begin{eqnarray}
\beta(m_{0,1}^2) = \beta(\overline{m}_0^2) \qquad\qquad
F(m_{0,1}^2) < F(\overline{m}_0^2),
\end{eqnarray}
where $F(m_0^2)$ is the free energy (\ref{free-energy}). 
Thus, $\overline{m}_0^2$ is not 
the relevant solution and simply represents a metastable state.

We start by observing that, if ${\cal N}$ is even, $\overline{m}_0^2$
is a local maximum or minimum of the function $\beta(m_0^2)$. 
Since $\beta(m_0^2)$ vanishes for $m_0^2\to \infty$, 
diverges for $m_0^2\to 0$, and is always positive under the 
assumption that $W'(x)$ is positive for $1\le x \le 2$, 
the function $\beta(m_0^2)$ must behave as in Fig.~\ref{figura-app}. 
If it is a local maximum, there exists $m_{0,1}^2$ such that
$m_{0,1}^2 < \overline{m}_0^2$, $\beta(m_{0,1}^2) = \beta(\overline{m}_0^2)$
and $\beta(m_0^2) < \beta(m_{0,1}^2)$ for 
$m_{0,1}^2 < m_0^2 < \overline{m}_0^2$. Similar considerations 
can be made if $\overline{m}_0^2$ is a local minimum  
(see Fig.~\ref{figura-app}). Then, consider the free energy
(\ref{free-energy}), a straightforward calculation gives 
\begin{equation}
{dF\over dm_0^2} = - 2 W(\tau) {d\beta\over dm_0^2}.
\label{dF}
\end{equation}
Consider first case $(a)$ of Fig.~\ref{figura-app}. Using eq.~(\ref{dF})
we can write
\begin{eqnarray}
&&F(\overline{m}_0^2) - F(m_{0,1}^2) = 
   - 2 \beta(\overline{m}_0^2)\, [W(\tau(\overline{m}_0^2) ) - 
                                W(\tau(m_{0,1}^2) )]
\nonumber \\
&& \quad\qquad 
   + 2 \int_{m_{0,1}^2}^{\overline{m}_0^2} dm_0^2\,
    W'(\tau(m_0^2)) \beta(m_0^2) {d\tau\over dm_0^2}.
\end{eqnarray}
Now $W'(\tau)>0$, $d\tau/ dm_0^2 < 0$, 
 $\beta(m_0^2) < \beta(\overline{m}_0^2)$ in the interval, so that 
\begin{eqnarray} 
\int_{m_{0,1}^2}^{\overline{m}_0^2} dm_0^2\,
    W'(\tau(m_0^2)) \beta(m_0^2) {d\tau\over dm_0^2} &>& 
 \beta(\overline{m}_0^2) \int_{m_{0,1}^2}^{\overline{m}_0^2} dm_0^2\,
    W'(\tau(m_0^2)) {d\tau\over dm_0^2} 
\nonumber \\
&= &
  \beta(\overline{m}_0^2) [W(\tau(\overline{m}_0^2) ) -
                                W(\tau(m_{0,1}^2) )].
\end{eqnarray}
If follows 
\begin{equation} 
F(\overline{m}_0^2) - F(m_{0,1}^2) > 0,
\end{equation}
as required. In the case $\overline{m}_0^2$ is a local minimum the 
analysis is identical.
If the disequalities  (\ref{eq-punto-critico}) (\ref{eq-punto-multicritico})
have opposite sign one can repeat the discussion reported it is
straightforward to recover the previous conclusion strictly following 
the case (a) of Fig.\ \ref{figura-app}.

\section{Equation of state and scaling fields} \label{chapter1-sec3}

In this section we want to describe the phase diagram of the model outlining
the mean field behavior. This is done 
in sec.\ \ref{sec-enne-1} developing the
gap-equation $\beta(\h)$ near the critical point 
(\ref{eq-punto-critico}). In sec.\ \ref{sec-enne-more}
 we generalize
the discussion for a multi-critical 
point (\ref{eq-punto-multicritico}).

\subsection{$\N=1$}\label{sec-enne-1}

We wish now to parametrize the singular behavior for $\beta \to \beta_c$ 
and $p \to p_c$. Expanding the gap equation (\ref{puntosella2})
near the critical point
we obtain
\begin{equation}
\beta - \beta_c = \sum_{nm} a_{nm} (p-p_c)^n (m_0^2 - m_{0c}^2)^m.
\label{expansion-gap}
\end{equation}
Because of the definition of $\beta_c$ we have $a_{00} = 0$. 
Moreover, eq.~(\ref{eq-punto-critico}) implies
that $a_{01} = a_{02} = 0$, $a_{03} < 0$. For $p=p_c$ we see that $m_0^2$ has the leading 
behavior 
\begin{equation}
    m_0^2 - m_{0c}^2 \approx 
   \left( {\beta - \beta_c \over a_{03}}\right)^{1/3},
\end{equation}
while for $\beta = \beta_c$, we have 
\begin{equation}
    m_0^2 - m_{0c}^2 \approx 
   \left( -{a_{10} (p-p_c) \over a_{03}}\right)^{1/3}.
\end{equation} 
The nonanalytic behavior with exponent 1/3 is observed along any straight 
line approaching the critical point, except that satisfying 
$\beta - \beta_c - a_{10} (p-p_c) = 0$. Therefore, the correct 
linear scaling field is 
\begin{equation}
  u_h \equiv \beta - \beta_c - a_{10} (p-p_c),
\label{scalingfield-uh}
\end{equation}
and we have $m_0^2 - m_{0c}^2 \sim u_h^{1/3}$ whenever $u_h\not = 0$. 
To include the case $u_h = 0$, we write the general scaling equation
\begin{equation}
m_0^2 - m_{0c}^2 = u_h^{1/3} f(x) ,
\label{m0-scaling}
\end{equation}
where $f(x)$ is a scaling function and $x$ is a scaling variable to be 
determined. Then, we use again the gap equation (\ref{expansion-gap}). 
Keeping only the leading terms we obtain 
\begin{equation}
u_h = a_{20} (p-p_c)^2 + a_{11}(p-p_c) u_h^{1/3} f(x) + 
      a_{03} u_h f(x)^3,
\end{equation}
so that 
\begin{equation} 
   a_{03} f(x)^3 + a_{11}(p-p_c) u_h^{-2/3} f(x) - a_{20} (p-p_c)^2 u_h^{-1} 
    - 1 = 0
\end{equation}
Since $(p-p_c)^2 u_h^{-1} = [(p-p_c) u_h^{-2/3}]^2 u_h^{1/3}$, 
the third term can be neglected. Thus, we may take
(the prefactor has been introduced for later convenience)
\begin{equation}
x \equiv a_{11} (p-p_c) |u_h|^{-2/3}. 
\label{chapter1-def-x}
\end{equation}
The function $f(x)$ satisfies 
\begin{equation}
a_{03} f(x)^3 + x f(x) - 1 = 0.
\label{eq-for-f}
\end{equation}
Such an equation is exactly the mean-field equation for the magnetization.
Indeed, if we consider the mean-field Hamiltonian 
\begin{equation} 
{\cal H} = - h M + {t\over2} M^2 + {u\over 24} M^4,
\end{equation}
the stationarity condition gives
\begin{equation}
- h + t M + {u\over6} M^3 = 0,
\end{equation}
which is solved by $M = h^{1/3} \hat{f}(t |h|^{-2/3})$, where 
$\hat{f}(x)$ satisfies eq.~(\ref{eq-for-f}) with\footnote{
Notice that the sign of $a_{03}$ implies that the
low temperature phase is obtained --as expected-- for $p>p_c$.
However the sign of $a_{03}$ is fundamental in 
order to guarantees the stability of the effective theory
for the zero mode in chapter \ref{chapter3}.} $a_{03} = u/6$. 
It is thus clear that the scaling field (\ref{scalingfield-uh}) corresponds 
to $h$, while $p$ corresponds to the temperature. Note that this 
identification is not unique, since only the line $H=0$ is 
uniquely defined by the singular behavior.
For instance, in the usual Ising case,
we could define $t' = t + a h$ without changing the scaling equation of state,
since $t |h|^{-2/3} = t' |h|^{-2/3} + a h^{1/3}$. Since the scaling 
limit is taken with $h\to0$, $t\to 0$ at fixed $t |h|^{-2/3}$, we see that 
$t |h|^{-2/3} \approx t' |h|^{-2/3}$. In the Ising case, however, 
there is exact $\mathbb{Z}_2$ symmetry and thus the natural
$t$ variable is defined so that it is invariant under the symmetry.
In our case we could define $u_t = p - p_c + A (\beta - \beta_c)$
and fix $A$ by requiring the leading correction on any line (except $u_h = 0$)
to be of order $u_h^{2/3}$ instead of order $u_h^{1/3}$, recovering in this way
an approximate $\mathbb{Z}_2$ symmetry. For our purposes this is 
irrelevant and thus we will use $p-p_c$ as thermal scaling field.

Eq.~(\ref{m0-scaling}) gives the leading behavior. It is also possible to 
compute the subleading corrections. We obtain for the leading one
\begin{equation}
m_0^2 - m_{0c}^2 = u_h^{1/3} f(x) + u_h^{2/3} g(x) + O(u_h),
\end{equation}
with 
\begin{equation}
g(x) = - {a_{20} x^2 + a_{12} a_{11} x f(x)^2 + a_{04} a_{11}^2 f(x)^4 \over 
        a_{11}^2 (x + 3 a_{03} f(x)^2)}.
\end{equation}
Finally, let us discuss the singular behavior of the energy.
We have 
\begin{eqnarray} 
    E = 2 W(\overline{\tau},p).
\end{eqnarray}
Such a function is regular in $m_0^2$ and $p$. Since 
$p - p_c\sim x |u_h|^{2/3}$ and $m_0^2 - m_{0c}^2 \sim u_h^{1/3}$, 
the leading term is obtained by expanding the previous equation in powers of 
$m_0^2 - m_{0c}^2$. Thus
\begin{equation}
   E = 2 W_c + 2 W'_c \left.{B_1^2 - B_2\over 4 B_1^2}\right|_{m_0 = m_{0c}}
    u_h^{1/3} f(x)  + O(u_h^{2/3}).
\end{equation}
where $W' = \partial W(x,p)/\partial x$ with $x = \overline{\tau}$ 
and the suffix $c$ indicates that all quantities must be computed at the 
critical point.

\subsection{The general $\N$ odd case}\label{sec-enne-more}

Following the same step of the previous section \ref{sec-enne-1}
we want now the expand the gap equation (\ref{puntosella2}) near 
a multicritical point (\ref{eq-punto-multicritico}).
First we present the $\N=3$ case because it requires to consider
non-linear terms in the definitions of the scaling fields,
 the generalization is then
straightforward. We have
\begin{equation}
\label{state}
\beta-\beta_c=\sum_{\ell,i_1 i_2 i_3} a_{i_1 i_2 i_3}^{(\ell)}
          (p_1 - p_{1c})^{i_1} (p_2 - p_{2c})^{i_2} (p_3 - p_{3c})^{i_3} 
          (m_0^2 - m_{0c}^2)^\ell.
\end{equation}
with $a_{000}^{(\ell)}  = 0$ for $0 \le \ell \le 4$ because of 
eq.~\reff{eq-punto-critico}. To simplify the notations 
we introduce a multi-index $A \equiv (i_1,i_2,i_3)$, rewrite the previous 
equation as 
\begin{equation}
\beta-\beta_c=\sum_{\ell,A} a_A^{(\ell)} 
   (p - p_c)^A (m_0^2 - m_{0c}^2)^\ell. 
\end{equation}
and define the linear scaling fields as
\begin{eqnarray}
&& u_h = \beta - \beta_c - \sum_{A:[A]=1} a_A^{(0)} (p - p_c)^A, \\
&& u_{1} = \sum_{A:[A]=1} a_A^{(1)} (p - p_c)^A, \\
&& u_{2} = \sum_{A:[A]=1} a_A^{(2)} (p - p_c)^A, \\
&& u_{3} = \sum_{A:[A]=1} a_A^{(3)} (p - p_c)^A, 
\end{eqnarray}
where $[A]=a$ indicates indices $i_1,i_2,i_3$ such that 
$i_1+i_2+i_3 = a$. The gap equation becomes 
\begin{eqnarray}
u_h &=& \sum_{\ell \ge 5} a_{000}^{(\ell)} (m_0^2 - m_{0c}^2)^\ell + 
    (m_0^2 - m_{0c}^2) u_1  + 
    (m_0^2 - m_{0c}^2)^2 u_2  + 
    (m_0^2 - m_{0c}^2)^3 u_3  
\label{gap-scaling} \\
&&    + \sum_{\ell \ge 4} \sum_{A:[A]=1} a_A^{(\ell)}
   (p - p_c)^A (m_0^2 - m_{0c}^2)^\ell
   + \sum_{\ell=0 } \sum_{A:[A]\ge 2} a_A^{(\ell)}
   (p - p_c)^A (m_0^2 - m_{0c}^2)^\ell.
\nonumber
\end{eqnarray}
From this equation one can read the correct scalings. We have 
$m_0^2 - m_{0c}^2 \sim u_h^{1/5}$ and, as a consequence,
$u_j \sim u_h (m_0^2 - m_{0c}^2)^{-j}\sim u_h^{1-j/5}$. 
Moreover, expressing $p_i - p_{ic}$ in terms of $u_i$, we obtain
$p_i - p_{ic} \sim u_3 \sim u_h^{2/5}$.
Thus, close to the multicritical point we expect 
\begin{eqnarray}
   m_0^2 - m_{0c}^2 = u_h^{1/5} f(x_1,x_2,x_3),
\label{eq-State}
\end{eqnarray}
where the scaling variables $x_j$ are defined by $x_j \equiv u_j u_h^{j/5-1}$. 
If we now substitute this Ansatz in eq.~\reff{gap-scaling} we find 
however a difficulty. All terms in the equation should vanish close 
to the MCP as $u_h$ or faster. On the other hand, the terms 
with $[A] = 2$ and $\ell = 0$ vanish only as $u_h^{4/5}$. Thus, they are 
the dominant ones and make the Ansatz (\ref{eq-State}) inconsistent. 
There is a very simple way out of this problem. It is enough to include these 
terms in the definition of $u_h$ and consider the nonlinear scaling field
\begin{equation}
u_h = \beta - \beta_c - \sum_{A:[A]=1,2} a_A^{(0)} (p - p_c)^A .
\end{equation}
With this definition the dangerous terms are no longer present in 
eq.~\reff{gap-scaling} and 
in the scaling limit, $u_h\to0$, $u_i\to 0$, at fixed $x_i$, we obtain
\begin{equation}
a_{000}^{(5)} f^5 + x_3 f^3 + x_2 f^2 + x_1 f - 1 = 0, 
\end{equation}
which is exactly the mean-field scaling equation of state for 
a generic $\phi^6$ theory with Hamiltonian
\begin{equation}
{\cal H} = - \phi + {x_1\over2} \phi^2 + 
         {x_2\over3} \phi^3 +
         {x_3\over4} \phi^4 +
         {a_{000}^{(5)}\over6} \phi^6.
\end{equation}
Note the absence of the $\phi^5$ term, which, if present, can always be 
eliminated by performing an appropriate shift of the fields. 

The generalization to generic values of $\cal N$ is straightforward. 
We define    
\begin{eqnarray}
u_h &\equiv & \beta - \beta_c - \sum_{A:[A]\le m_0} a_A^{(0)} (p - p_c)^A
\label{gap-near-critical-point}
 \\
u_{i} &\equiv & \sum_{A:[A]\le m_i} a_A^{(i)} (p - p_c)^A 
\label{scaling-fields-classici}
\\
x_i &\equiv & u_i u_h^{i/({\cal N}+2) - 1},
\label{scaling-classico}
\end{eqnarray}
$i = 1,\ldots,{\cal N}$, which imply 
$(p_i - p_{ic})\sim u_h^{2/({\cal N}+2)}$ and 
$(m_0^2 - m_{0c}^2) \sim  u_h^{1/({\cal N}+2)}$.
The integers $m_i$
must be determined by requiring that 
\begin{equation}
  \sum_{A:[A]> m_\ell} a_A^{(\ell)} (p - p_c)^A (m_0^2 - m_{0c}^2)^\ell
\end{equation}
is of order $u_h^\alpha$ with $\alpha > 1$ for 
$0\le \ell \le {\cal N}$. It follows 
\begin{equation}
{2 m_\ell + 2 + \ell \over {\cal N} + 2} > 1, 
\end{equation}
so that 
\begin{equation}
m_\ell = \left\lfloor {{\cal N} - \ell \over 2}\right\rfloor + 1,
\end{equation}
where $\lfloor x\rfloor$ indicates the largest integer that is smaller than
or equal to $x$. In terms of these variables we have therefore
in the scaling limit
\begin{equation}
m_0^2 - m_{0c}^2 = u_h^{1/({\cal N} + 2)}
   f(x_1,\ldots,x_{\cal N}),
\end{equation}
where the scaling function satisfies the mean-field equation of state 
of a $\phi^{3 + \cal N}$ theory:
\begin{equation}
a_{0,\ldots,0}^{(2 + {\cal N})} f^{2 + {\cal N}} + 
    \sum_{j=1}^{\cal N} x_j f^j - 1 = 0.
\end{equation}
This discussion also clarifies why only the case $\cal N$ odd is relevant.
If $\cal N$ is even we obtain a $\phi^n$ theory with $n$ odd, 
which is unstable, providing another indication for the instability of 
the corresponding solution.

\section{The $1/N$ calculation: propagator and effective vertices} \label{chapter1-sec4}

In order to develop the standard $1/N$ expansion 
we insert the fields expansion (\ref{field-exp}) into 
the action for the auxiliary fields (\ref{azione-ausiliari}).
Taking a short-hand notation  we define a five-component field $\Psi_A$
\begin{equation}
\Psi = (\hat{\mu}, \hat{\lambda}_1, \hat{\lambda}_2,
        \hat{\rho}_1, \hat{\rho}_2),
\end{equation}
then ${\cal H}$ can be written as 
\begin{eqnarray}
{\cal H} &=& {1\over 2} \int_{\mathbf{p}} \sum_{A_1A_2} 
   \Psi_{A_1}(-\mathbf{p}) P_{A_1A_2}^{-1}(\mathbf{p}) \Psi_{A_2}(\mathbf{p})
\label{effexp} \\
  &+& \sum_{n=3} {1\over n!} {1\over N^{n/2-1} }
  \int_{\mathbf{p}_1} \cdots \int_{\mathbf{p}_n} 
    \delta\left(\sum_i \mathbf{p}_i\right) \; 
   \sum_{A_1,\ldots,A_n} 
   V^{(n)}_{A_1,\ldots,A_n} (\mathbf{p}_1,\ldots,\mathbf{p}_n)
   \Psi_{A_1}(\mathbf{p}_1)\cdots \Psi_{A_n}(\mathbf{p}_n),
\nonumber 
\end{eqnarray}
where the indices $A_i$ run from 1 to 5 and we have neglected
the constant part $F(\h)$ (\ref{free-energy}). Notice that 
in (\ref{effexp}) the linear term disappears due to the
gap equations (\ref{def-gammabar}-\ref{def-alphabar}-\ref{def-taubar}).
The propagator can be explicitly written as\footnote{
It is useful to write the Fourier transform of $\hat{\lambda}_x$ as 
$\hat{\lambda}(\mathbf{p}) = 
e^{-i p_\mu/2} \sum_x e^{-ip\cdot x} \hat{\lambda}_x$. This makes all vertices 
and propagators real.}
\begin{equation}
\mathbf{P^{-1}(p)}= - {1\over {W'}^2}
\left( \begin{array}{ccccc}
\fr{1}{2}A_{0,0} &-\fr{1}{2} A_{1,0} &-\fr{1}{2}A_{0,1} & 0 & 0 \\
-\fr{1}{2}A_{1,0} &\fr{1}{2} A_{2,0} &\fr{1}{2}A_{1,1} &
-\frac {1}{2}\beta {W'}^2 & 0 \\
-\fr{1}{2}A_{0,1} & \fr{1}{2}A_{1,1} &\fr{1}{2} A_{0,2} & 0 &
 -\frac{1}{2}\beta {W'}^2\\
0 & - \frac{1}{2}\beta {W'}^2 & 0 & \beta W'' {W'}^2 & 0 \\
0 & 0 & - \frac{1}{2}\beta {W'}^2 & 0 & \beta W''{W'}^2
\end{array} \right),
\label{propagator}
\end{equation} 
where $W$ should always be intended as a function of $\bar{\tau}(m_0^2)$,
\begin{equation}
A_{i,j}(\mathbf{p},m_0^2) \equiv \int_\mathbf{q}
\frac{  \cos^i q_x \cos^j q_y }{
     [m_{0}^2 + \widehat{(q + \frac{p}{2})}^2 ]
     [m_{0}^2 + \widehat{(q - \frac{p}{2})}^2]  },
\end{equation}
and $\hat{p}^2 \equiv 4 (\sin^2 p_x/2 + \sin^2 p_y/2)$.  

For $p\to0$, by using the algebraic algorithm described in App.\ A of
Ref.~\cite{CP-95}, it is easy to express the integrals $A_{i,j}(0,m_{0}^{2})$
in terms of the integrals $B_n(m_0^2)$ defined in eq.~\reff{defBn} 
with $n=1,2$. Explicitly we have
\begin{eqnarray}
&& A_{00}(\mathbf{0},m_0^2) = B_2,
\nonumber \\
&& A_{10}(\mathbf{0},m_0^2) = A_{01}(\mathbf{0},m_0^2) =
    \left(1 + {m_0^2\over 4}\right) B_2 - {1\over4} B_1 ,
\nonumber \\  
&& A_{11}(\mathbf{0},m_0^2) =
  - {1\over8}(4+m_0^2) B_1 + {1\over8} (8 + 8 m_0^2 + m_0^4) B_2,
\nonumber \\
&& A_{20}(\mathbf{0},m_0^2) = A_{02}(\mathbf{0},m_0^2) =
 {1\over8} + B_2 - {1\over8}(4 + m_0^2) B_1.
\label{relAijBn}
\end{eqnarray}
Vertices are analogously computed. It is easy to check that the only 
nonvanishing contributions for which some $A_i$ is equal to 4 or 5 are 
\begin{equation}
V_{4\ldots 4}^{(n)}(\mathbf{p}_1,\ldots,\mathbf{p}_n) =  
V_{5\ldots 5}^{(n)}(\mathbf{p}_1,\ldots,\mathbf{p}_n) =  
   - \beta W^{(n)}(\bar{\tau}).
\end{equation}
If all indices satisfy $A_i\le 3$, then
\begin{eqnarray}
&& V_{A_1,\ldots,A_n}^{(n)}(\mathbf{p}_1,\ldots,\mathbf{p}_n) 
\delta(\sum_i \mathbf{p}_i) = 
\label{defVn}
\\
&& = 
 {(-1)^{n+1} \over [W'(\bar{\tau})]^n} \left\{
  \prod_{i=1}^n \left[ \int_{\mathbf{q}_i} 
\delta(\mathbf{q}_{i+1} - \mathbf{q}_i - \mathbf{p}_i) 
   {1\over \hat{q}_i^2 + m_0^2} R_{A_i} (\mathbf{p}_i,\mathbf{q}_i)
    \right] + {\rm permutations}\right\},
\nonumber
\end{eqnarray}
where
\begin{equation}
R_1(\mathbf{p},\mathbf{q}) = 1, \qquad 
R_2(\mathbf{p},\mathbf{q}) = - \cos(q_{x} + p_{x}/2), \qquad 
R_3(\mathbf{p},\mathbf{q}) = - \cos(q_{y} + p_{y}/2).
\end{equation}
The permutations should made the quantity in braces 
symmetric under any exchange of $(\mathbf{p}_i,A_i)$ [the total number 
of needed terms is $(n-1)!/2\,$]. 
As already discussed in Ref.~\cite{CP-02}, at the critical point 
the inverse propagator at zero momentum has a vanishing eigenvalue. 
Indeed, a straightforward computation gives 
\begin{eqnarray}
{\rm det}\, \mathbf{P}^{-1} (\mathbf{0}) &=& K_{0,\rm det} s_1,
\label{detPinv-p0}
\end{eqnarray}
where $K_{0,\rm det}$ is given by 
\begin{equation}
 K_{0,\rm det}\equiv
  - {B_1^3\over 128 {W'}^6}
  \left[4 B_1 W' - (1 - m_0^2 (8 + m_0^2) B_2) W''\right],
\end{equation}
and 
\begin{equation}
s_1 \equiv - {\partial \beta \over \partial m_0^2} =
{1\over 4 B_1 {W'}^2} (4 B_1 B_2 W' + B_1^2 W'' - B_2 W'').
\label{dbdm}
\end{equation}
Eq.~(\ref{detPinv-p0}) shows that the determinant vanishes at the
critical point---hence there is at least one vanishing eigenvalue---since 
there $\partial \beta/\partial m_0^2 = 0$. 
The corresponding eigenvector  can be written as 
\begin{equation}
\mathbf{z} = \left(2 {A_{01}(\mathbf{0},m_0^2)\over A_{00}(\mathbf{0},m_0^2)}, 
          1, 1, {1\over 2 W''},
          {1\over 2 W''} \right)
\label{defz}
\end{equation} 
computed at the critical point. Indeed,
\begin{equation}
\sum_{B = 1}^5 (P^{-1})_{AB}(\mathbf{0}) z_B  =
        {B_1 \over 4 B_2 W''} s_1 (0,1,1,0,0)_{A}.
\label{eqZ}
\end{equation}
Note that there is always only one zero mode. Indeed, at the critical 
point, we have from eq.~\reff{dbdm} 
\begin{equation}
W'' = - {4 B_1 B_2 \over B_1^2 - B_2} W',
\end{equation}
so that we can write at criticality
\begin{equation}
 K_{0,\rm det} = - {B_1^4 [B_1^2 - (8 + m_0^2) m_0^2 B_2^2]\over 
                    32 (B_1^2 - B_2)} {1\over {W'}^5}.
\end{equation}
We have verified numerically that the prefactor of $1/{W'}^5$ 
(that is independent of $W$ so that is the same for all the interactions
that one consider) is always finite
and negative, so that $K_{0,\rm det}$ is always nonvanishing. 
Thus, if we consider (multi)critical interaction, for $p_i = p_{ci}$ 
the determinant
${\rm det}\, \mathbf{P}^{-1} (\mathbf{0}) $ vanishes 
as $(m_0^2 - m_0^2)^{\N+1}$ that is  the scaling of
eigenvalue associated with the zero mode,
cf.~Eq.~\reff{eqZ}. Thus, there can only be a single eigenvector with zero
eigenvalue. This means that  the effective action
for the zero mode will be (see chapter \ref{chapter3}) a scalar 
interaction. However
choosing different $\N$ will change drastically the nature of
this.

\section{The failure of $1/N$ expansion}\label{chapter1-sec5}

Now we want to show as the appearance of a zero mode (see sec.\ 
\ref{chapter1-sec4}) invalidate the standard large-$N$ expansion.
Because of the zero mode, it is natural to express the fields in terms of a new
basis. For each $m_0^2$ and $p$,
given the inverse propagator $P^{-1}_{AB}(\mathbf{p})$, there exists
an orthogonal matrix $U(\mathbf{p};m_0^2,p)$ such that
$U^TP^{-1}U$ is diagonal. If 
$v_A(\mathbf{p};m_0^2,p) \equiv U_{A1}(\mathbf{p};m_0^2,p)$
is the eigenvector that correspond to the zero eigenvalue for $\mathbf{p} = 0$ 
at the critical
point and $Q_{Aa}(\mathbf{p};m_0^2,p) \equiv U_{A,a+1}(\mathbf{p};m_0^2,p)$,
$a = 1,\ldots 4$ are the other eigenvectors, we define new fields
$\Phi_A(\mathbf{p})$ by writing
\begin{equation}
\Psi_A(\mathbf{p}) \equiv  \sum_B U_{AB} (\mathbf{p}) \Phi_B(\mathbf{p}) =
  v_A(\mathbf{p}) \phi(\mathbf{p}) + \sum_a Q_{Aa} (\mathbf{p}) 
  \varphi_a(\mathbf{p}),
\label{depPhi}
\end{equation}
where $\Phi = (\phi,\varphi_a)$. Eq.~\reff{depPhi} defines the fields 
$\Phi_A$ up to a sign. For definiteness we shall assume $v_A(\mathbf{p})$ 
to be such that, at the critical point, 
\begin{equation}
   v_A(\mathbf{0}) = z_A/(\sum_B z_B^2)^{1/2}.
\end{equation}
We do not specify the sign of $\varphi_a$ since it will 
not play any role in the following.

The new field $\phi$ corresponds to the zero mode, while the four fields
$\varphi_a$ are the noncritical (massive) modes. 
The effective Hamiltonian for the fields $\Phi$ has an expansion analogous 
to that presented in eq.~\reff{effexp} for $\Psi$. The propagator 
$\hat{P}_{AB}(\mathbf{p})$ of $\Phi$ is 
\begin{eqnarray}
&& \hat{P}_{AB}(\mathbf{p}) = \sum_{CD} {P}_{CD}(\mathbf{p})
      U_{CA}(\mathbf{p}) U_{DB}(\mathbf{p}) ,
\end{eqnarray}
while the effective vertices are related to the previous ones by
\begin{equation} 
 \hat{V}_{A_i,\ldots,A_n}^{(n)}(\mathbf{p}_1,\ldots,\mathbf{p}_n) = 
\sum_{B_1,\ldots,B_n}    
  {V}_{B_1,\ldots,B_n}^{(n)}(\mathbf{p}_1,\ldots,\mathbf{p}_n)
  U_{B_1A_1}(\mathbf{p}_1;m_0,p) \cdots
  U_{B_nA_n}(\mathbf{p}_n;m_0,p).
\end{equation}
Note that $\hat{P}_{AB}(\mathbf{p})$ is diagonal by definition, i.e., 
$\hat{P}_{AB}(\mathbf{p}) = \delta_{AB} \hat{P}_{AA}(\mathbf{p})$.
Relation (\ref{detPinv-p0}) implies 
\begin{equation}
\hat{P}_{11}(\mathbf{0}) \sim s_1 \sim (p_i-p_{ic}),(m_0^2 - m_{0c}^2)^{\N+2}
\label{p-singular}
\end{equation}
close to the critical point.
Now we want to show that due to the singular behavior of $\hat{P}$
 (\ref{p-singular}) the standard $1/N$ expansion
fails near the critical point.
Suppose we want to compute the expectation value of the
$n$-th point function of the critical mode $\chi_n$
\begin{eqnarray}
\chi_n \equiv \int \di\mathbf{x}_2 \cdots\di\mathbf{x}_n
<\varphi(\mathbf{0})\varphi(\mathbf{x}_2)\cdots\varphi(\mathbf{x}_n)>_{
\mathrm{connected}}
\label{def-chi-n}
\end{eqnarray}
In order to do that one can
use the effective action for the zero mode ${\cal H}_{\mathrm{eff}}$, in 
which the massive mode $\varphi_a$ (\ref{depPhi})
have been integrated out. The study of ${\cal H}_{\mathrm{eff}}$ is the main
subject of the chapter \ref{chapter3}.
 There we will show that at the critical point
$p_i=p_{ci}$ the effective vertices $V^{(n)}$ behave as 
(in the infra-red limit, 
for $j=1\cdots \N+3$)
\begin{eqnarray}
V^{(n)}(\vp_i , \h ; p_i=p_{ci}) \approx l_n (\h-m_{0c}^2)^{k-n} + 
\sum_i \alpha^{(n)}_i\vp_i^2 
+\cdots
\label{vertex-first-chapter}
\end{eqnarray} 
where $k=\N + 3$ if one considers the zero mode defined in (\ref{depPhi})
or $k=\N+4$ if one translate the zero mode (\ref{depPhi}) by a constant
factor so that $l_{\N+2}=0$.\footnote{However in the considerations 
reported in this section we can take $k$ undefined.}
If we consider then a general $\ell$-loops graph $D_{n,\ell}$ entering
 into the computation of $\chi_n$ [eq.\ (\ref{def-chi-n})]
 with $N_j$ $j-$legs vertices ($j=3,\cdots,\N+1$) 
and $N_\mathrm{int}$ internal line,
neglecting ultraviolet divergences, by using eq.\ (\ref{p-singular})
and eq.\ (\ref{vertex-first-chapter}) the following
scaling relation holds\footnote{
In eq.\ (\ref{Eq.1.76})we neglect  contributions coming from 
$\N+2$-legs vertex, however
the general claim remains unchanged.}
\begin{eqnarray}
D_{n,\ell} & \sim & (\h-m_{0c}^2)^{-2n -2 N_\mathrm{int}
 + d \ell + \sum_{j=1}^{\N+1} (k-N_j)}
\nonumber\\
&=& (\h-m_{0c}^2)^{-2 n +k\N -1 -3 N_\mathrm{int} + (d+1) \ell},
\label{Eq.1.76}
\end{eqnarray}
$d$ being the dimension of the Euclidean space. In
(\ref{Eq.1.76}) we have used the topological relation
\begin{equation}
 N_\mathrm{int}= \ell + \sum_j N_j -1
\label{Eq.1.77} 
\end{equation}
and we have
neglected possible logarithmic corrections. The $1/N$ expansion
is a loop expansion around the saddle point solution. 
Consider for instance the $d=2$ case, from (\ref{Eq.1.76}-\ref{Eq.1.77}) 
\begin{equation}
D_{n,\ell} \sim (\h-m_{0c}^2)^{-2 n +k\N -1 -3 (\sum_j N_j -1)}.
\label{Eq.1.78}
\end{equation}
Looking at  eq.\ (\ref{Eq.1.78}) it is clear 
how if one considers higher order 
in the $1/N$ expansion new more severe algebraic infrared
divergences appear while in principle one expect 
only logarithmic contributions. 
This is the clear sign of the 
breakdown of the standard $1/N$ expansion. In order to 
solve the problem one has to consider a non-perturbative approach
based on the study of  weakly coupled scalar theories which are
studied in chapter \ref{chapter2} and will be applied to 
the effective zero mode interaction  ${\cal H}_\mathrm{eff}$
in chapter \ref{chapter3}.
One can guess that divergences appearing in (\ref{Eq.1.78})
cancel when explicitly Feynman diagrams are evaluated.
One can convince that this is not the case considering
for instance $\chi_2$ at one loop. There are three different diagrams
entering into the computation, one with a four-legs vertex 
($\sim m^{-4}\log m^2$) and two
built with a couple of three-legs vertices 
($\sim m^{-6}$, $\sim m^{-4}$)  so that there is no chance for any
cancellation (we have defined $m\equiv \h-m_{0c}^2$).
 On the other hand this $1/N$ contributions cannot
match with $1/N$ corrections of the $N=\infty$ result 
$\chi_2 \sim 1/(\h-m_{0c}^2)^2$
\begin{eqnarray}
\chi_2 &\sim& {1 \over \Big(\h-m_{0c}^2+{a\over N}\Big)^{2+b/N} }
\nonumber\\
&=& {1\over (\h-m_{0c}^2)^2}\Bigg(
1-{b\over 2 N} \log (\h-m_{0c}^2)^2 - {2 a \over N} {1\over (\h-m_{0c}^2)}
\Bigg).
\label{Eq.1.79}
\end{eqnarray}
In the previous equation (\ref{Eq.1.79}) we generate
 (with respect to the leading 
order) only $\Delta m^{-1}$ and $\log \Delta m$ divergences, 
while in principle, using the scaling arguments reported
above, one would match $m^{-4}$ and $m^{-2}$ divergences.

The problem of Infra 
Red Divergences  is well known in
large $N$ expansion \cite{MZ-03} and it is usually addressed
as a failure of $1/N$ expansion. They are not peculiar of the
Heisenberg models introduced above (\ref{HNLSigma}) but are
also present in other models like Gross-Neveu and Yukawa theory at 
finite temperature
\cite{BMW-02} \cite{MZ-03} \cite{SF-78}. It is important to 
 stress that for the previous
models at zero temperature $1/N$ expansion works well \cite{MZ-03}
\cite{Zinn-Justin} \cite{HKK-93}. In this case, concerning
universal coefficients for instance, one finds the $N=\infty$ 
value and   then standard perturbation theory is able to get 
$1/N$ corrections (e.g.\ \cite{HKK-93}). This works 
exactly in the same way of $1/N$ expansion of the vector 
$\lambda \Phi^4$ model: one finds for instance\footnote{
With $\nu$ commonly one refers to the critical exponent related to the 
critical behaviour of the order parameter $M$ near the critical point
$M\sim (-t)^{\nu}$, $t$ being the reduced temperature
$t=(T-T_c)/T_c$.} for $N=\infty$
$\nu=1/(d-2)$ (e.g.\ \cite{A-72} \cite{M-72}) 
and 
fluctuations around the saddle point solution can be resummed 
giving a $1/N$ correction
\cite{BW-73} \cite{M-74} ($\epsilon\equiv 4-d$)
\begin{eqnarray}
 \nu &=& {1\over d-2} - {2(3-\epsilon)\over N (2-\epsilon)(4-\epsilon)} X_1
+O\Bigg({1\over N^2} \Bigg)
\nonumber\\
X_1 &=& {4 \sin(\pi \epsilon/2) \Gamma(2-\epsilon)
\over \pi \Gamma(1-\epsilon/2) \Gamma(2-\epsilon/2)}
\nonumber
\end{eqnarray}
However when the temperature is finite 
the Infra Red Divergences become
more severe so that the perturbative series 
cannot be useful \cite{Y-97} 
\cite{K-89}. This can be understood 
also looking at the standard perturbation
theory in which  nonanalytic terms  
appear if the temperature is not zero \cite{Y-97}.
For a discussion  of the 
finite temperature see
also  sec.\ 10 of \cite{CM-02}.
For the case we have investigated we suggest that one cannot
correct the $N=\infty$ critical behavior because critical exponents
are discontinuous for $N=\infty$. However one observes a crossover
between the two different critical region that will be illustrated
in the next chapter.

%%%%%%%%%%%%%%%%%%%%%%%%%%%%%%%%%%%%%%%%%%%%%%%%%%%%%%%%%%%%%%%%%
%%%%%%%%%%%%%%%%%%%%%%%%%%%%%%%%%%%%%%%%%%%%%%%%%%%%%%%%%%%%%%%%%
%%%%%%%%%%%%%%%%%%%%%%%%%%%%%%%%%%%%%%%%%%%%%%%%%%%%%%%%%%%%%%%%%
%%%                                                           %%%
%%%                                                           %%%
%%% CHAPTER 2 - CHAPTER 2 - CHAPTER 2 - CHAPTER 2 - CHAPTER 2 %%%
%%%                                                           %%%
%%%                                                           %%%
%%%%%%%%%%%%%%%%%%%%%%%%%%%%%%%%%%%%%%%%%%%%%%%%%%%%%%%%%%%%%%%%%
%%%%%%%%%%%%%%%%%%%%%%%%%%%%%%%%%%%%%%%%%%%%%%%%%%%%%%%%%%%%%%%%%
%%%%%%%%%%%%%%%%%%%%%%%%%%%%%%%%%%%%%%%%%%%%%%%%%%%%%%%%%%%%%%%%%

\chapter{Crossover in critical phenomena} \label{chapter2}

In the past decades a huge amount of work
(experimental, theoretical and numerical) has been 
devoted to the study of the nature of critical points
in physical system characterised by a large number of elementary components
(since now ``spins'').
  In particular --from the theoretical point of view--
the possibility to apply  field theoretic methods (e.g.\ \cite{Zinn-Justin},
\cite{ItzDr}) to implement  
basic physical assumptions like 
{\em scaling invariance} of the {\em universal}
 quantities at the critical point (e.g. \cite{DombGreen}),
has allowed to obtain accurate predictions for these 
observables  in the scaling region $\tau \ll 1$ 
[$\tau\equiv (T-T_c)/T_c$, $T$ being the temperature].\footnote{For
 a review on the argument with a complete list of theoretical and 
experimental references see \cite{PV-02}.}
Outside the scaling region\footnote{Instead of  ``scaling region'' we 
could use also ``critical region''.} (physically dominated by correlations 
between the degrees of freedom of the system) one expects to recover
the physical picture given by  mean-field approximations which
is supposed to be dominated by the short range interaction
of the spins.
Several works 
\cite{BB-84a} \cite{BB-84b} \cite{BB-85}
\cite{BBMN-87} \cite{BB-02} \cite{PRV-99}
\cite{LBB-96} \cite{PRV-98} \cite{CCPRV-01}
\cite{LBB-97} \cite{LB-98}
have investigated the nature of the crossover
between the mean field region and the non trivial critical region
(for instance Ising in simple fluid).
Using the language of Wilson renormalization group theory, 
the main effort
of the theoretical works  is to resum the infinity set of irrelevant 
operators 
 that becomes important as soon as one leaves the scaling region 
(for instance $\phi^6$, $\phi^8$, ... in the Ising theory),
tuning a minimal set of adjustable non universal parameters 
\cite{BB-84a} \cite{BB-84b}. This description is valid because the introduction
of the irrelevant operators leads simply to a multiplicative
renormalization of this set of free parameters.
 Accordance with experiments
is under investigations \cite{CD-85} \cite{DC-88}
\cite{APKS-95} \cite{BB-95} \cite{APKS-96}. 
For simple liquid it seems to
work, however for complex fluid the
 field theoretic methods is questionable.\footnote{In condensed matter system (contrarily to field theory) the cut-off 
$\Lambda^{-1}$ has a physical origin related to some size of the basic 
constituents. In simple fluid we have $\Lambda \approx 1$
and one can reach the condition $\xi/\Lambda^{-1}\gg 1$ that is
necessary in order to apply RG.
In complex fluid $\Lambda^{-1}$ is not simply related to inter-molecular
distance (think for instance to a polymer solution) and the previous 
condition could not be satisfied \cite{APKS-96}. Maybe the understanding of
 these problems
(for which  techniques presented in this work could be useful) deserves
future investigations.}

In this chapter we will investigate the {\em universal} crossover behavior
that is observed for a family of scalar models in the weakly coupled
limit. We do that because similar interactions will
be obtained  considering the effective interactions for the zero mode
in  large-$N$ expansions affected by infrared
singularities that are the main topic of this work.
In order to understand the nature of the problem,
let us consider for instance the following interaction 
\begin{eqnarray}
{\cal H} &=& {1\over 2} (\partial \phi)^2 + {r\over 2} \phi^2
+ {u\over 4!} \phi^4.
\label{eq1.1}
\end{eqnarray} 
Tuning both $u\to 0$ or $t\equiv r -r_c\to 0$ in (\ref{eq1.1}),  
the two universal behaviour (Ising for $r=r_c$, Mean Field for $u=0$)
enter into competition; 
keeping a proper scaling variable $x\equiv t u^{-\alpha}$ fixed  one expect
to compute the observables (in particular critical exponents)
as functions of $x$
\begin{eqnarray}
\langle{\cal O}\rangle_{\cal H}
 \sim f_{\cal O}(t u^{-\alpha})
\nonumber
\end{eqnarray} 
and to recover Ising (Mean Field) behavior by taking
$x\to 0$ ($x\to\infty$).
The previous statements have been demonstrated in a series of theoretical 
works  \cite{BB-85} \cite{BBMN-87}. More interesting it has been shown
 that the crossover functions $f_{\cal O}$  are independent of the 
regularization used (i.e.\ they are universal). 
This means that
 in principle one can compute
this functions using 
e.g.\ field theory \cite{BB-85}, spin system  with medium range interaction 
\cite{PRV-99} \cite{CCPRV-01} \cite{LBB-97} \cite{LB-98},
or  polymer systems \cite{CCPRV-01}.

In this chapter we consider more general interactions than (\ref{eq1.1})
or that considered in \cite{BB-85} \cite{BBMN-87}. The reason is that
 in our large-$N$ computation we will obtain an effective action for the zero
mode without $\mathbb{Z}_2$ symmetry [$\phi\to -\phi$ in eq. (\ref{eq1.1})],
so that we need to include also odd-legs vertices and higher than four-legs
vertices. Our goals will be to 
show that including proper counterterms, in the limit in which our
crossover is defined, the crossover functions remain unchanged with
respect to the symmetric case.
In some sense we recover algebraically the $\mathbb{Z}_2$ symmetry that
is then dynamically broken.\footnote{This is similar to what happen when one
wants to investigate chiral symmetry using Wilson fermion.} 
This will permit us to investigate
the nature of the Ising Mean-Field crossover in $O(N)$ models with 
a critical zero mode (chapter \ref{chapter3}). The presentation given 
follows \cite{CMP-05}. In section \ref{1.2} we will also consider 
 the case in which
 the four- and five-legs vertices are tuned to zero at the
critical point so that one needs to generalize (\ref{eq1.1}) including
also a   $\phi^6$  operator. This will permit us to describe
(see chapter \ref{chapter4})
the crossover  in $O(N)$ models with multicritical zero mode which
effective theory will be described by an interaction similar to 
that presented in sec.\ \ref{1.2}. We stress that the basic point
that permits us to apply the results presented in this chapter
 to our large-$N$ models, is that 
the appearance of a zero mode [i.e. $r\to 0$ in eq.\ (\ref{Heff-ccl})]
is accompanied by the relations $V^{(3)}(\mathbf{0},\mathbf{0},\mathbf{0})=0$
(and similar relations, involving higher order vertices,
in the case of multicritical zero mode).

\section{Critical crossover limit}\label{chapter2-1}

We wish now to discuss the critical behavior of 
the following generic Hamiltonian
\begin{eqnarray}
{\cal H}_{\rm eff} &=& H \varphi(\mathbf{0}) + 
      {1\over 2} \int_\mathbf{p} [K(\mathbf{p}) + r] 
      \varphi(\mathbf{p}) \varphi(-\mathbf{p}) 
\nonumber \\ 
&& +  {\sqrt{u}\over 3!} \int_\mathbf{p} \int_\mathbf{q}
                V^{(3)}(\mathbf{p},\mathbf{q},-\mathbf{p}-\mathbf{q}) 
                \varphi(\mathbf{p}) \varphi(\mathbf{q}) 
                \varphi(-\mathbf{p}-\mathbf{q}) 
\nonumber \\ 
&& + 
      {u\over4!} \int_\mathbf{p} \int_\mathbf{q} \int_\mathbf{s}
         V^{(4)}(\mathbf{p},\mathbf{q},\mathbf{s},
                 -\mathbf{p}-\mathbf{q}-\mathbf{s}) 
           \varphi(\mathbf{p}) \varphi(\mathbf{q}) 
           \varphi(\mathbf{s}) \varphi(-\mathbf{p}-\mathbf{q}-\mathbf{s}) 
\label{Heff-ccl}
\end{eqnarray}
in some particular range of parameters that we will specify
in the following.
We use  a square lattice to regularize (\ref{Heff-ccl}), 
however most of the considerations reported in the following
easily apply also to  sharp-cut off regularization (used in chiral 
models) simply replacing the integration
over the first Brillouin zone with an integration over $\vp<\Lambda$.
On the other hand the magnetic and thermal regularization $h_c(u)$
$r_c(u)$ depend of the regularization chosen. 
One is free to impose both $V^{(3)}(\mathbf{0},\mathbf{0},\mathbf{0})=0$
and $K(\vp)\approx \vp^2$
normalizing and translating $\varphi$ by a constant factor
$\varphi(\vp)\to \alpha \varphi(\vp) + k \delta (\vp)$ while
 $V^{(4)}(\mathbf{0},\mathbf{0},\mathbf{0}, \mathbf{0} )=1$ 
can be obtained normalizing $u$.
In the models to which we have applied our formalism, 
$H$ and $r$ are functions of
 the two tunable parameter
(for instance $p$ and $T$ in Heisenberg models introduced in chapter 
\ref{chapter1}, or $M$ and $T$
in Yukawa models chap.\ \ref{chapter5}), 
while   the coupling parameter $u$ goes to zero like
$1/N$ in the spherical limit. Simple considerations show
that for $u \to 0$ Hamiltonian
(\ref{Heff-ccl}) exhibits a nontrivial limit
only if both $H$ and $r$ go to zero.

Let us first review the case $V^{(3)}(\mathbf{0},\mathbf{0},\mathbf{0})=0$.
There is an interesting critical
limit, the so-called Critical Crossover Limit (CCL) 
\cite{BB-84a} \cite{BB-84b} \cite{BB-85}.
 Indeed, one can show that,
by properly defining a function $r_c(u)$, in the limit
$u\to 0$, $t \equiv r - r_c(u)\to 0$, and $H\to 0$ at fixed $H/u$ and $t/u$,
the $n$-point susceptibility $\chi_n$, i.e.\ the zero-momentum 
$n$-point connected correlation function
\begin{eqnarray}
\chi_n &=& \langle \varphi^n(\mathbf{0})\rangle_{\mathrm{connected}}
\nonumber 
\end{eqnarray}
 has the following scaling form
\begin{equation}
  \chi_{n} \approx u t^{-n} f_n^{\rm symm} (t/u,H/u).
\label{scaling-cross-lim}
\end{equation}
The scaling function $f_n^{\rm symm}(x,y)$ is universal, i.e. it does 
not depend on the explicit form of $K(\mathbf{p})$ and of 
$V^{(4)}(\mathbf{p},\mathbf{q},\mathbf{r},\mathbf{s})$
as long as the normalization conditions fixed above are satisfied.
The definition of $r_c(u)$ has been discussed in Ref.~\cite{BB-85} for the 
continuum model and in Ref.~\cite{PRV-99} for the more complex case of 
medium-range models. Developing the standard expansion in $u$ one
can compute $r_c(u)$ which is obtained requiring that
$\chi_2$  scales
according to eq.~(\ref{scaling-cross-lim}) in the critical crossover limit.
If $r_c(u) = r_1 u + O(u^2)$, then at one loop we have
\begin{equation}
\chi_2 = {1\over t} - {u\over t^2} \left[r_1 + {1\over2} 
       \int_\mathbf{p} 
   {V^{(4)}(\mathbf{0},\mathbf{0},\mathbf{p},-\mathbf{p})\over 
       K(\mathbf{p}) + t}\right] + 
       O(u^2).
\end{equation}
For $t\to 0$
\begin{equation}
\int_\mathbf{p}  {V^{(4)}(\mathbf{0} ,\mathbf{0} ,\mathbf{p},-\mathbf{p})\over 
   K(\mathbf{p}) + t} \approx
  \int_\mathbf{p} {1\over \hat{p}^2 + t} + {\rm constant}\, 
  = - {1\over 4\pi} \ln t + {\rm constant},
\end{equation}
where $\hat{p}^2 = 4 \sin^2(p_1/2) + 4 \sin^2(p_2/2)$. Thus, we obtain 
\begin{equation} 
\chi_2 = {1\over t} - {u\over t^2}\left[
    r_1 - {1\over 8\pi} \ln u - {1\over 8\pi} \ln {t\over u} + 
    {\rm constant}\, \right] + O(u^2).  
\end{equation} 
Therefore, if we define
\begin{equation}
r_c(u) = {u\over 8\pi} \ln u,
\label{rcu-phi4}
\end{equation}
the susceptibility $\chi_2$ scales according to eq.~(\ref{scaling-cross-lim}).
Actually one can show that (\ref{rcu-phi4}) is sufficient to regularize the 
theory to all order in the perturbation theory. Indeed in the continuum
two-dimensional theory there is only one primitively divergent graph, 
the one-loop tadpole, and therefore only a one-loop 
mass counterterm is needed to make the theory finite. However the exact
knowledge (to all order) of $r_c(u)$ is a non-perturbative problem
and leaves some ambiguity in the definition. This will be discussed 
in sec.\ \ref{sec6.5}.

In this work we will consider the more general case in which  
also  the three-leg vertex is present in (\ref{Heff-ccl}). 
We will show 
that, if one properly defines functions $r_c(u)$ and $h_c(u)$, then
in the limit 
$u\to 0$, $t \equiv r - r_c(u)\to 0$, and $h \equiv H - h_c(u)\to 0$ 
at fixed $h/u$ and $t/u$,
the $n$-point susceptibility $\chi_n$ has the scaling form
(\ref{scaling-cross-lim}) with $h$ replacing $H$ and 
with the same scaling functions $f^{\mathrm{symm}}_n$
of the symmetric case. The previous
statement easily follows  from the condition $V^{(3)}(\mathbf{0},\mathbf{0},
\mathbf{0})=0$. Indeed considering the continuum limit of the 
action (\ref{Heff-ccl})
\begin{eqnarray}
{\cal H}_{\mathrm{eff}} &=& \int \di^2 x {1\over 2 } \varphi(x) ( \Box + 
r )\varphi(x)
+{v_3 \sqrt u\over 3!} \varphi(x)^2 \Box \varphi(x) + 
{u\over 4!} \varphi(x)^4
\label{exp-cont-a} 
\end{eqnarray}
and making the change of variable $x = y u^{-1/2}$, defining
$\phi(x)= \varphi(x /\sqrt u)$, we obtain
\begin{eqnarray}
{\cal H}_{\mathrm{eff}} & \approx & \int \di^2 y {1\over 2 } \phi(y) ( \Box + 
r/u )\phi(y)
+{v_3 \sqrt u \over 3!} \phi(y)^2 \Box \phi(y) + 
{1\over 4!} \phi(y)^4 \, .
\label{exp-cont-b} 
\end{eqnarray}  
In the previous equation (\ref{exp-cont-b}) it is clear how in the critical
crossover limit ($u\to 0$) the three legs vertex can be neglected, so that
the critical crossover coincide with that of the symmetric theory 
(\ref{scaling-cross-lim}) . However
in the previous naive discussion we do not have taken into account 
divergences that come from this contribution. This divergences must be 
carefully taken into account by proper counter-terms that define
$h_c(u)$ introduced above.

We will first prove this result at two loops and then we will give a general
argument that applies to all perturbative orders.
Notice that it is enough to consider the case $h = 0$. Indeed, if 
eq.~(\ref{scaling-cross-lim}) is valid for $h=0$ and any $n$, then
\begin{eqnarray}
\chi_n(h) &=& \sum_{m=0} {1\over m!} \chi_{n+m}(h=0) h^m \approx 
            \sum_{m=0} {1\over m!} u t^{-n-m} f_{n+m}^{\rm symm}(t/u,0) h^m = 
\nonumber \\
          &=& u t^{-n} \sum_{m=0} 
           {1\over m!} f_{n+m}^{\rm symm}(t/u,0) (t/u)^{-m} (h/u)^m.
\end{eqnarray}
which proves eq.~(\ref{scaling-cross-lim}) for all values of $h$.

\begin{figure}[t]
\centerline{\epsfig{height=10truecm,file=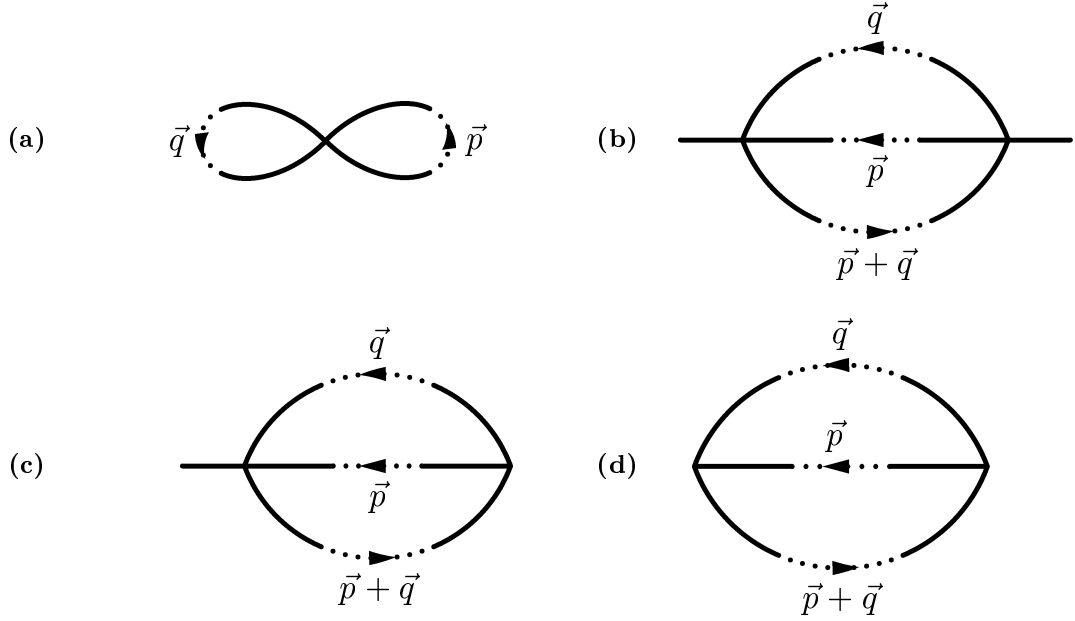}}
\caption{The four topologies appearing at two loops. Dots indicate 
parts of the graphs with additional external legs. }
\label{figgraph}
\end{figure}

\subsection{Explicit two-loop perturbative calculation} \label{twoloop-ccl}

At tree level, all contributions to $\chi_n$ that contain three-leg vertices
vanish because $V^{(3)}(\mathbf{0},\mathbf{0},\mathbf{0}) = 0$. 
Therefore, $\chi_n$ is identical to 
$\chi_n$ in the $\mathbb{Z}_2$-symmetric 
$\varphi^4$ theory; in particular $\chi_{2n+1} = 0$. 
At one-loop order, graphs contributing to $\chi_n$ are formed by a single loop
made of $a$ three-leg vertices and of $b$ four-leg vertices, 
with $a + 2 b = n$. Each of them contributes a term of the form
\begin{equation}
{u^{n/2}\over t^n} 
   \int_\mathbf{p} {V_3^a(\mathbf{p}) V_4^b(\mathbf{p})\over 
                   [K(\mathbf{p}) + t]^{a+b}},
\end{equation}
where $V_3(\mathbf{p}) \equiv V^{(3)}(\mathbf{0},\mathbf{p},-\mathbf{p})$ and 
$V_4(\mathbf{p}) \equiv V^{(4)}(\mathbf{0},\mathbf{0},\mathbf{p},-\mathbf{p})$.
The leading contribution for $t\to 0$ is obtained by replacing 
each quantity with its small-$\mathbf{p}$ behavior, i.e. 
$V_4(\mathbf{p})\approx 1$, 
$V_3(\mathbf{p}) \sim \hat{p}^2$, and $K(\mathbf{p}) \approx \hat{p}^2$,
with $\hat{p}^2 = 4 \sin^2(p_1/2) + 4 \sin^2(p_2/2)$. Then, we obtain 
a contribution proportional to 
\begin{equation}
{u^{n/2}\over t^n} \int_\mathbf{p}
     {(\hat{p}^2)^a \over (\hat{p}^2 + t)^{a+b}}. 
\end{equation} 
Now, for $t\to 0$ we have 
\begin{equation} 
\int_\mathbf{p}
     {(\hat{p}^2)^a \over (\hat{p}^2 + t)^{a+b}}
   \sim \cases{1              & for $b=0$ \cr
            \ln t         & for $b=1$ \cr
            t^{1-b}        & for $b\ge 2$.
           }
\label{int-1loop}
\end{equation}
Therefore, for $t\to 0$, $u\to 0$ at fixed $t/u$, we have 
\begin{equation}  
t^n \chi_n/u \sim \cases{u^{n/2-1} & for $b=0$ \cr  
                         u^{n/2-1} \ln t & for $b=1$ \cr
                         u^{n/2-b} = u^{a/2} & for $b\ge 2$.
           }
\end{equation}
Thus, all contributions vanish except those with: (i) $n=1$, $a=1$, $b=0$;
(ii) $n=2$, $a=2$, $b=0$; 
(iii) $n$ even, $b=n/2$, $a=0$. Contributions (iii) are those 
that appear in the standard theory without $\varphi^3$ interaction. 
Let us now show that contributions (i) and (ii) can be eliminated 
by redefining $r_c(u)$ and $h_c(u)$. Consider first $\chi_1$. At one loop 
we have 
\begin{equation}
{t\over u} \chi_1 = - {h_c(u)\over u} - {1\over 2 \sqrt{u}} 
\int_\mathbf{p} {V_3(\mathbf{p})\over K(\mathbf{p}) + t} + O (\sqrt{u}).
\label{chi1-oneloop}
\end{equation}
For $t\to 0$ we have 
\begin{equation}
\int_\mathbf{p} {V_3(\mathbf{p})\over K(\mathbf{p}) + t} = 
\int_\mathbf{p} {V_3(\mathbf{p})\over K(\mathbf{p})} + O(t\ln t).
\end{equation}
Thus, if we define 
\begin{equation}
h_c(u) = - {1\over2} \sqrt{u} \int_\mathbf{p} {V_3(\mathbf{p})\over 
        K(\mathbf{p})}, 
\label{def-hc-App}
\end{equation}
then 
$t\chi_1/u \sim O(t \ln t/\sqrt{u},\sqrt{u}) \to 0$ in the critical crossover 
limit.

Now, let us consider the two-point function. At one loop we have 
\begin{eqnarray}
{t^2\over u}\chi_2 \approx {t\over u} - {r_c(u)\over u} - 
 {1\over 2} \int_\mathbf{p} \left[
   {V_4(\mathbf{p})\over  K(\mathbf{p}) + t} - 
   {V_3(\mathbf{p})^2 \over (K(\mathbf{p}) + t)^2} \right].
\label{chi2-oneloop}
\end{eqnarray}
The first one-loop term is the contribution of the tadpole that has to be 
considered in the pure $\varphi^4$ theory and which requires an appropriate 
subtraction to scale correctly, cf. eq.~(\ref{rcu-phi4}).
The second one is due to the 
three-leg vertex and is finite for $t\to 0$. Therefore, if we define 
\begin{equation}
r_c(u) = r_c^{\rm symm}(u) + 
     {u\over 2} \int_\mathbf{p}
     {V_3(\mathbf{p})^2 \over K(\mathbf{p})^2},
\label{def-rc-App}
\end{equation}
where $r_c^{\rm symm}(u)$ is given by eq.~(\ref{rcu-phi4}), we cancel all 
contributions of the $\varphi^3$ vertex. 

Now, let us repeat the same discussion at two loops, 
in order to understand the 
general mechanism. There are four different topologies that must be considered,
see Fig. \ref{figgraph}.

{\em Topology (a).}

A contribution to $\chi_n$ is proportional to 
\begin{equation}
{u^{n/2+1}\over t^n} 
  \int_\mathbf{p} \int_\mathbf{q} 
   V^{(4)}(\mathbf{p},-\mathbf{p},\mathbf{q},-\mathbf{q}) 
  {V_4(\mathbf{p})^a V_4(\mathbf{q})^b V_3(\mathbf{p})^c V_3(\mathbf{q})^d
      \over (K(\mathbf{p})+t)^{a+c+1} (K(\mathbf{q})+t)^{b+d+1}},
\end{equation}
where $2a+2b+c+d=n$. The leading contribution for $t\to 0$ 
is obtained by setting
$V^{(4)}(\mathbf{p},-\mathbf{p},\mathbf{q},-\mathbf{q}) \approx 
 V^{(4)}(\mathbf{0},\mathbf{0},\mathbf{0},\mathbf{0}) = 1$. Then,
the integral factorizes and
we can use eq.~(\ref{int-1loop}). Ignoring logarithmic terms, we see that 
the corresponding contribution to $t^n \chi_n/u$ scales as 
\begin{equation}
  \left({u\over t}\right)^{n/2} t^{n/2-a-b} = 
  \left({u\over t}\right)^{n/2} t^{(c+d)/2}\; .
\end{equation}
Thus, a non-vanishing contribution is obtained only for $c+d=0$. Three-leg 
vertices can be neglected in the critical crossover limit.

{\em Topology (b).}

A contribution to $\chi_n$ has the form
\begin{eqnarray}
&& {u^{n/2+1}\over t^n}
  \int_\mathbf{p} \int_\mathbf{q} 
   [V^{(4)}(\mathbf{0},\mathbf{p},\mathbf{q},-\mathbf{p}-\mathbf{q})]^2
\nonumber \\
  && \qquad \times 
    {V_4(\mathbf{p})^a V_4(\mathbf{q})^b V_4(\mathbf{p}+\mathbf{q})^c 
     V_3(\mathbf{p})^d V_3(\mathbf{q})^e V_3(\mathbf{p}+\mathbf{q})^f \over 
   (K(\mathbf{p})+t)^{a+d+1} (K(\mathbf{q})+t)^{b+e+1} 
   (K(\mathbf{p}+\mathbf{q})+t)^{c+f+1}},
\end{eqnarray}
where $2(a+b+c)+d+e+f=n-2$.
The leading infrared contribution is obtained by approximating all 
expressions with their small-$p$ behavior. Therefore, we can write for $t\to 0$
\begin{equation}
{u^{n/2+1}\over t^n}
  \int_\mathbf{p} \int_\mathbf{q}
  {(\hat{p}^2)^d (\hat{q}^2)^e (\widehat{p+q}^2)^f \over
   (\hat{p}^2+t)^{a+d+1} (\hat{q}^2+t)^{b+e+1} (\widehat{p+q}^2+t)^{c+f+1}}.
\end{equation} 
Integrals of this type can be easily evaluated. By writing 
$\hat{p}^2 = (\hat{p}^2 + t) - t$ in the numerator, we obtain integrals
\begin{equation} 
I_{mnr} \equiv  \int_\mathbf{p} \int_\mathbf{q}
  {1\over 
  (\hat{p}^2+t)^{m} (\hat{q}^2+t)^{n} (\widehat{p+q}^2+t)^{r}},
\label{int-mnr}
\end{equation}
with $m > 0$, $n > 0$, $r > 0$. Then,
we can rescale 
$\mathbf{p}\to t^{1/2} \mathbf{p}$, $\mathbf{q} \to t^{1/2} \mathbf{q}$ and 
extend the 
integration over all $\mathbb{R}^2$. This is possible since the corresponding 
continuum integral is finite. As a consequence the 
integral scales as $t^{2-m-n-r}$ and
the corresponding contribution to $t^n \chi_n/u$ scales as 
\begin{equation}
  \left({u\over t}\right)^{n/2} t^{n/2-a-b-c-1} = 
  \left({u\over t}\right)^{n/2} t^{(d+e+f)/2}.
\end{equation}
Thus, a non vanishing contribution is obtained only for $d+e+f=0$. Three-leg 
vertices can be neglected in the critical crossover limit.

{\em Topology (c).}

A contribution to $\chi_n$ has the form
\begin{eqnarray}
&& {u^{n/2+1}\over t^n}
  \int_\mathbf{p} \int_\mathbf{q} 
   V^{(4)}(\mathbf{0},\mathbf{p},\mathbf{q},-\mathbf{p}-\mathbf{q})
    V^{(3)}(\mathbf{p},\mathbf{q},-\mathbf{p}-\mathbf{q})
\nonumber \\
&& \quad\qquad \times 
  {V_4(\mathbf{p})^a V_4(\mathbf{q})^b V_4(\mathbf{p}+\mathbf{q})^c 
   V_3(\mathbf{p})^d V_3(\mathbf{q})^e V_3(\mathbf{p}+\mathbf{q})^f \over 
   (K(\mathbf{p})+t)^{a+d+1} (K(\mathbf{q})+t)^{b+e+1} 
   (K(\mathbf{p}+\mathbf{q})+t)^{c+f+1}},
\label{topc}
\end{eqnarray}
where $2(a+b+c)+d+e+f=n-1$ and we assume without loss of generality 
$a\ge b \ge c$. We can now repeat the analysis performed for 
topology (b). We replace each quantity with its small-momentum behavior.
In particular, we can replace 
$V^{(3)}(\mathbf{p},\mathbf{q},-\mathbf{p}-\mathbf{q})$ with 
$(\hat{p}^2 + \hat{q}^2 + \widehat{p+q}^2)$. Then, 
we rewrite each contribution in terms of the integrals 
$I_{mnr}$, cf. eq.~(\ref{int-mnr}). However, in this case it is possible that 
one (and only one) of the indices vanishes. If this the case, 
$I_{mnr}$ factorizes and we can use the one-loop result (\ref{int-1loop}).
A careful analysis shows 
that in all cases the integral scales as $t^{-a-b-c}$ for $t\to 0$ 
except when $b=c=0$. In this case, if $a\not= 0$ the integral scales as 
$t^{-a} \ln t$, while for $a=0$ it scales as $\log^2 t$.
Thus, ignoring logarithms all 
integrals scale as $t^{-a-b-c}$. Therefore, 
the corresponding contribution to $t^n \chi_n/u$ scales as 
\begin{equation}
  \left({u\over t}\right)^{n/2} t^{n/2-a-b-c} = 
  \left({u\over t}\right)^{n/2} t^{(d+e+f+1)/2}.
\end{equation}
These contributions always vanish. 

{\em Topology (d).}

A contribution to $\chi_n$ has the form
\begin{equation}
{u^{n/2+1}\over t^n}
  \int_\mathbf{p} \int_\mathbf{q} 
  [V^{(3)}(\mathbf{p},\mathbf{q},-\mathbf{p}-\mathbf{q})]^2
  {V_4(\mathbf{p})^a V_4(\mathbf{q})^b V_4(\mathbf{p}+\mathbf{q})^c 
   V_3(\mathbf{p})^d V_3(\mathbf{q})^e V_3(\mathbf{p}+\mathbf{q})^f \over 
   (K(\mathbf{p})+t)^{a+d+1} (K(\mathbf{q})+t)^{b+e+1} 
   (K(\mathbf{p}+\mathbf{q})+t)^{c+f+1}},
\end{equation}
where $2(a+b+c)+d+e+f=n$, $a\ge b\ge c$. 
We repeat the analysis done for topology (b) and (c). We find that the 
integral scales as 
\begin{equation}
t^{-a-b-c+1} \cases{\ln t/t   &  for $a=b=c=0$;  \cr
                    1/t        &  for $b=c=0$, $a \ge 1$; \cr
                    \ln t     &  for $c=0$, $b=1$, $a \ge 1$; \cr
                    {\rm constant }  &  otherwise.}
\end{equation}
Thus, if $b\not=0$ and $c\not=0$, ignoring logarithms, the contribution to 
$t^n\chi_n/u$ scales as 
\begin{equation}
\left({u\over t}\right)^{n/2} t^{n/2-a-b-c+1} = 
\left({u\over t}\right)^{n/2} t^{(d+e+f)/2+1} 
\end{equation}
These contributions therefore always vanish in the critical crossover limit. 
If, however, $b=c=0$, then 
\begin{equation}
\left({u\over t}\right)^{n/2} t^{n/2-a-b-c} =
\left({u\over t}\right)^{n/2} t^{(d+e+f)/2}
\end{equation} 
and thus one may have a finite (or a logarithmically divergent 
if $a = 0$) contribution for $d=e=f=0$. Let us now focus on this last
case in which $a=n/2 >0$ 
with $n$ even. Let us now show that these contributions are cancelled by
the one-loop counterterm due to $r_c(u)$. Indeed, at two loops we obtain 
a contribution to $\chi_n$ of the form 
\begin{equation}
r_c(u) {u^{n/2}\over t^n} \int_\mathbf{p}
       {V^b_3(\mathbf{p}) V_4^a(\mathbf{p}) \over [K(\mathbf{p}) + t]^{a+b+1}} 
  = 
    r_c(u) {u^{n/2}\over t^n} \cases{\ln t  & for $b=0$ \cr 
                                     t^{-a} & for $b\ge 1$ 
        }
\end{equation} 
where $2 a + b =n$. Thus, contributions to $t^n\chi_n/u$ scale as 
$t^{n/2-a} = t^{b/2}$ and thus vanish unless $b=0$. 
Thus, for each even $n$ there are two contributions that should be 
considered: one with topology (d) and one associated with the counterterm
$r_c(u)$. Taking properly into account the combinatorial factors,
their sum is given by (of course, 
we only consider here the contribution to $r_c(u)$ due to the three-leg 
vertices)
\begin{equation} 
{u^{n/2+1}\over t^n} 
    \int_\mathbf{p} {1\over (K(\mathbf{p}) + t)^{a+1}} 
    \int_\mathbf{q} \left[
    {V^{(3)}(\mathbf{p},\mathbf{q},-\mathbf{p}-\mathbf{q})^2 \over 
      (K(\mathbf{q}) + t)(K(\mathbf{p}+\mathbf{q})+t)} - 
    {V^{(3)}(0,\mathbf{q},-\mathbf{q})^2 \over (K(\mathbf{q}) + t)^2}\right]
\end{equation}
For $a\ge 1$, the subtracted term improves the infrared behavior, and 
indeed the integral scales as $t^{-a+1} \ln t$  and is therefore 
irrelevant in the critical crossover limit. 

\subsection{Higher powers of the fields}

It is interesting to consider also Hamiltonians with 
higher powers of the field $\varphi$. 
The standard scaling argument indicates that
 these additional terms 
 are irrelevant for the 
critical behavior, but in principle they can contribute 
to the renormalization constants like $h_c$ and $r_c$. For this purpose let us 
suppose that the Hamiltonian contains also terms of the form 
\begin{equation}
\Delta{\cal H}_{\rm eff} = \sum_{k>4} {u^{(k-2)/2}\over k!}
    \int_{\mathbf{p}_1}\cdots \int_{\mathbf{p}_k}
\delta(\mathbf{p}_1 + \cdots \mathbf{p}_k) \, 
    V^{(k)}(\mathbf{p}_1,\cdots,\mathbf{p}_k) 
   \varphi(\mathbf{p}_1)\cdots \varphi(\mathbf{p}_k).
\end{equation}
Note the particular dependence on the coupling constant $u$, which is 
motivated by the large-$N$ calculation and is crucial in the argument 
reported below.

Let us consider the contributions of these additional terms. At tree level
we have an additional contribution to $\chi_n$ given by
\begin{equation}
\Delta\chi_n = {u^{(n-2)/2}\over t^n} V^{(n)}
    (\mathbf{0},\ldots,\mathbf{0})
\end{equation}
Thus, $t^n \Delta\chi_n/u \sim u^{(k-4)/2}$, that vanishes as $u\to 0$ for 
$k > 4$. At one-loop order there are additional contributions of the 
form 
\begin{equation}
   {u^{n/2}\over t^n} \int_\mathbf{p} (K(\mathbf{p}) + t)^{-\sum a_i} 
   V_3(\mathbf{p})^{a_3} \cdots V_{n+2}(\mathbf{p})^{a_{n+2}},
\end{equation}
where $V_k(\mathbf{p}) = 
   V^{(k)}(-\mathbf{p},\mathbf{p},\mathbf{0},\ldots,\mathbf{0})$ and 
$\sum_k(k-2)a_k = n$. 
Proceeding as in the previous section, using eq.~\reff{int-1loop}, and 
noting that $\sum_{k>4} a_k > 0$ by hypothesis, we obtain
\begin{equation}
t^n \Delta\chi_n/u \sim u^{n/2-1} t^{1 - \sum_{k > 3} a_k} \sim
   u^{{1\over2} a_3 + {1\over2} \sum_{k \ge 5} (k-4) a_k}.
\end{equation}
Here, possible logarithmic terms have been neglected. Then, since some 
$a_k$ with $k\ge 5$ is non-vanishing by hypothesis, we find that this
correction vanishes in the crossover limit. 
Therefore, no contribution survives at one loop. The same is expected at any
perturbative order.

\subsection{The general argument}\label{general-argument}

The discussion reported above shows that at two loops one can 
define a critical crossover limit with crossover functions that are identical 
to those of the symmetric theory. This is expected to be a general result
since formally the added interaction is irrelevant. 
This result can be understood diagrammatically. 

Consider the continuum theory with Hamiltonian
\begin{equation}
{\cal H} = \int \di^2 x\, {1\over2}\sum_\mu (\partial_\mu \phi(x))^2 + 
           {t\over2} \phi(x)^2 + {\sqrt{u}\over 3!} \phi(x)^2 \Box \phi(x) + 
           {u\over 4!} \phi(x)^4.
\end{equation}
Given an $l$-loop diagram contributing to the zero-momentum $n$-point 
irreducible correlation function, we can compute  the
superficial degree of divergence of Feynman integrals $D \sim \Lambda^\delta$
as  
\begin{equation}
D \sim \Lambda^{d \ell+2 N_3 - 2 N_i}
\label{scalingD}
\end{equation}
obtained rescaling each momenta with $\Lambda$. In (\ref{scalingD})
$\ell$ is the number of loop of the diagram, $N_3$ ($N_4$) is the
number of the three (four) legs vertices and $N_i$ is the number
of the internal lines. Using the topological relations
\begin{eqnarray}
n+2 N_i &=& 3 N_3 + 4 N_4
\nonumber
\\
N_3 + 2 N_4 &=& n + 2 \ell -2
\nonumber
\end{eqnarray}
we obtain for $d=2$
\begin{eqnarray}
D &\sim & \Lambda^{2(1-N_4)} 
\label{scalingI}
\end{eqnarray}
Thus, there are primitively divergent diagrams for any $n$: those with 
$N_4=0$ (and correspondingly $N_3 = n + 2l - 2$) are quadratically 
divergent, while those with $N_4 = 1$ (and $N_3 = n + 2l - 4$) are 
logarithmically divergent. 
The previous considerations suggest that in the evaluation of the
generic Feynman diagram with $n$ external legs $\Lambda^2$
and $\log \Lambda^2$ divergences will be generated. 
To regularize the theory a counterterm $\delta_n \cdot \phi(\vp)^n$ 
is introduced. However in principle this could not be
sufficient. Indeed we observe that
other divergences comes from 
second derivatives of the $n$-point irreducible correlation functions with
respect to the external momenta $\phi(\vp)^{n-1} \Box_p \phi(\vp)$ . Following
 the 
same step done before we find 
\begin{eqnarray}
\Box_p\langle\phi(\vp=\mathbf{0}) \phi(\mathbf{0})^{n-1} \rangle \sim \Lambda^{-2 N_4}
\end{eqnarray}
which 
diverge logarithmically for every $n$  when $N_4 = 0$, so that also 
these operators
must be included in the renormalized Hamiltonian which will contain
an infinite number of counterterms 
\begin{equation}
{\cal H}^{\rm ren} = {\cal H} + 
   \sum_n [Z_n(\Lambda) \phi^n + \zeta_n(\Lambda) \phi^{n-1}\Box\phi].
\end{equation}
Now, let us show that all counterterms except those 
computed in the previous section can be neglected in the critical 
crossover limit. Suppose that we wish to compute $Z_n(\Lambda)$ at 
$l$ loops. Keeping into account that the divergence may be quadratic or 
logarithmic, we expect the $l$-loop divergent contribution to $\chi_n$ to be 
of the form 
\begin{equation}
{u^{n/2+l-1}\over t^n} (a_1 \Lambda^2 + t P_1[\ln (\Lambda^2/t)] + 
      P_2[\ln (\Lambda^2/t)] ). 
\end{equation}
where $P_1(x)$ and $P_2(x)$ are polynomials and $a_1$ a constant.
Therefore, the contribution to $t^n\chi_n/u$ vanishes unless 
$n/2 + l - 2\le 0$. The only two cases satisfying this condition 
(of course $n\ge 1$ and $l\ge 1$) are $l=n=1$, $l=1$ and $n=2$, which are 
the cases considered before.
Let us now consider the contributions to $\bar{\chi}_{n,i}$, which is 
the first derivative of the $n$-point connected correlation function
with respect to the square of an external momentum $\mathbf{p}$ computed
at zero momentum. Since momenta scale as $t^{1/2}$, in the critical
crossover limit we should have 
$\bar{\chi}_{n,i}\approx u t^{-n-1} \bar{f}_n(t/u,H/u)$. 
The divergent contributions are logarithmic (diagrams with $N_4 = 0$) 
and therefore we expect an $l$-loop contribution of the form
\begin{equation}
{u^{n/2+l-1}\over t^n} (P[\ln (\Lambda^2/t)] + \hbox{\rm finite  terms}).
\end{equation}
Considering $t^{n+1} \bar{\chi}_{n,i}/u$, we see that this contribution 
always vanishes in the critical crossover limit. Therefore, 
the renormalization constants $\zeta_n(\Lambda)$ can be neglected. 
Thus, the only renormalizations needed are those that we have 
considered. Finally, let us show that correlation functions computed in 
the renormalized theory have the correct scaling behavior. Indeed, in
the renormalized theory diagrams scale canonically with possible logarithmic 
corrections. Therefore, $D$ introduced at the beginning of this section
 scales as 
$u^{N_4 + N_3/2} t^{1-N_4} \times$logs, so that the contribution of 
$D$ to $t^n \chi_n/u$ scales as 
\begin{equation}
   {t^n\over u} \times {1\over t^n} \times u^{N_4 + N_3/2} t^{1 - N_4} 
   \sim u^{N_3/2}.
\end{equation}
Therefore, the only non-vanishing diagrams have $N_3=0$, confirming the 
claim that three-leg vertices do not play any role.

\subsection{A unique definition for the renormalization functions
$h_c(u)$ and $r_c(u)$} \label{sec6.5}

In this section we wish to discuss again the definition of $r_c(u)$
and $h_c(u)$. It is obvious that these functions are not uniquely defined,
since one can add a term proportional to $u$ without modifying the 
scaling behavior. We wish now to fix this ambiguity by 
requiring that $t = h = 0$ corresponds to the critical point. 

It is easy to see that no modifications are needed for $h_c(u)$. 
Indeed, with the choice (\ref{def-hc-App}) one obtains the 
correlation functions of the symmetric theory and in this case the 
critical point is uniquely defined by $h = 0$ by symmetry. The proper
definition of $r_c(u)$ requires more care, since we must perform a 
non-perturbative calculation in order to identify the critical point. 
For this purpose we will use the fact that in the critical crossover limit 
the perturbative expansion in powers of $u$ is equivalent to the 
perturbative expansion in the continuum $\phi^4$ theory once a proper 
mass renormalization is performed. 

In the continuum theory, if $\tilde{t}\equiv t_{\rm cont}/u_{\rm cont}$ 
is the a-dimensional reduced temperature defined so that $\tilde{t} = 0$ 
corresponds to the critical point, we have at one loop, 
cf. eq.~(2.10) of Ref.~\cite{PRV-99}, 
\begin{equation}
u_{\rm cont} \chi_{2,\rm cont} = 
   {1\over \tilde{t}} + {1\over 8 \pi \tilde{t}^2} \left(
   \ln {8 \pi \tilde{t}\over 3} + 3 + 8 \pi D_2\right) + 
   O(\tilde{t}^{-3} \ln \tilde{t}^2),
\label{chi2-cont}
\end{equation}
where $D_2$ is a non-perturbative constant that can be expressed in terms 
of renormalization-group functions, cf.~eq.~(2.11) of Ref.~\cite{PRV-99}. 
By using the four-loop perturbative results of Ref.~\cite{BNGM-77},
Ref.~\cite{PRV-99} obtained the estimate $D_2 = - 0.0524(2)$. 
It is not clear whether the error can really be trusted, since in two 
dimensions the resummation of the perturbative expansion is
not well behaved due to 
non-analyticities of the renormalization-group functions at the 
fixed point \cite{CCCPV-00,CPV-01}; still, the estimate should provide the 
correct order of magnitude. 

The expansion \reff{chi2-cont} should be compared with the perturbative 
expansion of $\chi_2$ in the lattice model. We write $r_c(u)$ as 
\begin{eqnarray}
r_c(u) &=& {u\over 8\pi} \ln u + {u\over2} \int_\mathbf{p}
      {V_3(\mathbf{p})^2\over K(\mathbf{p})^2} + A u ,
\nonumber\\
A &=& - D_2 - {1\over 8\pi} \ln {256\pi\over3} - {3\over 8\pi} - 
    {1\over2} \int_\mathbf{p} \left[
          {V_4(\mathbf{p})\over K(\mathbf{p})} - {1\over \hat{p}^2}\right].
\label{defrc-const}
\end{eqnarray}
where $A$ has been determined comparing  $\chi_2$
with a lattice regularisation eq.~\reff{chi2-oneloop} 
\begin{equation}
u \chi_2 = {u\over t} + {u^2\over 8 \pi t^2} 
   \left\{ \log{t\over 32 u} - 8 \pi A - 
          4\pi \int_\mathbf{p} \left[ 
          {V_4(\mathbf{p})\over K(\mathbf{p})} - 
          {1\over \hat{p}^2}\right]\right\} + 
          O(u^3).
\end{equation}
with $\chi_2$ in the continuum limit given by
 eq.~\reff{chi2-cont}. 
If we use definition \reff{defrc-const}, the critical point 
corresponds to $\tilde{t} = 0$.

%%%%%%%%%%%%%%%%%%%%%%%%%%%%%%%%%%%%%%%%%%%%%%%%%%%%%%%%%%%%%%%%%%%%
%
\subsection{Higher than two dimension}\label{chapter2-higher-dimension}
%
%%%%%%%%%%%%%%%%%%%%%%%%%%%%%%%%%%%%%%%%%%%%%%%%%%%%%%%%%%%%%%%%%%%%

In this section we want to generalize the previous results 
(obtained for $d=2$) to higher dimension $2\leq d<4$,
for which Ising criticality is still present. 
Cosidering the continuum action (\ref{general-cont})
one is able to predict the general scaling relations
(\ref{provv-chapter1}) and (\ref{scalingrel}).
On the other hand the critical parameters
$h_c(u)$ and $r_c(u)$ will include more contributions
with respect to the two dimensional case \cite{CMP-05}.
This is simple related to the fact that in three dimension
for instance, the mass counterterm must be computed up to two 
loops. However in the case in which the theory is not symmetric,
anomalous terms appear also for $\chi_3$; these bad
contributions cannot be deleted by using  
$h_c(u)$ or $r_c(u)$. In this section we show that
the expected scaling relation (\ref{scalingrel})
still survive if the critical field is translated
by a constant factor\footnote{The normalization of
$k_R$ by using $u^{1/2}$, has been taken for convenience.} (i.e.\ 
$\varphi \to \varphi + u^{1/2} k_R$), so that the anomalous three
legs diagrams can be deleted fixing $k_R$.

Let we start considering the 
continuum interaction in $d<4$ dimension 
\begin{equation}
{\cal S}_{\rm cont}[\varphi] = 
\int d^d\mathbf{r}\,
\left[ H \varphi(\mathbf{r}) + {{1\over2}} 
\Big(\partial\varphi(\mathbf{r})\Big)^2 +
    {{r\over2}} \varphi(\mathbf{r})^2 +
{\sqrt u v_3\over 3!} \varphi(\mathbf{r})^2 \Box \varphi(\mathbf{r})
  +  {{u\over 4!}}  \varphi(\mathbf{r})^4\right],
\label{general-cont}
\end{equation}
generalizing the considerations done in (\ref{exp-cont-a}),
i.e.\ making a change of variable $\mathbf{r} = 
\mathbf{y} u^{-1/(4-d)}$ and defining a new bosonic field
\begin{equation}
\psi(\mathbf{x}) = u^{(d-2)/[2(d-4)]}
    \varphi(\mathbf{x} u^{-1/(4-d)}) .
\end{equation}
we can rewrite (\ref{general-cont}) as
\begin{equation}
{\cal S}_{\rm cont}[\psi] =
\int d^d\mathbf{s}\,
\left[ \tilde{H} \psi(\mathbf{s}) + {1\over2} 
\Big(\partial\psi(\mathbf{s})\Big)^2 +
    {\tilde{r}\over2} \psi(\mathbf{s})^2 +
{ u^{d\over 4-d} v_3\over 3!} \psi(\mathbf{s})^2 \Box \psi(\mathbf{s})
+    {1\over 4!}  \psi(\mathbf{s})^4\right],
\end{equation}
where  we have defined. 
\begin{equation}
\tilde{H} \equiv  H u^{-(d+2)/[2(4-d)]}, 
\qquad\qquad
\tilde{r} \equiv  r u^{-2/(4-d)}.
\label{provv-chapter1}
\end{equation}
Thus, formally, once the action is expressed in terms of 
$\psi$, the bare parameters appear only in the combinations 
$\tilde{H}$ and $\tilde{r}$ while the three fields coupling
can be neglected. 
Then, consider the zero-momentum connected 
correlation function $\chi_n$,
neglecting the three-legs vertex, we have
\begin{eqnarray}
\chi_n &\equiv &
   \int d^d\mathbf{r}_2\ldots  d^d\mathbf{r}_n\,
   \langle \varphi(\mathbf{0})\varphi(\mathbf{r}_2) \ldots
            \varphi(\mathbf{r}_n) \rangle^{\rm conn}
\nonumber \\
   &=&u^{[2d-n(2+d)]/[2(4-d)]} \int d^d\mathbf{s}_2\ldots  d^d\mathbf{s}_n\,
   \langle \psi(\mathbf{0})\psi(\mathbf{s}_2) \ldots
            \psi(\mathbf{s}_n) \rangle^{\rm conn}
\nonumber \\
      &=& u^{[2d-n(2+d)]/[2(4-d)]} f_n(\tilde{H},\tilde{r}),
\label{scalingrel}
\end{eqnarray}
i.e. $u^{-[2d-n(2+d)]/[2(4-d)]} \chi_n$ is a scaling function that is universal
given the normalization conditions for 
$\tilde{H}$ and $\tilde{r}$. 
The above-reported discussion is valid only at the formal level 
because we have not regularized the theory, or in other words
a cut-off regularization breaks scale invariance.

Let us first discuss the symmetric case because it will be used
in this work.
In a similar way of what happens for $d=2$, the previous 
relation (\ref{scalingrel}) remains true
if proper counterterms are included in the theory.
For $d < 4$ we have to include only a mass counterterm $r_c(u)$
[$h_c(u)=0$ due to the symmetry].
 Having done this
the correlation functions $\chi_n$ 
satisfy the scaling relations (\ref{scalingrel})  
with 
$\tilde{t} = t u^{-2/(4-d)}$, $t \equiv r - r_c(u)$ replacing $\tilde{r}$,
and $\tilde h = h u^{-(d+2)/[2(4-d)]} $, $h\equiv H- h_c(u)$
replacing $\tilde H$
\begin{equation}
  \chi_n = u^{[2d-n(2+d)]/[2(4-d)]} f_n(\tilde{h},\tilde{t}) 
\label{scalingrel2}
\end{equation}
However in more than two dimensions more than one loop terms
enter into the definition of $r_c(u)$ and $h_c(u)$. This is
related to the fact that in the continuum theory, 
divergent diagrams are present beyond the one  loop order.
For instance in three dimension ($h_c(u)=0$) we have\footnote{The form of
$r_c$ and $h_c$ depends on the explicit regularization
used. Here we use a cut-off because it will be used 
for a fermionic model investigated in this work.}
\begin{equation}
r^{(\Lambda)}_c(u) 
= - {\Lambda\over 4 \pi^2} u + {u^2\over 96\pi^2} \ln u + K u^2
\label{rcd3}
\end{equation}
where $K$ is  chosen so that 
$t = 0$ corresponds to the critical point (see sec.\ \ref{sec6.5}).

The non-symmetric case demands a more accurate analysis.
Following the discussion of sec.\ \ref{general-argument}
we have that a generic contribution to $\chi_n$ scales as\footnote{
$G_n(\Lambda)$ for $d=3$ computed up to $\ell$ loops diverges 
as $\Lambda^{\ell+2}$. Indeed generalizing  (\ref{scalingI}) we have that 
for a generic $\ell$ loops diagram $D\sim \Lambda^{\ell+2(1-N_4)}
=\Lambda^{3-N_4+N_3/2-n/2}$ 
}
$\chi_n\sim G(\Lambda,t) t^{-n} u^{n/2+l-1}$. Using the scaling
relation (\ref{scalingrel2}) $t\sim u^{2/(4-d)}$ we observe that
\begin{eqnarray}
u^{n(2+d)-2d\over 2(4-d)}\chi_n &\sim& u^{{n-4\over 4-d}+\ell}
G(\Lambda,u^2)
\end{eqnarray}
so that the anomalous diagram can be obtained
by studying
${n-4\over 4-d}+\ell \le 0$. If $d=3$ there are
the following dangerous contributions:
a) $n=1$, $\ell=1,2,3$ (so that $h_c$ can be obtained 
by a three loops computation), 
b) $n=2$ $\ell=1,2$ [cancelled in the definition of $r_c(u)$], but
also c) $n=3$ $\ell=1$  that in principle give us some problems.
Now we show that the one loop $n=3$ anomalous contributions can be
deleted by introducing a new bosonic 
field $\phi(\vp)$\footnote{
This operation is exactly what we do when we obtain the effective
action for the zero mode in the large $N$ limit where in principle
the condition $V^{(3)}(\mathbf{0},\mathbf{0},\mathbf{0})=0$ is not
satisfied. For $d=3$ one needs to take into account also radiative 
corrections.
}
\begin{eqnarray}
\varphi(\vp) &=& \phi(\vp) - u^{1/2}k_R \delta(\vp)  
\label{k-ren}
\end{eqnarray}
and taking $k_R$ so that the effective three-legs vertex [computed
up to $O(u)$ in $d=3$] for the new bosonic fields is zero. Defining
$\ol V^{(n)}(\vp_1,\cdots \vp_n)$ the $n-$legs vertex 
for the translated 
fields, we have
\begin{eqnarray}
\ol V^{(3)}(\vp_1,\vp_2,\vp_3) &=& V^{(3)}(\vp_1,\vp_2,\vp_3)
-u k_R V^{(4)}(\vp_1,\vp_2,\vp_3,\mathbf{0})
\\
\ol V^{(2)}(\vp_1,\vp_2) &=&  V^{(2)}(\vp_1,\vp_2) 
-u k_R V^{(3)}(\vp_1,\vp_2,\mathbf{0}) 
\nonumber\\
&&+ {u^2{k_R}^2\over 2}
V^{(4)}(\vp_1,\vp_2,\mathbf{0},\mathbf{0})  
\label{termal-3d}
\\
\ol H &=& H -u k_R  V^{(2)}(\mathbf{0},\mathbf{0}) 
+{u^2{k_R}^2\over 2} V^{(3)}(\mathbf{0},\mathbf{0},\mathbf{0}) 
\nonumber\\
&&- {u^3{k_R}^3\over 6}
V^{(4)}(\mathbf{0},\mathbf{0},\mathbf{0},\mathbf{0})
\label{magnetic-3d}
\end{eqnarray}
Imposing that the first radiative correction for the three-legs vertex
is null we get the result\footnote{Notice that both the integrals
are infra red finite in three dimensions.}
\begin{eqnarray}
 k_R &=&{1\over V_4(\mathbf{0})}\Bigg[ {1\over 2}\int \di^{3} \vp 
{V_3(\vp) V_4(\vp)\over k(\vp)^2}
-{1\over 6}\int \di^{3} \vp {V_3(\vp)^3\over k(\vp)^3} \Bigg]
\label{kr-esplicito}
\end{eqnarray}
that delete the the anomalous c) contributions. We have regularized
the anomalous a) and b) contributions introducing respectively
$h_c(u)$ and $r_c(u)$ similarly to what done for
$d=2$ (but with more diagram). But now new anomalous contributions
 appear due to the new terms (proportional to $k_R$) in 
(\ref{termal-3d}) (\ref{magnetic-3d}). It is straightforward to
understand that this terms can be deleted recomputing
$h_c(u)$ and $r_c(u)$ as a function of the
 new vertices $\ol V^{(n)}$. 
Indeed this is possible because the leading 
behavior of the vertices remain
the same $\ol V_n(\mathbf{0}) \sim V_n(\mathbf{0}) $.

\section{Multicritical crossover limit}\label{1.2}

Now we want to generalize the previous considerations,
considering  more general interactions than (\ref{Heff-ccl}). 
We want to take into account interactions in which
we tune to zero at the same time all $V_j(\mathbf{0})$ for 
$j=0,1\cdots \N+2$ (with $\N$ odd),
while $V_{\N+3}(\mathbf{0})$ remain finite and positive.\footnote{
The positive condition guarantees us that the stability of the
theory.}
In the continuum limit we want to study the following
Hamiltonian
\begin{eqnarray}
{\cal H} = H \chi + {1\over 2} \sum_\mu (\partial_\mu\chi)^2 +
   \sum_{n=2}^{\N+3} {a_n\over n!} {1\over N^{n/2-1}} \chi^n,
\end{eqnarray}
with $a_{{\cal N} + 2} = 0$. Here we have disregarded the momentum 
dependence of the vertices---it is formally irrelevant in the 
continuum limit---and higher-order terms in the kinetic term.
We are interested in the limit in which $a_i \to 0$ for $i=2,\cdots \N+1$, 
so that the leading 
non-vanishing term is $a_{{\cal N} + 3} \chi^{{\cal N} + 3}$. 
In analogy to what have been done in the previous section \ref{chapter2-1}
 we wish now to 
compute the crossover from mean-field to multicritical behavior.
The analysis is quite complex and thus we only consider explicitly the case 
${\cal N} = 3$. 

Using the lattice regularization
we start from the following Hamiltonian  
\begin{eqnarray}
{\cal H} &=& u_{01} \phi(\mathbf{0})+
    {1\over 2}\int_{\vp }
\LL(K(\mathbf{p})+ u_{02}\RR)\phi(\mathbf{p})\phi(-\mathbf{p})
\nonumber\\
&&+{1\over 3!}  \LL(\prod_{i=1}^3\int_{\vp_i } 
 \,\phi({\mathbf p}_i)\RR)
\delta\LL(\sum_{i=1}^3 {\mathbf p}_i\RR) \left[u_{03} + u_6^{1/4}
V^{(3)}(\{ {\mathbf p}_i\})\right]
\nonumber\\
&&+{1\over 4!} \LL(\prod_{i=1}^4\int_{\vp_i } 
 \,\phi({\mathbf p}_i)\RR)
\delta\LL(\sum_{i=1}^4 {\mathbf p}_i\RR) \left[u_{04} + u_6^{1/2}
V^{(4)}(\{\mathbf{p_i}\})\right]
\nonumber\\
&& +{u_6^{3/4}\over 5!}\LL(\prod_{i=1}^5\int_{\vp_i } 
 \,\phi({\mathbf p}_i)\RR) 
\delta\LL(\sum_{i=1}^5 {\mathbf p}_i\RR)
V^{(5)}(\{ {\mathbf p}_i\}) 
\nonumber\\
&&+{u_6\over 6!}\LL(\prod_{i=1}^6 \int_{\vp_i } 
\,\phi( {\mathbf p}_i)\RR)
\delta\LL(\sum_{i=1}^6 {\mathbf p}_i\RR) 
\left[ 1+V^{(6)}(\{ {\mathbf p}_i\})\right]
\label{starting}
\end{eqnarray}
where we have defined
\begin{equation}
V^{(i)}({\mathbf 0},\ldots,{\mathbf 0} )=0
\end{equation}
for $i=1,\cdots,6$. 
We have written each vertex as 
a constant plus a function of the momenta multiplied by a power of 
$u_{6}$. This is indeed the expression that one would obtain 
from the effective Hamiltonian for the zero mode by setting 
$u_6 \sim 1/N^2$. Moreover, we have dropped the contributions of 
vertices with more than six fields. We will discuss them in 
sec.~\ref{sec5.5}. The normalization of the field is fixed by requiring:
\begin{eqnarray} 
K(\vp) &\approx& \vp^2 + O(\vp^4),
\label{normalization-1}
\end{eqnarray}
while we introduce $\alpha_i$ to name the leading
contribution in $\mathbf{p}$ for higher order 
than two vertices
\begin{eqnarray}
V^{(i)}(\vp,-\vp,\mathbf{0},\cdots\mathbf{0}) &\approx& \alpha_i \vp^2
+ O(\vp^4).
\label{normalization-2}
\end{eqnarray}

We wish now to show
that four critical parameters ($u_{ci}$, $i=1,\ldots 4$)
can be defined, such that 
for $u_i \equiv  u_{0i} - u_{ci} \to 0$, ($i=1,\ldots 4$), 
$u_6\to 0$, the $n$-point susceptibility has the scaling form:
\begin{equation}
\chi_n = t^{1-n} f_n(x_1,x_3,x_4,x_6),
\label{scalingeq}
\end{equation}
where we write for notational convenience $t \equiv u_2$ and we set
$x_i \equiv u_i/t$. Moreover, we will show that the function 
$f_n$ is independent of the vertices $V^{(i)}$ and it is thus 
a universal scaling function of the $\phi^6$ theory. The 
non-universal features of the model appear only in the functions
$u_{ci}$, which can be computed in perturbation theory 
up to an additive constant. In our case a two-loop calculation
provides the full answer. Notice that, as pointed
out in section \ref{chapter2-1} for the $\N=1$ case,
in order to prove eq.~\reff{scalingeq} it is enough
 to consider the 
case in which $u_1 = 0$. This is the case we
 shall consider below.

\subsection{One-loop computation}

We begin by considering 
a generic one-loop diagram with $n_3$ three-leg vertices
and so on up to $n_6$ six-leg vertices contributing to 
$\chi_n$. For $t \to0$ its contribution is 
\begin{equation}
\chi_n\sim t^{-n} 
\int_{\mathbf{p}} {1\over [K(\vp)+ t]^{\sum n_i}} 
\prod_{i=3}^6 [u_i + w_i V_i (\vp)]^{n_i}
\end{equation}
where $V_i(\vp) \equiv V^{(i)}(\vp,-\vp,\mathbf{0},\ldots)$, 
$w_j = u_6^{(j-2)/4}$ ($j = 3,4,5,6$), and $u_5 = 0$.
Expanding we obtain contributions of the form
\begin{equation}
\chi_n\sim t^{-n} \prod_{i=3}^6 u_i^{n_i-k_i} w_i^{k_i}
\int_{\mathbf{p}} {\prod_i [V_i (\vp)]^{k_i} \over
[K(\vp) + t]^{\sum n_i}},
\end{equation}
where $0\le k_i \le n_i$ and $k_5 = n_5$.
If $\sum (n_i - k_i)\ge 2$ the integral behaves 
as $t^{-A}$ with 
$A = - 1 + \sum (n_i - k_i)$, so that 
\begin{equation}
t^{n-1} \chi_n\sim \prod_i x_i^{n_i - k_i} w_i^{k_i}.
\end{equation}
Thus, in the scaling limit, the contributions of the formally-irrelevant
vertices  can be neglected and thus we can set $k_i = 0$. 
The remaining contributions 
(with $k_3 = k_4 = k_6 = n_5 = 0$) scale according to 
eq.~\reff{scalingeq}. Note that in this case the integral can be computed 
in the continuum, by replacing $K(\vp)$ with $p^2$ and by integrating over
the whole plane. Therefore, the result is independent of the explicit form
of $K(\vp)$.

If $\sum (n_i - k_i) = 1$, we have similarly
\begin{equation}
t^{n-1} \chi_n\sim x_i^{n_i - k_i} w_i^{k_i} \log t.
\end{equation}
The formally-irrelevant vertices
can still be neglected. However, 
in this case there is an additional $\log t$ that breaks the scaling law 
for $k_i = 0$, i.e. for $n_3 + n_4 + n_6 = 1$.
Anomalous contributions appear therefore in $\chi_1$, $\chi_2$, and 
$\chi_4$.

Finally, if $\sum (n_i - k_i) = 0$
(this implies $n_3 = k_3$, $n_4 = k_4$, $n_6 = k_6$) the integral
is constant as $t\to 0$. Therefore, we have 
\begin{equation}
t^{n-1} \chi_n \sim {1\over t} \prod_i w_i^{n_i} \sim
     {u_6^{n/4}\over t}.
\end{equation}
In the scaling limit this contribution is irrelevant except when
$n \le 4$.

In conclusion, at one loop, 
the only contributions that scale anomalously
in $\chi_n$ are that whith $n \le 4$. 
We shall now show that we
can absorb these anomalous terms in the definition of the functions
$u_{ci}$ (in this section contributions to $u_{ci}$ given by one-loop
diagrams will be indicated as $u^{1l}_{ci}$).

Let us begin by considering the one-loop expression of $\chi_1$:
\begin{eqnarray}
\chi_1^{1l} &=& -{u^{1l}_{c1}\over t} - {1\over 2 t}\int_{\mathbf{p}}
{u_3+ u_6^{1/4} V_3(\mathbf{p})\over
K(\vp) +t }
\nonumber\\
&=& -{u^{1l}_{c1}\over t} - {u_3\over 2 t} 
  \left[- {1\over 4 \pi} \log {t\over 32} + K_1 \right] -
{u_6^{1/4} \over 2 t}\int_{\mathbf{p}} {
   V_3(\mathbf{p})\over K(\vp)}
+ O(u_6^{1/4} \log t, u_3\log t),
\end{eqnarray}
with
\begin{equation}
K_1 = \int_\vp \left({1\over K(\vp)} - {1\over \hat{p}^2}\right)
\end{equation}
and $\hat{p}^2 = 4 \sum_i \sin^2(p_i/2)$.
Thus, if we define
\begin{eqnarray}
u^{1l}_{c1} &=& {u_3\over 8 \pi } \left(\log u_6  - 4 \pi K_1\right) -
   {u_6^{1/4} \over 2}\int_{\mathbf{p}} {
   V_3(\mathbf{p})\over K(\vp)},
\label{uc1-1loop}
\\
&\equiv&U_1 + U_2
\nonumber
\end{eqnarray}
the anomalous contributions in $\chi_1$ cancel out
 and the result is 
independent of $V^{(j)}$ and of $K(\vp)$
\begin{equation}
\chi^{1l}_1 = - {x_3 \over 8 \pi} \log (32 x_6).
\end{equation}
In order to render easier the analysis of 
the theory at two loop level,
in (\ref{uc1-1loop}) we have introduced $U_1$ and $U_2$,
defined as
\begin{eqnarray}
U_1 &=&{u_3\over 8 \pi } \left(\log u_6  - 4 \pi K_1\right)
\\
U_2 &=&-   {u_6^{1/4} \over 2}\int_{\mathbf{p}} {
   V_3(\mathbf{p})\over K(\vp)}.
\end{eqnarray}
Let us now consider the susceptibility with $n=2$. We have 
\begin{eqnarray}
t \chi_2^{1l} &=& -{u^{1l}_{c2}\over t}- {1\over 2 t}\int_{\mathbf{p}}
{u_4+ w_4 V_4(\mathbf{p})\over
K(\vp) +t } + 
{1 \over 2 t}  \int_{\mathbf{p}}
   {[u_3 + w_3 V_3(\mathbf{p})]^2 \over 
   [K(\vp) + t]^2 }
\\
&=& -{u^{1l}_{c2}\over t} - {u_4\over 2 t}
  \left[- {1\over 4 \pi} \log {t\over 32} + K_1 \right] +
  {u_3^2\over 8 \pi t^2} + {u_6^{1/2} \over 2 t}
  \int_{\mathbf{p}} \left[
          {V_3(\mathbf{p})^2\over K(\vp)^2} - 
          {V_4(\mathbf{p})\over K(\vp)} \right] + 
  O(u_6^{1/4} \log t). \nonumber 
\end{eqnarray}
Thus, if we define
\begin{eqnarray}
u^{1l}_{c2} &\equiv& D_1 + D_2 + D_3
\nonumber\\
D_1 &=& {u_4\over 8 \pi } \left(\log u_6  - 4 \pi K_1\right) 
\nonumber\\
D_2 &=&    {u_6^{1/2} \over 2}\int_{\mathbf{p}} 
   {V_3(\mathbf{p})^2\over K(\vp)^2} 
\nonumber\\
D_3 &=& -{u_6^{1/2} \over 2}\int_{\mathbf{p}} 
          {V_4(\mathbf{p})\over K(\vp)} 
\label{uc2-1loop}
\end{eqnarray}
we obtain 
\begin{equation}
t \chi_2^{1l}  = {x_3^2\over 8 \pi} - {x_4\over 8 \pi} \log (32 x_6),
\end{equation}
that is independent of the vertices and of $K(\vp)$.

Now let us consider $\chi_3$. At one loop we have
\begin{eqnarray}
t^2\chi_3^{1l} &=& -{u^{1l}_{c3}\over t} 
- {1\over t} \int_\vp {[u_3 + w_3 V_3(\vp)]^3\over [K(\vp) + t]^3} 
\nonumber \\
&& 
+ {3\over 2t} \int_\vp {[u_3 + w_3 V_3(\vp)] [u_4 + w_4 V_4(\vp)]\over 
      [K(\vp) + t]^2}  
- {1\over 2t} \int_\vp {w_5 V_5(\vp)\over K(\vp) + t} 
\nonumber\\
&=& -{u^{1l}_{c3}\over t} - {x_3^3 - 3 x_3 x_4\over 8\pi} 
  - {u_6^{3/4}\over 2 t} \int_\vp 
  \left[ {2 V_3(\vp)^3 \over K(\vp)^3} - 
         {3 V_3(\vp) V_4(\vp) \over K(\vp)^2} +
         {V_5(\vp) \over K(\vp)} \right] + O(u_6^{3/4} \log t).
\nonumber\\
\end{eqnarray}
Thus, the correct scaling is obtained by setting
\begin{eqnarray}
u^{1l}_{c3} &\equiv & T_1+T_2+T_3
\nonumber\\
T_1 &=& - {u_6}^{3/4}  \int_\vp
           {2 V_3(\vp)^3 \over K(\vp)^3} 
\nonumber\\
T_2 &=& {3 {u_6}^{3/4}\over 2} \int_\vp
         { V_3(\vp) V_4(\vp) \over K(\vp)^2} 
\nonumber\\
T_3 &=& -{{u_6}^{3/4}\over 2} \int_\vp
         {V_5(\vp) \over K(\vp)} 
\label{uc3-1loop}
\end{eqnarray}
Finally, let us consider the four-point function. There are five graphs 
contributing to it. It is easy to see that an anomalous contribution
arises only from the insertion of the 6-leg vertex. However, if we wish that
$\chi_4$ does not depend on any lattice detail we should also subtract
other contributions that arise from the other graphs.
We define therefore
\begin{eqnarray}
u^{1l}_{c4} &\equiv& Q_1+Q_2 +Q_3 + Q_4 +Q_5 +Q_6
\nonumber\\
Q_1 &=& {u_6\over 8 \pi } \left(\log u_6  - 4 \pi K_1\right) 
\nonumber \\
Q_2 &=& 3 u_6 \int_{\mathbf p}{ V_3(\mathbf{p})^4\over K(\vp)^4}
 \nonumber\\
Q_3 &=& -6 u_6\int_{\mathbf p}{V_4(\mathbf{p}) V_3(\mathbf{p})^2\over K(\vp)^3}
\nonumber\\
Q_4 &=& {3 u_6\over 2} \int_{\mathbf p} { V_4(\mathbf{p})^2\over K(\vp)^2}
\nonumber\\
Q_5 &=& 2 u_6 \int_{\mathbf p} {V_5(\mathbf{p}) V_3(\mathbf{p})\over K(\vp)^2}
\nonumber\\
Q_6 &=& -{u_6\over 2}\int_{\mathbf p}   { V_6(\mathbf{p})\over K(\vp)}
\label{uc4-1loop}
\end{eqnarray}

\subsection{Two loop computation}

\begin{figure}
\begin{center}   
%% \vspace{-9cm}
\includegraphics[angle=0,scale=0.75]{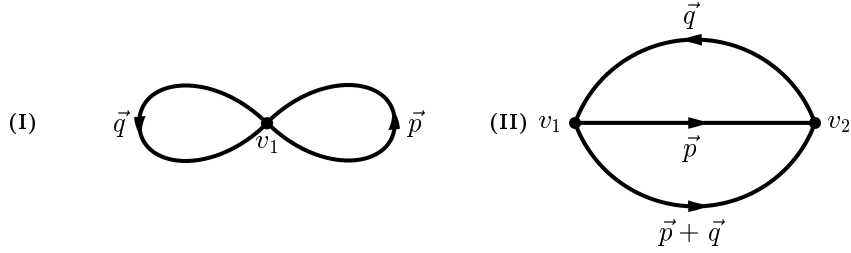}
%% \vspace{-9cm}
\caption{The general two loop topologies for $\chi_n$. We report only the 
vertex ($v_1$ and $v_2$) for which more than two internal legs are 
contracted.}\label{twoloop}
\end{center}
\end{figure}

At two loops there many graphs contributing to $\chi_n$. In general
they can be grouped in two different topologies reported in Fig.~\ref{twoloop}.
Graphs are obtained from these topologies by adding external legs to vertices
$v_1$ and $v_2$ and to the internal lines.
We will then distinguish the contributions in different classes, according
to the nature of the vertex appearing in $v_1$ and $v_2$.
We will thus consider contributions (IA) and (IB): in the first case
we consider at the vertex $v$ the zero-momentum part of the vertex,
while in the opposite case we consider the remaining part.
Analogously we distinguish contributions (IIA), (IIB), (IIC).

\subsubsection{Case (IA)} 

In this case the contribution to $\chi_n$ can be written as 
\begin{eqnarray}
\chi_n &\sim& u_j t^{-n} \int_\mathbf{p} 
   [K(\vp) + t]^{-\sum n_i - 1}\prod_i [u_i + w_i V_i(\vp)]^{n_i} 
\nonumber \\
&& \qquad \int_\mathbf{q}
   [K(\vq) + t]^{-\sum m_i - 1}\prod_i [u_i + w_i V_i(\vq)]^{m_i}
\end{eqnarray}
Expanding we obtain a sum of terms of the form
\begin{eqnarray}
\chi_n &\sim& u_j t^{-n} \prod_i u_i^{k_i + h_i}
    w_i^{n_i + m_i - k_i - h_i}
   \int_\mathbf{p} \int_\mathbf{q}
   [K(\vp) + t]^{-\sum n_i - 1}\prod_i [ V_i(\vp)]^{n_i - k_i} 
\nonumber \\
  && \qquad
   [K(\vq) + t]^{-\sum m_i - 1}\prod_i [ V_i(\vq)]^{m_i - h_i}.
\end{eqnarray}
The integral scales as 
$t^{-A}$ with $A = \sum_i (k_i + h_i)$ apart from logarithmic factors. 
Therefore,
\begin{eqnarray}
t^{n-1} \chi_n \sim x_j \prod_i x_i^{k_i + h_i} w_i^{n_i + m_i - k_i - h_i}.
\label{eqIA}
\end{eqnarray}
Thus, the contributions of the additional vertices can be neglected
and we should only consider the case $n_i = k_i$, $m_i = h_i$. 
In this case, if there are no additional terms proportional to $\log t$,
these contributions scale according to eq.~\reff{scalingeq}. 
Logarithmic terms occur when $\sum n_i = 0$ or $\sum m_i = 0$, i.e.\
when there are tadpoles. Thus, an anomalous contribution occurs for 
the terms of the form
\begin{eqnarray}
 u_j t^{-n}  \prod_i u_i^{m_i} \int_\mathbf{p} {1\over K(\vp) + t}
  \int_\vq {1\over [K(\vq) + t]^{\sum m_i + 1}}
\end{eqnarray}
with $n = j + m_3 + 2 m_4 + 4 m_6 - 4$. However, at the same time
we must consider the contributions of the one-loop counter-terms that give
rise to terms of the form
\begin{eqnarray}
\chi_n \sim u^{1l}_{cj} t^{-n}  
  \int_\vq [K(\vq) + t]^{-\sum m_i - 1}
  \prod_i [u_i + w_i V_i(\vq)]^{m_i}
\end{eqnarray}
The analysis presented above can be repeated for these terms here,
showing that one can neglect the contributions of the vertices $V_i$.
Thus, we have a contribution 
\begin{eqnarray}
\chi_n \sim u^{1l}_{cj} t^{-n} \prod_i  u_i
  \int_\vq [K(\vq) + t]^{-\sum m_i - 1}
\end{eqnarray}
Now,  $u^{1l}_{cj}$ is the sum of several contributions: 
we only consider here the logarithmic terms [$U_1$, $D_1$ and $Q_1$ in
(\ref{uc1-1loop},\ref{uc2-1loop},\ref{uc4-1loop})]. The additional ones 
will be relevant for case (IIC) below. Taking into account the 
combinatorial factors we obtain
\begin{eqnarray}
\chi_n \sim t^{-n}  u_{j} \prod_i  u_i 
  \left[ \int_\vp {1\over K (\vp) + t} + {1\over 4\pi} \log u_6-K_1 \right]
  \int_\vq [K(\vq) + t]^{-\sum m_i - 1}
\end{eqnarray}
The subtracted term replaces the $\log t$ term that arises from the 
integration over $\vp$ with $\log (t/u_6)$. Thus, if no other logarithmic 
terms are generated by the integration over $\vq$ this term scales correctly.
However, when $m_i = 0$, another $\log t$ is generated. This occurs 
for $n = 2$ and $j = 6$. 
Thus taking into consideration this new anomalous contribution to 
$\chi_2$ (and neglecting other contributions):
\begin{eqnarray}
t{\chi_2}^{\mathrm{2l};\mathrm{IA}} &=& -{u_6\over 4t}\Big[
\int_{\mathbf{p}}{1\over K(\mathbf{p})+t}+{1\over 4\pi}\log u_6 - K_1
\Big] \int_{\mathbf{q}}[K(\mathbf{q})+t]^{-1}
\nonumber\\
&=& -{u_6\over 4 t }\LL[{1\over 4\pi}\log32 x_6 
\RR]\LL[{1\over 4 \pi}\log 32 x_6-{1\over 4\pi}\log u_6+K_1\RR]
\end{eqnarray}
In order to cancel the previous anomalous contribution to 
$\chi_2$, a correction to $u_{c2}$ will be introduced in sec.\
\ref{correction_critical}.

\subsubsection{(IB)}
  
The general contribution scales according to
\begin{eqnarray}
\chi_n &\sim& w_j t^{-n} \prod_i u_i^{k_i + h_i}
    w_i^{n_i + m_i - k_i - h_i}
   \int_\mathbf{p} \int_\mathbf{q}
    V^{(j)}(\vp,-\vp,\vq,-\vq,\mathbf{0},\ldots)
\nonumber \\
&&
   [K(\vp) + t]^{-\sum n_i - 1}\prod_i [ V_i(\vp)]^{n_i - k_i} 
   [K(\vq) + t]^{-\sum m_i - 1}\prod_i [ V_i(\vq)]^{m_i - h_i}.
\label{IB-general}
\end{eqnarray}
Keeping into account the fact that 
$ V^{(j)}(\vp,-\vp,\vq,-\vq,\mathbf{0},\ldots) \sim (p^2 + q^2)$ for 
$p,q\to 0$, we find that, if $\sum_i k_i \not= 0$ and $\sum_i h_i \not= 0$,
the integral scales as 
$t^{-A}$ with $A = \sum_i (k_i + h_i) - 1$ apart from logarithmic factors. 
Thus
\begin{equation}
t^{n-1} \chi_n \sim w_j \prod_i x_i^{k_i + h_i}
    w_i^{n_i + m_i - k_i - h_i};
\end{equation}
such a term always vanishes in the critical crossover limit.

Let us now consider the case in which $\sum_i k_i=0$ but $\sum_i h_i\neq 0$
There are several anomalous diagrams that come from this contributions. 
We are going to show now that all these non-scaling terms are eliminated by
one loop diagrams with insertion of counter terms defined in the one loop
computation. We define:
\begin{equation}\label{vertice_due} 
V^{(j)}(\mathbf{p},-\mathbf{p},\mathbf{q},-\mathbf{q},\mathbf{0},\cdots)
\equiv V_j(\mathbf{p}) + V_{j}(\mathbf{q})+\tilde V_j (\mathbf{p},\mathbf{q}) 
\end{equation}
then inserting (\ref{vertice_due}) into (\ref{IB-general}) and following the discussion 
of the $\sum_i h_i\neq 0$ and $\sum_i k_i\neq 0$ case, 
we find that the anomalous
terms only come from  $V_j(\cdot)$ in
(\ref{vertice_due}), while $\tilde V_j$ terms are regular.
 At this point the integrals factorize
and the discussion reduces to one loop computation of the previous section
with the properly counter terms introduced.
Explicitly we have
\begin{equation}
t^{n-1}\chi_n \sim {w_j\over t} \prod_i {x_i}^{h_i}{w_i}^{n_i+m_i-h_i}
\end{equation}
so that we have to take care all the diagram for which:
\begin{equation}
\sum_{i=3}^6 {(i-2)\over 4}(n_i+d_i)+{j-2\over 4}\leq 1 
\end{equation}
having defined $d_i=m_i-h_i$ and with $j=4,5,6$.\\
If $j=6$ we find a scaling diagram with $d_1=0$ and $n_i=0$. This
non-universal contribution is cancelled by $Q_6$ term of 
$u^{1l}_{c4}$ (\ref{uc4-1loop}),
indeed for this terms (including combinatorial factors):
\begin{eqnarray}
t^{n-1}\chi_n &\sim& {w_j\over t}\prod_i {x_i}^{h_i}{w_i}^{n_i+m_i-h_i}
\int_{\mathbf{p}} {V_{6}(\mathbf{p})\over K(\mathbf{p})+t} - {V_6(\mathbf{p})
\over K(\mathbf{p})}
\nonumber\\
&\sim& w_j\prod_i {x_i}^{h_i}{w_i}^{n_i+m_i-h_i}\log t
\end{eqnarray} 
that can be neglected in the CCL limit. In a very similar manner
using $T_3$ (\ref{uc3-1loop}), 
we can delete the terms with $j=5$, $d_i=0$, $n_i=0$ 
and $j=5$, $n_i=0$, $d_i=\delta_{i,3}$; while using $D_3$ 
(\ref{uc2-1loop}) the diagrams
with $j=4$ and $n_i=0$ are cancelled (anomalous one are $d_i=0$, $d_i
=\delta_{i,3}$, $d_i=2\delta_{i,3}$ and $d_i=\delta_{i,4}$). 
The last diagrams with $j=5$ we have to discuss 
is the $n_3=1$, $d_i=0$. 
This diagrams are regularized by
one loop-insertion of $Q_5$  (\ref{uc4-1loop}). Indeed we have:
\begin{eqnarray}
t^{n-1}\chi_n &\sim& {w_j\over t} w_3 \prod_i {x_i}^{h_i}
\int_{\mathbf{p}} {V_3(\mathbf{p})V_5(\mathbf{p})\over
(K(\mathbf{p})+t)^2}-{V_3(\mathbf{p})V_5(\mathbf{p})\over K(\mathbf{p})^2}
\nonumber\\
&\sim & w_j w_3 \prod_i {x_i}^{h_i} \log t
\end{eqnarray}
\\
In a very similar way the remaining anomalous diagrams for  $j=4$ are
regularized by $T_2$ ($n_3=1$ and $d_i=0$, $n_3=1$ and $d_i=\delta_{i,3}$),
$Q_4$ ($n_4=1$ and $d_i=0$) and $Q_3$ ($n_3=2$ and $\delta_i=0$). In the discussion above we do not have used $Q_2$, $T_1$ and 
$D_2$, we expect that these terms enter into topologies of type (II).

In the case in which $\sum_i k_i=\sum_i h_i=0$ other counterterms must
be included in order to regularize the theory. In this case (neglecting
$\log$):
\begin{equation}
t^{n-1}\chi_n \sim {w_j\over t}\prod_i {w_i}^{n_i+m_i}
={{u_6}^{n+2\over 4}\over t},
\end{equation}
so that we have to take care $\chi_1$ and $\chi_2$ in (IB).
From now we take the following short-hand notation
for the propagator
\begin{equation}
\Delta(\mathbf{p}) \equiv  K(\mathbf{p})+t,
\end{equation}
then
\begin{equation}
 {\chi_1}^{\mathrm{2l};\mathrm{IB}} 
= -{u_6^{3/4}\over 8 t } \int_{\mathbf{p},\mathbf{q}}
{V_5(\mathbf{p},\mathbf{q})\over \Delta(\mathbf{p})\Delta(\mathbf{q})}
+{u_6^{3/4}\over 4t}\int_{\mathbf{p},\mathbf{q}}
{V_3(\mathbf{p})V_4(\mathbf{p},\mathbf{q})\over \Delta(\mathbf{p})^2
\Delta(\mathbf{q})}
\end{equation}
Therefore in the CCL (i.e. for $t\to 0$):
\begin{eqnarray}
{\chi_1}^{\mathrm{2l};\mathrm{IB}} &=& {u_6^{3/4}\over  t }
\LL({\cal A}^{(1)}_{\mathrm{2l};\mathrm{IB}}-{{\cal B}^{(1)}_{\mathrm{2l};\mathrm{IB}}
\over 4\pi}
\log{t\over 32} \RR)
\\
{\cal A}^{(1)}_{\mathrm{2l};\mathrm{IB}} 
&=& -{1\over 8  } \int_{\mathbf{p},\mathbf{q}}
{\tilde V_5(\mathbf{p},\mathbf{q})\over K(\mathbf{p})K(\mathbf{q})}
+{1\over 4  }\int_{\mathbf{p},\mathbf{q}}{\tilde V_4(\mathbf{p},
\mathbf{q})V_3(\mathbf{p})\over K(\mathbf{p})^2 K(\mathbf{q})}
-{\alpha_3 \over 16\pi  }\int_{\mathbf{q}}{V_4(\mathbf{q})\over
K(\mathbf{q})}
\nonumber\\
&&-{1\over 4}\int_{\mathbf{p},\mathbf{q}}{V_5(\mathbf{p})\over
K(\mathbf{p})}\LL({1\over K(\mathbf{q})}-{1\over \wi{\mathbf{q}}^2}\RR)
+{1\over 4 }\int_{\mathbf{p},\mathbf{q}}{V_3(\mathbf{p})V_4(\mathbf{p})\over K(\mathbf{p})^2}\LL({1\over K(\mathbf{q})}-{1\over \wi{\mathbf{q}}^2}\RR)
\nonumber\\
 &&+{1\over 4 }\int_{\mathbf{p},\mathbf{q}}{V_4(\mathbf{p})\over K(\mathbf{p})} \LL( {V_3(\mathbf{q})\over K(\mathbf{q})}{1\over K(\mathbf{q})}-{\al_3\over {\wi{\mathbf{q}}}^2} 
\RR)
\\
{\cal B}^{(1)}_{\mathrm{2l};\mathrm{IB}}&=&{1\over 4}\int_{\mathbf{p}}{V_3(\mathbf{p})V_4(\mathbf{p})\over K(\mathbf{p})^2}
 -{1\over 4}\int_{\mathbf{p}}{V_5(\mathbf{p})\over K(\mathbf{p})}
+{\al_3\over 4} \int_{\mathbf{p}}{V_4(\mathbf{p})\over K(\mathbf{p})}
\end{eqnarray}
where $\alpha_i$ was defined in (\ref{normalization-2}).
Now we compute the IB diagrams that enter into the computation
of $\chi_2$:
\begin{eqnarray}
t {\chi_2}^{\mathrm{2l};\mathrm{IB}} &=& -{u_6 \over 8 t}\int_{\mathbf{p},\mathbf{q}}{ 
V_6(\mathbf{p},\mathbf{q})\over \Delta(\mathbf{p})\Delta(\mathbf{q})}
+{u_6\over 2t}\int_{\mathbf{p},\mathbf{q}}{V_5(\mathbf{p},\mathbf{q})
V_3(\mathbf{p})\over \Delta(\mathbf{p})^2\Delta(\mathbf{q})}
-{u_6\over 2t}\int_{\mathbf{p},\mathbf{q}}{V_4(\mathbf{p},\mathbf{q})
V_3(\mathbf{p})^2\over \Delta(\mathbf{p})^3\Delta(\mathbf{q})}
\nonumber\\
&&-{u_6\over 4t}\int_{\mathbf{p},\mathbf{q}}
{V_3(\mathbf{p})V_3(\mathbf{q})V_4(\mathbf{p},\mathbf{q})
\over \Delta(\mathbf{p})^2\Delta(\mathbf{q})^2}
+{u_6\over 4t}\int_{\mathbf{p},\mathbf{q}}{V_4(\mathbf{p})
V_4(\mathbf{p},\mathbf{q})\over\Delta(\mathbf{p})^2\Delta(\mathbf{q}) }
\end{eqnarray}
Therefore in the CCL:
\begin{eqnarray}
t{\chi_2}^{\mathrm{2l};\mathrm{IB}} &=& {u_6\over t}\LL({\cal A}^{(2)}_{\mathrm{2l};\mathrm{IB}}-{{\cal B}^{(2)}_{\mathrm{2l};\mathrm{IB}}
\over 4\pi}
\log{t\over 32}\RR)
\\
{\cal A}^{(2)}_{\mathrm{2l};\mathrm{IB}} &=&\int_{\mathbf{p},\mathbf{q}}\Big[
-{1\over 8}{\tilde V_6(\mathbf{p},\mathbf{q})\over K(\mathbf{p}) K(\vq)}
+{1\over 2}{\tilde V_5(\mathbf{p},\mathbf{q}) V_3(\mathbf{p})
\over K(\mathbf{p})^2 K(\mathbf{q})}
-{1\over 2}{V_3(\mathbf{p})^2\tilde V_4(\mathbf{p},\mathbf{q})
\over K(\mathbf{p})^3K(\mathbf{q})}
+{1\over 4}{V_4(\mathbf{p})\tilde V_4(\mathbf{p},\mathbf{q})\over
K(\mathbf{p})^2K(\mathbf{q})}
\nonumber\\
&&-{1\over 4}{V_3(\mathbf{p})V_3(\mathbf{q})\tilde V_4(\mathbf{p},\mathbf{q})
\over K(\mathbf{p})^2K(\mathbf{q}) }
\Big]-{1\over 4}\int_{\mathbf{p},\mathbf{q}}{V_6(\mathbf{p})\over K(\mathbf{p})}\LL({1\over K(\mathbf{q})}-{1\over \wi{\mathbf{q}}^2}\RR)
+{1\over 2}\int_{\mathbf{p},\mathbf{q}}{V_5(\mathbf{p})\over K(\mathbf{p})}
\nonumber\\
&&\LL({V_3(\mathbf{p})\over K(\mathbf{p})}{1\over K(\mathbf{q})}+{V_3(\mathbf{q})\over K(\mathbf{q})}{1\over K(\mathbf{q})} -{V_3(\mathbf{p})\over K(\mathbf{p})}{1\over \wi{\mathbf{q}}^2}-\al_3 {1\over {\wi{\mathbf{q}}}^2}
 \RR)-{\al_3\over 8\pi}\int_{\mathbf{q}}{V_5(\mathbf{q})\over K(\mathbf{q})}
\nonumber\\
&&-{1\over 2}\int_{\mathbf{p},\mathbf{q}}{V_4(\mathbf{p})\over K(\mathbf{p})}
\Bigg( {V_3(\mathbf{p})^2\over K(\mathbf{p})^2}\Big({1\over K(\mathbf{q})}
-{1\over \wi{\mathbf{q}}^2} \Big)
+{V_3(\mathbf{q})^2\over K(\mathbf{q})^2}{1\over K(\mathbf{q})}-
{{\al_3}^2\over\wi{\mathbf{q}}^2}
\Bigg)+
{3{\al_3}^2\over 16\pi}\int_{\mathbf{q}}{V_4(\mathbf{q})\over K(\mathbf{q})}
\nonumber\\
&&-{1\over 2}\int_{\mathbf{p},\mathbf{q}}{V_3(\mathbf{p})V_4(\vp)\over K(\mathbf{p})^2}
\Bigg({V_3(\mathbf{q})\over K(\mathbf{q}) }-{\al_3\over \wi{\mathbf{q}}^2}
\Bigg)+{\al_3\over 8\pi}\int_{\mathbf{q}}{V_3(\mathbf{q})V_4(\mathbf{q})
\over K(\mathbf{q})^2}-{\al_4\over 16\pi}\int_{\mathbf{q}}{V_4(\mathbf{q})\over K(\mathbf{q})}
\nonumber\\
&&+{1\over 4}\int_{\mathbf{p},\mathbf{q}}
{V_4(\mathbf{p})^2\over K(\mathbf{p})^2}
\Bigg({1\over K(\mathbf{q})}-{1\over\wi{\mathbf{q}}^2}\Bigg)
+{1\over 4}\int_{\mathbf{p},\mathbf{q}}{V_4(\mathbf{p})\over K(\mathbf{p})}
\Bigg({V_4(\mathbf{q})\over K(\mathbf{q})^2}-{\al_4\over\wi{\mathbf{q}}^2}\Bigg)
\\
{\cal B}^{(2)}_{\mathrm{2l};\mathrm{IB}} &=& -{1\over 4}\int_{\mathbf{p}}
{V_6(\mathbf{p})\over K(\mathbf{p})}+{1\over 2}\int_{\mathbf{p}}
{V_5(\mathbf{p})\over K(\mathbf{p})}\Bigg(
\al_3 + {V_3(\mathbf{p})\over K(\mathbf{p})}
\Bigg)-{1\over 2}\int_{\mathbf{p}}
{V_4(\mathbf{p})\over K(\mathbf{p})}\Bigg({\al_3}^2 + {V_3(\mathbf{p})^2\over K(\mathbf{p})^2}
\Bigg)
\nonumber\\
&&-{\al_3 \over 2} \int_{\mathbf{q}}{V_3(\mathbf{q})V_4(\vq)\over
 K(\mathbf{q})^2}
+{1\over 4}
\int_{\mathbf{p}}{V_4(\mathbf{p})\over K(\mathbf{p}) }
\Bigg({V_4(\mathbf{p})\over K(\mathbf{p}) }+\al_4\Bigg)
\end{eqnarray}

\subsubsection{Case (IIA)}

In this case the contributions of vertices $v_1$ and $v_2$ are simply
$u_{j_1} u_{j_2}$. The contribution to $\chi_n$ has the
general form
\begin{eqnarray}
\chi_n &\sim& t^{-n} u_{j_1} u_{j_2} \prod_i [u_i^{k_i + h_i + l_i} 
    w_i^{n_i + m_i + p_i - k_i - h_i - l_i} ]
\nonumber \\
&&   \int_{\mathbf p} \int_{\mathbf q} 
   (t+K(\vp))^{-1 - \sum n_i} 
   (t+K(\vq))^{-1 - \sum m_i} 
   (t+K(\vp + \vq))^{-1 - \sum p_i}  
\nonumber \\
   && \qquad \prod_i [V_i(\vp)]^{n_i - k_i} [V_i(\vq)]^{m_i - h_i} 
             [V_i(\vp + \vq)]^{p_i - l_i},
\end{eqnarray}
with the topological relation $n = j_1 + j_2 - 6 + 
\sum_\ell (\ell - 2) (n_\ell + m_\ell + p_\ell)$.
The analysis of this 
contribution is completely analogous to that presented in 
sec.\ \ref{twoloop-ccl}
for the two-loop topology (b). The integral scales always as 
$t^{-A}$ with $A = 1 + \sum_i (k_i + h_i + l_i)$ in the infrared limit. Thus,
we have
\begin{equation}
t^{n-1} \chi_n \sim 
x_{j_1} x_{j_2} \prod_i [x_i^{k_i + h_i + l_i} 
    w_i^{n_i + m_i + p_i - k_i - h_i - l_i} ]
\end{equation}
In the scaling limit, the only non-vanishing contributions are those with 
$n_i = k_i$, $m_i = h_i$, $p_i = l_i$: the formally irrelevant vertices 
can be disregarded. The remaining terms scale correctly.

\subsubsection{Case (IIB)}

In this case the vertex $v_1$ enters into the diagrams
with the zero momentum part of a vertex
$u_{j_1}$, while $v_2$ gives  
the contribution $V^{(j_2)}(\vp,\vq,-\vp-\vq,\mathbf{0},\ldots)$. Then $\chi_n$ has the following form
\begin{eqnarray}
\chi_n &\sim& t^{-n} u_{j_1} w_{j_2} \prod_i [u_i^{k_i + h_i + l_i} 
    w_i^{n_i + m_i + p_i - k_i - h_i - l_i} ]
   \int_{\mathbf p} \int_{\mathbf q} 
    V^{(j_2)}(\vp,\vq,-\vp-\vq,\mathbf{0},\ldots)
\nonumber \\
&& \qquad
  (t+ K(\vp))^{-1 - \sum n_i} 
  (t+ K(\vq))^{-1 - \sum m_i} 
  (t+ K(\vp + \vq))^{-1 - \sum p_i}  
\nonumber \\
   && \qquad \prod_i  [V_i(\vp)]^{n_i - k_i} [V_i(\vq)]^{m_i - h_i} 
             [V_i(\vp + \vq)]^{p_i - l_i} .
\end{eqnarray}
Here $n = j_1 + j_2 - 6 + \sum_\ell (\ell - 2) (n_\ell + m_\ell + p_\ell)$ 
as before. The analysis follows from the analogous one presented in sec.\ \ref{twoloop-ccl}
 for topology (c). Disregarding logarithmic factors
the integral scales as 
$t^{-A}$ with $A = \sum_i (k_i + h_i + l_i)$ in the infrared limit. Thus,
we have 
\begin{equation}
t^{n-1} \chi_n \sim 
x_{j_1} w_{j_2} \prod_i [x_i^{k_i + h_i + l_i} 
    w_i^{n_i + m_i + p_i - k_i - h_i - l_i} ],
\end{equation}
which shows that these contributions can always be neglected.

\subsubsection{Case (IIC)}
Using a similar notation of case IIB,
we have that $\chi_n$ has the following form
\begin{eqnarray}
\chi_n &\sim& t^{-n} w_{j_1} w_{j_2} \prod_i [u_i^{k_i + h_i + l_i} 
    w_i^{n_i + m_i + p_i - k_i - h_i - l_i} ]
\nonumber \\
&&   \int_{\mathbf p} \int_{\mathbf q} 
    V^{(j_1)}(\vp,\vq,-\vp-\vq,\mathbf{0},\ldots)
    V^{(j_2)}(\vp,\vq,-\vp-\vq,\mathbf{0},\ldots)
\nonumber \\
&&   (t+K(\vp))^{-1 - \sum n_i} 
   (t+K(\vq))^{-1 - \sum m_i} 
   (t+K(\vp + \vq))^{-1 - \sum p_i}
\nonumber\\  
&&\qquad   \prod_i \left\{[V_i(\vp)]^{n_i - k_i} [V_i(\vq)]^{m_i - h_i} 
             [V_i(\vp + \vq)]^{p_i - l_i} \right\}.
\end{eqnarray}
Here $n = j_1 + j_2 - 6 + \sum_\ell (\ell - 2) (n_\ell + m_\ell + p_\ell)$ 
as before. The analysis follows from the analogous one presented in
sec.\ \ref{twoloop-ccl} for topology (d). Assuming without loss of generality
that $\sum k_i \le \sum h_i \le \sum l_i$, we should consider three cases:
(a) $k_i = h_i = l_i = 0$; (b) $k_i = h_i = 0$, $\sum l_i > 0$; 
(c) $\sum h_i > 0$ and $\sum l_i > 0$.

In case (c) the integral behaves as $t^{-A}$, with 
 with $A = \sum_i (k_i + h_i + l_i) - 1$ in the infrared limit. Thus,
\begin{equation}
t^{n-1} \chi_n \sim 
w_{j_1} w_{j_2} \prod_i [x_i^{k_i + h_i + l_i} 
    w_i^{n_i + m_i + p_i - k_i - h_i - l_i} ],
\end{equation}
which can always be neglected.
In case (a) the integral behaves as $\log t$, while in case 
(b) it behaves as $t^{-\sum l_i}$. Therefore,
\begin{equation}
t^{n-1}\chi_n \sim {1\over t} w_{j_1} w_{j_2}  
     \prod_i w_i^{n_i + m_i + p_i - l_i} x_i^{l_i}
   \sim {u^{n/4+1/2}\over t} w^{-\sum l_i},
\end{equation}
so that we have to discuss some cases.
First let us discuss the anomalous diagrams that can be cancelled by
one-loop insertions. If $d_i = p_i-l_i$, they are the following
(with $l_1\ge 1$):
(A) $d_i=\delta_{i,3}$, $n_i=m_i=0$, $j_1=j_2=3$ ($t^{n-1}\chi_n\sim {u_6}^
{-1/4}$); 
(B) $d_i=2 \delta_{i,3}$, $n_i=m_i=0$, $j_1=j_2=3$ ($t^{n-1}\chi_n\sim {u_6}^0
$);
(C) $d_i= \delta_{i,4}$, $n_i=m_i=0$, $j_1=j_2=3$ ($t^{n-1}\chi_n\sim {u_6}
^0$);
(D) $d_i=  \delta_{i,3}$, $n_i=m_i=0$, $j_1=3$, $j_2=4$ ($t^{n-1}\chi_n\sim {u_6}^0$);
(E) $d_i=  \delta_{i,3}$, $n_i=\delta_{i,3}$, $m_i=0$, $j_1=3$, $j_2=3$ ($t^{n-1}\chi_n\sim {u_6}^0$). Diagrams (A), (B) and (C) are regularized by insertion
on respectively one-loop diagrams of $D_2$ counterterm of $u_{c2}$
 (\ref{uc2-1loop}).
Indeed, including combinatorial factors (let us consider A):
\begin{eqnarray}\label{previous}
t^{n-1}\chi_n &\sim& {{u_6}^{3/4}\over t}\prod_i {u_i}^{l_i}
\int_{\mathbf{p},\mathbf{q}}\Big[
{V_3(\mathbf{p}) V^{(3)}(\mathbf{p},\mathbf{q},-\mathbf{p}-\mathbf{q})^2
\over \Delta(\mathbf{p})^{\sum l_i+2}\Delta(\mathbf{p}+\mathbf{q})
\Delta(\mathbf{q})}
-{V_3(\mathbf{p}) V_3(\mathbf{q})^2\over \Delta(\mathbf{p})^{\sum l_i+2}
K(\mathbf{q})^2}
\Big].
\end{eqnarray}
If we write $V^{(3)}(\mathbf{p},\mathbf{q},-\mathbf{p}-\mathbf{q})
=:V_3(\mathbf{p})+V_3(\mathbf{q})+\tilde V_3(\mathbf{p},\mathbf{q})$ 
in the first term in the r.h.s. of the previous expression 
(\ref{previous}), we find that only  $V_3(\mathbf{q})$ terms are relevant
[indeed for the other contributions the integral scale as
$t^{-A}$ with $A=\sum_i l_1 -1$  as in the case (c)] but these are regularized
by the second term on the r.h.s. of (\ref{previous}). The same happens for (B)
and (C). Similar considerations follow for
(D) and (E) using respectively $T_2$ and $T_1$ of $u_{c3}$
 (\ref{uc3-1loop}).

Now we have to compute exactly the previous two loop contributions to
$\chi_1$ and $\chi_2$. We define:
\begin{equation}
V^{(j)}(\mathbf{p},\mathbf{q},-\mathbf{p}-\mathbf{q},\mathbf{0},\cdots)
= {V_j(\mathbf{p})+V_j(\mathbf{q})+V_j(\mathbf{p}+\mathbf{q})\over 2}
+\ol V_j(\mathbf{p},\mathbf{q}).
\end{equation}
We notice in particular that $\ol V_j=\mathbf{0}$ for $\mathbf{p}=\mathbf{0}$, $\mathbf{q}=\mathbf{0}$ or
$\mathbf{p}+\mathbf{q}=\mathbf{0}$. 
One-point function is given by the following
two diagrams:
\begin{eqnarray}
\chi_1 &=& - {{u_6}^{3/4}\over 4t}\int_{\mathbf{p},\mathbf{q}}
{V_3(\mathbf{p})V^{(3)}(\mathbf{p},\mathbf{q},-\mathbf{p}-\mathbf{q})^2
\over \Delta(\mathbf{p})^2\Delta(\mathbf{q})\Delta(\mathbf{p}+\mathbf{q})}
\nonumber\\
&&+{{u_6}^{3/4}\over 6t}\int_{\mathbf{p},\mathbf{q}}
{V^{(4)}(\mathbf{p},\mathbf{q},-\mathbf{p}-\mathbf{q},\mathbf{0})
V^{(3)}(\mathbf{p},\mathbf{q},-\mathbf{p}-\mathbf{q})\over
\Delta(\mathbf{p})\Delta(\mathbf{q})\Delta(\mathbf{p}+\mathbf{q})}
\end{eqnarray}
then in the CCL:
\begin{eqnarray}
{\chi_1}^{\mathrm{2l};\mathrm{IIC}} &=& {{u_6}^{3/4}\over t}\LL(
{\cal A}^{(1)}_{\mathrm{2l};\mathrm{IIC}} 
-{{\cal B}^{(1)}_{\mathrm{2l};\mathrm{IIC}}\over 4\pi}\log{t\over 32}
\RR)
\\
{\cal B}^{(1)}_{\mathrm{2l};\mathrm{IIC}} &=&-{1\over 4}
\int_{\mathbf{p}}\Bigg(2{V_3(\mathbf{p})^3\over K(\mathbf{p})^3}+\al_3 {V_3(\mathbf{p})^2\over K(\mathbf{p})^2}
\Bigg)
+{1\over 2}\int_{\mathbf{p}}{V_4(\mathbf{p}) V_3(\mathbf{p})\over K(\mathbf{p})^2}.
\end{eqnarray}
The explicit expression for ${\cal A}^{(1)}_{\mathrm{2l};\mathrm{IIC}}$ is quite
long and we not report here.

Two points function computation give us:
\begin{eqnarray}
t \chi_2 &=& {u_6\over t}\int_{\mathbf{p},\mathbf{q}}V^{(3)}(\mathbf{p},\mathbf{q},-\mathbf{p}-\mathbf{q})^2
\Bigg[ {1\over 2}{V_3(\mathbf{p})V_3(\mathbf{q})\over \Delta(\mathbf{p})^2
 \Delta(\mathbf{q})^2 \Delta(\mathbf{p}+\mathbf{q}) }
+ {1\over 2}{V_3(\mathbf{p})^2 \over \Delta(\mathbf{p})^3
 \Delta(\mathbf{q})^2 \Delta(\mathbf{p}+\mathbf{q}) }
\nonumber\\
&&-{1\over 4}{V_4(\mathbf{p}) \over \Delta(\mathbf{p})^2
 \Delta(\mathbf{q}) \Delta(\mathbf{p}+\mathbf{q}) }
\Bigg]-{V^{(3)}(\mathbf{p},\mathbf{q},-\mathbf{p}-\mathbf{q})V^{(4)}(\mathbf{p},\mathbf{q},-\mathbf{p}-\mathbf{q},\mathbf{0}) V_3(\mathbf{p})\over\Delta(\mathbf{p})^2\Delta(\mathbf{q})\Delta(\mathbf{p}+\mathbf{q}) }
\nonumber\\
&&+{1\over 6} { V^{(4)}(\mathbf{p},\mathbf{q},-\mathbf{p}-\mathbf{q},
\mathbf{0})^2\over \Delta(\mathbf{q})
\Delta(\mathbf{p}+\mathbf{q})\Delta(\mathbf{p}) }
+{1\over 6} {V^{(3)}(\mathbf{p},\mathbf{q},-\mathbf{p}-\mathbf{q})V^{(5)}(\mathbf{p},\mathbf{q},-\mathbf{p}-\mathbf{q},\mathbf{0},\mathbf{0}) \over\Delta(\mathbf{p})\Delta(\mathbf{q})\Delta(\mathbf{p}+\mathbf{q}) }
\end{eqnarray}
and in the CCL:
\begin{eqnarray}
t{\chi_2}^{\mathrm{2l};\mathrm{IIC}} &=& {u_6\over t}\LL({\cal A}^{(2)}_{\mathrm{2l};\mathrm{IIC}}-{{\cal B}^{(2)}_{\mathrm{2l};\mathrm{IIC}}
\over 4\pi}
\log{t\over 32}\RR)
\\
\\
{\cal B}^{(2)}_{\mathrm{2l};\mathrm{IIC}} &=& {1\over 2}\int_{\mathbf{p}}{V_3(\mathbf{p})^2\over K(\mathbf{p})^2}\Bigg({\al_3}^2+ 2 \al_3{V_3(\mathbf{p})\over K(\mathbf{p})} + 3 {V_3(\mathbf{p})^2\over K(\mathbf{p})^2} \Bigg)
-\int_{\mathbf{p}}{V_3(\mathbf{p})\over K(\mathbf{p})}
\Bigg({5\over 2}{V_4(\vp)V_3(\mathbf{p})\over K(\mathbf{p})^2}
\nonumber\\
&&+\al_3{V_4(\vp)\over K(\vp)}
+{1\over 4}\al_4{V_3(\vp)\over K(\vp)}
\Bigg)
+{1\over 2}\int_{\vp} {V_4(\vp)^2\over K(\vp)^2}+{1\over 2}\int_{\vp} {V_3(\vp) V_5(\vp)\over K(\vp)^2}
\end{eqnarray}
also in this case we have reported explicitly only the $log$ term of 
the previous expression.

\subsection{Correction to $u_{1c}$ and $u_{2c}$ at two loop}
\label{correction_critical}

In order to complete the two loop analysis we have to compute the 
contributions to $\chi_1$ and to $\chi_2$ of one loop diagrams with
the insertion of counterterms $u_{ci}$. Then corrections to $u_{c1}$
and $u_{c2}$ are needed in order to delete non-scaling or 
non-universal terms. In the analysis of IA we have just used the 
logarithmic part of $u_{ci}$ ($U_1$ and $D_1$) that will be neglected
in this section.
Using (\ref{uc1-1loop}) (\ref{uc2-1loop}) (\ref{uc3-1loop}) (\ref{uc4-1loop}) 
we have:
\begin{eqnarray}
\chi_1^{\mathrm{1l};C} &=& -{u_{3c}\over 2t}\int_{\mathbf{p}}
{1\over K(\mathbf{p})+t}
+{u_6}^{1/4}
{D_2+D_3\over 2t}\int_{\mathbf{p}}{V_3(\mathbf{p})\over \LL(K(\mathbf{p})+t\RR)^2}
\end{eqnarray}
and in the MCCL:
\begin{eqnarray}
{\chi_1}^{\mathrm{1l};C} &=& {{u_6}^{3/4}\over t}\LL({\cal A}^{(1)}_{\mathrm{1l};C}-{{\cal B}^{(1)}_{\mathrm{1l};C}
\over 4\pi}
\log{t\over 32}\RR)
\\
{\cal A}^{(1)}_{\mathrm{1l};C} &=& {K_1\over 4}\int_{\mathbf{p}} 2{ V_3(\mathbf{p})^3\over K(\mathbf{p})^3}-3 { V_3(\mathbf{p})V_4(\mathbf{p})\over K(\mathbf{p})^2}+{V_5(\mathbf{p})\over k(\mathbf{p})}
\nonumber\\
&&+{\alpha_3\over 4}\int_{\mathbf{p}}\LL({ V_3(\mathbf{p})^2\over K(\mathbf{p})^2}-{V_4(\mathbf{p})\over K(\mathbf{p})}\RR)
\int_{\mathbf{q}}\LL({V_3(\mathbf{q})\over K(\mathbf{q})^2}-{\alpha_3\over \wi{\mathbf{q}}^2} \RR)
\\
{\cal B}^{(1)}_{\mathrm{1l};C} &=&{1\over 4}\int_{\mathbf{p}} 2{ V_3(\mathbf{p})^3\over K(\mathbf{p})^3}-3 { V_3(\mathbf{p})V_4(\mathbf{p})\over K(\mathbf{p})^2}+{V_5(\mathbf{p})\over k(\mathbf{p})}
+\alpha_3 { V_3(\mathbf{p})^2\over K(\mathbf{p})^2}-\alpha_3 {V_4(\mathbf{p})
\over K(\mathbf{p})}
\end{eqnarray}
For $\chi_2$ we have:
\begin{eqnarray}
{\chi_2}^{\mathrm{1l};C} &=& -{\sum_{i=2}^6 Q_i\over 2t^2}\int_{\mathbf{p}} {1\over \Delta(\mathbf{p})}+{u_{c3}u_6^{1/4}\over t^2}\int_{\mathbf{p}}{V_3(\mathbf{p})\over \Delta(\mathbf{p})^2}-{(D_2+D_3)u_6^{1/2}\over t^2}\int_{\mathbf{p}}{V_3(\mathbf{p})^2 \over \Delta(\mathbf{p})^3}
\nonumber\\
&&+{(D_2+D_3)u_6^{1/2}\over t^2}\int_{\mathbf{p}}{V_4(\mathbf{p}) \over \Delta(\mathbf{p})^2}
\end{eqnarray}
and in the MCCL limit:
\begin{eqnarray}
t {\chi_2}^{\mathrm{1l};C} &=& {{u_6}\over t}\LL({\cal A}^{(2)}_{\mathrm{1l};C}-{{\cal B}^{(2)}_{\mathrm{1l};C}
\over 4\pi}
\log{t\over 32}\RR)
\\
{\cal A}^{(2)}_{\mathrm{1l};C} &=& -{\sum_{i=2}^6 Q_i \over 2 {u_6}}K_1 +{u_{c3}\over {u_6}^{3/4}}\int_{\mathbf{p}}{V_3(\mathbf{p})\over K(\vp)^2}-{\al_3\over\wi{\vp}^2}-
{\al_3 u_{c3}\over 8\pi {u_6}^{3/4}} -{D_2+D_3\over{u_6}^{1/2}}\int_{\vp}{V_3(\vp)^2\over K(\vp)^3}-{{\al_3}^2\over\wi{\vp}^2}
\nonumber\\
&&+{3\over 8\pi}{(D_2+D_3){\al_3}^2\over {u_6}^{1/2}}
+{D_2 + D_3\over 2{u_6}^{1/2}}\int_{\vp}{V_4(\vp)\over K(\vp)^2}-{\al_4\over \wi{\vp}^2}-{\al_4\over 8\pi}{D_2+D_3\over {u_6}^{1/2}}
\\
{\cal B}^{(2)}_{\mathrm{1l};C} &=&-{\sum_{i=2}^6 Q_i \over 2 {u_6}}+{u_{c3}\,\alpha_3\over {u_6}^{3/4}} -{(D_2+D_3)\,{\alpha_3}^2\over{u_6}^{1/2}}+{(D_2+D_3)\,\alpha_4\over 2} 
\end{eqnarray}
where we have used (ref. to $u_{ci}$)
At the end the only $log$ correction comes from IA, while 
\begin{eqnarray}
 u^{2l}_{c2} &=&-u_6\LL({\cal A}^{(2)}_{\mathrm{2l};\mathrm{IB}}+ {\cal A}^{(2)}_{\mathrm{2l};\mathrm{IIC}}+{\cal A}^{(2)}_{\mathrm{1l};\mathrm{C}}\RR) 
-{u_6\over 16\pi}\log 32 x_6 \LL(K_1-{1\over 4\pi}\log u_6\RR)
\\
 u^{2l}_{c1} &=& -{u_6}^{3/4}\LL({\cal A}^{(1)}_{\mathrm{2l};\mathrm{IB}}+ {\cal A}^{(1)}_{\mathrm{2l};\mathrm{IIC}}+{\cal A}^{(1)}_{\mathrm{1l};\mathrm{C}}\RR)
\end{eqnarray}
where we have used the fact that:
\begin{eqnarray}
0&=&{\cal B}^{(2)}_{\mathrm{2l};\mathrm{IB}}+ {\cal B}^{(2)}_{\mathrm{2l};\mathrm{IIC}}+{\cal B}^{(2)}_{\mathrm{1l};\mathrm{C}}
\\
0&=&{\cal B}^{(1)}_{\mathrm{2l};\mathrm{IB}}+ {\cal B}^{(1)}_{\mathrm{2l};\mathrm{IIC}}+{\cal B}^{(1)}_{\mathrm{1l};\mathrm{C}}
\end{eqnarray}
The previous equalities are expected, indeed $\log$ terms come from the factorization of two loops integrals, so that the one-loop result is recovered.

\subsection{Higher powers of the fields}\label{sec5.5}

We want to show that higher power of the fields in the starting interaction
(\ref{starting}) does not affect the MCCL: indeed we want to give evidence
of the fact that including higher power vertices  the scaling equations
(\ref{scalingeq}) and the critical parameters $u_{ci}$ do not change.
In order to do that we have to show that every diagram, in which a
more than six leg vertex appear, is suppressed in the MCCL.
 Let us consider an extra term in the 
Hamiltonian (\ref{starting}):
\begin{eqnarray}
\Delta {\cal H} =\sum_{j>6} {u_6}^{(j-2)/4} \int {\di \vp_1\over (2\pi)^2}
\cdots \int {\di \vp_j\over (2\pi)^2}\delta(\vp_1+\cdots \vp_j)
[v_j+V^{(j)}(\vp_1,\cdots\vp_j)]
\end{eqnarray}
At tree-level we have the following additional contributions 
$\Delta \chi_n$ to $\chi_n$:
\begin{eqnarray}
t^{n-1} \Delta \chi_n \sim {{u_6}^{(j-2)/4}\over t } \sim {u_6}^{(j-6)/4}
\end{eqnarray}
so that the contributions are null in the MCCL.

The generalization to one and two loops diagrams is direct.
Suppose to have a diagram ${\cal G}_n$ with n-external legs and
 \emph{at least} a vertex $V^{(j)}$ with $j>6$.  $\tilde {\cal G}^*_{n-j+6}$
is the diagram obtained replacing $V^{(j)}$ in ${\cal G}_n$ with 
$V^{(6)}$ in such a way that
the topology remains unchanged.\footnote{
This can be done cutting $j-6$ external lines of $V^{(j)}$
in ${\cal G}$. We note that this operation is always possible
up to two loops.} 
In the previous section we have shown that the
general behaviour for a diagram in which a six-legs vertex is present 
$\tilde {\cal G}^*_{n-j+6}$ is:
\begin{eqnarray}
\tilde {\cal G}^*_{n-j+6} \sim t^{1-(n-j+6)}(\tilde A+\tilde B\log u_6)
\end{eqnarray}
where $\tilde A$ and $\tilde B$ are  constants. 
Otherwise the loop-structure of
${\cal G}_n$ is exactly the same as the loop structure of 
$\tilde {\cal G}^*_{n-j+6}$; it follows that
\begin{eqnarray}
t^{n-1}{\cal G}_n \sim  {u_6}^{(j-6)/4} (A + B\log u_6) 
\end{eqnarray}
that shows that graphs with higher order vertex 
are suppressed in the MCCL.

In this section we have obtained similar results 
of sec.\
\ref{chapter2-1} for Multicritical interactions
eq.\ (\ref{starting}). In particular several critical parameters
can be defined (sec.\ \ref{correction_critical}) so that 
taking the multicritical crossover limit (\ref{scalingrel})
the connected n-point correlation functions $\chi_n$
behave in a universal way (\ref{provv-chapter1}). The universality
property is not affected by lattice details
[$V^{(j)}(\vp,\cdots)$ in eq.\ (\ref{starting})] or by
higher order vertex sec.\ \ref{sec5.5}, but can be defined
using only $\phi^{\N+3}$ field theory.

%%%%%%%%%%%%%%%%%%%%%%%%%%%%%%%%%%%%%%%%%%%%%%%%%%%%%%%%%%%%%%%%%
%%%%%%%%%%%%%%%%%%%%%%%%%%%%%%%%%%%%%%%%%%%%%%%%%%%%%%%%%%%%%%%%%
%%%%%%%%%%%%%%%%%%%%%%%%%%%%%%%%%%%%%%%%%%%%%%%%%%%%%%%%%%%%%%%%%
%%%                                                           %%%
%%%                                                           %%%
%%% CHAPTER 3 - CHAPTER 3 - CHAPTER 3 - CHAPTER 3 - CHAPTER 3 %%%
%%%                                                           %%%
%%%                                                           %%%
%%%%%%%%%%%%%%%%%%%%%%%%%%%%%%%%%%%%%%%%%%%%%%%%%%%%%%%%%%%%%%%%%
%%%%%%%%%%%%%%%%%%%%%%%%%%%%%%%%%%%%%%%%%%%%%%%%%%%%%%%%%%%%%%%%%
%%%%%%%%%%%%%%%%%%%%%%%%%%%%%%%%%%%%%%%%%%%%%%%%%%%%%%%%%%%%%%%%%

\chapter{Large-$N$ expansion of $O(N)$ models:
the critical zero mode}\label{chapter3}

In sec.\ \ref{chapter1-sec5} we have shown as the standard $1/N$
expansion fails for certain $O(N)$ models 
at the finite temperature
phase transition. Indeed in the $1/N$ expansion severe 
infra-red divergences appear that cannot be resummed including
corrections to the leading ($N=\infty$) order. 
This means that one 
--in principle-- needs to consider all the $1/N$ orders. 
However the failure
of the expansion is suggested because, using a  symmetry
argument  \cite{BGH-02}, one expects an Ising behaviour 
(if $\N=1$) for every $N$ finite. 
However, we have just pointed out, as 
 in principle there is also the
problem of the $1/N$ expansion: in \cite{SS-00} the study
of  $O(N)$ models in one dimension points out that 
the $N=\infty$ limit
is affected by unphysical phase transitions
 that strictly disappear
when $N$ is taken finite.\footnote{A phase transition (i.e.\ a singularity
in the partition function) can appear only if infinite degrees of
freedom appear in the system (Yang-Lee Theorem
\cite{YL-52} \cite{LY-52}).
This can be realised both with a (physical) thermodynamic limit
$V\to \infty$ or taking the (unphysical) $N\to\infty$ limit.}

This chapter is organised in the following way. In sec.\ (\ref{sec-hnl-5})
we discuss the effective interaction of the models
with a critical mode (i.e.\ $\N=1$). We will find an interaction
similar to what introduced in the past chapter 
\ref{chapter2}, so that
introducing proper scaling fields, we 
will be able to describe the crossover
between the Ising criticality with the Mean Field one.
More interesting,
we will present also some numerical results (for the correction
to the $N=\infty$ critical temperature) 
for a pair of model considered in
literature \cite{BGH-02} \cite{MR-87}.
For the mixed O($N$)-RP$^{N-1}$ model \cite{MR-87}, there
are no numerical evidences of the existence of a phase transition
for $N=3$; however we predict the possibility that
for $N<N_c$ ($N_c\approx 100$, see sec.\ \ref{sec9}), the
critical point disappears. On the other hand, for the model
simulated in \cite{BGH-02}, there are strong evidence 
that for $N=3$ the system undergoes a phase transition.
Our numerical prediction for the non universal critical
constant $p_c$ seems to have the same magnitude order 
of that measured in \cite{BGH-02}, however the accordance
is not very good due to the big difference between 
$p_c(\infty)$ ($\approx 5$, see \cite{CP-02}) and
$p_c(3)$ ($\approx 20$, see \cite{BGH-02}). More stringent 
check will come from other $\N=1$ models 
(see chapter \ref{chapter5}), for which
simulations with several $N$ are available \cite{KSS-98}. 
We stress that this could be an important test 
for the scheme presented in this work, and due to its
universal setting, it could be applied to all the model
with Mean Field crossover.

\section{Effective Hamiltonian for the zero mode}\label{sec-hnl-5}

We have discussed in the first Chapter 
that, for the class of phase transitions
we are interested in, the propagator $P$ has 
a zero mode at the critical point. In order to understand
the role of $1/N$ fluctuations, a careful
treatment of the zero mode is required. For this 
purpose, we are now going to 
integrate out the massive modes, obtaining an effective 
Hamiltonian for the 
critical field $\phi$. More precisely, we define
\begin{equation}
e^{-{\cal H}_{\rm eff}[\phi]} = 
   \int \prod_{xa} d\varphi_{xa} e^{-{\cal H}[\Phi]}.
\end{equation}
The effective Hamiltonian ${\cal H}_{\rm eff}[\phi]$ 
has the following expansion 
\begin{eqnarray}
{\cal H}_{\rm eff}[\phi] &=& {1\over \sqrt{N}} \tilde{H} \phi(\mathbf{0}) + 
        {1\over2} \int_{\mathbf{p}} 
   \phi(-\mathbf{p}) \tilde{P}^{-1}(\mathbf{p}) \phi(\mathbf{p})
\label{effexp2} \\
  &+& \sum_{n=3} {1\over n!} {1\over N^{n/2-1} }
  \int_{\mathbf{p}_1} \cdots \int_{\mathbf{p}_n} 
\delta(\sum_i \mathbf{p}_i) \; 
   \tilde{V}^{(n)} (\mathbf{p}_1,\ldots,\mathbf{p}_n)
   \phi(\mathbf{p}_1)\cdots \phi(\mathbf{p}_n),
\nonumber 
\end{eqnarray}
where vertices and propagator also depend on $N$ and have an expansion 
of the form
\begin{eqnarray}
  && \tilde{H} = \sum_{m=0} {1\over N^m} \tilde{H}_m, \nonumber \\
  && \tilde{P}^{-1}(\mathbf{p}) = 
       \sum_{m=0} {1\over N^m} \tilde{P}_m^{-1}(\mathbf{p}), \nonumber \\
   && \tilde{V}^{(n)} (\mathbf{p}_1,\ldots,\mathbf{p}_n) = 
     \sum_{m=0} {1\over N^m} 
    \tilde{V}^{(n)}_m (\mathbf{p}_1,\ldots,\mathbf{p}_n).
\end{eqnarray}
We report here the explicit expressions that we shall need in the following:
\begin{eqnarray}
&& \tilde{H}_0 = {1\over2} \int_{\mathbf{p}} \sum_{ab}
         \hat{V}_{1ab}(\mathbf{0},\mathbf{p},-\mathbf{p}) 
                       \hat{P}_{ab}(\mathbf{p}), 
\\
&& \tilde{P}^{-1}_0 (\mathbf{p})= \hat{P}_{11}^{-1} (\mathbf{p}),
\\
&& \tilde{P}^{-1}_1(\mathbf{p}) = {1\over2} \int_{\mathbf{q}}
   \left[ \sum_{ab} 
    \hat{V}_{11ab}^{(4)}(\mathbf{p},-\mathbf{p},\mathbf{q},-\mathbf{q}) 
        \hat{P}_{ab}(\mathbf{q}) -
   \right. 
\nonumber \\
&& \qquad\qquad\qquad  
       \sum_{abcd} \hat{V}_{11a}^{(3)}(\mathbf{p},-\mathbf{p},\mathbf{0})
       \hat{P}_{ab}(\mathbf{0})
       \hat{V}_{bcd}^{(3)}(\mathbf{0},\mathbf{q},-\mathbf{q}) \
               \hat{P}_{cd}(\mathbf{q}) -
\nonumber \\
&& \qquad\qquad\qquad  \left.
       \sum_{abcd}
       \hat{V}_{1ab}^{(3)}(\mathbf{p},\mathbf{q},-\mathbf{p}-\mathbf{q})
       \hat{V}_{1cd}^{(3)}(\mathbf{p},\mathbf{q},-\mathbf{p}-\mathbf{q})
       \hat{P}_{ac}(\mathbf{q}) \hat{P}_{bd}(\mathbf{p}+\mathbf{q})\right],
\\
&& \tilde{V}^{(3)}_0(\mathbf{p},\mathbf{q},\mathbf{r}) = 
 \hat{V}^{(3)}_{111}(\mathbf{p},\mathbf{q},\mathbf{r}) ,
\\
&& \tilde{V}^{(4)}_0(\mathbf{p},\mathbf{q},\mathbf{r},\mathbf{s}) = 
 \hat{V}^{(4)}_{1111}(\mathbf{p},\mathbf{q},\mathbf{r},\mathbf{s}) -
\nonumber \\
&& \qquad
   \sum_{ab} \hat{V}_{11a}^{(3)}(\mathbf{p},\mathbf{q},-\mathbf{p}-\mathbf{q})
          \hat{V}_{11b}^{(3)}(\mathbf{r},\mathbf{s},-\mathbf{r}-\mathbf{s})
          \hat{P}_{ab}(\mathbf{p}+\mathbf{q})
  + \hbox{\rm two permutations},
\\
&& \tilde{V}^{(5)}_0(\mathbf{p},\mathbf{q},\mathbf{r},\mathbf{s},\mathbf{t}) = 
 \hat{V}^{(5)}_{1111}(\mathbf{p},\mathbf{q},\mathbf{r},\mathbf{s},\mathbf{t}) -
\nonumber \\
&& \qquad \left[
   \sum_{ab} \hat{V}^{(3)}_{11a}(\mathbf{p},\mathbf{q},-\mathbf{p}-\mathbf{q})
  \hat{V}^{(4)}_{111b}(\mathbf{r},\mathbf{s},\mathbf{t},
         -\mathbf{r}-\mathbf{s}-\mathbf{t})
          \hat{P}_{ab}(\mathbf{p}+\mathbf{q}) + 
      \hbox{\rm 9 permutations}\right] + 
\nonumber \\
&& \qquad \left[
   \sum_{abcd} \hat{V}^{(3)}_{11a}(\mathbf{p},\mathbf{q},-\mathbf{p}-\mathbf{q})
  \hat{V}^{(3)}_{1bc}(\mathbf{r},\mathbf{p}+\mathbf{q},\mathbf{s}+\mathbf{t})
  \hat{V}^{(3)}_{11d}(\mathbf{s},\mathbf{t},-\mathbf{s}-\mathbf{t})
  \times \right. 
\nonumber \\
&&  \qquad\qquad \left. \times 
          \hat{P}_{ab}(\mathbf{p}+\mathbf{q}) 
          \hat{P}_{cd}(\mathbf{s}+\mathbf{t}) 
       + 
      \hbox{\rm 14 permutations}\right],
\end{eqnarray}
where $a$, $b$, $c$, $d$ run from 1 to 4 over the massive modes. 

The vertices introduced above are not independent,
but near the critical point can be related to the scaling fields
introduced to parametrize the gap equation eq.\ (\ref{puntosella2}).
The identities presented in sec.~\ref{AppA} allow us to derive several 
relations among the effective vertices at the critical point. We have 
for $p = p_c$ and $m_0^2 = m_{0c}^2$
\begin{eqnarray}
&& \tilde{P}^{-1}_0 (\mathbf{0}) = 
  {\partial \tilde{P}^{-1}_0 (\mathbf{0})\over \partial m_0^2} = 0,
\label{id-P-CP}
\\
&& \tilde{V}_0^{(3)} (\mathbf{0},\mathbf{0},\mathbf{0}) = 0, 
\label{id-V3-CP}
\\
&& \left[{\partial\over \partial m_0^2}\tilde{V}_0^{(3)} 
      (\mathbf{0},\mathbf{0},\mathbf{0})\right]^2 = 
  {\partial^2 \tilde{P}^{-1}_0 (\mathbf{0})\over \partial (m_0^2)^2} 
 \tilde{V}_0^{(4)} (\mathbf{0},\mathbf{0},\mathbf{0},\mathbf{0}) .
\label{id-V4-CP}
\end{eqnarray}
Relations (\ref{id-P-CP}), (\ref{id-V3-CP}) and (\ref{id-V4-CP})
are the first step to show the effective interaction for the zero
mode is a weakly coupled interaction studied in the previous
chapter. On the other hand
(\ref{id-P-CP}) clearly confirm that the standard $1/N$ expansion 
fails close to the critical point. Indeed, outside the critical point, 
$\tilde{P}^{-1}_0 (\mathbf{p})$ is nonsingular and for large $N$ it is 
enough to 
expand the interaction Hamiltonian in powers of $1/N$. On the other hand, this
not possible at the critical point. Since also the three-leg vertex vanishes
at zero momentum in this case, cf.~eq.~\reff{id-V3-CP}, the zero-momentum
leading term is the quartic one. 
Since the coupling constant is proportional 
to $1/N$, the model effectively corresponds to a weakly
coupled $\phi^4$ theory. 
In order to have a stable $\phi^4$ theory, we must also have 
$\tilde{V}_0^{(4)} (\mathbf{0},\mathbf{0},\mathbf{0},\mathbf{0}) > 0$.
For a generic solution of the gap equations satisfying 
eq.~\reff{eq-punto-critico}, this is not {\em a priori} guaranteed.
Note, however, that if 
$\tilde{V}_0^{(4)} (\mathbf{0},\mathbf{0},\mathbf{0},\mathbf{0}) < 0$,
then, for $p=p_c$, we have 
$\tilde{P}^{-1}_0(\mathbf{0}) \approx a (m_0^2 - m_{0c}^2)^2$,
with $a < 0$, as a consequence of eq.~\reff{id-V4-CP}: the propagator
has a {\em negative} mass for $N=\infty$. We believe---but we have 
not been able to prove---that such a phenomenon signals the fact that
the solution we are considering is not the relevant one. We expect 
the existence of another solution of the gap equation 
\reff{puntosella2} with a lower free energy.

In the previous chapter we have seen that the weakly
coupled $\phi^4$ theory shows an interesting crossover limit. If 
one neglects fluctuations it corresponds to tune $p$ and $m_0^2$ so that 
$\tilde{H}$ and $\tilde{P}^{-1}(\mathbf{0})$ go to zero as $N\to \infty$,
in such a way that $\tilde{H} N$ and $\tilde{P}^{-1}(\mathbf{0}) N$
remain constant. In this limit, Ising behavior is observed when 
the two scaling variables go to zero, while mean-field behavior is 
observed in the opposite case. 
Fluctuations change the simple scaling 
forms reported above and one must consider two additive renormalization constants
($ h_c(u)$ and $ r_c(u)$,
 in chap.\ \ref{chapter2}).

However in order to strictly apply the results of the previous
chapter (including also the computation of the renormalization constants with the lattice regularization) we need to impose the condition $V^{(3)}(\mathbf{0},\mathbf{0},
\mathbf{0})=0$ for every values of the parameters, 
while in the case under investigation this happens 
only near the critical 
point (\ref{id-V3-CP}).\footnote{However the fact that 
$V^{(3)}(\mathbf{0},\mathbf{0},\mathbf{0})$ goes to zero
at the critical point is a fundamental topic in order
to apply the following considerations. In particular
this implies that $k$, defined in (\ref{eq_3.13}), can
be computed perturbatively (\ref{k-enpassant}) 
near the critical point.}
In order to do that we change 
again the definition of the field so that the 
effective zero-momentum 
three-leg vertex vanishes for all $p$ and $m_0^2$ in the 
limit $N\to \infty$.
For this purpose we now define a new field
\begin{equation}
\alpha \chi(\mathbf{p}) = \phi(\mathbf{p}) + k \delta(\mathbf{p}),
\label{eq_3.13}
\end{equation}
where $\alpha$ and $k$ are functions of $p$ and $m_0^2$ to be fixed. 
The function $k$ is fixed by requiring that the 
large-$N$ zero-momentum three-leg vertex vanishes. 
Apparently, all $\tilde{V}^{(n)}$ contribute in this calculation. 
However, because of eq.~\reff{id-V3-CP}, 
$\tilde{V}^{(3)}_0 (\mathbf{0},\mathbf{0},\mathbf{0}) $
vanishes at the critical point. As we already mentioned the 
interesting limit corresponds to considering 
$\Delta_m\equiv m_0^2 - m_{0c}^2\to 0$ and 
$\Delta_p\equiv p - p_c  \to 0$ together with $N\to \infty$. We will 
show in sec.~\ref{sec7} that this limit should be 
taken keeping fixed $\Delta_m N^{1/3}$ and $\Delta_p N$, so that 
$\tilde{V}^{(3)}_0 (\mathbf{0},\mathbf{0},\mathbf{0}) $ is effectively of 
order $N^{-1/3}$. Therefore, the equation defining $k$ 
can be written in a compact form as 
\begin{equation}
{a_3 \over N^{5/6}} + 
  \sum_{n\ge 4} {a_n k^{n-3}\over N^{n/2 - 1}} = 0,
\label{eqaa}
\end{equation}
where $a_n \approx a_{n0} + a_{n1}/N$ is the contribution of the 
$n$-point vertex at zero momentum. Eq.~(\ref{eqaa}) can be rewritten as 
\begin{equation}
\sum_{n=1}^\infty {a_{n+3}\over N^{(n-1)/3}}
        \left( {k\over N^{1/6}}\right)^n = - a_3 
\end{equation}
which shows that $k$ has an expansion of the form
\begin{equation} 
k = k_0 N^{1/6} [1 + k_1 N^{-1/3} + O(N^{-2/3})].
\end{equation}
The leading constant $k_0$ depends only on the three- and four-leg 
vertex, the constant $k_1$ depends also on the five-leg vertex, and 
so on. The explicit calculation gives
\begin{eqnarray}
k &=& \sqrt{N} {\tilde{V}^{(3)}_0 (\mathbf{0},\mathbf{0},\mathbf{0}) \over 
       \tilde{V}^{(4)}_0 (\mathbf{0},\mathbf{0},\mathbf{0},\mathbf{0}) } + 
\nonumber \\
&& +    {\sqrt{N}\over2} 
  {\tilde{V}^{(3)}_0 (\mathbf{0},\mathbf{0},\mathbf{0})^2 
\tilde{V}^{(5)}_0 (\mathbf{0},\mathbf{0},\mathbf{0},\mathbf{0},\mathbf{0})
     \over
  [\tilde{V}^{(4)}_0 (\mathbf{0},\mathbf{0},\mathbf{0},\mathbf{0})]^3 } + 
    O(N^{-1/2}).
\label{k-enpassant}
\end{eqnarray}
By performing this rescaling, the
$k$-leg $\phi$-vertex that scales with $N$ as $N^{1 - k/2}$ 
(except for $k\le 3$) 
gives a contribution to the $m$-leg $\chi$ vertex (of course $m\le k$) of order 
$N^{1 - m/2 + (m-k)/3}$, which is therefore always subleading. 
Taking this result into account we obtain 
\begin{eqnarray}
{\cal H}_{\rm eff} &=& H \chi(\mathbf{0}) + 
     {1\over 2} \int_{\mathbf{p}} \chi(\mathbf{p}) \chi(-\mathbf{p}) 
        \bar{P}^{-1} (\mathbf{p}) + 
\nonumber \\
&& +  {1\over 3! \sqrt{N}} \int_{\mathbf{p},\mathbf{q}}
      \bar{V}^{(3)}(\mathbf{p},\mathbf{q},-\mathbf{p}-\mathbf{q})
      \chi(\mathbf{p}) \chi(\mathbf{q}) \chi(-\mathbf{p}-\mathbf{q})
\nonumber \\
  &&  + {1\over 4! N} \int_{\mathbf{p},\mathbf{q},\mathbf{r}}
      \bar{V}^{(4)}(\mathbf{p},\mathbf{q},\mathbf{r},
           -\mathbf{p}-\mathbf{q}-\mathbf{r})
      \chi(\mathbf{p}) \chi(\mathbf{q}) \chi(\mathbf{r})
      \chi(-\mathbf{p}-\mathbf{q}-\mathbf{r}) + \ldots
\label{caleffchi}
\end{eqnarray}
where 
\begin{eqnarray} 
&& H={\alpha \tilde{H}_0\over \sqrt{N}} - 
    \alpha k \left[\tilde{P}_{0}^{-1}(\mathbf{0}) + 
                  {1\over N} \tilde{P}_{1}^{-1}(\mathbf{0})\right]+
    {\alpha k^2\over 2\sqrt{N}} \tilde{V}^{(3)}_{0}
         (\mathbf{0},\mathbf{0},\mathbf{0}) - 
    {\alpha k^3\over 6N} \tilde{V}^{(4)}_{0}
         (\mathbf{0},\mathbf{0},\mathbf{0},\mathbf{0})  
\nonumber \\
&& \qquad\qquad 
    + {\alpha k^4\over 24{N}^{3/2}} \tilde{V}^{(5)}_{0}
         (\mathbf{0},\mathbf{0},\mathbf{0},\mathbf{0},\mathbf{0}) 
+ O(N^{-7/6}),
\nonumber \\
&& \bar{P}^{-1} (\mathbf{p}) = 
   \alpha^2 \tilde{P}_{0}^{-1}(\mathbf{p}) + 
   {\alpha^2\over N} \tilde{P}_{1}^{-1}(\mathbf{p}) -
   {\alpha^2 k\over \sqrt{N}} \tilde{V}^{(3)}_0
         (\mathbf{0},\mathbf{p},-\mathbf{p}) + 
   {\alpha^2 k^2\over 2 N} 
     \tilde{V}^{(4)}_{0}(\mathbf{0},\mathbf{0},\mathbf{p},-\mathbf{p}) 
\nonumber \\ 
 && \qquad\qquad
   - {\alpha^2 k^3\over 6 N^{3/2}}  
     \tilde{V}^{(5)}_{0}(\mathbf{0},\mathbf{0},\mathbf{0},
               \mathbf{p},-\mathbf{p}) + O(N^{-4/3}), 
\nonumber \\ [3mm]
&& \bar{V}^{(3)}(\mathbf{p},\mathbf{q},\mathbf{r}) = 
   \alpha^3 \tilde{V}^{(3)}_{0}(\mathbf{p},\mathbf{q},\mathbf{r}) - 
   {\alpha^3 k\over \sqrt{N}} 
     \tilde{V}^{(4)}_{0}(\mathbf{0},\mathbf{p},\mathbf{q},\mathbf{r}) + 
     O(N^{-2/3}), 
\nonumber \\ [3mm]
&& \bar{V}^{(4)}(\mathbf{p},\mathbf{q},\mathbf{r},\mathbf{s}) =
   \alpha^4 \tilde{V}^{(4)}_0(\mathbf{p},\mathbf{q},\mathbf{r},\mathbf{s})
     + O(N^{-1/3}), 
\end{eqnarray} 
where $\alpha = \alpha_0 + {\alpha_1 N^{-2/3}} + O(N^{-4/3})$ 
is fixed by requiring that 
\begin{equation}
\bar{P}^{-1}(\mathbf{p}) - \bar{P}^{-1}(\mathbf{0}) \equiv 
K(\mathbf{p}) \approx  \mathbf{p}^2,
\end{equation}
for $\mathbf{p} \to 0$.

\subsection{Identities among effective vertices} \label{AppA}

In this section we wish to derive a general set of identities 
for the zero-momentum vertices in order to justify the relations introduced
above (\ref{id-P-CP},\ref{id-V3-CP},\ref{id-V4-CP}).
 For this purpose we first rewrite
the propagator and the vertices by considering the constants 
$\alpha$, $\gamma$, and $\tau$ appearing in eq.~\reff{field-exp} 
as independent variables {\em without} assuming the saddle-point relations. 
We define the integrals 
\begin{equation}
{\cal A}_{ij,n}(\alpha,\gamma) = 
  \int_\mathbf{q} 
    {\cos^i q_x \cos^j q_y\over (\gamma - \alpha \sum_\mu \cos q_\mu)^n}.
\end{equation}
It is easy to verify that, if $\alpha$ and $\gamma$ are replaced by 
their saddle-point values, $\alpha = 2 W'(\bar{\tau})$ and 
$\gamma = (4 + m_0^2) W'(\bar{\tau})$, then 
\begin{equation}
{\cal A}_{ij,n}(\alpha,\gamma) = 
  {1\over [W'(\overline{\tau})]^n} 
   \int_\mathbf{q}
 {\cos^i q_x \cos^j q_y\over (\hat{q}^2 + m_0^2)^n}.
\end{equation}
In terms of $\alpha$, $\gamma$, and $\tau$ the propagator is simply obtained
by using eq.~\reff{propagator}
and replacing $A_{i,j}(0,m_0^2)/[W'(\overline{\tau})]^2$ with 
${\cal A}_{ij,2}(\alpha,\gamma)$. Let us now consider a generic 
$n$-leg vertex $V^{(n)}_{A_1,\ldots,A_n}$ at zero momentum. It is easy to 
verify that the only nonvanishing terms with $A_i = 4$ or $A_i = 5$ 
for some $i$ are (in this section we do not explicitly write the momentum 
dependence since in all cases we are referring to zero-momentum 
quantities)
\begin{equation}
V^{(n)}_{4,\ldots 4} = V^{(n)}_{5,\ldots 5} = - \beta W^{(n)}(\tau)\; .
\end{equation}
If all $A_i \le 3$, eq.~\reff{defVn} gives
\begin{equation}
V^{(n)}_{A_1,\ldots,A_n} = {1\over2} (-1)^{i_2+i_3+n+1} (n-1)! 
    {\cal A}_{i_2 i_3,n}(\alpha,\gamma),
\end{equation}
where $i_2$ (resp. $i_3$) is the number of indices equal to 2 (resp. to 3).
These expressions allow us to obtain simple recursion relations for the 
vertices. 
\begin{eqnarray}
&& V^{(n+1)}_{1,A_1,\ldots,A_n} = 
   {\partial\over \partial\gamma} V^{(n)}_{A_1,\ldots,A_n},
\nonumber \\
&& V^{(n+1)}_{2,A_1,\ldots,A_n}  + V^{(n+1)}_{3,A_1,\ldots,A_n} = 
   {\partial\over \partial\alpha} V^{(n)}_{A_1,\ldots,A_n}, 
\nonumber \\  
&& V^{(n+1)}_{4,4,\ldots,4}  = 
   {\partial\over \partial\tau} V^{(n)}_{4,\ldots,4},
\nonumber \\  
&& V^{(n+1)}_{5,5,\ldots,5}  = 
   {\partial\over \partial\tau} V^{(n)}_{5,\ldots,5},
\end{eqnarray}
where $\alpha$, $\gamma$, $\tau$, and $\beta$ are considered as independent 
variables.
These relations also apply to the case $n = 2$, once we identify 
$V^{(2)}_{AB} = P^{-1}_{AB}$. 

Let us consider the projector on the zero mode $v_A$, cf.~eq.~\reff{depPhi}. 
Keeping into account that the symmetry of the matrix at zero momentum 
implies $v_2 = v_3$ and $v_4 = v_5$, we obtain 
\begin{eqnarray}
\sum_B v_B V^{(n+1)}_{B,A_1,\ldots,A_n} = 
   \left( v_1 {\partial\over \partial\gamma} + 
          v_2 {\partial\over \partial\alpha} + 
          v_4 {\partial\over \partial\tau} \right)
    V^{(n)}_{A_1,\ldots,A_n}.
\end{eqnarray}
Close to the critical point we can write, cf. eq.~\reff{defz},
\begin{equation}
  v_A = \hat{z}_A + O(s_1)
\end{equation}
with
\begin{eqnarray}
   \hat{z}_A = K \left( 2 {{\cal A}_{01,2}\over {\cal A}_{00,2}}, 1, 1,
   {1\over 2 W''(\tau)}, {1\over 2 W''(\tau)} \right),
\end{eqnarray}
where $K$ is such to have $\sum_B \hat{z}_B^2 = 1$. 
Now, let us note that at the saddle point we have
(we now think of $\alpha$, $\tau$, and 
$\gamma$ as functions of $m_0^2$ and $p$)
\begin{eqnarray}
&& {\partial \gamma\over \partial m_0^2} = 
   W'' {\partial\tau\over \partial m_0^2} 
     \left(4 + m_0^2 - {B_1\over B_2}\right) + O(s_1) = 
   {2 W''\over K} {\partial \tau\over \partial m_0^2} \hat{z}_1 + O(s_1),
\nonumber \\
&& {\partial \alpha\over \partial m_0^2} = 
   2 W'' {\partial \tau\over \partial m_0^2} = 
   {2 W''\over K}  {\partial \tau\over \partial m_0^2} \hat{z}_2 + O(s_1),
\label{what-is-zero}
\end{eqnarray}
where we have used eqs.~\reff{dbdm} and \reff{relAijBn}.
Thus, if we define 
\begin{equation}
   C \equiv \left( {2 W''\over K}  {\partial \tau\over \partial m_0^2}
            \right)^{-1},
\end{equation}
we can rewrite
\begin{eqnarray}
\sum_B v_B V^{(n+1)}_{B,A_1,\ldots,A_n} &=& C
   \left( {\partial \gamma\over \partial m_0^2}
          {\partial\over \partial\gamma} + 
          {\partial \alpha\over \partial m_0^2}{\partial\over \partial\alpha} + 
         {\partial \tau\over \partial m_0^2}{\partial\over \partial\tau} 
         \right) 
    V^{(n)}_{A_1,\ldots,A_n}
    + O(s_1) 
\nonumber \\ 
   &= &
    C {\partial\over \partial m_0^2} 
    V^{(n)}_{A_1,\ldots,A_n} + O(s_1),
\label{identity1}
\end{eqnarray}
where in the last term $\alpha$, $\gamma$, $\tau$, and $\beta$ take their
saddle-point values in terms of $m_0^2$ and $p$.

To go further we compute the derivative of $\hat{z}_A$ with respect to $m_0^2$. 
Since 
\begin{equation}
   \sum_B P^{-1}_{AB} \hat{z}_B = O(s_1),
\end{equation}
we have 
\begin{equation}
\sum_B P^{-1}_{AB} {\partial \hat{z}_B\over \partial m_0^2} = 
    - \sum_B {\partial P^{-1}_{AB} \over \partial m_0^2} \hat{z}_B + O(s_2),
\end{equation}
where $s_2 = \partial^2 \beta / \partial (m_0^2)^2$ also vanishes at 
the critical point. 

Since $\sum_A \hat{z}_A^2 = 1$, $\partial \hat{z}_B/ \partial m_0^2$ 
belongs to the subspace orthogonal to $\hat{z}_A$. 
Thus, if $P^\bot_{AB}$ is the inverse of $P^{-1}$ in the massive subspace,
we obtain
\begin{equation}
 {\partial \hat{z}_A\over \partial m_0^2} 
  = - \sum_{BC} P^\bot_{AB} {\partial P^{-1}_{BC} \over \partial m_0^2} 
   \hat{z}_C + O(s_2) = 
  - {1\over C} \sum_{BCD} P^\bot_{AB}  V^{(3)}_{BCD} \hat{z}_C \hat{z}_D + 
    O(s_2).
\label{dhatz}
\end{equation}
Finally, let us compute 
\begin{eqnarray}
\hat{V}_{1,\ldots 1}^{(n+1)} &=& C 
   \sum_{A_1,\ldots,A_n} z_{A_1} \ldots z_{A_n} 
     {\partial \over \partial m_0^2} 
       V^{(n)}_{A_1,\ldots,A_n}  + O(s_1)
\nonumber  \\
   & = & C {\partial \over \partial m_0^2}  \hat{V}_{1,\ldots 1}^{(n)} - 
      n C \sum_{A_1,\ldots,A_n} z_{A_1} \ldots z_{A_{n-1}} 
          V^{(n)}_{A_1,\ldots,A_n} {\partial z_{A_n}\over \partial m_0^2} 
       + O(s_1)
\nonumber  \\
   & = & C {\partial \over \partial m_0^2}  \hat{V}_{1,\ldots 1}^{(n)}
     + n \sum_{ab} \hat{V}_{11a}^{(3)} \hat{P}_{ab} 
          \hat{V}_{b1\ldots 1}^{(n)} + O(s_2),
\end{eqnarray}
where the indices $a$ and $b$ run from 1 to 4 over the massive subspace.
Since $s_2$ is of order $p-p_c$ and $m_0^2 - m_{0c}^2$ this relation 
implies an identity only at the critical point. 
However, for $n=2$ we can obtain another relation.
Since $V^{(2)}_{AB} = P^{-1}_{AB}$ and $\sum_B P^{-1}_{AB} z_B = O(s_1)$
we do not need eq.~\reff{dhatz} and obtain directly
\begin{eqnarray}
\hat{V}_{111} = C {\partial \over \partial m_0^2} P^{-1}_{11} + O(s_1).
\label{identity3}
\end{eqnarray}
The presence of corrections of order $s_1$ gives rise 
to two critical-point identities. Since $P^{-1}_{11}\sim s_1$ we obtain
\begin{eqnarray}
 \hat{V}_{111} & = & 0, \nonumber \\
 {\partial \over \partial m_0^2} \hat{V}_{111} & = & 
    C {\partial^2 \over \partial (m_0^2)^2} P^{-1}_{11} ,
\end{eqnarray}
where all quantities are computed at the critical point.

%%%%%%%%%%%%%%%%%%%%%%%%%%%%%%%%%%%%%%%%%%%%%%%%%
%
\section{Critical crossover limit in the large-$N$ case} \label{sec6.6}
%
%%%%%%%%%%%%%%%%%%%%%%%%%%%%%%%%%%%%%%%%%%%%%%%%%

The main result of the previous section \ref{sec-hnl-5} is the
identification of the zero mode Hamiltonian with the weakly
coupled theory introduced in the previous chapter \reff{Heff-ccl}.
One can work in order to apply the same results
of chapter \ref{chapter2} by using the following
identifications 
\begin{eqnarray}
\mathrm{chapter\,\, \ref{chapter2}}&&
\mathrm{chapter\,\, \ref{chapter3}}
\nonumber\\
u &=& {1\over N}{\ol V^{(4)}(\mathbf{0},\mathbf{0},\mathbf{0},\mathbf{0})}
\label{u-HNL}
\\
V^{(n)}(\vp_1,\cdots \vp_n) &=&
{\ol V^{(n)}(\vp_1\cdots \vp_n)  \over  
\ol V^{(4)}(\mathbf{0},\mathbf{0},\mathbf{0},\mathbf{0})^{n-2\over 2} 
}
\label{Vn-HNL}
\\
K(\vp) &=& \ol P(\vp) - \ol P(\mathbf{0})
\label{K-HNL}
\\
r &=& \ol P(\mathbf{0}) 
\label{r-HNL}
\\
H &=& \ol H
\label{H-HNL}
\end{eqnarray}
so that the normalization conditions assumed in sec.\ 
(\ref{chapter2-1}) are
satisfied. In particular one can obtain the mass- and the magnetic-term
regularisation  ($ r_c$ and $ h_c$) in the large $N$ limit, 
using the function
$h_c(u)$ and $r_c(u)$ obtained in the previous 
chapter\footnote{
$r_c(u)$ and $h_c(u)$ are given
by eq.\ (\ref{defrc-const})  and by eq.\ (\ref{def-hc-App}).
}
 for the lattice regularisation using  $u$ defined in eq.\
(\ref{u-HNL}).

However in our 
large-$N$ case  we have to take into the mind that the propagator
and the vertices of the theory are function of 
$\Delta_m \equiv m_0^2 - m^2_{0c}$ and 
of $\Delta_p \equiv p - p_c$, and moreover all quantities, beside $r$ and $H$,
depend on these two variables.  We have seen in the previous
chapter that the Critical Crossover Limit (CCL) is defined
having fixed two scaling variables $x_t$ and $x_h$
\begin{eqnarray}
x_t & = & N{ P(\mathbf{0}) - r_c(u) 
\over \ol V^{(4)}(\mathbf{0},\mathbf{0},\mathbf{0},\mathbf{0})}
\label{xt-HNL}
\\
x_h & = & N{ \ol H - h_c(u) 
\over \ol V^{(4)}(\mathbf{0},\mathbf{0},\mathbf{0},\mathbf{0})}
\label{xh-HNL}
\end{eqnarray}
Expanding the previous couple of equations (\ref{xt-HNL},\ref{xh-HNL})
 near the critical point one can obtain  $\Delta_m$ and $\Delta_p$ 
as function of $x_t$ $x_h$ and $N$. 
As we will show in the next section
$\Delta_m$ and $\Delta_p$ scale  respectively as $1/N^{1/3}$ 
and $1/N$, so that we can assume an additional dependence on $u^{1/3}$
for the vertices, the propagator and the counterterms of the 
theory: 
i.e.\ we consider $K(\mathbf{p};u^{1/3})$ and 
$V^{(k)}(\mathbf{p}_1,\ldots, \mathbf{p}_k;u^{1/3})$. Note that, by definition,
$K(\mathbf{0};u^{1/3}) = 0$ for all values of $u$. 
Moreover, we assume, as in the case
of interest, that $V^{(3)}(\mathbf{0},\mathbf{0},\mathbf{0};u^{1/3}) = 0$ 
for all values of $u$. 
Following the calculation presented in sec.~\ref{twoloop-ccl}, it is easy to 
realize that the dependence of the vertices on $u$ is irrelevant
except in $h_c(u)$. In this case, eq.~\reff{chi1-oneloop} becomes
\begin{equation}
{t\over u} \chi_1 = - {h_c(u)\over u} - {1\over 2 \sqrt{u}} 
\int_\mathbf{p} {V_3(\mathbf{p};u^{1/3})\over K(\mathbf{p};u^{1/3}) + t} + 
O (\sqrt{u}).
\end{equation}
Because of the prefactor $1/\sqrt{u}$ we must take here into account the 
first correction proportional to $u^{1/3}$. Thus, if 
$V_3(\mathbf{p};u^{1/3}) \approx V_{3,0}(\mathbf{p}) + 
                        u^{1/3}  V_{3,1}(\mathbf{p})$ and 
$K(\mathbf{p};u^{1/3}) \approx K_{0}(\mathbf{p}) + u^{1/3}  K_{1}(\mathbf{p})$,
we obtain 
\begin{equation}
{t\over u} \chi_1 = - {h_c(u)\over u} - {1\over 2 \sqrt{u}} 
\int_\mathbf{p}
\left[{V_{3,0}(\mathbf{p}) + u^{1/3}  
       V_{3,1}(\mathbf{p})\over K_0(\mathbf{p}) + t} - 
       {u^{1/3} V_{3,0}(\mathbf{p}) K_1(\mathbf{p})\over 
         (K_0(\mathbf{p}) + t)^2}\right] + O(u^{1/6}).
\end{equation}
Now, $t$ can be set to zero without generating infrared divergences, 
neglecting corrections of order $u^{-1/2} t\ln t\sim u^{1/2} \ln u$.
It follows
\begin{equation}   
h_c(u) = - {\sqrt{u}\over 2} 
  \int_\mathbf{p} \left[ {V_{3,0}(\mathbf{p})\over K_0(\mathbf{p})} + 
    u^{1/3} {V_{3,1}(\mathbf{p}) K_0(\mathbf{p}) -  
             V_{3,0}(\mathbf{p}) K_1(\mathbf{p})\over K_0(\mathbf{p})^2}\right].
\label{hc-largeN}
\end{equation}
Eq.~\reff{hc-largeN} represents the only equation in which the explicit 
dependence of the vertices on $\Delta_m$ should be considered 
($\Delta_p$ is proportional to $u$ and thus it can always be set to zero).
In all other cases, we can simply set $\Delta_m = \Delta_p = 0$. 

\section{Crossover between mean-field and Ising behavior} \label{sec7}

In this section we wish to apply the above-reported results
for the critical crossover limit to the Hamiltonian \reff{caleffchi} 
of the zero mode.
With the identification (\ref{u-HNL}-\ref{H-HNL})
the effective Hamiltonian \reff{effexp2} 
corresponds to the Hamiltonian
discussed in sec.~\ref{chapter2-1}.

We begin by reporting the additive renormalization constants. 
The mass renormalization constant is given by, 
cf. eqs.~\reff{defrc-const},
\begin{eqnarray}
r_c(N) &=& {u\over 8\pi} \ln \left( {3 u\over 256\pi}\right) - 
      {u\over 8\pi} (3 + 8 \pi D_2) 
\nonumber \\
   && + {u\over2} \int_\mathbf{p}\left\{
    {[\bar{V}^{(3)}(\mathbf{0},\mathbf{p},-\mathbf{p})]^2 \over 
     \bar{V}^{(4)}(\mathbf{0},\mathbf{0},\mathbf{0},\mathbf{0}) K(\mathbf{p})^2}
    -{\bar{V}^{(4)}(\mathbf{0},\mathbf{0},\mathbf{p},-\mathbf{p}) \over 
     \bar{V}^{(4)}(\mathbf{0},\mathbf{0},\mathbf{0},\mathbf{0}) K(\mathbf{p})} 
    + {1\over \hat{p}^2} \right\},
\label{rc1}
\end{eqnarray}
where $D_2$ is a nonperturbative constant defined in Ref.~\cite{PRV-99}
(numerically $D_2 \approx  - 0.052$). As discussed in sec.~\ref{sec6.6}, 
we can compute all quantities appearing in eq.~\reff{rc1} at the critical
point and keep only the leading terms for $N\to \infty$. We thus obtain
\begin{eqnarray}
r_c(N) &=& {u\over 8\pi} \ln \left( {3 u\over 256\pi}\right) -
      {u\over 8\pi} (3 + 8 \pi D_2)
\\
   && + {\alpha^2\over 2 N} 
    \int_\mathbf{p}\left\{
    [\tilde{V}^{(3)}_0(\mathbf{0},\mathbf{p},-\mathbf{p}) 
     \tilde{P}_0(\mathbf{p})]^2 - 
     \tilde{V}^{(4)}_0(\mathbf{0},\mathbf{0},\mathbf{p},-\mathbf{p}) 
     \tilde{P}_0(\mathbf{p}) + 
     {\alpha^2\over \hat{p}^2} 
     \tilde{V}^{(4)}_0(\mathbf{0},\mathbf{0},\mathbf{0},\mathbf{0}) \right\},
\nonumber 
\end{eqnarray}
where all quantities are computed at the critical point. 

Analogously, we should introduce a counterterm for the magnetic field: 
\begin{equation}
h_c(N) = - {1\over2} \sqrt{u} 
    \int_\mathbf{p} {\bar{V}^{(3)}(\mathbf{0},\mathbf{p},-\mathbf{p})\over 
    [\bar{V}^{(4)}(\mathbf{0},\mathbf{0},\mathbf{0},\mathbf{0})]^{1/2} 
     K(\mathbf{p})}.
\end{equation}
As discussed in sec.~\ref{sec6.6}, such a quantity should not be simply 
computed at the critical point, but one should also take into account the 
additional corrections of order $N^{-1/3}$. Since $k\sim O(N^{1/6})$
in the critical crossover limit, we have 
\begin{eqnarray}
h_c(N) &=& - {\alpha\over 2\sqrt{N}} 
   \int_\mathbf{p} \left[ 
     \tilde{V}^{(3)}_0(\mathbf{0},\mathbf{p},-\mathbf{p}) - 
     {\tilde{V}^{(3)}_0(\mathbf{0},\mathbf{0},\mathbf{0})\over 
      \tilde{V}^{(4)}_0(\mathbf{0},\mathbf{0},\mathbf{0},\mathbf{0})} 
      \tilde{V}^{(4)}_0(\mathbf{0},\mathbf{0},\mathbf{p},-\mathbf{p})
     \right] {1\over \tilde{P}^{-1}_0(\mathbf{p}) - 
                \tilde{P}^{-1}_0(\mathbf{0}) } 
\nonumber \\
   && - {\alpha\over 2\sqrt{N}} 
     {\tilde{V}^{(3)}_0(\mathbf{0},\mathbf{0},\mathbf{0})\over
      \tilde{V}^{(4)}_0(\mathbf{0},\mathbf{0},\mathbf{0},\mathbf{0})}
      \int_\mathbf{p} 
    [\tilde{V}^{(3)}_0(\mathbf{0},\mathbf{p},-\mathbf{p}) 
     \tilde{P}_0(\mathbf{p})]^2 .
\end{eqnarray}
The second integral should be computed at the critical point, while the
first one and the prefactor of the second one should be expanded 
around the critical point. Since, as we shall show below, 
$\Delta_m \equiv m^2_0 - m_{0c}^2 \sim N^{-1/3}$ and $p - p_c \sim N^{-1}$, 
it is enough to compute the first correction in $\Delta_m$. 
In practice, we find that the renormalization terms have the form
\begin{eqnarray}
 r_c(N) &=& {1\over N} (r_{c0} \ln N + r_{c1}), \nonumber \\
 h_c(N) &=& {1\over \sqrt{N}} (h_{c0} + \Delta_m h_{c1}).
\label{rchc-exp}
\end{eqnarray}
Once $r_c(N)$ and $h_c(N)$ are computed, we can define the scaling variables
$x_t \sim t/u$ and $x_h \sim h/u$. Choosing the normalizations appropriately 
for later convenience, we define 
\begin{eqnarray}
x_t &\equiv& {N\over \alpha^2}  [\bar{P}^{-1}(0) - r_c(N)], 
\nonumber \\
x_h &\equiv& {N\over \alpha} [H - h_c(N)].
\label{xtxhdef}
\end{eqnarray}
Now, the critical crossover limit is obtained by tuning $p$ and $m_0^2$ 
around the critical point in such a way that $x_t$ and $x_h$ are 
kept constant, i.e. $p\to p_{\rm crit}$, $m_0^2\to m_{0,\rm crit}^2$, 
$N\to \infty$ at fixed $x_t$ and $x_h$. Note that here $p_{\rm crit}$ and 
$m_{0,\rm crit}^2$ correspond to the position of the critical point 
as a function of $N$ (thus $p_{\rm crit}\to p_c$ and 
$m_{0,\rm crit}^2\to m_{0c}^2$ as $N\to \infty$) and are obtained by 
requiring $x_t=x_h=0$ (cf.~sec.~\ref{sec6.5}). 

In order to compute the relation between $p$, $m_0^2$ and 
$x_t$, $x_h$ we set
$\Delta_m \equiv m^2_{0} - m^2_{0c}$ and 
$\Delta_p \equiv p - p_c$ and expand eq.~\reff{xtxhdef} in 
powers of $\Delta_m$ and $\Delta_p$. In the following we shall show that 
$\Delta_m \sim N^{-1/3}$ and $\Delta_p\sim N^{-1}$,
so that the relevant terms are
\begin{eqnarray}
x_t &=& N (b_{t0} \Delta_p + b_{t1} \Delta_p \Delta_m + 
         b_{t2} \Delta_m^3) + d_1 \ln N + d_2 
\label{eqxt}
\\
x_h &=& N^{3/2} (b_{h0} \Delta_p^2 + b_{h1} \Delta_p \Delta_m + 
               b_{h2} \Delta_p \Delta_m^2 + b_{h3} \Delta_m^3 + 
               b_{h4} \Delta_m^4) + 
\nonumber \\ 
&& + N^{1/2} (d_{30} + d_{31} \Delta_m).
\label{eqxh}
\end{eqnarray}
The constants are obtained by expanding $\bar{P}^{-1}(\mathbf{0})$ and 
$H$ around the critical point and by using the expansions
\reff{rchc-exp}. All quantities are analytic in 
$\Delta_m$ and $\Delta_p$ and several terms are absent because of 
identities (\ref{id-P-CP}), (\ref{id-V3-CP}), and (\ref{id-V4-CP}). 
In particular, a term proportional to $\Delta_m^2$ is absent in the equation
for $x_t$. This is a consequence of eq.~(\ref{id-V4-CP}). 
Indeed, we have
\begin{eqnarray}
{1\over \alpha^2} \bar{P}^{-1}(\mathbf{0}) &=& 
       \tilde{P}_0^{-1}(\mathbf{0}) - 
       {[\tilde{V}_0^{(3)}(\mathbf{0},\mathbf{0},\mathbf{0})]^2 \over 
        2 \tilde{V}_0^{(4)}(\mathbf{0},\mathbf{0},\mathbf{0},\mathbf{0})} 
      + O(\Delta_m^3,N^{-1}) 
\nonumber \\
   &=& {1\over 2} \left[
    {\partial^2 \tilde{P}_0^{-1}(\mathbf{0})\over \partial (m_0^2)^2} - 
     {1\over \tilde{V}_0^{(4)}(\mathbf{0},\mathbf{0},\mathbf{0},\mathbf{0})}
    \left( {\partial\over \partial m_0^2} 
     \tilde{V}_0^{(3)}(\mathbf{0},\mathbf{0},\mathbf{0}) \right)^2\right]
    \Delta_m^2 + O(\Delta_m^3,N^{-1}) 
\nonumber \\
    &=& 0 + 
    O(\Delta_m^3,N^{-1}),
\end{eqnarray}
where in the last step we have used eq.~(\ref{id-V4-CP}).

We wish now to determine the behavior of $\Delta_m$ and $\Delta_p$ that is 
fixed by eqs.~(\ref{eqxt}) and (\ref{eqxh}). We assume 
\begin{equation}
\Delta_m = {\delta_{m0}\over N^{\alpha}}, \qquad\qquad 
\Delta_p = {\delta_{p0}\over N^{\beta}}, 
\end{equation}
where $\delta_{m0}$ and $\delta_{p0}$ are nonvanishing constants and 
$\alpha$ and $\beta$ exponents to be determined. If 
$\beta < 1$, eq.~\reff{eqxt} implies 
\begin{equation}
b_{t0} \delta_{p0} N^{1-\beta} + b_{t2} \delta_{m0} N^{1-3\alpha} = 
  o(N^{1-\beta}),
\end{equation}
which requires $\beta = 3\alpha$. Now, consider eq.~(\ref{eqxh}).
The term $N^{3/2} \Delta_m^3$ is of order $N^{3/2 - \alpha}$ and cannot 
be made to vanish. Therefore, we must have $\beta \ge 1$. 
Considering again eq.~(\ref{eqxh}), it is easy to see that all terms 
containing $\Delta_p$ cannot increase as fast as $N^{1/2}$. Therefore, 
cancellation of the term $d_{30} N^{1/2}$ requires $\alpha = 1/3$. 
Consideration of eq.~\reff{eqxt} implies finally $\beta = 1$. 
This analysis can be extended to the subleading corrections, obtaining 
an expansion of the form
\begin{eqnarray}
\Delta_m &=& {\delta_{m0}\over N^{1/3}} + {\delta_{m1}\over N^{2/3}} + 
             {\delta_{m2}\over N^{5/6}},
           + O(N^{-1}) \nonumber \\
\Delta_p &=& {\delta_{p0}\over N} + 
             {\delta_{p1}\over N^{7/6}}
           + O(N^{-4/3}).
\end{eqnarray}
The coefficients are given by
\begin{eqnarray}
\delta_{m0} &=& - \left({d_{30}\over b_{h3}} \right)^{1/3} ,
\nonumber \\
\delta_{m1} &=& {\delta_{m0}^2 \over 3 b_{t0} d_{30}} 
   \left[ b_{t0} d_{31} - b_{h1} (d_1 \ln N + d_2) \right]
   + {b_{h1} \delta_{m0}^2\over 3 b_{t0} d_{30}} x_t + 
\nonumber \\
        && + {\delta_{m0}^5 \over 3 b_{t0} d_{30}}
   (b_{h4} b_{t0} - b_{h1} b_{t2}),
\nonumber \\ 
\delta_{p0} &=& - {1\over b_{t0}} \left( \delta_{m0}^3 b_{t2} +
             d_1 \ln N + d_2 - x_t\right).
\end{eqnarray}
We are not able to compute $\delta_{m2}$ and $\delta_{p1}$ but we can
however compute a relation between these two quantities. We obtain 
\begin{equation}
\delta_{p1} = - {3 \delta_{m0} \delta_{m2} b_{h3} \over b_{h1}} + 
       {1\over  b_{h1} \delta_{m0}} x_h.
\end{equation}
Correspondingly, by using eq.~(\ref{expansion-gap}), we can compute the 
expansion of $u_h$:
\begin{eqnarray}
u_h &=& {u_{h0}\over N} + {u_{h1}\over N^{4/3}} + {u_{h2}\over N^{3/2}},
\end{eqnarray}
where $u_{h1}$ depends on $x_t$ and $u_{h2}$ on $x_h$. 
We thus define a new scaling field by requiring that no term proportional
to $x_t N^{-4/3}$ is present. For this purpose we set
\begin{equation}
\hat{u}_h = u_h + {x_{\rm mix} \over N^{1/3}} (p - p_c)  
\end{equation}
where the coefficient $x_{\rm mix}$ is given by
\begin{equation}
x_{\rm mix} = {(a_{03} b_{h1} - a_{11} b_{h3}) \delta_{m0} \over 
     b_{h3}}.
\end{equation}
The new scaling field has an expansion of the form 
\begin{equation}
\hat{u}_h= {\hat{u}_{h0}\over N} + {\hat{u}_{h1}\over N^{4/3}} + 
             {\hat{u}_{h2}\over N^{3/2}},
\end{equation}
where 
\begin{eqnarray}
\hat{u}_{h0} &= &
     a_{03} \delta_{m0}^3,
\\
\hat{u}_{h1} &=& {\delta_{m0}\over b_{h3}}
        [ - a_{03} d_{31} + 
          (a_{04} b_{h3} - a_{03} b_{h4}) \delta_{m0}^3  ]
\\
\hat{u}_{h2} &=& {a_{03}\over b_{h3} } x_h .
\end{eqnarray}
Interestingly enough, in $\hat{u}_{h2}$ all terms proportional to 
the unknown quantity $\delta_{m2}$ cancel.
At this point we can easily compute the $1/N$ expansion of the critical point
$p_{\rm crit}$, $\beta_{\rm crit}$.
It is enough to set $x_h$ = $x_t = 0$ in the previous expansions,
obtaining 
\begin{eqnarray}
   p_{\rm crit} &=& p_c + 
          \left. {\delta_{p0}\over N}\right|_{x_t = 0} + O(N^{-7/6}),
\\
  \beta_{\rm crit} &=& \beta_c 
         + {\hat{u}_{h0} +
            a_{10} \delta_{p0}|_{x_t = 0}\over N}  + O(N^{-7/6}).
\end{eqnarray}
It follows that 
\begin{eqnarray}
p - p_{\rm crit} &\approx &  \left(-  {2\over3} 
   {b_{t1} b_{h1} \delta_{m0}^2 \over d_{30}} + 1\right) {x_t\over b_{t0} N},
\label{eq:scaldp}
\\
\hat{u}_h - \hat{u}_{h,\rm crit} &\approx& 
           {3 a_{03} b_{t0} - a_{11} b_{t1} \over
                3 b_{h3} b_{t0} - b_{h1} b_{t1}}
           {x_h\over N^{3/2}},
\label{eq:scaluh}
\end{eqnarray}
where $\hat{u}_{h,\rm crit}$ is the value of $\hat{u}_h$ at the critical point.
Therefore, the $1/N$ corrections modify the position of the critical
point and change the magnetic scaling field which should now be identified with 
$\bar{u}_h = \hat{u}_h - \hat{u}_{h,\rm crit}$ (as we already discussed the 
thermal magnetic field is not uniquely defined and we take again
$p - p_{\rm crit}$). Eqs.~(\ref{eq:scaldp}) and (\ref{eq:scaluh}) also indicate
which are the correct scaling variables. Ising behavior is observed only if 
$N(p - p_{\rm crit})$ and $N^{3/2}(\hat{u}_h - \hat{u}_{h,\rm crit} )$ 
are both small; in the opposite case mean-field behavior is observed.
Therefore, as $N$ increases the width of the critical region 
decreases, and no Ising behavior is observed at $N=\infty$ exactly. 

\section{Critical behavior of $\langle\bsigma_x\cdot \bsigma_{x+\mu}\rangle$
} \label{sec8}

In this section we wish to compute the large-$N$ behavior of 
\begin{equation}
 \overline{E} = \< \bsigma_x \cdot \bsigma_{x+\mu}\> .
\end{equation}
Such a quantity does not coincide with the energy. 
However, as far as the critical behavior is concerned, there should not be 
any significant difference. In order to perform the computation, 
note that the equations of motion for the field $\lambda_{x\mu}$ give
\begin{equation}
\< \rho_{x\mu} \> = 1 + \overline{E }.
\end{equation}
Thus, we have
\begin{equation}     
\overline{E} = \tau - 1 + {1\over \sqrt{N}} \< \hat{\rho}_{x\mu} \> = 
    \tau - 1 + {1\over \sqrt{N}} \sum_B U_{4B}(\mathbf{0}) 
    \< \Phi_{Bx} \>.
\end{equation}
Therefore, we need to compute the correlations $\<\phi\>$ and 
$\<\varphi_a\>$. For the zero mode we have immediately 
\begin{equation}
\< \phi_x\> = \alpha \< \chi_x \> - k = 
   {\alpha \over x_t} f_1^{\rm symm}(x_t,x_h) - k,
\end{equation}
where $f_1^{\rm symm}(x_t,x_h)$ is the crossover function for the 
magnetization in the Ising model and the constant $k$ diverges as $N^{1/6}$, modulo some obvious normalizations.

Let us now consider $\<\varphi_a \>$ and show that such a correlation vanishes
as $N\to\infty$ in the critical crossover limit. Indeed, we can write
\begin{eqnarray}
\<\varphi_a \> &=& {1\over \sqrt{N}} 
 \hat{P}_{ab}(\mathbf{0}) 
   \int_{\mathbf{p}} [\hat{V}_{bcd}(\mathbf{0},\mathbf{p},-\mathbf{p})
             \hat{P}_{cd}(\mathbf{p}) + 
      (\hat{V}_{b11}(\mathbf{0},\mathbf{p},-\mathbf{p}) - 
      \hat{V}_{b11}(\mathbf{0},\mathbf{0},\mathbf{0})) \hat{P}_{11}(\mathbf{p})]
\nonumber \\
       && + {1\over \sqrt{N}} \hat{P}_{ab}(\mathbf{0}) 
            \hat{V}_{b11}(\mathbf{0},\mathbf{0},\mathbf{0}) \<\phi^2_x\>.
\end{eqnarray}
Note that in the last term we have replaced 
\begin{equation}
\int_\mathbf{p} \hat{P}_{11}(\mathbf{p}) = \<\phi^2_x\>,
\end{equation}
a necessary step, since the integral diverges at the critical point and 
therefore should be computed in the effective theory for the zero mode. 
Now, we have 
\begin{equation}
\<\phi^2_x\>  = k^2 - 2 \alpha k \< \chi_x \> + \alpha^2 \< \chi_x^2 \>.
\end{equation}
In the critical crossover limit, the two expectation values are replaced 
by the crossover functions for the magnetization and the energy and by 
a regular term. Therefore, for $N\to \infty$, the leading behavior is 
$\<\phi^2_x\>  = k^2 \sim N^{1/3}$. It follows that 
$\<\varphi_a\> \sim N^{-1/6}$. In conclusion we can write
\begin{eqnarray}
\overline{E} &= &
    \tau - 1 + {1\over \sqrt{N}} U_{41}(\mathbf{0}) 
    \left[
   {\alpha\over x_t} f_1^{\rm symm}(x_t,x_h) - k \right]
\nonumber \\
 &=& \overline{E}_{\rm reg} + {1\over \sqrt{N}} U_{41}(\mathbf{0})
     {\alpha\over x_t} f_1^{\rm symm}(x_t,x_h) + O(N^{-2/3}),
\label{E-largeN}
\end{eqnarray}
where $\overline{E}_{\rm reg}$ is the regular part of $\overline{E}$:
\begin{eqnarray} 
  \overline{E}_{\rm reg} &= & \tau - 1 - {1\over \sqrt{N}} U_{41}(\mathbf{0}) k
\nonumber \\
   &=& \tau_0 - 1 + [\tau_1 - k_{0} U_{41}(\mathbf{0})] \delta_{m0} 
      N^{-1/3} + O(N^{-2/3}),
\end{eqnarray}
where we have written $\tau \approx \tau_0 + \tau_1 \Delta_m$ and 
$k \approx k_{0} \Delta_m \sqrt{N}$. 

Equation \reff{E-largeN} shows that the singular part of $\overline{E}$ 
behaves as the magnetization
in the Ising model. For $x_t \ll 1$ and $x_h\ll 1$ one observes Ising behavior
and thus 
\begin{eqnarray}
\overline{E} - \overline{E}_{\rm reg} \sim |x_t|^{\beta_I} && \qquad \hbox{for}\quad x_h = 0, 
      \hbox{low $t$ phase}
\nonumber \\
\overline{E} - \overline{E}_{\rm reg} \sim |x_h|^{1/\delta_I} && \qquad \hbox{for}\quad x_h \not= 0,
\end{eqnarray}
where $\beta_I = 1/8$ and $\delta_I = 15$. On the other hand, in the opposite
limit we have
\begin{eqnarray}
\overline{E} - \overline{E}_{\rm reg} \sim |x_t|^{\beta_{MF}} && \qquad \hbox{for}\quad x_h = 0, 
           |x_t|\to \infty,
      \hbox{low $t$ phase}
\nonumber \\
\overline{E} - \overline{E}_{\rm reg} \sim |x_h|^{1/\delta_{MF}} && \qquad \hbox{for}\quad 
           |x_h|\to \infty
\end{eqnarray}
with $\beta_{MF} = 1/2$ and $\delta_{MF} = 3$. Note that the limit 
$|x_t|\to \infty$ and $|x_h|\to\infty$ should always be taken close to the 
critical limit. Therefore, $|x_h|\to\infty$ means that we should consider 
$N\to \infty$, $\hat{u}_h\to \hat{u}_{h,\rm crit}$, $p\to p_{\rm crit}$
in such a way that $N^{3/2}(\hat{u}_h - \hat{u}_{h,\rm crit})\to \infty$,
i.e. $N$ should increase much faster that the rate of approach to the 
critical point.

\section{Numerical results for selected Hamiltonians} \label{sec9}

In this section we present some numerical results for some selected 
Hamiltonians. First,
as in Ref. \cite{BGH-02}, we consider 
\begin{equation}
W(x) = {2\over p}\left( {x\over2}\right)^p.
\label{Ham1}
\end{equation}
Second, we consider the mixed O($N$)-RP$^{N-1}$ model with 
Hamiltonian \cite{MR-87} 
\begin{equation}
{\cal H} = - N \beta_V \sum_{x\mu} (\bsigma_x\cdot \bsigma_{x+\mu}) -
{N\beta_T\over2} \sum_{x\mu} (\bsigma_x\cdot \bsigma_{x+\mu})^2.
\end{equation}
This Hamiltonian corresponds to the function 
\begin{equation}
W(x) = p x + {1\over 4} (1 - p) x^2,
\label{Ham2}
\end{equation}
where we set $\beta_V = (1 + p) \beta/2$ and $\beta_T = (1 - p) \beta/2$. 
This Hamiltonian is ferromagnetic for $p > -1$. 
Note that for $p = -1$ we obtain the $RP^{N-1}$ Hamiltonian 
\cite{HM-82}, \cite{MR-87}, \cite{Ohno_90}, \cite{KZ}, \cite{Butera_92}, 
\cite{CEPS_RP}, \cite{Hasenbusch_96}, \cite{Niedermayer_96},
\cite{Catterall_98}, \cite{SS-01}
\begin{equation}
{\cal H} = - {N \beta \over 2} \sum_{x\mu} (\bsigma_x\cdot\bsigma_{x+\mu})^2,
\end{equation}
that has the additional gauge invariance $\bsigma_x \to \epsilon_x \bsigma_x$,
$\epsilon_x = \pm 1$. Under standard assumptions, the large-$N$
analysis should apply also to this last model: the local gauge invariance 
should not play any role.\footnote{
The irrelevance of the $\mathbb{Z}_2$ symmetry for the large-$\beta$ 
behavior of $RP^{N-1}$ models have been discussed in detail in 
Ref.~\cite{Hasenbusch_96,Catterall_98,Niedermayer_96}. Thus, in spite of 
the additional local invariance, $RP^{N-1}$ models are 
expected to be asymptotically free and to be described by the 
perturbative renormalization group.}

In the large-$N$ limit \cite{CP-02}, the first Hamiltonian has a critical 
point at $\beta_c\approx 1.335 $ and $p_c \approx 4.538 $. 
By using the numerical results reported in Table~\ref{numerical}
we can compute the first corrections to the critical parameters. 
We obtain
\begin{eqnarray}
\beta_c  &\approx&  1.335 + {1\over N} \left( 36.127 +2.093 \ln N \right),
\\
p_c  &\approx& 4.538 + {1\over N}\left(87.92+4.80 \ln N \right).
\label{pc-W1}
\end{eqnarray}
Note the presence of a $\ln N/N$ correction due to the nonanalytic 
nature of the renormalization counterterms. The correction terms are 
quite large, indicating that the large-$N$ results are
quantitatively predictive only for large values of $N$. This is not 
totally unexpected since \cite{BGH-02} $p_c \approx 20$ for $N=3$, 
that is quite far from the large-$N$ estimate $p_c \approx 4.538$. 
If we substitute $N=3$ in eq.~\reff{pc-W1}, we obtain 
$p_c \approx 35$, which shows that the corrections have the 
correct sign and give at least the correct order of magnitude. 

\begin{table}[t]
\begin{center}
\begin{tabular}{ccc}
\hline\hline
& $W_1$ & $W_2$ \\
\hline
$\beta_c$   &     1.334721915850    &     0.9181906464057     \\
$p_c$       &     4.537856778637    &  $-$0.9707166650184     \\
$m_{0c}^2$  &     0.1501849439193   &     0.8657494320430     \\
$a_{10} $   &     0.4359516292302   &  $-$1.1359750388653     \\
$a_{11} $   &     0.5042522341176   &  $-$0.6248171602172     \\
$a_{12} $   &  $-$1.2793594495686   &     0.1122363376128     \\
$a_{03} $   &  $-$1.3015714087645   &  $-$0.0061080440602     \\
$a_{04} $   &     67.512019516378   &     0.2847340288532     \\
$b_{t0} $   &  $-$0.04261881236908  &     0.21169346791995    \\
$b_{t1} $   &  $-$0.0043954143923   &     0.05334711842197    \\
$b_{t2} $   &     1.85335235385232  &     0.00015229016444    \\
$b_{h0} $   &  $-$0.0077917231312   &  $-$0.664558775414698   \\
$b_{h1} $   &     0.085888253027    &  $-$0.094120307536470   \\
$b_{h2} $   &  $-$0.5785580673083   &     0.049596987743194   \\
$b_{h3} $   &  $-$0.221693999401    &  $-$0.000920094744508   \\
$d_1    $   &     0.205             &     0.2678             \\
$d_2    $   &     0.614             &  $-$0.808514            \\
$d_{30} $   &     0.374             &     1.1692              \\
$\alpha^{-2}$ &     0.0315969         &     0.00933367          \\
\hline\hline
\end{tabular}
\end{center}
\caption{Numerical estimates for interaction  \reff{Ham1} ($W_1$)
and \reff{Ham2} ($W_2$).}
\label{numerical}
\end{table}

For $N=\infty$ the second Hamiltonian has a critical point
at \cite{MR-87} $\beta_c\approx 0.918 $ and $p_c \approx  -0.971 $. 
Including the first correction we obtain
\begin{eqnarray}
\beta_c  &=& 0.9182-{1\over N}\left(11.062-1.437 \ln N\right),
\\
p_c & =& -0.9707 +{1\over N}\left( 2.905 - 1.265\ln N \right).
\end{eqnarray}
In this case the corrections are smaller. However, for 
$N \le 99.4$ they predict $p_c < -1$, although only slightly. 
This is of course not possible, since for $p < -1$ the system 
is no longer ferromagnetic. Thus, we expect the transition to disappear 
for some $N = N_c$, with $N_c \approx 100$ (this is of course a very 
rough estimate). Since $p_c \approx - 1$, we also predict 
$RP^{N-1}$ models to show a very weak first-order transition for large
$N$.

\section{Conclusions} \label{sec10}

In this chapter we have explicitly presented our technique to solve
the problem of Infra Red Divergences in large
$N$ expansion, giving a detailed description
of a class of one-parameter $O(N)$ models that present
this kind of problem. This models
 show a line of first-order 
finite-$\beta$ transitions. We focus on the endpoint of the 
first-order transition line where energy-energy correlations 
become critical, while the spin-spin correlation length 
remains finite, in agreement with Mermin-Wagner theorem \cite{MW-66}. 
In sec.\ \ref{chapter1-sec5}
we showed that, at the critical point, the standard $1/N$ expansion
breaks down, since the inverse propagator of the auxiliary fields 
has a zero eigenvalue. A careful treatment shows that the zero mode,
i.e.\ the field associated with the vanishing eigenvalue, has an effective 
Hamiltonian that corresponds to a weakly coupled one-component $\phi^4$ theory.
Thus, the phase transition belongs to the Ising universality class for any
$N$, in agreement with the argument of Ref.~\cite{BGH-02}. In Ref.~\cite{CP-02}
it was shown that for $N=\infty$ the transition has mean-field exponents.
We reconcile here these two results. If $u_t$ and $u_h$ are 
the linear thermal and magnetic scaling fields, in the critical limit 
a generic long-distance quantity $\cal O$
has a behavior of the form
\begin{equation}
\langle {\cal O} \rangle_{\rm sing} \approx 
  u_t^\sigma f_{\cal O}(u_t N, u_h N^{3/2}),
\end{equation}
where $f(x,y)$ is a crossover function. Only if 
$u_t N \ll 1$ and $u_h N^{3/2} \ll 1$ does one observe Ising behavior.
In the opposite limit one observes mean-field criticality. 
Therefore, the width of the Ising critical region goes to zero as $N\to\infty$
and, even if the transition is an Ising one for any $N$, only 
mean-field behavior is observed for $N=\infty$. The behavior observed 
at the critical point for $N\to\infty$ resembles very closely what 
is observed in medium-range models 
\cite{LBB-97}, \cite{LB-98}, \cite{PRV-98},
\cite{CCPRV-01}, with $N$ 
playing the role of the interaction range. Our analysis fully confirms
the conclusions of Ref.~\cite{BGH-02}.

From a more quantitative point of view, we give explicit expressions 
for the critical values $\beta_c$ and $p_c$ and for the nonlinear scaling 
fields to order $1/N$. Numerical results are given for the 
Hamiltonian introduced in Ref.~\cite{BGH-02} and for mixed 
$O(N)$-$RP^{N-1}$ models \cite{MR-87}. We plan to compute
corrections to the critical parameters for other model for
which more available data are present (for instance
Yukawa models, chap.\ \ref{chapter5} and ref.\ \cite{KSS-98}).
This would provide a more evident check  
for the scheme proposed in this work.

The main result of this paper, i.e. the fact
that the width of the Ising critical region goes to zero as $N\to\infty$, 
holds in any $d < 4$. However, in generic dimension $d$ the natural 
scaling variables are $(p - p_{\rm crit}) N^{2/(4-d)}$ and 
$(\hat{u}_h - \hat{u}_{h,\rm crit}) N^{3/(4-d)}$,\footnote{
 Thus, for $d> 2$ the 
Ising critical region shrinks faster as $N$ increases: in three 
dimensions as $N^{-2}$ and $N^{-3}$ in the thermal and magnetic directions
respectively.} while the scaling equation eq.\ (\ref{scaling-cross-lim})
is now replaced by
\begin{equation}
\chi_n = u^{d/(4-d)} t^{-n(d+2)/4}
   f_{d;n}^{\rm symm}(u t^{-(4-d)/2}, u h^{-2(4-d)/(2+d)}).
\label{scaling-high-dimension}
\end{equation}
 The claim simply follows from the fact that
for any $d$ the zero mode has a Hamiltonian that corresponds to 
a weakly coupled $\phi^4$ theory. Thus, for $d < 4$, the phase transition
is always Ising-like. One only needs to take into account the
different definition of the scaling fields (\ref{scaling-high-dimension}).
Also the expressions for the renormalization constants 
$r_c$ and $h_c$ should be changed: for instance, in three dimensions 
we also expect contributions from two-loop graphs, as it happens in
medium-range models \cite{PRV-99}. More important, following the discussion
of sec.\ (\ref{chapter2-higher-dimension}) in order to get
the previous scaling equation (\ref{scaling-high-dimension})
one need to correct the definition of the zero mode (\ref{k-ren}) 
to impose that the three-leg vertex renormalized must be equal
to zero unless contributions that are irrelevant in the 
Critical Crossover Limit.

%%%%%%%%%%%%%%%%%%%%%%%%%%%%%%%%%%%%%%%%%%%%%%%%%%%%%
%%%%%%%%%%%%%%%%%%%%%%%%%%%%%%%%%%%%%%%%%%%%%%%%%%%%%
%%%%%%%%%%%%%%%%%%%%%%%%%%%%%%%%%%%%%%%%%%%%%%%%%%%%%
%%%                                               %%%
%%%                                               %%%
%%% CHAPTER 4 - CHAPTER 4 - CHAPTER 4 - CHAPTER 4 %%%
%%%                                               %%%
%%%                                               %%%
%%%%%%%%%%%%%%%%%%%%%%%%%%%%%%%%%%%%%%%%%%%%%%%%%%%%%
%%%%%%%%%%%%%%%%%%%%%%%%%%%%%%%%%%%%%%%%%%%%%%%%%%%%%
%%%%%%%%%%%%%%%%%%%%%%%%%%%%%%%%%%%%%%%%%%%%%%%%%%%%%

\chapter{Large-$N$ expansion of $O(N)$ models:
the  multicritical zero mode}\label{chapter4}

In this chapter we want to generalize the considerations
 of the previous chapter
to the case in which a multicritical zero mode appear $\N>1$.\footnote{
Referring to the notation introduced in the first chapter, we remember
that only the $\N$ odd case is investigated, the other case being
unstable for $N=\infty$.}  
  In this case the study of the effective interaction for the zero mode 
would require a more involved algebra which gives 
several non-trivial cancellations in the expansion of the scaling
fields near the multicritical point. 
This cancellations generalize what found in 
chapter \ref{chapter3} [where for instance the $\Delta_m^2$ term
disappears in the expansion of $x_t$ eq.\ 
(\ref{eqxt})] and appear as soon as one consider
the translated theory (\ref{eq_3.13}).
The main effects of this cancellations will be the identification
of the leading order of the effective vertices for the
translated theory with the scaling fields for $N=\infty$
--i.e.\ without $1/N$ corrections-- (\ref{scaling-fields-classici})
introduced in order to parametrize the gap equation near the critical point
(\ref{gap-near-critical-point}).\footnote{This identification
 will
 be exact apart some contribution that 
are irrelevant in the scaling limit taken in the past chapters
(also called Critical Crossover Limit --CCL--).}
Following the notations of the previous chapter, 
if we define
$\tilde V^{(j)}(\mathbf{p})\equiv \tilde V^{(j)}(\mathbf{0},\cdots \mathbf{0},
\vp,-\vp)$ the vertices of the 
zero mode, the first result will be to 
show that near the multicritical point
\begin{eqnarray}
\tilde V^{(j)}(\mathbf{0}) & \sim & {u_h}^{\N+2-j\over \N+2} 
\nonumber\\
&\sim& u_{j-1} ,\, \Delta_m u_{j}, \, {\Delta_m}^2 u_{j+1}
\cdots {\Delta_m}^{\N-j+3}
\label{eq_4.1} 
\end{eqnarray}
for $j=0,\cdots \N$. In the second line of (\ref{eq_4.1}) we have written all
the contributions that scale in the same way in the 
$N=\infty$ critical limit defined by (\ref{scaling-fields-classici})
\begin{eqnarray}
u_j\sim {u_h}^{\N+2-j\over \N + 2} &\qquad &
\Delta_m \sim {u_h}^{1\over \N+2}.
\nonumber
\end{eqnarray}
In the next section we will show that introducing a proper
translation for the zero mode [similar to what introduced 
in (\ref{eq_3.13})], one is able to cancel in eq.\ (\ref{eq_4.1})
all the contribution that factorize $\Delta_m$ so that
$\ol V^{(j)} (\mathbf{0}) \sim u_{j-1}$.\footnote{ As
in the previous chapter $\ol V^{(j)}$
refers to the vertices for the translated theory.}
This result is very important in order
to include $1/N$ fluctuations to the $N=\infty$ theory
 of chapter \ref{chapter1}. Indeed  in chapter
\ref{chapter2} we have obtained that the scaling fields
describing the crossover to the multicritical point are given by
\begin{eqnarray}
u_i(N) \equiv \ol V^{(i+1)}(\mathbf{0}) - u_{ci} 
\label{eq_4.2}
\end{eqnarray}
where $u_{ci}$ regularise the theory and cancel the 
nonuniversal details. If  the $\Delta_m$ contributions of
the second line of eq.\ (\ref{eq_4.1}) do not cancel,
inserting (\ref{eq_4.1}) in (\ref{eq_4.2}) one would obtain
an expression similar to the following
\begin{eqnarray}
u_i(N) &=& u_i(\infty) + \sum_{j=i+1}^{\N} c^{(i)}_j u_{j}(\infty) + O\Big(
{1\over N}\Big)
\label{eq_4.3}
\end{eqnarray}
If $c^{(i)}_j \neq 0 $ for {\em every} $i$
and  $j>i$, then the scaling fields 
would be discontinuous for $N=\infty$, so that our mechanism 
to explain crossover between the two different critical
regime would fail. We remember that there is some freedom in the definition
of the scaling fields $u_i$ (\ref{scaling-fields-classici}). Indeed also
\begin{eqnarray}
u'_i = u_i + \sum_{j=0}^{i-1}a_j u_j
\label{eq_4.4}
\end{eqnarray}
 satisfy the scaling relations
of the $N=\infty$ theory; however eq.\ (\ref{eq_4.4}) 
can never match with eq.\ (\ref{eq_4.3}). We
stress how in principle the require to cancel
$\N(\N-1)/2$ parameters [$c^{(j)}_i$ in eq.\ (\ref{eq_4.3})]
tuning a single parameter
(the translation constant of the zero mode)
looks an impossible work! The answer to this paradox relies
on the fact that the vertices of the 
effective theory are not independent but are strictly
related each other and at the end to the 
gap equation. In the first part of this chapter we want
to elucidate this relations by using similar techniques
of sec.\ \ref{AppA}, while in the last part of this
chapter we  will describe the crossover behaviour
 to mean field in a similar way of what done
for the $\N=1$ case, chapter \ref{chapter2}.
In particular we will describe (in a general way),
how to obtain the scaling field in the large $N$ limit
and how to include $1/N$ corrections to their
$N=\infty$ definitions computing explicitly the
$\N=3$ case.

\section{Effective Hamiltonian for the zero mode}\label{sec-four}

\subsection{Basic definitions}

As in the previous chapter the starting point is the definition
of the effective interaction for the zero mode through 
the integration of the massive mode of the theory.
However in this case, being interested in general relations,
we need to take a more general point of view.
If $v_A(\mathbf{p}; m_0^2, p_i)$ 
is the normalized eigenvector that corresponds to the zero eigenvalue for 
$\mathbf{p} = 0$ at the MCP we have that
we define the critical field (rescaled)
\begin{equation} 
 \varphi(\mathbf{p}) = {1\over \sqrt{N}}
 \sum_A v_A(\mathbf{p}; m_0^2, p_i) \Psi_A,
\label{def-varphi}
\end{equation}
while the effective Hamiltonian becomes
\begin{equation} 
{\cal H}_{\rm eff}[\varphi] = - {1\over N} \ln {Z[\varphi]\over Z[0]},
\label{App:Heff}
\end{equation}
where 
\begin{equation}
Z[\varphi] = 
\int \prod_{x,A} d\Psi_{xA}\, 
\delta(\Psi\cdot v - \varphi \sqrt{N}) e^{-{\cal H}},
\label{App:Z}
\end{equation}
where the functional $\delta$ function guarantees the identification
(\ref{def-varphi}) and  $\Psi\cdot v = \sum_A \Psi_A v_A$. 
In this expression we assume that $v_A(\mathbf{p}; m_0^2, p_i)$ is 
normalized, i.e., $\sum_A v_A^2 = 1$. 
Note the additional factor $\sqrt{N}$ in the definition of 
$\varphi$. It guarantees that, in the large-$N$ limit,
the effective Hamiltonian becomes $N$ independent. 

The effective Hamiltonian can be expanded as before:
\begin{eqnarray}
{\cal H}_{\mathrm{eff}}[\varphi]
&=&\tilde{H} \varphi + {1\over 2}\int_\vp \tilde{P}^{-1}(\vp)\varphi (\vp)
\varphi(-\vp)
\\
&&+\sum_{n=3}{1\over n!}
\int_{\vp_1}\cdots\int_{\vp_n} \delta\LL(\sum_i \vp_i\RR)
\tilde{V}^{(n)}(\vp_1,\cdots,\vp_n)\varphi(\vp_1)\cdots\varphi(\vp_n)\ .
\nonumber
\end{eqnarray}
The magnetic field, the propagator, and the vertices acquire now an 
explicit $N$ dependence and we can write
\begin{eqnarray}
\tilde{H} &=& \sum_{n=1} \tilde{H}_n N^{-n}, \\
\tilde{P}^{-1}(\vp) &=& \sum_{n=0} \tilde{P}_n^{-1}(\vp) N^{-n}, \\
\tilde{V}^{(j)}(\vp_1,\ldots,\vp_j) &=& \sum_{n=0} 
   \tilde{V}_n^{(j)}(\vp_1,\ldots,\vp_j) N^{-n}.
\end{eqnarray}

\subsection{Identities among the vertices of the effective Hamiltonian}\label{identities}

Let us now derive some important relations concerning the zero-momentum
vertices $\tilde{V}_0^{(j)} (\mathbf{0},\ldots,\mathbf{0})$ and 
propagator $\tilde{P}_0^{-1}(\mathbf{0})$
[that below will be denoted by $\tilde{V}_0^{(2)} (\mathbf{0})$]. 
Since we will discuss only zero-momentum quantities, in this section
we will not write the explicit momentum dependence.

The leading behavior of ${\cal H}_{\rm eff}[\varphi]$ for $N\to\infty$ 
is obtained by considering 
the tree-level diagrams that arise from the perturbative expansion of 
\reff{App:Heff} in powers of $N$. 
If we are only interested in zero-momentum correlations
of the field $\varphi$ at leading order in $1/N$ expansion\footnote{
In order to  consider $1/N$ corrections to the vertices, one have to
include radiative contributions in which   vertices
appear  at every momenta.  },
in $\cal H$ we can simply consider the zero-momentum vertices 
and in $Z$ we need only to integrate over the five zero-momentum fields
$\Psi$. In the following we take the point of view of sec.\
\ref{AppA}, in particular taking the short-hand notation
\begin{eqnarray}
f_A &\equiv& (\gamma, \alpha, \alpha, \tau, \tau)
\nonumber
\end{eqnarray}
we have that, using the gap-equations (\ref{def-gammabar},\ref{def-alphabar},\ref{def-taubar},\ref{puntosella2}), the vertices of the theory that in principle
depend on $\beta$ and $f_A$ are parametrized only by $\h$. In particular we
can write\footnote{Here and in the following we are neglecting the dependence
on $W$ or $p_i$.}
\begin{eqnarray}
V^{(n)}_{A_1 \cdots A_n} =  V^{(n)}_{A_1 \cdots A_n}(f_A(\h), \beta(\h))
\label{vertices}
\end{eqnarray}
In (\ref{vertices})  $f_A(\h)$ and $\beta(\h)$ are given in
(\ref{def-gammabar},\ref{def-alphabar},\ref{def-taubar},\ref{puntosella2}).
We are not interested in the explicit form of $ V^{(n)}_{A_i}$ that is
given in \cite{CMP-05}, otherwise in the following we need only
the following relations:
\begin{eqnarray}
 V^{(n+1)}_{A,A_1,\cdots A_n} &=& {\partial  V^{(n)}_{A_1,\cdots A_n}
 \over \partial f_A}
\nonumber\\
{ \partial^2   V^{(n)}_{A_1,\cdots A_n}  \over  \partial \beta^2 } &=& 0
\label{second-v}
\end{eqnarray}
in particular the (\ref{second-v}) easily follows from the fact that the action
of the auxiliary fields is linear in $\beta$ \cite{CP-02}. In 
section \ref{AppA},
it was shown that the derivative of $f_A$ with
respect to $\h$ near the multicritical point is related to the zero mode 
$v_A$ of the theory 
\begin{eqnarray}
{\di f_A (\h) \over \di \h} &=& {1\over C(\h,p_i)}\left(
v_A(\h,p_i) + {\cal O}(m_0^2,p_i) \xi_A(m_0^2,p_i)
\right)
\nonumber\\
{\cal O}(m_0^2,p_i) &\sim & {\partial \beta (m_0^2,p_i)\over \partial 
m_0^2} \equiv s_1(\h,p_i)
\label{relation-oldpaper}
\end{eqnarray}
$\xi_A$ is a no-more 
specified vector while $v_A$ is the zero mode at zero external momentum.
 ${\cal O}$ is a function of $p_i$ and $\h$
that goes to zero at the MCP (\ref{relation-oldpaper}).
Since now we could neglect the explicit dependence on $m_0^2$ and $p_i$
of ${\cal O}$, $\xi_A$, $v_A$ and $C$. Using (\ref{relation-oldpaper})
and (\ref{second-v}) we recover the following equality
\begin{eqnarray}
{\di \over \di m_0^2} V^{(n)}_{A_1,\ldots,A_n} &=& 
   {1\over C } \sum_B \Big( v_B+ {\cal O} \xi_B \Big) 
         V^{(n+1)}_{B,A_1,\ldots,A_n} + s_1 {\partial \over \partial \beta}
V^{(n)}_{A_1,\ldots,A_n},
\label{identity}
\end{eqnarray}
Identity (\ref{identity}) can also be written in the compact form 
\begin{equation}
{\di {\cal H}\over \di m_0^2} = 
{\sqrt{N}\over C} \sum_B \Big( v_B+{\cal O} \xi_B \Big) \left[
    {\partial {\cal H}\over \partial \Psi_B} - \sum_A V^{(2)}_{BA}\Psi_A
    \right] +s_1 {\partial {\cal H} \over \partial \beta} 
\label{identity2}
\end{equation}
We begin by computing $d{\cal H}_{\rm eff}/dm_0^2$. We have
\begin{eqnarray}
{d{\cal H}_{\rm eff}\over dm_0^2} &=& 
   - {1\over NZ[\varphi]} \left[
   \int \prod d\Psi\, \delta'(\Psi\cdot v - \varphi \sqrt{N}) 
        {\Psi\cdot {dv\over dm_0^2}} e^{-{\cal H}} \right. 
\nonumber \\
  && \qquad \left. - 
   \int \prod d\Psi\, \delta(\Psi\cdot v - \varphi \sqrt{N}) 
        {d{\cal H}\over dm_0^2} e^{-{\cal H}} \right] - (\varphi \to 0).
\label{dHvarphi-over-dm}
\end{eqnarray}
The second term can be related to $d{\cal H}_{\rm eff}/d\varphi$, by 
using identity \reff{identity2}. Indeed defining
$\langle \cdot \rangle$ as  
\begin{equation}
\langle {\cal F}[\Psi] \rangle \equiv 
   {1\over Z[\varphi]} 
   \int d\Psi\, \delta(\Psi\cdot v - \varphi \sqrt{N}) 
        {\cal F}[\Psi]  e^{-{\cal H}} .
\end{equation}
we find
\begin{eqnarray}
{d{\cal H}_{\rm eff}\over d\varphi} &=& 
  \tilde{P}_0^{-1}\varphi + 
   {C\over N} \langle {d{\cal H}\over dm_0^2}\rangle
-{C s_1 \over N}\langle {\partial {\cal H}\over \partial \beta} \rangle
+{{\cal O} \over \sqrt N }  \sum_B \xi_B
\langle \sum_A V^{(2)}_{BA} \Psi_A-{\partial {\cal H}\over \partial \Psi_B}
\rangle
\label{dHvarphi-over-dvarphi}
\end{eqnarray}
A simple calculation allows also to write
\begin{equation}
   - {1\over NZ[\varphi]} 
   \int d\Psi\, \delta'(\Psi\cdot v - \varphi \sqrt{N}) 
        {\Psi\cdot {dv\over dm_0^2}} e^{-{\cal H}} 
   = {1\over N^{3/2}} {d\over d\varphi} \langle 
        \Psi\cdot {dv\over dm_0^2} \rangle - 
     {1\over N^{1/2}} {d{\cal H}_{\rm eff}\over d\varphi} \langle 
        \Psi\cdot {dv\over dm_0^2} \rangle,
\label{equation47}
\end{equation}
Thus, we obtain the following representation
\begin{eqnarray}
{d{\cal H}_{\rm eff}\over d m_0^2}   &=&  
  {1\over C}  {d{\cal H}_{\rm eff}\over d\varphi}    +
      {1\over N^{3/2}} {d\over d\varphi} \langle 
        \Psi\cdot {dv \over dm_0^2} \rangle - 
     {1\over N^{1/2}} {d{\cal H}_{\rm eff}\over d\varphi} \langle 
        \Psi\cdot {dv \over dm_0^2} \rangle 
\nonumber\\ 
&&-{ {\cal O}\over C\sqrt{N}}\sum_B \xi_B\langle \sum_A
V^{(2)}_{BA}\Psi_A-{\partial {\cal H}\over \partial \Psi_B}\rangle
- { {\tilde P}_0 \varphi\over C}   
+{s_1 \over N} \langle {\partial {\cal H} \over \partial\beta}\rangle
\label{resultHvarphi}
\end{eqnarray}

Now, let us show that the second term in the first line of 
(\ref{resultHvarphi}) is of order $1/N$ and can therefore 
be dropped. Since $v \cdot {d v /dm_0^2} = 0$, 
$dv/ dm_0^2$ projects over the massive subspace, i.e. 
the field $\Psi$ appearing in the mean value is a massive field. 
Thus, the leading contribution to the coefficient of $\varphi^n$ 
is due to tree-level diagrams with $n$ lines corresponding to the 
field $\varphi$ and one line to the massive field 
$\Psi\cdot {dv /dm_0^2}$. The
$N$-dependence of a tree-level graph contributing to the coefficient of 
$\varphi^n$ is given by $N^{n/2 - a}$ with 
\begin{equation}
a = {1\over 2} \sum_{k=3}^\infty (k-2)n_k,
\end{equation}
where $n_k$ is the number of $k$-leg vertices appearing in the diagram. 
Now we use the fact that, for a tree-level diagram, $N_{\rm ext} = 2 
+ \sum_{k=3} (k-2) n_k$, where $N_{\rm ext}$ is the number of 
external legs. Since $N_{\rm ext} = n + 1$, we obtain 
$a = N_{\rm ext}/2 - 1 = (n-1)/2$, independently of the diagram. 
Thus, 
\begin{equation}
  \langle \Psi\cdot {dv\over dm_0^2} \rangle  \sim N^{1/2},
\end{equation}
independently of the number of $\varphi$ legs. This proves that at leading
order in $1/N$ the 
second term can be dropped, while the third one must be kept. 
Consider now the first term of the second line of (\ref{resultHvarphi}):
\begin{equation}
{ {\cal O}\over C}\sum_{n\geq 2}{1\over n! N^{n/2 }}\langle \sum_{B\{A_i\}}V^{(n+1)}_{B A_1 \cdots A_n} \xi_B \Psi_{A_1}\cdots \Psi_{A_n}
\rangle
\label{os1} 
\end{equation}
In ~\reff{os1} we decompose
$\Psi$ on the eigenvector of $\tilde P$ (one of which is the critical
mode $v_A$, while the other four are massive modes $v^{(m)}$ with 
$m=1,2\cdots 4$) 
\begin{eqnarray}
\Psi_A &=& (v\cdot\Psi) v_A +\sum_{m=1}^4 (v^{(m)}\cdot \Psi)v^{(m)}_A
\label{evecP}
\end{eqnarray}
Considering only mass-less component in ~\reff{os1} we find the 
following $\varphi^n$ contribution:
\begin{eqnarray}
{{\cal O}\over C}\sum_{n\ge 2} \varphi^n \sum_{B\{A_i\}} V^{(n+1)}_{B A_1
\cdots A_n}\xi_B v_{A_1} \cdots v_{A_n}
\label{eq-53}
\end{eqnarray}
Otherwise inserting $l$ ($1\le l\le n$) massive vectors into ~\reff{os1}
we have a contribution that looks like:
\begin{eqnarray}
{{\cal O}\over C}\sum_{n\ge 2}{\varphi^{n-l}\over N^{l/2} }
\sum_{B\{A_i\}}V^{(n+1)}_{B A_1\cdots A_n}\xi_B v_{A_1} \cdots v_{A_{n-l}}
v^{(m_1)}_{A_{n-l+1}} \cdots v^{(m_l)}_{A_n}
\langle (v^{(m_1)}\cdot \Psi)\cdots (v^{(m_l)}\cdot \Psi) \rangle 
\label{eq-54}
\end{eqnarray}
A leading order graph that enters into the evaluation of ~\reff{eq-54} 
is a tree graph with 
$l$ external massive legs and an arbitrary number of massless one. 
It behaves independently of the massless legs like
$N^{-l/2+1}$ so that we recover an $O(N^0)$ terms in ~\reff{eq-54}
taking $l=1$. As a consequence of (\ref{eq-53}-\ref{eq-54})
(up to $1/N$) we have that the first term in the second line of 
~\reff{resultHvarphi} goes as  $\varphi^2$ for $\varphi\to 0$. 
The same considerations follows for $\partial {\cal H}/N \partial \beta $  
in (\ref{resultHvarphi}), indeed following the same previous steps one can show
that the leading order is $O(N^0)$ and that for $\varphi \to 0$ this 
behaves like $\varphi^2$.
In the following we will write:
\begin{eqnarray}
\sum_B {{\cal O} \xi_B\over C \sqrt N}\langle \sum_A
 {\partial {\cal H}\over \partial \Psi_B}-V^{(2)}_{BA}\Psi_A \rangle
+{s_1 \over N} \langle{\partial {\cal H} \over \partial \beta} \rangle
&=&\sum_{k\ge 2} {o_k \over k!}  \varphi^k + O\left(1\over N\right)
\label{def-o-piccoli}
\end{eqnarray}
with $o_k\sim s_1$ near the MCP.
Collecting all the previous results, 
at leading order in $1/N$ we have
\begin{eqnarray}
{d{\cal H}_{\rm eff}\over d m_0^2} &=& 
  {1\over C} {d{\cal H}_{\rm eff}\over d\varphi}  -
     {1\over N^{1/2}} {d{\cal H}_{\rm eff}\over d\varphi} \langle 
        \Psi\cdot {dv \over dm_0^2} \rangle 
+\sum_{n\geq 2} {o_n \over n!} \varphi^n
-{\tilde P^{-1}_0 \varphi
\over C}+O\Big({1\over N}\Big)
\label{resultHvarphi-1}
\end{eqnarray}
Since $ \langle \Psi\cdot {dv /dm_0^2} \rangle  \sim \varphi^2$ 
for $\varphi \to 0$ 
we obtain a set of differential equation for the vertices
\begin{eqnarray}
{d \tilde{V}_0^{(n)}\over dm_0^2} &=& 
  {1-\delta_{n,1}\over C} \tilde{V}_0^{(n+1)}  -
 n! \sum_{k=2}^{n-1} {c^{(0)}_{n-1-k} \over (k-1)!} \tilde{V}_0^{(k)} + o_n
\label{Main}
\end{eqnarray}
where  coherently with (\ref{def-o-piccoli}) we have 
implicitly assumed
$o_1 = o_2 =0$.
In (\ref{Main}) we have defined
\begin{eqnarray}
{1\over \sqrt N}\langle \Psi\cdot {dv \over dm_0^2} \rangle &=& \sum_{k\geq 0} c^{(0)}_k \varphi^{k+2}+O\Big({1\over N}\Big).
\label{eq-58}
\end{eqnarray}

Eq.~\reff{Main} allows us to compute the behavior close to the 
MCP of the zero-momentum vertices $\tilde{V}_0^{(k)}$ for 
$k \le {\cal N} + 3$. Using eq.~\reff{state} and the definition of the 
scaling fields (\ref{scaling-classico}) we have 
\begin{eqnarray}
{d\beta\over dm_0^2} &=& 
  \sum_{k=1}^{\cal N} k u_k {\Delta_m}^{k-1} + 
   a_{\bf 0}^{({\cal N} + 2)} ({\cal N} + 2) {\Delta_m}^{{\cal N} + 1} 
\nonumber \\
&& + O\Big[{\Delta_m}^{{\cal N} + 2},
   \Big{\{}{\Delta_m}^{i-1} \prod_{l=1}^{m_i+1} u_{k_l} \Big{\}}_{i=1,2\cdots\N},
 \{{\Delta_m}^{\N}u_k\}_{k=1,\cdots\N} \Big],
\label{dboverdm}
\end{eqnarray}
where ${\Delta_m} \equiv  m_0^2 - m_{0c}^2$. All the terms in the first line
of (\ref{dboverdm}) go as $u_h$ near the MCP; the next to leading
contribution ($\sim {u_h}^{\N+3\over \N+2}$) is obtained considering the 
three terms of the second line of (\ref{dboverdm}) with $i$ even and 
$u_{k_l}=u_{\N}$ concerning the second term, and with $u_{k}=u_{\N}$
in the third term.
We will now show that for $2\le k \le {\cal N} + 3$  
zero-momentum vertices can be expanded close to the MCP as 
\begin{eqnarray}
\tilde V_0^{(k)} = a_k {\Delta_m}^{{\cal N} + 3 - k} + 
   \sum_{j=1}^{k-2} b_j^{(k)} u_j + 
    \sum_{j=k-1}^{\cal N} b_j^{(k)} u_j {\Delta_m}^{j-k+1} + \ldots 
\label{expansion-general}
\end{eqnarray}
For $k=2$ the second  term is absent.

The (\ref{expansion-general}) can be obtained 
starting from the following equality
\begin{equation}
\tilde V^{(2)}_0 = \tilde{P}^{-1}_0 = 
   f(m_0^2,p_i) {d\beta\over dm_0^2}
\label{two}
\end{equation}
(that was demonstrated in \cite{CMP-05} and recovered in chapter
\ref{chapter1}) and using recursively
(\ref{Main}).
We notice that  for $k=1$ the r.h.s. of (\ref{Main}) cancels in
agreement with the condition $\tilde V^{(1)}_0= 0$ given by the gap
equation.
For $k=2$ expansion \reff{expansion-general} follows from 
eq.~\reff{dboverdm} and it is enough to identify 
$a_2 = f(0,0) a_{000}^{({\cal N} + 2)} ({\cal N} + 2)$, 
$b_j^{(2)} = j f(0,0)$. For $k=3$ the expansion follows 
from eq.~\reff{Main} and indeed all coefficients can be related 
with those appearing in the expansion of $\tilde V^{(2)}_0$ and 
${\cal O}$. 
We will need the explicit expression
for $a_k$ and $b_j^{(k)}$ for $j\ge k-1$. Applying eq.~\reff{Main}
recursively we obtain
\begin{eqnarray}
a_k &=& {({\cal N} + 1)! \over ({\cal N} - k + 3)!} C^{k-2} a_2 
\\
b_j^{(k)} &=& {(j-1)! \over (j - k + 1)!} C^{k-2} b_j^{(2)} ,
\label{relations-vertici}
\end{eqnarray}
where the second relation applies only for $j\ge k-1$. 
For $j< k-1$  the expression for $ b_j^{(k)}$ is not so simple, indeed
it will appear also contributions coming from ${\cal O}$ and 
higher derivative of $f$.
  However we are not interested
in the explicit expression of $ b_j^{(k)}$ for $j\ge k-1$ because this
coefficients are not involved in the cancellations of 
the translated
theory. This  will be
clarified in the next section \ref{zero-mode}.

\subsection{Translated zero-mode Hamiltonian}\label{zero-mode}

In the previous section we have shown that all zero-momentum
vertices with $k < {\cal N}+ 3$ vanish at the MCP. We shall now perform
a field redefinition in order to have 
$V^{({\cal N} + 2)}(\mathbf{0},\ldots,\mathbf{0}) = 0$ for all 
values of the parameters. Proceeding in a similar
way of what done in chapter \ref{chapter3} eq.\ (\ref{eq_3.13}) 
we write
\begin{equation}
\varphi(\vp) = \alpha \chi(\vp) + k \delta(\vp).
\label{def-kappa}
\end{equation} 
In terms of $\chi$, the effective Hamiltonian has the expansion
\begin{eqnarray}
{\cal H}_{\mathrm{eff}}[\chi]
&=&\ol H \chi + {1\over 2}\int_\vp \ol {P}^{-1}(\vp)\chi (\vp)
\chi(-\vp) \\
&&+\sum_{n=3}{1\over n!}
\int_{\vp_1}\cdots\int_{\vp_n} \delta\LL(\sum_i \vp_i\RR)
\ol V^{(n)}(\vp_1,\cdots,\vp_n)\chi(\vp_1)\cdots\chi(\vp_n).\nonumber
\end{eqnarray}
The parameters $\alpha$ and $k$ are determined by requiring 
that 
\begin{eqnarray}
&& \ol V^{({\cal N} + 2)}(\mathbf{0},\ldots,\mathbf{0}) = 0,
\label{cond-shift}
\\
&& \ol P^{-1}(\vp)-\ol P^{-1}(\mathbf{0}) = \mathbf{p}^2 +
O(\mathbf{p}^4)
\end{eqnarray}
for all values of the parameters.
Condition (\ref{cond-shift}) is satisfied if 
\begin{eqnarray}
\sum_{j=0}^\infty 
   {k^j \over j!}
   \tilde{V}^{({\cal N} + 2 + j)} (\mathbf{0},\ldots,\mathbf{0}) = 0.
\end{eqnarray}
Since $\tilde{V}^{({\cal N} + 2)} (\mathbf{0},\ldots,\mathbf{0})$ vanishes
at the MCP for $N\to\infty$, close to the MCP we can expand $k$ as 
\begin{equation}
k = \sum_{n=1}^\infty k_n \left[
   \tilde{V}^{({\cal N} + 2)} (\mathbf{0},\ldots,\mathbf{0}) \right]^n.
\end{equation}
The first term is easily computed:
\begin{equation}
k_1 = - {1 \over 
         \tilde{V}^{({\cal N} + 3)}_0 (\mathbf{0},\ldots,\mathbf{0})} + 
         O(1/N).
\end{equation}
by using eq.~\reff{Main} and eq.~\reff{expansion-general} 
it is easy to see that 
\begin{equation}
k = - {1\over C} {\Delta_m} + O[{\Delta_m}^2, u_i,N^{-1}].
\label{k1exp}
\end{equation}
Since $k$ vanishes at the MCP for $N\to \infty$, we have at the MCP
\begin{equation}
\ol V^{(j)}(\mathbf{0},\ldots,\mathbf{0}) = 
\tilde{V}^{(j)}(\mathbf{0},\ldots,\mathbf{0}) + O(1/N),
\end{equation}
for any $j \ge {\cal N} + 3$. 
In principle one can observe that due to the leading behaviour of $k$
(\ref{k1exp}) $\ol V^{(j)}$ will have the same general expansion of 
$\tilde V^{(j)}$ (\ref{expansion-general}), for $j \le {\cal N} + 2$:
\begin{eqnarray}
\ol V^{(j)} &=&\sum_{i=1}^{j-2}\ol b^{(j)}_i u_i+ 
\sum_{i=j-1}^{\N}\ol b^{(j)}_i u_{i}{\Delta_m}^{i-j+1}
+\ol a_j {\Delta_m}^{\N-j+3}
+\cdots
\label{exp-naive}
\end{eqnarray}
 On the other hand in this case 
  an important cancellation
occurs (for $j \le {\cal N} + 2$). 
Indeed one can show that $\ol b^{(j)}_i=0$ for $i\ge j$ and that $\ol a_j=0$
so that at leading order in $u_h$ one has $\ol V^{(j)}\sim u_{j-1}$. This 
implies that the translated vertices vanishes faster as ${\Delta_m} \to 0$.

We start from 
\begin{eqnarray}
\overline{V}^{(j)} = \alpha^j \sum_{n=0}^{{\cal N} + 3 - j}
   {k^n \over n!} \tilde{V}_0^{(n + j)}  + 
   O(k^{{\cal N} + 4 - j},N^{-1}).
\end{eqnarray}
The leading contribution independent of the scaling fields is given by
[using the notations given in (\ref{expansion-general})]
\begin{eqnarray}
&& \alpha^j \sum_{n=0}^{{\cal N} + 3 - j} {k^n a_{n+j} \over n!}
   ({\Delta_m})^{{\cal N} + 3 - j - n} + 
   O[({\Delta_m})^{{\cal N} + 4 - j},N^{-1}]  = 
\nonumber \\
&& \alpha^j {a_2 ({\cal N} + 1)! C^{j-2} \over ({\cal N} + 3 - j)!}
    (C k + {\Delta_m})^{{\cal N} + 3 - j} + 
    O[({\Delta_m})^{{\cal N} + 4 - j} ,N^{-1}]  =
    O[({\Delta_m})^{{\cal N} + 4 - j},N^{-1}],
\nonumber\\
\label{can1}
\end{eqnarray}
where we have used eqs.~\reff{relations-vertici} and \reff{k1exp}.
As for the contributions proportional to the scaling fields $u_i$,
no cancellation occurs for $i \le j - 2$. If $i \ge j - 1$
the coefficient of $u_i$ is given by
\begin{eqnarray}
&& \alpha^j \sum_{n=0}^{i - j + 1} {k^n b^{(n+j)}_i \over n!}
   ({\Delta_m})^{i - j - n + 1} + 
   O[({\Delta_m})^{i - j + 2},N^{-1}]  = 
\nonumber \\
&& \alpha^j {b_i^{(j)} (i-1)! C^{j-2} \over (i - j + 1)!}
    (C k + {\Delta_m})^{i - j + 1} + 
    O[({\Delta_m})^{i - j + 2},N^{-1}]  =
    O[({\Delta_m})^{i - j + 2},N^{-1}]
\label{can2}
\end{eqnarray}
The previous equations (\ref{can1}-\ref{can2}) show that $\ol a_j=0$ and
$\ol b^{(j)}_i=0$ if $i\ge j$ in (\ref{exp-naive}) confirming the starting 
claim.
Otherwise other cancellations appear  at next to leading order, so
that the general structure of the translated vertices is not given by
the corrections of (\ref{exp-naive}). This is investigated in the next 
sub-section \ref{translated} and in \ref{appB} where the 
structure of $k$ is investigated in detail, in particular it is
shown that the ${\Delta_m}^2$ correction of $k$ in (\ref{k1exp}) is due
only to contribution coming from $C$.
The strategy of the next sub-section \ref{translated} is to generalize
the set of differential equations for $\tilde V^{(j)}$ (\ref{Main}) to
 to the translated one $\ol V^{(j)}$. Near the MCP this set of differential
equations decouple so that we can easily recover the cancellations discovered
in this section. Otherwise using the new set of equations one is able 
to discover other cancellations. One can think to have in mind the final 
results for the translated vertices (\ref{result},\ref{result-a},
\ref{result-b}) and in particular the translated vertices for the 
$\N =3$ case (\ref{Three-1},\ref{Three-2},\ref{Three-3}) that will
be used in the final section of this chapter \ref{sec-six}, where
$1/N$  corrections for the scaling fields and the scaling 
variables
will be explicitly built for the ${\cal N}=3$ case.

\subsection{Structure of the translated vertices: another derivation}\label{translated}
For convenience in this section we will take $\alpha =1$
in eq.~\reff{def-kappa} (at the end  $\ol V^{(j)}$  must
 be rescaled by $\alpha^j$).
One can try to solve in a more direct way  the structure 
of the translated vertices $\ol V^{(j)}$. 
In a similar way of what done in previous sections we define
\begin{eqnarray}
Z[\chi] &=& \int \di \Psi \delta({\bf \psi} \cdot v-\sqrt N \chi-\sqrt N k) e^{-{\cal H}}
\\
{\cal H}_{\mathrm{eff}} &=& -{1\over N}\log {Z[\chi]\over Z[0]}
\end{eqnarray}
and we follow the same steps of sec \ref{identities}. The first change
is in  equation ~\reff{dHvarphi-over-dm}. Using an obvious definition
for $\langle\cdot\rangle$ and ~\reff{equation47} we have:
\begin{eqnarray}
{\di{\cal H}_{\mathrm{eff}}\over \di \h}& =& 
{1\over N}\langle{\di {\cal H} \over \di\h} \rangle
-{1\over \sqrt N}{\di {\cal H}_{\mathrm{eff}} \over \di\chi }\langle
\Psi\cdot{\di v\over \di \h } \rangle
+{\di {\cal H}_{\mathrm{eff}} \over \di\chi }{\di k\over \di \h}
+O\left(1\over N\right)
\nonumber\\
&&- \Big[\,\chi \to 0 \,\Big]
\label{equation47Tr}
\end{eqnarray}
On the other hand ~\reff{dHvarphi-over-dvarphi} does not change
apart the substitution $\varphi \to \chi + k $ (notice however that the
explicit $k$ dependence disappear in ~\reff{equation47Tr} because the subtraction in the second line).
 So we generalize ~\reff{resultHvarphi-1} in:
\begin{eqnarray}
{\di {\cal H}_{\mathrm{eff}}\over \di \h} &=&\Bigg( {1\over C}+{\di k\over \di \h}\Bigg) {\di {\cal H}_{\mathrm{eff}}\over \di \chi} - {\tilde V^{(2)}_0\over C} \chi +
\sum_n {o_n\over n!} (\chi + k)^n-{1\over \sqrt N}{\di {\cal H}_{\mathrm{eff}} \over \di\chi }\langle
\Psi\cdot{\di v\over \di \h } \rangle
\nonumber\\
&&+O\Bigg({1\over N} \Bigg) - \Big[\,\chi \to 0 \,\Big]
\end{eqnarray}
Using ~\reff{eq-58} we are now able to generalize  ~\reff{Main} for
the translated vertices:
\begin{eqnarray}
k_s(\h)\ol V^{(n+1)} &=&  {\di \ol V^{(n)}\over \di \h} +n!\Bigg[ {c^{(0)}_0 \ol V^{(n-1)}\over (n-2)!}+
{c^{(0)}_1 \ol V^{(n-2)}\over (n-3)!}+ \cdots+{c^{(0)}_{n-2} \ol V^{(1)}}
\nonumber\\
&&\qquad\qquad +k \Bigg(2 {c^{(0)}_0 \ol V^{(n)}\over(n-1)!}+3
{c^{(0)}_1 \ol V^{(n-1)}\over(n-2)!}+\cdots + (n+1)c^{(0)}_{n-1}\ol V^{(1)} \Bigg)
\nonumber\\
&&\qquad\qquad+{k^2\over 2} \Bigg(2 {c^{(0)}_0 \ol V^{(n+1)}\over n!}+6
{c^{(0)}_1 \ol V^{(n)}\over(n-1)!}+\cdots + (n+1)n c^{(0)}_{n-1}\ol V^{(1)} \Bigg)
\nonumber\\
&&\qquad \qquad +O(k^3)\Bigg]
\nonumber\\
o_j \mathrm{\,with\,} j\ge 2
&& +\Big[ o_n +(n+1)  o_{n+1} k +(n+2)(n+1)
o_{n+2} {k^2\over 2} + O(k^3)
\Big]
\nonumber\\
&& +{\tilde V^{(2)}_0 \over C} \delta_{n,1}
\label{MainTr}
\end{eqnarray}
In (\ref{MainTr}) we have defined
\begin{eqnarray}
k_s(\h) &=& {1\over C} + {\di k\over \di \h}.
\end{eqnarray}
$k_s$ is studied in detail in sec.\ \ref{appB}. There we have 
shown that
\begin{eqnarray}
k_s(\h)
&=& {\alpha c^{(0)}_0(\N+2)\over A_0 C} u_\N 
+{c^{(0)}_0{\Delta_m}^2\over  C^2}
 + O\left( {\Delta_m}^3,{u_h}^{3\over \N+2}\right)
\label{ks}
\\
k_s(\h)-c^{(0)}_0 k^2 &=& {\alpha c^{(0)}_0(\N+2)\over A_0 C} u_\N 
+O\left( {\Delta_m}^3,{u_h}^{3\over \N+2}\right)
\label{kss}
\end{eqnarray}
in (\ref{ks}) the ${\Delta_m}$ term cancels in a non trivial way, while in
(\ref{kss}) also the ${\Delta_m}^2$ term disappears.

We have just noticed that 
 $\ol V^{(j)}$ in principle has the same structure of $\tilde V^{(j)}$
(\ref{exp-naive}) near the MCP. Otherwise 
we notice that at leading order in $u_h$ due
to the relations (\ref{ks}) and (\ref{kss}), eq.\  (\ref{MainTr}) becomes 
\begin{eqnarray}
{\di \ol V^{(j)}\over \di \h} = 0 &\Longrightarrow& \ol V^{(j)} \sim u_{j-1} 
\label{MainTr-leading}
\end{eqnarray}
so that we recover the cancellations discovered in the
previous section \ref{zero-mode}:
\begin{eqnarray}
\ol b^{(j)}_i = 0 &\qquad\qquad & i=j,j+1,\cdots \N
\label{b-traslated}
\\
\ol a_j =0. &&
\label{a-traslated}
\end{eqnarray}
We can complete ~\reff{b-traslated} with
\begin{eqnarray}
\ol b^{(j)}_{i}=b^{(j)}_{i} &\qquad\qquad & i=1,\cdots j-1
\end{eqnarray}
that follows from the fact that $k$ goes to zero
at the MCP.
If $n=1$  (\ref{MainTr}) becomes:
\begin{eqnarray}
{\di \ol V^{(1)}\over\di\h } = -{1\over C}\tilde V^{(2)}_0- 2 o_2 k 
\approx -{1\over C}\tilde V^{(2)}_0
&\qquad&  \ol V^{(1)} = b^{(1)}_0 u_h
\label{eq_4.62}
\end{eqnarray}
where we have used the gap equation (in the following $u_0\equiv u_h$).
Eq.\ (\ref{eq_4.62}) is expected if compared
with similar results for higher order vertices eq.\ 
(\ref{MainTr-leading}).

Now we consider corrections to ~\reff{MainTr-leading}. 
In particular we have to compute at what order (in ${\Delta_m}$)
$u_j$ enters into $\ol V^{(k)}$ for $j>k-1$, i.e.\ we want
to compute $\chi^{(j)}$ and $\chi^{(j)}_i$ ($j\le i \le \N$) in the 
following expression
\begin{eqnarray}
\ol V^{(j)} = \sum_{i=1}^{j-1}b^{(j)}_{i} u_{i}+ {\cal V}^{(j)}_0 
{\Delta_m}^{\chi^{(j)}}
+ \sum_{i=j}^{\N} {\cal V}^{(j)}_i 
u_i {{\Delta_m}}^{\chi^{(j)}_i}.
\label{result}
\end{eqnarray}
Writing~\reff{MainTr} for $n=\N+2$ (taking into account the constraint
$\ol V^{(\N+2)}=0$) we have:
\begin{eqnarray}
(k_s(\h)-k^2)\ol V^{(\N+3)}&=&c^{(0)}_0(\N+2)(\N+1)(1+O(k))\ol V^{(\N+1)}
\nonumber\\
&&+c^{(0)}_1(\N+2)(\N+1)\N(1+O(k))\ol V^{(\N)}+\cdots
\label{eq83}
\end{eqnarray}
this show the importance of cancellations that happen in~\reff{ks}
and~\reff{kss}, indeed the previous equations tell us that the ${\Delta_m}^2$
term in $\ol V^{(\N+1)}$ disappears according with (\ref{a-traslated}).
However in principle solving ~\reff{eq83} one is able to compute the 
following expression
\begin{eqnarray}
k_s(\h)-c^{(0)}_0 k^2 = \sum_{i=1}^{\N} {\cal K}_i u_i
 + {\cal K}_0 {\Delta_m}^3
\label{develop-eta}
\end{eqnarray}
that can be recovered simply using the leading behaviour of the
translated field. Then one can try to find $\chi^{(j)}_i$.
If we consider $n=1$ case of (\ref{MainTr})
[defining $\eta(\h)\equiv k_s(\h)-c^{(0)}_0 k(\h)^2 $]
\begin{eqnarray}
\eta \ol V^{(2)}={\di \ol V^{(1)}\over \di \h }+ \alpha \tilde V^{(2)}
\end{eqnarray} 
where $\alpha$ is an appropriate constant that can be obtained including
all the proper terms in (\ref{MainTr}).
Remembering that $\tilde V^{(2)}= f(\h) s_1$, we easily obtain
that the correction to the leading behavior $\ol V^{(1)}\sim u_h$ is of order
${\Delta_m} $ and comes from $f(\h)$ so that
\begin{eqnarray}
\chi^{(1)}_i  = i+1 &\qquad & \chi^{(1)}= \N+3. 
\label{chi-1}
\end{eqnarray} 
Considering the $n=2$ case of (\ref{MainTr}):
\begin{eqnarray}
\eta \ol V^{(3)}={\di \ol V^{(2)}\over \di \h }
+o_2 \Big[ 1 +O(k)\Big]
+2 c^{(0)}_0
\ol V^{(1)} +2 k c^{(0)}_0 \ol V^{(2)}.
\end{eqnarray}
The leading corrections comes from $o_2$ terms so we find
\begin{eqnarray}
\chi^{(2)}_i = i &\qquad & \chi^{(2)}= \N+2.
\label{chi-2}
\end{eqnarray}
Using  (\ref{MainTr}) is then quite easy to compute the terms that
appear in the expansion (\ref{result}). Indeed we can rewrite the 
differential equation (\ref{MainTr}) including only the terms we are
interested in as
\begin{eqnarray}
{\di \overline V^{(n)}_0 \over \di \h}= o_n + t_n {\Delta_m}^3 \ol V^{(n+1)}_0
+\sum_{j=n+2}^{\N+3} t_{j,n }{\Delta_m}^{j+1-n} \ol V^{(j)}_0.
\end{eqnarray} 
Integrating the previous equation,
using $\ol V^{(j)}_0 \sim u_{j-1}$, we get
\begin{eqnarray}
\ol V^{(n)}_0({\Delta_m})-\ol V^{(n)}_0(0) &=& \Big[\int_{0}^{{\Delta_m}}o_n \Big] 
+ {t_n\over 4} {\Delta_m}^4 u_n +\sum_{j=n+1}^{\N} {t_{j+1,n }\over j+3-n} {\Delta_m}^{j+3-n} u_j 
\nonumber\\
&&+ {t_{\N+3}\over \N+5-n} \ol V^{(\N+3)}_0 {\Delta_m}^{\N+5-n}. 
\label{prepre}
\end{eqnarray} 
Using the previous equation (\ref{prepre}) 
we are able to give our
final result (for $j>2$)
\begin{eqnarray}
\chi^{(j)}_j &=& \mathrm{Min}[j,4]
\nonumber\\
\chi^{(j)}_i &=& \mathrm{Min}[i,i+3-j] \qquad j\neq i
\label{result-a}
\\
\chi^{(j)} &=& \mathrm{Min}[\N+2,\N+5-j].
\label{result-b}
\end{eqnarray}

To finish this section we give an explicit expression of the translated 
vertices for the $\N=3$ case that will be studied in detail in the rest of
this chapter:
\begin{eqnarray}
\ol V^{(1)} &=& b^{(1)}_0 u_h +{\cal V}^{(1)}_1{\Delta_m}^2 u_1 
+ {\cal V}^{(1)}_2{\Delta_m}^3 u_2+{\cal V}^{(1)}_3{\Delta_m}^4 u_3
+{\cal V}^{(1)} {\Delta_m}^6
\label{Three-1}
\\
\ol V^{(2)} &=& b^{(2)}_1 u_1 +{\cal V}^{(2)}_2 u_2 {\Delta_m}^2 
+{\cal V}^{(2)}_3 u_3 {\Delta_m}^3 + {\cal V}^{(2)} {\Delta_m} ^5  
\label{Three-2}
\\
\ol V^{(3)} &=& b^{(3)}_1 u_1 + b^{(3)}_2 u_2
+{\cal V}^{(3)}_3 u_3 {\Delta_m}^3 + {\cal V}^{(3)} {\Delta_m} ^5
\label{Three-3}
\\
\ol V^{(4)} &=& b^{(4)}_1 u_1 + b^{(4)}_2 u_2 + b^{(4)}_2 u_3 
+{\cal V}^{(4)}{\Delta_m}^4. 
\label{Three-4}
\end{eqnarray}

\subsection{$k(\h)$ near the MCP}\label{appB}

%%%%%%%%%%%%%%%%%%%%%%%%%%%%%%%%%%%%%%%%%%%%%%%%%%%%%%%%%%%%%%%%%%%%%%%%%%%%%
%%%%%%%%%%%%%%%%%%%%%%%%%%%%%%%%%%%%%%%%%%%%%%%%%%%%%%%%%%%%%%%%%%%%%%%%%%%%%
%%%%%%%%%%%%%%%%%%%%%%%%%%%%%%%%%%%%%%%%%%%%%%%%%%%%%%%%%%%%%%%%%%%%%%%%%%%%%

In this section we want to investigate the structure of $k(\h)$
(\ref{def-kappa}) as a power of ${\Delta_m}$ in order to recover
the claims of the previous section [see eqs.\ (\ref{ks}) and
(\ref{kss})].
 This is the fundamental
issue in order to investigate the algebraic structure of the translated 
vertices $\overline V^{(j)}$. In particular we want to show that the 
${\Delta_m}$ and ${\Delta_m}^2$ terms that enter into the expression of 
$k_s(\h)$ near the MCP are due only to the $C(\h)$ expansion near the MCP.
 We want to investigate only the 
structure in ${\Delta_m}$ [notice that the scaling fields $u_i$ trivially appear
linearly in $k(\h)$] so in the following we can
 think to put $u_i=0$. This
 allows us to think to the expansion (\ref{kappa-app}) as an expansion
in ${\Delta_m}$. Otherwise the considerations are general in the MCCL, the
expression (\ref{kappa-app}) being an expansion in $u_h$.

 An easy computation gives us:
\begin{eqnarray}
k(\h)&=& -{ \tilde V^{(\N+2)} \over \tilde V^{(\N+3)} }-{1\over 2}{\tilde V^{(\N+4)}\over \tilde V^{(\N+3)}} \left(
\tilde V^{(\N+2)} \over \tilde V^{(\N+3)} \right)^2+ \left( {1\over 6}{\tilde V^{(\N+5)}\over  \tilde V^{(\N+3)}}-
{1\over 2} \left[ { \tilde V^{(\N+4)}\over  {\tilde V}^{(\N+3)} } \right]^2 
\right)\left(
\tilde V^{(\N+2)} \over \tilde V^{(\N+3)} \right)^3  
\nonumber\\
&&+O\left( {\Delta_m}^4 \right) 
\label{kappa-app}
\end{eqnarray}
In this subsection we limit ourself  to $\N>1$. In this 
case we have
that $\di \beta / \di {m_0}^2$ is taken into account in  the 
${\Delta_m}^4$ term in (\ref{kappa-app}). This allows us 
to neglect in 
(\ref{Main}) the terms that are order $s_1$ near the MCP
\begin{equation}
{\di \tilde V^{(n)}_0 \over \di {m_0}^2 } = {1\over C(\h)} \tilde V^{(n+1)}_0
-n(n-1)c^{(0)}_0 \tilde V^{(n-1)}_0 + \cdots
\label{Main-simplified}
\end{equation}
where the dots mean that in the following the corrections can be
neglected. Using (\ref{Main-simplified}) and (\ref{kappa-app}) we get
a very easy expression
\begin{eqnarray}
{\di k(\h) \over \di m_0^2} &=& -{1\over C(\h)} + c^{(0)}_0(\N^2+3\N+2)
{\tilde V^{(\N+1)}_0 \over \tilde V^{(\N+3)}_0 } 
- c^{(0)}_0{\N^2+3\N\over 2} \Bigg[{\tilde V^{(\N+2)}_0 \over
V^{(\N+3)}_0 } \Bigg]^2
\nonumber\\
&&+O\left({\Delta_m}^3,{1\over N}\right)
\end{eqnarray}
In the previous expression we have taken the convention used in this 
work
\begin{eqnarray}
\tilde V^{(n)} &=& \tilde V^{(n)}_0 + O\left( 1\over N \right)
\nonumber
\end{eqnarray}
As a straightforward consequence of (\ref{Main-simplified}),
 we have that
\begin{eqnarray}
{\tilde V^{(\N+1)}_0 \over \tilde V^{(\N+3)}_0 }  &=&
{1\over 2 C(\h)^2} {\Delta_m}^2 + O({\Delta_m}^3)
\\
\Bigg[{\tilde V^{(\N+2)}_0 \over
V^{(\N+3)}_0 } \Bigg]^2 &=&
{1\over  C(\h)^2} {\Delta_m}^2 + O({\Delta_m}^3),
\end{eqnarray}
so that we recover the wanted result 
\begin{eqnarray}
{\di k(\h) \over \di m_0^2} &=& -{1\over C(\h)} + {c^{(0)}_0\over C(\h)^2} {\Delta_m}^2
+O({\Delta_m}^3)
\label{appb-1}
\\
k_s(\h) &\equiv& {1\over C(\h)} + {\di k(\h) \over \di \h}
\nonumber\\
&=&{c^{(0)}_0{\Delta_m}^2\over  C(\h)^2} + O({\Delta_m}^3),
\end{eqnarray}
and finally
\begin{eqnarray}
k_s(\h)-k(\h)^2 &=& O({\Delta_m}^3).
\label{appb-2}
\end{eqnarray}
In (\ref{appb-1}) the ${\Delta_m}$ is due only to $C(\h)$, this disappears
with the ${\Delta_m}^2$ term in the (\ref{appb-2}).

\section{Scaling Fields and the large $N$ limit}\label{sec-six}

In the previous section we have obtained 
a detailed description of the zero mode Hamiltonian
near the MCP. We are now in a position to apply the results
of chapter \ref{chapter2} concerning the multicritical points.
In particular we have verified that at the ${\cal N}$-th order 
 multicritical point 
all the vertices $\tilde V_0^{(j)}(\mathbf{0})$ (with $j=1,\cdots\N+2$)
go to zero in a proper way that has been careful characterized 
above.\footnote{We are neglecting $1/N$ corrections 
and the counterterms that one needs to include 
in the effective action
in order to regularize
the theory. This contributions will be deleted including corrections
to the $N=\infty$ multicritical point.}
This is a necessary condition in order to apply the results
of chapter \ref{chapter2} concerning the crossover
to multicritical points.

Taking a pedagogical point of view we will give a detail description of the $\N=3$ case 
for which one can apply the  results of 
sec.\ \ref{1.2}. Following the same step of sec.\
\ref{sec6.6}, we want to investigate {\em i}) the scaling variables 
that describe
the crossover between classical behavior to Mean Field behavior
eq.\ (\ref{scalingeq}), {\em ii}) the correction to the critical
 point $p_i
=p_{ic}$ and $\beta=\beta_c$ given by (\ref{eq-punto-multicritico})
and {\em iii}) the $1/N$ corrections to the classical
scaling fields defined in (\ref{scaling-fields-classici}) and
to the classical scaling relations (\ref{scaling-classico}). 
We have just pointed out at the beginning of this chapter that the
cancellations we have found in the previous sections are very important
to solve the point {\em iii}).

%%%%%%%%%%%%%%%%%%%%%%%%%%%%%%%%%%%%%%%%%%%%%%%%%%%%%%%%%%%%%%%%%%%%%%%%%%%
%%                N=3
%%%%%%%%%%%%%%%%%%%%%%%%%%%%%%%%%%%%%%%%%%%%%%%%%%%%%%%%%%%%%%%%%%%%%%%%%%%

\subsection{${\cal N}=3$ case in two dimension}

For the ${\cal N}=3$ the translated vertices 
near the MCP has been explicitly given in eqs.\
(\ref{Three-1}), (\ref{Three-2}) and (\ref{Three-3}).
Using them we are now in a position to apply the result 
of sec.\ \ref{1.2}, in particular the results for the Hamiltonian 
(\ref{starting}) can be used with the following identification
(for $i=1,2,3,4$)
\begin{eqnarray}
u_{0i} = \ol V^{(i)}(\mathbf{0})
&\qquad & u_6 = {\ol V^{(6)}(\mathbf{0})\over N^{2}}.
\label{identification-multicritico}
\end{eqnarray}
Notice in particular that the normalization conditions assumed in
eq.\ (\ref{starting}) are satisfied due to the definition 
of $u_6$ (\ref{identification-multicritico}) and to eq.\ 
(\ref{def-kappa}) that ensures $\ol V^{(5)}(\mathbf{0})=0$ 
and $K(\vp)\approx\vp^2$.
We will consider only the two dimensions. Otherwise, as pointed out
for ${\cal N}=1$, for $d>2$ one has to translate the theory to impose
that the renormalized five legs vertex is zero. 
The scaling variable are then simply given by
\begin{eqnarray}
x_h &=& -{ N^2 \over \alpha }(\ol H - h_c) 
\label{eq0}
\\
x_1 &=&{ N^2 \over \alpha^2 } (\ol V^{(2)}(\mathbf{0}) - u_{c1})
\label{eq1}
\\
x_2 &=&{ N^2 \over \alpha^3 }(\ol V^{(3)}(\mathbf{0})-u_{c2})
\label{eq2}
\\
x_3 &=&{ N^2 \over  \alpha^4 }(\ol V^{(4)}(\mathbf{0}) - u_{c3})
\label{eq3}
\end{eqnarray}
where $u_{ci}$ have been computed using a lattice regularization
in chapter \ref{chapter4}. We are not interested in the exact
form of this variable but we will develop them in the most
general way around the MCP.
Notice that in the computation of  $u_{c1}$ and $u_{c3}$ we are
not able to give $1/N^2$ terms: indeed it would require an hard fine tuning
not available until now. On the other hand $h_c$ and $u_{c1}$ can be 
obtained imposing $\mathbb{Z}_2$ symmetry at the transition.

{\bf Scaling fields equations.} 
Using the results of the tree-level translated theory we have:
\begin{eqnarray}
x_h &=& N^{5/2}\Big(  b^{(1)}_0 u_h +{\cal V}^{(1)}_1{\Delta_m}^2 u_1 
+ {\cal V}^{(1)}_2{\Delta_m}^3 u_2+{\cal V}^{(1)}_3{\Delta_m}^4 u_3
+{\cal V}^{(1)} {\Delta_m}^6 +\cdots \Big)
\nonumber\\
&&+N^{3/2}(H_0+ H_1{\Delta_m}+\cdots)+N^{1/2}\cdots
\label{eqexp0}
\\
x_1 &=& N^{2}  \Big( b^{(2)}_1 u_1 +{\cal V}^{(2)}_2 u_2 {\Delta_m}^2 
+{\cal V}^{(2)}_3 u_3 {\Delta_m}^3 + {\cal V}^{(2)} {\Delta_m} ^5  +\cdots \Big) 
\nonumber\\
&&+N(U_0+U_1 {\Delta_m}+\cdots)
\label{eqexp1}
\\
x_2 &=&  N^{3/2}  \Big(b^{(3)}_1 u_1 + b^{(3)}_2 u_2
+{\cal V}^{(3)}_3 u_3 {\Delta_m}^3 + {\cal V}^{(3)} {\Delta_m} ^5   +\cdots \Big)
\nonumber\\
&&+N^{1/2}(D_0+D_1{\Delta_m} +\cdots)
\label{eqexp2}
\\
x_3 &=&  N  \Big(b^{(4)}_1 u_1 + b^{(4)}_2 u_2 + b^{(4)}_3 u_3 
+{\cal V}^{(4)}{\Delta_m}^4+\cdots   \Big)
\nonumber\\
&&+(T_0+T_1{\Delta_m}) 
\label{eqexp3}
\end{eqnarray}
In the previous expression we do not have considered the possibility
of cancellations at one loop order, this is not necessary for the
 considerations we are going to do.
 Otherwise if one is searching for particular symmetries
some cancellation could be model dependent.\footnote{
For instance if $\N=1$ and the action is $\mathbb{Z}_2$ symmetric, 
then in chapter \ref{chapter5} we will show that some cancellations
are present also at one loop order. These cancellations give
the obvious result $h_c = 0$ (i.e.\ explicit symmetry is not
broken).
}

We notice that at the classical level, the scaling fields $u_i$ 
are not defined in a unique way. Indeed we are able to re-define:
\begin{eqnarray}
u'_i &=&k_i u_i + \sum_{j=0}^{i-1} f_j u_j  
\end{eqnarray}
so that we can eventually resum \emph{at classical level} the $b^{(i)}_j$
contribution in (\ref{eqexp0}-\ref{eqexp3}). On the other hand 
in order to cancel
the ${\cal V}^{(i)}_l$ terms in (\ref{eqexp0}-\ref{eqexp3}) we have to
include $1/N$ correction in the definition of the scaling fields.

{\bf Correction to the critical point.} Solving (\ref{eqexp1}-\ref{eqexp3}) in $u_i$ and putting the result into
(\ref{eqexp0}) we argue that ${\Delta_m} \sim m_0 / N^{1/5}$  in the MCCL.
Using then (\ref{eqexp1},\ref{eqexp2},\ref{eqexp3}) we find:
\begin{eqnarray}
u_{i} \sim  {u^{(0)}_{i}\over N} + O\left(1\over N^{6/5} \right) 
&\qquad& i=1,2
\label{corr-CP1}
\\
u_{3} \sim {u^{(0)}_{3}\over N^{4/5}}+ O\left(1\over N\right) &&
\label{corr-CP2}
\end{eqnarray}
with:
\begin{eqnarray}
{m_0}^5 &=&-{H_0\over b^{(1)}_0 a^{(5)}_{000}}
\label{corr-CP3}
\\
u^{(0)}_1 &=&-{{\cal V}^{(2)}{m_0}^5+U_0\over b^{(2)}_1 }
\label{corr-CP4}
\\
u^{(0)}_2 &=&{({\cal V}^{(2)}-{\cal V}^{(3)}){m_0}^5+U_0-D_0\over b^{(3)}_2 }
\label{corr-CP5}
\\
u^{(0)}_3 &=&-{{\cal V}^{(4)}{m_0}^4\over b^{(4)}_3}
\label{corr-CP6}
\end{eqnarray}
The previous equations (\ref{corr-CP1}-\ref{corr-CP6}) are the leading
correction to the position of the critical point (notice they are independent
of $x_i$). Otherwise the position of the critical point ($u_{ci}$)
can be obtained
studying (\ref{eqexp0}-\ref{eqexp3}) with $x_i=0$. It is not hard to
see that for $u_{ci}$ we have the following  developments in 
powers of $N^{-1/5}$: 
\begin{eqnarray}
{\Delta_m}_c &=&{m_{0c} \over N^{1/5} } +{m_{1c} \over N^{2/5} } +
{m_{3c}\over N^{3/5} } +\cdots 
\label{CP-0}
\\
u_{1c} &=& {u^{(0)}_{1c}\over N } +{u^{(1)}_{1c}\over N^{6/5} } +
{u^{(2)}_{1c}\over N^{7/5} } +\cdots 
\label{CP-1}
\\
u_{2c} &=&{u^{(0)}_{2c}\over N } +{u^{(1)}_{2c}\over N^{6/5} } +
{u^{(2)}_{2c}\over N^{7/5} } +\cdots 
\label{CP-2}
\\
u_{3c} &=&{u^{(0)}_{3c}\over N^{4/5} } +{u^{(1)}_{3c}\over N } +
{u^{(2)}_{3c}\over N^{6/5} } +\cdots 
\label{CP-3}
\end{eqnarray}
in particular $m_{0c}=m_0$ and $u^{(0)}_{ic}=u^{(0)}_{i}$ 
 (\ref{corr-CP3}-\ref{corr-CP6}).

{\bf Redefinitions of the scaling fields.} Now we 
take $x_i\neq 0$ in (\ref{eqexp0}-\ref{eqexp3}).
Defining $\tilde m\equiv {\Delta_m} - {\Delta_m}_c$  
$\tilde u_i\equiv u_i - u_{ic}$, taking into account 
 (\ref{CP-0}-\ref{CP-3}) the (\ref{eqexp0}-\ref{eqexp3}) now become:

\begin{eqnarray}
{x_h \over b^{(1)}_0 N^{5/2}} &=& {m_{0c}\over N^{1/5}}\tilde u_1 + 
{{m_{0c}}^2\over N^{2/5}} \tilde u_2 + {{m_{0c}}^3\over N^{3/5}} \tilde u_3 +
 5 a^{(5)}_{000} {{m_{0c}}^4\over N^{4/5}}
 \tilde m
+\cdots
\label{eqexpN0}
\\
{x_1 \over N^2} &=& b^{(2)}_1 \tilde u_1+{{\cal V}^{(2)}_2{m_{0c}}^2\over N^{2/5} }\tilde u_2  + {{\cal V}^{(2)}_3{m_{0c}}^3\over N^{3/5} }\tilde u_3+5{{\cal V}^{(2)}{m_{0c}}^4\over N^{4/5} }\tilde m 
+\cdots
\label{eqexpN1}
\\
{x_2 \over N^{3/2}} &=& b^{(3)}_1 \tilde u_1+b^{(3)}_2\tilde u_2+{{\cal V}^{(3)}_3{m_{0c}}^3\over N^{3/5} }\tilde u_3 +5{{\cal V}^{(3)}{m_{0c}}^4\over N^{4/5} }\tilde m
+\cdots
\label{eqexpN2}
\\
{x_3\over N} &=& b^{(4)}_1 \tilde u_1+b^{(4)}_2\tilde u_2+b^{(4)}_3\tilde u_3
+{4{\cal V}^{(4)}{m_{0c}}^3\over N^{3/5}}\tilde m
+\cdots
\label{eqexpN3}
\end{eqnarray}
where dots indicate sub-leading contributions.
In principle one can solve the previous equations (\ref{eqexpN0}-\ref{eqexpN3})
obtaining the following expansion:
\begin{eqnarray}
\tilde m &=&  \sum_{i,j,k,l} {\cal M }_{i  j   k l} {x_3}^i
{x_2}^j  {x_1}^k {x_h}^l
\label{sviluppo-a}
\\
\tilde u_p &=&  \sum_{i,j,k,l} {\cal U }^{(p)}_{i  j   k l} {x_3}^i
{x_2}^j  {x_1}^k {x_h}^l
\label{sviluppo-c}
\end{eqnarray}
with ${\cal M},{\cal U}\to 0$ when $N\to \infty$. Following
chapter \ref{chapter3} and \cite{CMP-05}, then one tries to solve the
previous mixing defining new scaling fields that have to take into
account $1/N$ corrections respect the mean-field definition.
 As a first step we compute (at leading order) the linear
terms (in which only one $x_i$ appears) in
(\ref{sviluppo-a}-\ref{sviluppo-c}). We easily obtain the following
expressions:
\begin{eqnarray}
\tilde u_1 &=&  {\cal U}^{(1)}_{x_1}{x_1 \over N^{2} } +
 {\cal U}^{(1)}_{x_2}{x_2 \over N^{19/10} } +
 {\cal U}^{(1)}_{x_3}{x_3 \over N^{8/5} } +
 {\cal U}^{(1)}_{x_h}{x_h \over N^{5/2} }+\cdots
\nonumber\\
{\cal U}^{(1)}_{x_1}&\equiv& N^2 {\cal U }^{(1)}_{0 0 10} = {1\over b^{(2)}_1 }
+O\left(1\over N^{1/10}\right)
\nonumber\\
{\cal U}^{(1)}_{x_2}&\equiv& N^{19/10}{\cal U }^{(1)}_{0 100} =
       { {m_{0c}}^2 {\cal V}^{(2)}\over a^{(5)}_{000} b^{(3)}_2 b^{(2)}_1 } 
- {{m_{0c}}^2 {\cal V}^{(2)}_2\over b^{(3)}_2 b^{(2)}_1} + O\left(1\over N^{1/10}\right)
\nonumber\\
{\cal U}^{(1)}_{x_3} &\equiv& N^{8/5}{\cal U }^{(1)}_{1 000} =
      { {m_{0c}}^3 {\cal V}^{(2)}\over a^{(5)}_{000} b^{(4)}_3 b^{(2)}_1 } 
- {{m_{0c}}^3 {\cal V}^{(2)}_3\over b^{(4)}_3 b^{(2)}_1} +O\left(1\over N^{1/10}\right)
\nonumber\\
{\cal U}^{(1)}_{x_h} &\equiv& N^{5/2}{\cal U }^{(1)}_{0001} =
-{{\cal V}^{(2)}\over a^{(5)}_{000} b^{(2)}_1 b^{(1)}_0}  +O\left(1\over N^{1/10}\right)
\label{u1}
\end{eqnarray}

\begin{eqnarray}
\tilde u_2 &=&  {\cal U}^{(2)}_{x_1}{x_1 \over N^{2} } +
 {\cal U}^{(2)}_{x_2}{x_2 \over N^{3/2} } +
 {\cal U}^{(2)}_{x_3}{x_3 \over N^{8/5} } +
 {\cal U}^{(2)}_{x_h}{x_h \over N^{5/2} }+\cdots
\nonumber\\
{\cal U}^{(2)}_{x_1}&\equiv& N^2 {\cal U }^{(2)}_{0 0 10} = 
-{b^{(3)}_1\over b^{(2)}_1 b^{(3)}_2}+O\left(1\over N^{1/10}\right)
\nonumber\\
{\cal U}^{(2)}_{x_2}&\equiv& N^{3/2}{\cal U }^{(2)}_{0 100} =
{1\over b^{(3)}_2}
+O\left(1\over N^{1/10}\right)
\nonumber\\
{\cal U}^{(2)}_{x_3} &\equiv& N^{8/5}{\cal U }^{(2)}_{1 000} =
{b^{(3)}_1 {m_{0c}}^3 {\cal V}^{(2)}_3-b^{(2)}_1 {\cal V}^{(3)}_3 {m_{0c}}^3\over
b^{(2)}_1 b^{(3)}_2 b^{(4)}_3}
-{{m_{0c}}^3\over a^{(5)}_{000} b^{(2)}_1 b^{(3)}_2 b^{(4)}_3}
(b^{(3)}_1 {\cal V}^{(2)}-b^{(2)}_1  {\cal V}^{(3)} )
\nonumber\\
&&+O\left(1\over N^{1/10}\right)
\nonumber\\
{\cal U}^{(2)}_{x_h} &\equiv& N^{5/2}{\cal U }^{(2)}_{0001} =
{b^{(3)}_1 {\cal V}^{(2)}\over a^{(5)}_{000} b^{(1)}_0 b^{(2)}_1
b^{(3)}_2 }- {{\cal V}^{(3)}\over a^{(5)}_{000} b^{(1)}_0 b^{(3)}_2 } 
+O\left(1\over N^{1/10}\right)
\label{u2}
\end{eqnarray}

\begin{eqnarray}
\tilde u_3 &=&  {\cal U}^{(3)}_{x_1}{x_1 \over N^{2} } +
 {\cal U}^{(3)}_{x_2}{x_2 \over N^{3/2} } +
 {\cal U}^{(3)}_{x_3}{x_3 \over N } +
 {\cal U}^{(3)}_{x_h}{x_h \over N^{23/10} }+\cdots
\nonumber\\
{\cal U}^{(3)}_{x_1}&\equiv& N^{2} {\cal U }^{(3)}_{0 0 10} = 
 {4 {\cal V}^{(4)}\over  5 a^{(5)}_{000} b^{(2)}_1 b^{(4)}_3 }+
 {b^{(3)}_1 b^{(4)}_2 \over  b^{(2)}_1 b^{(3)}_2 b^{(4)}_3  }-
 {b^{(4)}_1 \over  b^{(2)}_1 b^{(4)}_3  }
+O\left(1\over N^{1/10}\right)
\nonumber\\
{\cal U}^{(3)}_{x_2}&\equiv& N^{3/2}{\cal U }^{(3)}_{0 100} =
 -{b^{(4)}_2 \over b^{(3)}_2 b^{(4)}_3 }
+O\left(1\over N^{1/10}\right)
\nonumber\\
{\cal U}^{(3)}_{x_3} &\equiv& N {\cal U }^{(2)}_{1 000} =
{1\over b^{(4)}_3}
+O\left(1\over N^{1/10}\right)
\nonumber\\
{\cal U}^{(3)}_{x_h} &\equiv& N^{23/10}{\cal U }^{(3)}_{0001} =
- {4 {\cal V}^{(4)} \over  5 a^{(5)}_{000} b^{(1)}_0 {m_{0c}} b^{(4)}_3 } 
+O\left(1\over N^{1/10}\right)
\label{u3}
\end{eqnarray}

\begin{eqnarray}
\tilde m &=&  {\cal U}^{(m)}_{x_1}{x_1 \over N^{14/10} } +
 {\cal U}^{(m)}_{x_2}{x_2 \over N^{11/10} } +
 {\cal U}^{(m)}_{x_3}{x_3 \over N^{4/5} } +
 {\cal U}^{(m)}_{x_h}{x_h \over N^{17/10} }+\cdots
\end{eqnarray}
\begin{eqnarray}
{\cal U}^{(m)}_{x_1}&\equiv& N^{14/10} {\cal M }_{0 0 10} = 
-{1 \over 5 a^{(5)}_{000} {m_{0c}}^3 b^{(2)}_1 }
+O\left(1\over N^{1/10}\right)
\nonumber\\
{\cal U}^{(m)}_{x_2}&\equiv& N^{11/10}{\cal M }_{0 100} =
-{1 \over 5 b^{(3)}_2 {m_{0c}}^2  a^{(5)}_{000} }
+O\left(1\over N^{1/10}\right)
\nonumber\\
{\cal U}^{(m)}_{x_3} &\equiv& N^{4/5} {\cal M }_{1 000} =
-{1\over 5 a^{(5)}_{000} m_{0c} b^{(4)}_3 }
+O\left(1\over N^{1/10}\right)
\nonumber\\
{\cal U}^{(m)}_{x_h} &\equiv& N^{17/10}{\cal M }_{0001} =
{1\over 5 a^{(5)}_{000} {m_0}^4 b^{(1)}_0}
+O\left(1\over N^{1/10}\right)
\label{m}
\end{eqnarray}

We see that at leading order in $1/N$ (\ref{u3}) becomes:
\begin{eqnarray}
\overline u_3 \equiv \tilde u_3 \sim {\cal U}^{(3)}_{x_3} {x_3 \over N}
\label{u3-sf}
\end{eqnarray}
so that the new scaling field ($\overline u_3$) does not 
need any corrections
with respect the older one ($\tilde u_3$). The same happen for 
$\overline u_2$:
\begin{eqnarray}
\overline u_2 \equiv \tilde u_2 \sim {\cal U}^{(2)}_{x_2} {x_2 \over N^{3/2}}
\label{u2-sf}
\end{eqnarray}
In the expression for $\tilde u_1$ (\ref{u1}) we see that the terms in
$x_2$ and $x_3$ are sub-leading with respect to $x_1$. 
So one has to include 
correction to this scaling fields in the following way [using
(\ref{u3-sf}) and (\ref{u2-sf})]:
\begin{eqnarray}
\overline u_1 \equiv \tilde u_1-{1\over N^{2/5}}{{\cal U}^{(1)}_{x_2}\over 
{\cal U}^{(2)}_{x_2}}\tilde u_2 - {1\over N^{3/5}}{{\cal U}^{(1)}_{x_3}\over 
{\cal U}^{(3)}_{x_3}}\tilde u_3 \sim {\cal U}^{(2)}_{x_1} {x_1 \over N^{2}}
\label{u1-sf}
\end{eqnarray}
Notice that in the previous expression 
${\cal U}^{(1)}_{x_2}/{\cal U}^{(2)}_{x_2}$
and
${\cal U}^{(1)}_{x_3}/{\cal U}^{(3)}_{x_3}$ need to be computed up
to corrections of order $N^{-3/5}$ and $N^{-2/5}$ with respect
the value reported in (\ref{u1},\ref{u2},\ref{u3}).
This implies that if one wants to compute exactly the mixing
coefficients in (\ref{u1-sf}) one needs to include in (\ref{eqexpN0}-\ref{eqexpN3}) 
almost $N^{-3/5}$ corrections to the leading behaviour, in particular
also radiative corrections will enter in the effective computation.
Now we have to discuss the effect of having discarded the non-linear terms
in (\ref{sviluppo-a}) and (\ref{sviluppo-c}). But it is easy to see that
[including contributions that correct (\ref{eqexpN1})] 
for instance the ${x_3}^2$
contribution scales as $1/N^{22/10}$. This shows that this contributions
can be neglected.

Now we pass to discuss the $\tilde u_h$ field. Using the gap-equation
and (\ref{eqexp0}) we include corrections to  (\ref{eqexpN0}) 
in the following way
\begin{eqnarray}
{x_h\over b^{(1)}_0 N^{5/2}} =\tilde u_h +{ {\cal H}_1\over N^{2/5}} \tilde u_1+
{ {\cal H}_2\over N^{3/5}} \tilde u_2 + 
{ {\cal H}_3\over N^{4/5}} \tilde u_3 + 
{ {\cal H}^{(1)}_m\over N} \tilde m+{ {\cal H}^{(2)}_m\over N^{4/5}}
\tilde m \tilde m 
\label{x-acca}
\end{eqnarray}
Notice that in the previous expansion (\ref{x-acca}), 
the $\tilde u_3 \tilde u_3$ term ($\sim {x_3}^2 N^{-2}$)
is not present; indeed
it has been adsorbed in the definition of 
$u_h$. At leading order in $N$ we have (\ref{eqexp0}-\ref{eqexp3})
\begin{eqnarray}
{\cal H}_1,{\cal H}_2,{\cal H}_3&=& {\cal V}^{(1)}_1 {m_{0c}}^2,{\cal V}^{(1)}_2 
{m_{0c}}^3,{\cal V}^{(1)}_3 {m_{0c}}^3  
\\
{\cal H}^{(1)}_m &=& 6 m_{0c}^5{\cal V}^{(1)}+H_1
\\
{\cal H}^{(2)}_m &=& 15 m_{0c}^4 {\cal V}^{(1)}
\end{eqnarray}
while in order to compute completely the mixing-coefficients many other 
contributions (in $1/N^{10}$) are needed. Using then (\ref{x-acca})
and (\ref{m}) we can define the new scaling field $\overline u_h$ 

\begin{eqnarray}
\overline u_h&=&\tilde u_h -{1\over {\cal U}^{(1)}_{x_1} } \Bigg( { {\cal H}_1 {\cal U}^{(1)}_{x_1}\over N^{2/5}} +{ {\cal H}^{(1)}_m {\cal U}^{(m)}_{x_1}
\over N^{2/5}} 
\Bigg)\overline u_1 -
{1\over {\cal U}^{(2)}_{x_2} }  
\Bigg({{\cal H}_1 {\cal U}^{(1)}_{x_2} \over N^{4/5}} +{{\cal H}_2 {\cal U}^{(2)}_{x_2} \over N^{3/5}}+{{\cal H}_3 {\cal U}^{(3)}_{x_2} \over N^{4/5}} 
\nonumber\\
&&+{ {\cal H}^{(1)}_m {\cal U}^{(m)}_{x_2}
\over N^{3/5}}
\Bigg)\overline u_2
-{1\over {\cal U}^{(3)}_{x_3} }  
\Bigg({{\cal H}_1 {\cal U}^{(1)}_{x_3} \over N} 
+{{\cal H}_2 {\cal U}^{(2)}_{x_3} \over N^{6/5}}
+{{\cal H}_3 {\cal U}^{(3)}_{x_3} \over N^{4/5}}
+{ {\cal H}^{(1)}_m {\cal U}^{(m)}_{x_3}
\over N^{4/5}}
\Bigg)
\overline u_3
\nonumber\\
&&
-{{\cal H}^{(2)}_m \over ({\cal U}^{(3)}_{x_3})^2}{1\over N^{2/5}} 
\overline u_3^2
\end{eqnarray}
so that we can write:
\begin{eqnarray}
\ol u_h &\sim& {x_h\over N^{5/2}}
\end{eqnarray}

\section{Conclusions}

In this chapter we have investigate the critical behavior of
$O(N)$ models with a generalized ferromagnetic nearest neighbour
interaction depending on $\N$ tunable parameters that undergo a phase
transition at finite temperature (see chapter \ref{chapter1}).
In sec.\ \ref{sec-four} we have studied in detail the algebraic structure
of the zero Mode Hamiltonian near the critical point, parametrizing 
the vertices using the scaling fields of the $N=\infty$ theory
(chapter \ref{chapter1}). Using the results of chapter \ref{chapter2} in
two dimension, in sec.\ \ref{sec-six} we have applied the theory
of Multicritical Crossover Limit to the specific case $\N=3$.
The approach we have given is only algebraic (for instance we 
do not have taken care of the explicit form of $u_{ci}$\footnote{
The explicit form of $u_{ic}$ has been computed in chapter
\ref{chapter2}} but we have 
taken the most general behavior near the MCP) but shows 
the steps that one need to follow in order to study the critical
limit when $N\to \infty$. In particular it shows the definition 
of the scaling fields and corrections to the critical point that in 
principle could be compared with available numerical results 
(in a similar way of what reported in sec.\ \ref{sec9}).
Using the scheme presented in this chapter we believe several physical
system could be investigate. For instance finite temperature 
Yukawa model with $N_f$ fermions (for $N_f\to \infty$)
has been investigate in chapter \ref{chapter5}. There we have shown
as the system, for every $N_f$ finite, undergoes an Ising phase transition 
that crosses on a Mean Field phase transition for $N_f=\infty$ in the
same way of the ${\cal N}=1$ models introduced in chapter \ref{chapter1}
and studied in chapter \ref{chapter3}. It is usually claimed that
if  one tune also a chemical  potential the phase diagram exhibit also
a tricritical point. We plan to apply the method developed in this chapter 
for the investigation of this multicritical point.

%%%%%%%%%%%%%%%%%%%%%%%%%%%%%%%%%%%%%%%%%%%%%%%%%%%%%%%%%%
%%%%%%%%%%%%%%%%%%%%%%%%%%%%%%%%%%%%%%%%%%%%%%%%%%%%%%%%%%
%%%%%%%%%%%%%%%%%%%%%%%%%%%%%%%%%%%%%%%%%%%%%%%%%%%%%%%%%%
%%%                                                    %%%
%%%                                                    %%%
%%% CHAPTER 5 - CHAPTER 5 - CHAPTER 5 - CHAPTER 5 - CH %%%
%%%                                                    %%% 
%%%                                                    %%%
%%%%%%%%%%%%%%%%%%%%%%%%%%%%%%%%%%%%%%%%%%%%%%%%%%%%%%%%%%
%%%%%%%%%%%%%%%%%%%%%%%%%%%%%%%%%%%%%%%%%%%%%%%%%%%%%%%%%%
%%%%%%%%%%%%%%%%%%%%%%%%%%%%%%%%%%%%%%%%%%%%%%%%%%%%%%%%%%

\chapter{ Fermionic Models with chiral symmetry}\label{chapter5}

In the previous chapters we have investigated several problems related
to the $1/N$ expansion in models with global symmetry. 
In order to investigate such  kinds of problems we 
have used generalized Heisenberg models. 
However we believe that the presentation
given is  general 
because in the effective action for the zero mode 
(translated or not) noone symmetries was assumed, but the most
general expansions was taken.\footnote{This is not 
completely correct. We have seen that in $O(N)$ models the sign of
the effective four leg vertex at the Critical Point has the same sign of
the mass term of the zero mode near the CP [the claim follows
from eq.\ (\ref{id-V4-CP})]. In principle this fact could 
not necessary holds for other models which share with Heisenberg
models the existence of a zero mode. } For this reason we have tried
to investigate the large $N$ limit of other Infra Red divergent models 
\cite{MZ-03}. In this chapter we present our analysis of fermionic
models with global symmetry at finite temperature. This kind
of studies could be significant in order to investigate the nature of the
QCD phase diagram.
The finite-temperature transition in QCD has been extensively studied in the
last twenty years and is becoming increasingly important because of the 
recent experimental progress in the physics of ultrarelativistic
heavy-ion collisions. Some general features of the transition, which is 
associated with the restoration of chiral symmetry, can be 
studied in dimensionally-reduced three-dimensional models 
\cite{PW-84,BPV-03}. However, a detailed understanding 
requires a direct analysis in QCD. Being the phenomenon
intrinsically nonperturbative, our present knowledge comes mainly from 
numerical simulations \cite{Karsch-02,KL-03}. Due to the many technical
difficulties---finite-size effects, proper inclusion of fermions, etc.
---results are not yet conclusive and thus it 
is worthwhile to study  simplified
models that show the same basic features but are significantly 
simpler.

\section{The model}

In this chapter we shall consider a Yukawa model in which
$N_f$ fermions are coupled with a scalar field through a Yukawa 
interaction.
\begin{eqnarray}
{\cal S} &=& D N_{f}
   \int\di^{d+1} \mathbf{x}\,\Big({1\over 2}
      (\partial \phi)^2 + {\mu \over 2} \phi^2 
       +{\lambda  \over 4!} \phi^4\Big) 
+\sum_{f=1}^{ N_f} \int\di^{d+1}\mathbf{x}\,
        \ol \psi_{f} \Big(\sc{\partial}+ g \phi+M\Big)\psi_{f}
\nonumber\\
\label{action}
\end{eqnarray}
where $\sc{\partial}\equiv \sum_{\mu=1}^{d+1} \gamma_\mu \partial_\mu$
and $\gamma_\mu$, $\mu=1,\cdots,d+1$ are the generator of a Clifford
Algebra satisfying the anti-commutation relations
$\{\gamma_\mu,\gamma_\nu \}=2\delta_{\mu,\nu}\mathbf{1}$
where
$\tr \gamma_\mu^2 =  D$,\footnote{One can chooses a representation 
so that $D = 2^{d/2}$ if $d$ is even, 
$D = 2^{(d+1)/2}$ if $d$ is odd} and $\mathbf{1}$ is the identity $D\times D$
matrix. The integration is over 
$\mathbb{R}^{d}\times [0,T^{-1}]$, and $\lambda \ge 0$ to ensure the 
stability of the quartic potential.  Along the
\emph{thermal} direction we take periodic boundary conditions for the
bosonic field $\phi$ and antiperiodic ones for the fermionic fields
$\psi_f$
\begin{eqnarray}
\phi(\mathbf{x},0)=\phi(\mathbf{x}, T^{-1})&\quad \psi(\mathbf{x},0)=-\psi(\mathbf{x}, T^{-1})\quad & \ol\psi(\mathbf{x},0)=-\ol\psi(\mathbf{x}, T^{-1}).
\label{periodic}
\end{eqnarray}
The theory must be properly regularized. We shall consider a 
sharp-cutoff regularization, restricting the momentum integrations in the 
\emph{spatial} directions to $p < \Lambda$. 
However, the discussion presented here can be extended without 
difficulty to any other 
regularization that maintains at least a remnant of chiral symmetry.
For instance an  explicit lattice calculation involving staggered fermions
has been considered. However all the computations strictly follow
the presentation of this chapter without any new consideration;\footnote{
The only significant differences are related to the doubling
problem (partially solved by staggered fermions) that
gives additional degrees of freedom.} in particular
the relations that protect the symmetry of the theory are
satisfied also for chiral fermions on the lattice
(in the following this fact will guarantee that $M_c=0$ for every $N_f$).

Given a representation of the Clifford Algebra in even 
dimension (see e.g.\ \cite{Zinn-Justin}), 
 $\gamma_{\cal C}$ is introduced
\begin{eqnarray}
\gamma_{\cal C}:=\gamma_1\gamma_2\cdots\gamma_d,
\label{gamma-five}
\end{eqnarray}
so that it  anti-commutes with
all the generators of the  algebra 
$\{\gamma_{\cal C}\,,\, \gamma_\mu  \} =0$,
$\mu=1,2\cdots d$.
On the other hand
an odd dimensional representation can be built by using
 the standard $d-1$-dimensional representation, plus
 $\gamma_{\cal C}$ (\ref{gamma-five})
 multiplied by a proper $\mathbb{C}$ number
\begin{displaymath}
\gamma_\mu(2d+1) = \left\{
\begin{array}{ll}
\gamma_u(2d) & \mathrm{if}\,\, \mu<2d+1
\\
\sqrt{(-1)^k}\gamma_{\cal C}(2d) & \mathrm{if}\,\, \mu=2d+1.
\end{array}
\right. 
\end{displaymath}

The interaction (\ref{action}) has a discrete symmetry for $M=0$. If 
$d$ is even this can be realized by the following transformations
\begin{eqnarray}
\psi \to  \gamma_{\cal C}\, \psi 
\qquad &
\ol\psi \to  -\ol\psi\,\gamma_{\cal C}
& \qquad
\phi \to - \phi
\label{symm_even}
\end{eqnarray}
In odd dimension $\gamma_{\cal C}$ is trivial $\gamma_{\cal C}
=(-1)^k\mathbf{1}$. Actually
the previous symmetry (\ref{symm_even})  can be realized 
using a $\gamma_i$ 
coupled  with a reflection in the Euclidean space along the 
$i$-th direction \cite{Zinn-Justin}:
\begin{eqnarray}
&\psi \to  \gamma_{i}\, \psi
\qquad \quad
\ol\psi \to  -\ol\psi\,\gamma_{i}
\qquad \quad
\phi \to - \phi &
\nonumber\\
& \{x_1,\cdots x_i, \cdots x_{d+1}\} \to \{x_1,\cdots - x_i, 
\cdots x_{d+1} \} &
\label{symm_odd}
\end{eqnarray}
We present an unified picture 
using (\ref{symm_odd}) with $i=d+1$. We are interested to study the 
reduced-dimensional action which is obtained integrating out the 
{\em thermal} component so that
 the minus sign appearing in (\ref{symm_odd})
can be neglected.
The previous symmetries (\ref{symm_even}-\ref{symm_odd}) protect
by fermions mass generation and can be investigate studying the expectation
value of the bosonic field $<\phi>$ that will be used as an order parameter
to investigate the phase diagram of the model (\ref{action}).

In the limit $N_f\to \infty$ this model can be solved analytically 
and one finds that there is a range of parameters in which 
it shows a transition analogous to that observed in QCD \cite{RSY-94,MZ-03}. 
It separates a low-temperature phase in which chiral symmetry
is broken from a high-temperature phase in which chiral symmetry is 
restored. For $N_f=\infty$ this transition shows mean-field behavior, 
in contrast with general arguments that predict the transition to 
belong to the Ising universality class. This apparent contradiction
was explained in Ref.~\cite{KSS-98} where, by means of scaling arguments,
it was shown that the width of the Ising critical region scales as 
a power of $1/N_f$, so that only mean-field behavior can be observed 
in the limit $N_f = \infty$.

This picture is strictly like what we have  
observed in a generalized $O(N)$ $\sigma$ model in
the first part of this work: for finite values of $N$ the 
transition was expected 
to be in the Ising universality class, while the $N=\infty$ solution
predicted mean-field behavior. In chap.\ \ref{chapter3} we performed a 
detailed calculation of the $1/N$ corrections, explaining the observed 
behavior in terms of a critical-region suppression.
The analytic technique presented in this work can be applied to 
model (\ref{action}). It allows us to obtain an analytic description of the 
crossover from mean-field to Ising behavior that occurs when $N_f$ is large
and to extend the discussion of Ref.~\cite{KSS-98} to the case $M\not = 0$.
More importantly, we are able to show that the phenomenon is universal.
In field-theoretical terms, it can be characterized as a crossover 
between two fixed points: the Gaussian fixed point and the Ising fixed point.
This implies that quantitative predictions for model (\ref{action}) 
can be obtained in completely different settings. One can use field theory
and compute the crossover functions by resumming the perturbative series
\cite{BB-85,BBMN-87,BB-02,PRV-99}. Alternatively, one can use the fact that 
the field-theoretical crossover is equivalent to the critical 
crossover that occurs in models with medium-range interactions
\cite{LBB-96,PRV-98,PRV-99,CCPRV-01,PV-02}. This allows one to use the 
wealth of results available for these spin systems
\cite{LBB-96,LBB-97,LB-98,PRV-98,PRV-99,PV-02}.
Finally, we should note that the phenomenon is quite general and occurs in any 
situation in which there is a crossover from the Gaussian fixed point to a 
nonclassical stable fixed point. For instance, similar considerations have been
recently presented for finite-temperature QCD in some very specific limit
\cite{Bringoltz-05}, while for other models the scenario presented above is
questionable \cite{CJ-06}.

In sec.~\ref{sec2} we review the 
behavior in the limit $N_f = \infty$. In sec.~\ref{sec3} we consider the 
$1/N_f$ fluctuations and determine the effective theory of the 
excitations that are responsible for 
the Ising behavior at the critical point. These modes are 
described by an effective weakly-coupled $\phi^4$ Hamiltonian.
These considerations are applied to the Yukawa model in sec.~\ref{sec4.2}
and in sec.~\ref{sec4.3}. We determine the relevant scaling variables 
and show how to compute the crossover behavior of the correlation functions.
In the appendix  \ref{App} we carefully 
discuss the relations among medium-ranged spin models, field theory,
and the Yukawa model. The presentation given  strictly
follows Ref.~\cite{CMP-05b} and  Ref.~\cite{CMP-06}.

%%%%%%%%%%%%%%%%%%%%%%%%%%%%%%%%
%                              %
% diagramma di fase N=infinito %
%                              %
%%%%%%%%%%%%%%%%%%%%%%%%%%%%%%%%

\section{Behavior for $N_f = \infty$} \label{sec2}

The solution of the model for $N_f = \infty$ is quite standard.
We briefly summarize here the main steps, following the presentation 
of Ref.~\cite{MZ-03}. As a first step we integrate the fermionic fields 
obtaining an effective action ${\cal S}_{\rm eff}[\phi]$ given by
\begin{eqnarray}
e^{-D N_{f}{\cal S}_{\rm eff} [\phi]} 
&=& \int \prod_{f=1}^{N_{f}}\di\ol\psi_{f}\di\psi_{f}\,
e^{-{\cal S}[\phi,\ol\psi,\psi]}, 
\end{eqnarray}
where
\begin{eqnarray}\label{S-chiral-fermions}
{\cal S}_{\rm eff}[\phi] &=& \int\di^{d+1} \mathbf{x}
\Big({1\over 2}(\partial\phi) ^2  + {\mu\over 2}  \phi ^2 
      +{ \lambda  \over 4!} \phi^4 \Big)
- {1\over D} \int\di^{d+1} \mathbf{x}\,
  \tr \log \Big(\sc{\partial} + g \phi+M\Big).
\end{eqnarray}
For $N_f\to \infty$ one can expand around the saddle point $\phi = \ol \phi$,
that is determined by the gap equation ${\delta{\cal S}_{\rm eff}[\ol\phi]/
\delta \phi =0 }$ 
\begin{equation}\label{saddle_a}
\ol \mu m+{\ol \lambda \over 6 }m^3 =  (m+M) T \sum_{n\in\mathbb{ Z}}
\int_{p < \Lambda} 
   {\di^{d}\vp \over (2 \pi)^{d}} 
\,{1\over {p}^2 + \omega_n^2 + (m+M)^2},
\end{equation}
where we define the {\em frequencies} $\omega_n \equiv (2 n + 1) \pi T$, and
\begin{eqnarray}
m \equiv  g\,\ol\phi\qquad
\ol\mu \equiv  \mu\,g^{-2}\qquad
\ol\lambda \equiv  \lambda\, g^{-4}\; .
\nonumber
\end{eqnarray}
The action corresponding to a saddle-point solution $m$ is:
\begin{eqnarray} 
\overline{\cal S}_{\rm eff} (m,M,T) &=& 
  {\ol\mu\over 2}m^2+{\ol \lambda\over 4!}m^4
-{T\over 2}\sum_{n\in \mathbb{Z}}\int_{p<\Lambda} 
{\di^{d} \mathbf{p}\over (2\pi)^d} 
   \log \left[{{p}^2 + \omega_n^2 +(m+M)^2 \over 
              {p}^2 + \omega_n^2} \right] ,
\label{eq5}
\end{eqnarray}
where we have added a mass-independent counterterm to regularize
the sum \cite{MZ-03}. Such a quantity has been chosen so that 
the action for $M=m=0$ vanishes.
Summations can be done analytically using the identity\footnote{
The (\ref{sum-continuum}) can be proved considering the identity
$0=\pi\int_{\cal C} \di z  {\mathrm{ctg} (\pi z)\over (z+1/2)^2 + x^2}$,  
that holds considering for instance ${\cal C}$ a 
rectangular in the complex plain with the vertex on semi-integer 
number [i.e.\ $v_i=(m_i+1/2, i n_i+1/2 )$, $m_i,n_i\in\mathbb{Z}$
and $i=1,2\cdots 4$],
and taking the limit $m_i,n_i\to\pm \infty$, with the proper sign
so that ${\cal C}$ encloses all the plane.
Using Cauchy theorem one can evaluate the series (\ref{sum-continuum}) 
computing the residues of the previous integrand at $z=-1/2 \pm i x$. } 
\begin{equation}
\sum_n {x\over (n+1/2)^2+x^2} = \pi \tanh (\pi x )
=\pi\Big(1-{2\over e^{2\pi x}+1}\Big).
\label{sum-continuum}
\end{equation}
The gap equation can then be written as 
\begin{eqnarray}\label{gap}
p(m) &=& (m+M)\, {\cal G}(m+M,T),
\end{eqnarray}
where the functions $p(m)$ and
${\cal G}(x,T)$ are defined by
\begin{eqnarray}
p(m) &\equiv &\ol\mu \,m+{\ol \lambda \over 6 }m^3 ,
\label{p} \\
{\cal G}(x,T) &\equiv & \int_{p < \Lambda}
{\di^{d}\vp\over (2\pi)^d} \,
 {1\over \sqrt{p^2 + x^2} }\LL({1\over 2}-{1\over e^{\sqrt{p^2 + x^2}/T}+1} 
  \RR) \; .
\label{defG}
\end{eqnarray}
Analogously we can rewrite eq.~(\ref{eq5}) as
\begin{eqnarray}
\overline{\cal S}_{\rm eff} (m,M,T) &=& 
    {\ol \mu\over 2 } m^2+{\ol \lambda\over 4! } m^4
\nonumber \\
&& -
  T \int_{p<\Lambda} {\di^{d} \mathbf{p}\over (2 \pi)^d}\,
    \left[
   \log \left(  \cosh {\sqrt{p^2 + (m + M)^2}\over 2T} \right) - 
   \log  \cosh {p \over 2T } \right]\; .
\label{ffe}
\end{eqnarray}
Using eqs.~(\ref{gap}) and (\ref{ffe}) 
we can determine the phase diagram of the model.
Given $\ol \mu$ and $\ol \lambda$, for each value of $T$ and $M$ we 
determine the solutions $m$ of the gap equations.
 When the solutions are more 
than one, the physical one is that with the lowest 
action $\overline{\cal S}_{\rm eff} (m,M,T)$.
Note that eqs.~(\ref{gap}) and (\ref{ffe}) are invariant 
under the transformations $m\to -m$ and $M\to -M$. Thus, we can limit our
study to the case $M\ge 0$.
In general, we can find either one solution or three different solutions 
$m_0$, $m_+$, and $m_-$ with $m_-\le m_0\le m_+$ (for some specific values 
of the parameters two of them may coincide). The previous claim can be proved
using the following identities (see app.\ \ref{app-G-cont}) 
\begin{eqnarray}\label{property-G}
{\di\over \di x }x\,{\cal G}(x,T) \ge 0 
&\qquad&
{\di^2 \over \di x^2}x\,{\cal G}(x,T) \theta(x) \le 0 \, , 
\end{eqnarray}
[$\theta(x)$ is the step function]
and straightforward consideration on the gap equation (\ref{gap}).
There are four different possibilities:
\begin{figure}
\begin{center}   
\includegraphics[angle=0,scale=0.9]{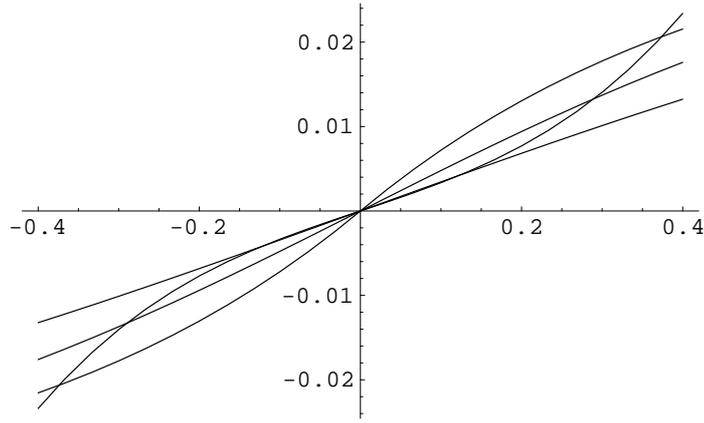}
\caption{Plot of $p(m)$ and $m{\cal G}(m)$ for $C_0 \Lambda^{d-1} > \ol \mu >0$ taking different temperature. The symmetry is broken at low temperature 
($m_+ \neq 0$)}
\label{GapCont3}
\end{center}
\end{figure}
\begin{itemize}
\item[(a)]
If $\ol \mu \ge {\cal G}(0,0) = C_0 \Lambda ^{d-1}$ with
\nonumber\\
\begin{eqnarray}
C_0 &\equiv & \Big[ 2^{d}\pi^{d/2}(d-1)\Gamma\Big({d\over 2} \Big)
\Big]^{-1},
\label{der}
\end{eqnarray}
then, for every $M\ge 0$, there is only one solution $m_+\ge 0$; for 
$M = 0$ we have $m_+=0$.
\item[(b)] 
If $0 < \ol \mu < C_0 \Lambda ^{d-1}$, there exists a critical temperature
$T_c(\ol \mu)$ (see fig.\ \ref{GapCont3}). For $T > T_c(\ol \mu)$ and any $M$ 
there is only one solution $m_+\ge 0$ (for $M = 0$ we have $m_+=0$).
For $T < T_c(\ol \mu)$ and $M < \tilde{M}$  there are three solutions 
$m_0$, $m_+$, and $m_-$ with $m_-\le m_0\le m_+$ and
$m_+\ge0$ and $m_-\le 0$. For $T < T_c(\ol \mu)$ and $M > \tilde{M}$ 
there is only one solution corresponding to $m_+$. 
The physical solution is always $m_+$ so that $\tilde{M}$ has no physical
meaning.
Moreover, for $T < T_c(\ol \mu)$ and $M = 0$, $m_+ > 0$. 
The critical temperature can be computed from the following relation:
\begin{eqnarray}
\ol \mu  = {\cal G}(0,T_c) &=&
  T_c \sum_{n\in\mathbb{ Z}}
\int_{p < \Lambda} 
   {\di^{d}\vp \over (2 \pi)^{d}} 
\,{1\over {p}^2 + \omega_{c,n}^2}
\nonumber \\ 
 &=&  \Lambda^{d-1}C_0 - 
  \Lambda^{d-1}\int_{p< 1}{\di^{d}\vp\over (2\pi)^d}\, 
{1\over p [e^{p/t_c}+1]},
\label{eq:Tc}
\end{eqnarray}
where $\omega_{c,n} \equiv (2 n + 1) \pi T_c$ and 
$T_c(\ol\mu)=t_c(\ol\mu)\Lambda$.
For $\ol\mu\to 0$, we have $T_c(\ol\mu)\to\infty$.
\item[(c)] If $ -C_1 < \ol{\mu} \le 0$
(see fig.\ \ref{GapCont1} left), with\footnote{$C_1$ is the solution 
of the equation $p(-x) = \lim_{M\to\infty} M {\cal G}(M,0)$ with
$x = (2 C_1/\overline{\lambda})^{1/2}$. The value $x$ corresponds to the 
position of a maximum of $p(m)$ for $\ol \mu = - C_1$.}
\begin{eqnarray}
C_1^{3/2} \equiv \lambda^{1/2}
{3 \Lambda^{d}\over 2^{d+2} \pi^{d/2} \sqrt{2} } \Gamma\Big({d+2\over 2}
\Big)^{-1},
\label{dis_mu}
\end{eqnarray}   
there is a critical mass $\tilde M$ such that there are
three solutions for  $M< \tilde M$, two of them coincide for $M = \tilde M$,
while for  $M> \tilde M$ the only solution
is $m_+$. The physical solution---the one with the lowest action---is 
always $m_+$ so that
$\tilde M $ has no physical meaning. Note that $m_+> 0$ for $M = 0$.
\item[(d)]  For  $ \ol{\mu} \le - C_1$ 
(see fig.\ \ref{GapCont3} right) there are three solutions 
for all values of $T$ and $M$. The relevant solution is always $m_+> 0$.
\end{itemize}
\begin{figure}
\begin{center}   
\includegraphics[angle=0,scale=0.6]{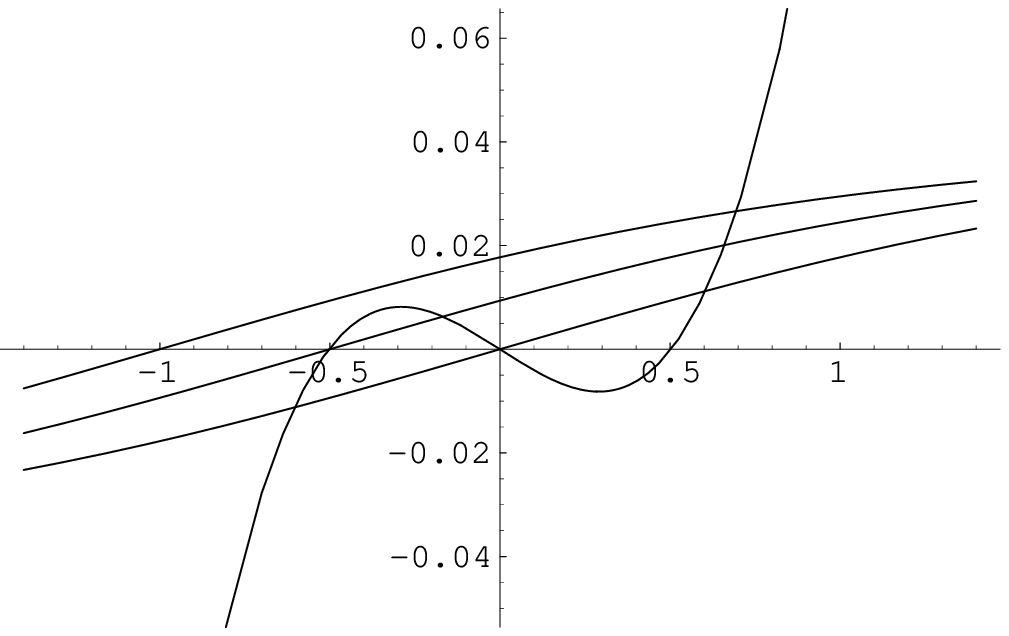}\quad\qquad\includegraphics[angle=0,scale=0.6]{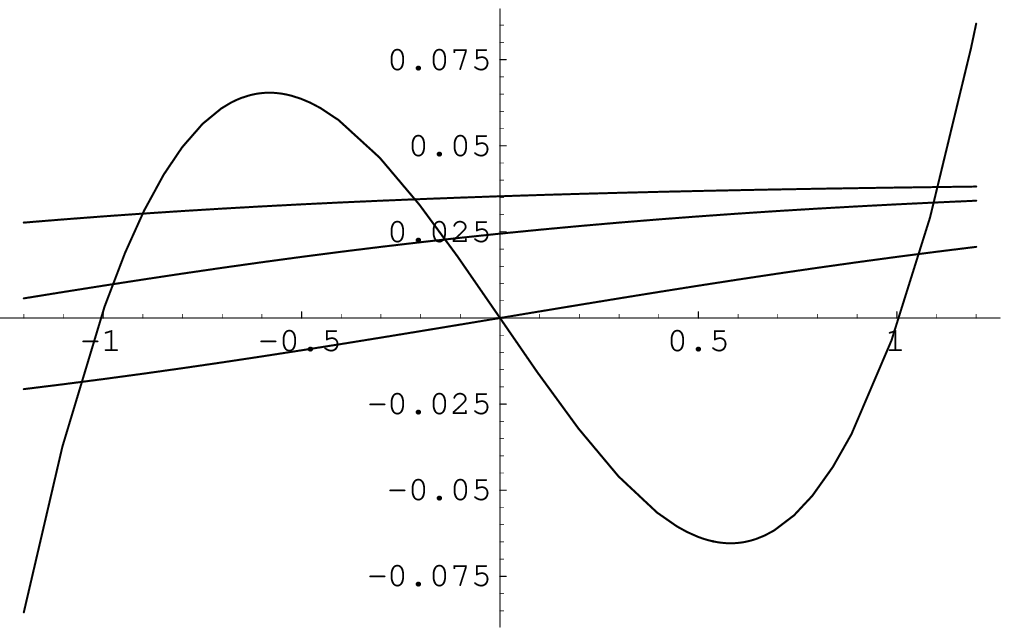}
\caption{Plot of $p(m)$ and $(m+M){\cal G}(m+M)$ for several $M$ and with $\ol \mu <0$ (the intersection points are the solutions of the gap equation \ref{gap}). (Left) Case (c): there could be one or three solution, 
however $m_+>0$ is always present and is always the more stable.
 (Right) Case (d): there are always  three solutions.}
\label{GapCont1}
\end{center}
\end{figure}
In order to compute the more stable solutions (if more than one are present)
one can use the previous consideration, the figs. \ref{GapCont1}
\ref{GapCont3} and\footnote{$c(M,T)$ can easily be computed using
eq.\ (\ref{eq5}), in particular $c(0,T)=0$.}
\begin{eqnarray}
S_{\mathrm{eff}}(m) &=& c(M,T)+\int_0^m \di x\,p(x) 
- (x+M){\cal G}(x+M,T)
\end{eqnarray}
to evaluate the more stable solution. Using the symmetry of $p(x)$ and 
$x {\cal G}(x)$ we get that 
in case (a) chiral symmetry is never broken, while in cases (c) and (d) 
chiral symmetry is never restored. Thus, the only case of interest---and 
the only one we shall consider in the following---is case (b), in which
there is a chirally-symmetric high-temperature phase and 
a low-temperature phase in which chiral symmetry is broken (see fig.
\ref{GapCont4}). 

\begin{figure}
\begin{center}   
\includegraphics[angle=0,scale=0.6]{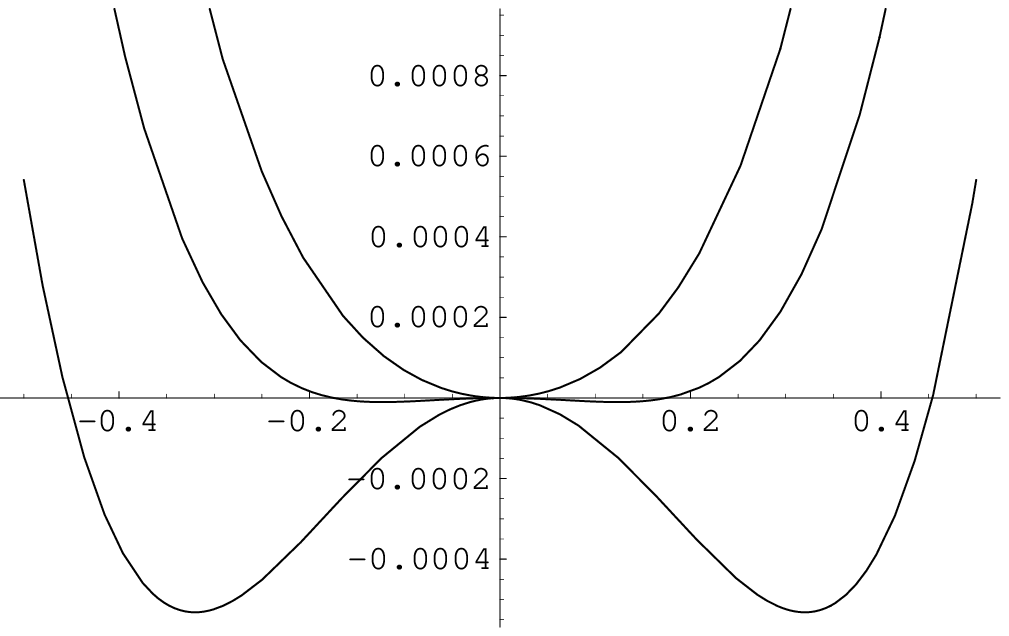}\qquad \quad
\includegraphics[angle=0,scale=0.6]{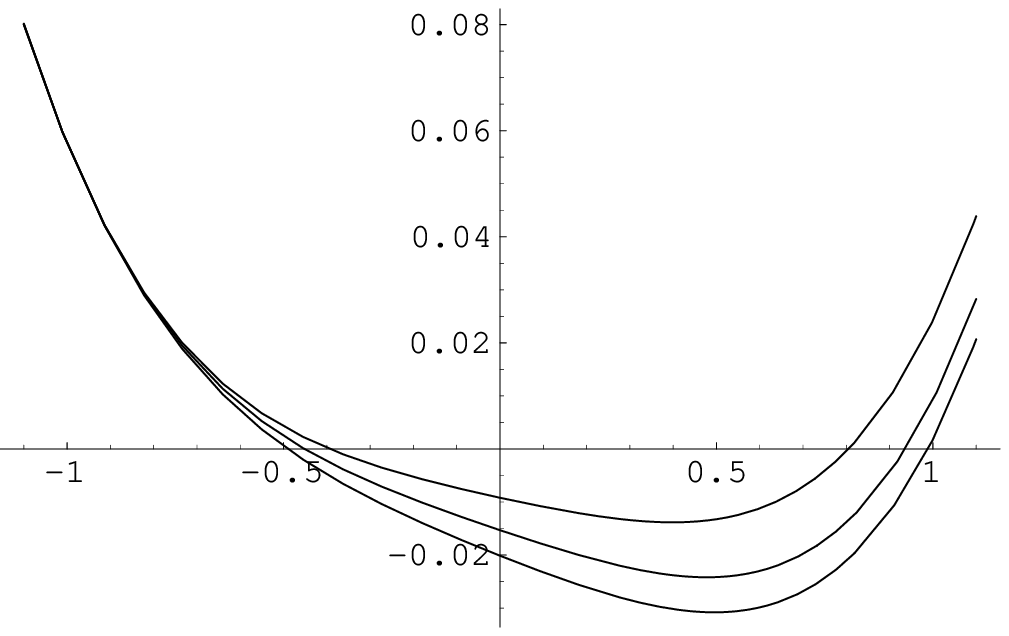}
\caption{$S_{\mathrm{eff}}$ for $M=0$ and $M>0$}
\label{GapCont4}
\end{center}
\end{figure}

The nature of the transition is easily determined.
We expand 
\begin{eqnarray}
{\cal G}(x,T) &=& \sum_{m,n} g_{mn}\, x^{2m} (T-T_c)^n,
\label{Gexp}
\end{eqnarray} 
where
\begin{eqnarray}
&& g_{00} = 
  T_c \sum_{n\in\mathbb{ Z}}
\int_{p < \Lambda} 
   {\di^{d}\vp \over (2 \pi)^{d}} 
\,{1\over {p}^2 + \omega_{c,n}^2},
\nonumber \\
&& g_{01} = 
  \sum_{n\in\mathbb{ Z}}
\int_{p < \Lambda} 
   {\di^{d}\vp \over (2 \pi)^{d}} 
\,{{p}^2 - \omega_{c,n}^2 \over ({p}^2 + \omega_{c,n}^2)^2}
\nonumber \\
&& \qquad = - {1\over T_c^2}
  \int_{p < \Lambda} {\di^{d}\vp \over (2 \pi)^{d}} 
  {e^{p/T_c}\over (e^{p/T_c} + 1)^2},
\nonumber \\
&& g_{10} = - T_c \sum_{n\in\mathbb{ Z}}
\int_{p < \Lambda} 
   {\di^{d}\vp \over (2 \pi)^{d}} 
\,{1\over ({p}^2 + \omega_{c,n}^2)^2},
\label{defgab}
\end{eqnarray}
and $\omega_{c,n} \equiv  (2 n + 1)\pi T_c$.
Since $g_{00}= {\cal G}(0,T_c) = \ol \mu$ [eq.~(\ref{eq:Tc})], 
the gap equation (\ref{gap}) becomes:
\begin{eqnarray}
{\ol \lambda \over 6 } m^3 &=& \ol\mu M + g_{01}(T-T_c)(m+M)+ g_{10}(m+M)^3
+\cdots 
\end{eqnarray}
where we have neglected subleading terms in $m+M$ and $T-T_c$. 
Defining
\begin{eqnarray}
u_h \equiv  {6 \ol \mu \over 6 g_{10} - \ol \lambda}\,  M
&\qquad & 
u_t\equiv {6 g_{01}\over 6 g_{10} - \ol \lambda}\, (T-T_c) ,
\end{eqnarray}
and taking the limit $u_h, u_t \to 0$ at fixed $x\equiv u_t/u_h^{2/3}$
we obtain the equation of state
\begin{eqnarray}
m &=& {u_h}^{1/3} f(x)
\\
0 &=& f(x)^3+x f(x) +1.
\label{eqst}
\end{eqnarray}
Note that the prefactor of $T - T_c$ in $u_t$ is always positive to ensure the 
existence of only one solution for $T > T_c$. 
Such an equation is exactly the mean-field equation of state that 
relates magnetization $\varphi$, magnetic field $h$, and 
reduced temperature $t$.
Indeed, if we consider the mean-field Hamiltonian
\begin{equation}
{\cal H} = h \varphi + {t\over2} \varphi^2 + {u\over 24} \varphi^4,
\end{equation}
the stationarity condition gives
\begin{equation}
h + t \varphi + {u\over6} \varphi^3 = 0,
\end{equation}
which is solved by $\varphi = A h^{1/3} {f}(B t |h|^{-2/3})$, where
${f}(x)$ satisfies eq.~(\ref{eqst}), and $A$ and $B$ are constants
depending on $u$.
This identification also shows that $M$ plays the role of an external
field, while $m\sim \overline{\phi}$ is the magnetization.

\section{Vertex, Propagator and  the failure of
$1/N$ expansion}\label{sec3}

In order to perform the $1/N_f$ calculation, we expand the field
$\phi$ around the saddle-point solution, 
\begin{equation}
\phi(\mathbf{x}) = \ol \phi + {1 \over g \sqrt{N}}  \hat{\phi}(\mathbf{x}),
\end{equation}
where $N \equiv  D N_f$, 
and $\hat{\phi}(\mathbf{x})$ in Fourier modes:
\begin{eqnarray}
\hat{\phi}(\mathbf{x}_d, x_{d+1}) = T
\sum_{n\in \mathbb{Z}} e^{2 i\pi n T x_{d+1} }
\int {\di^{d} \vp\over (2\pi)^d} \, \hat{\phi}_n(\vp) 
e^{i\vp\cdot\mathbf{x}_d} \; .
\end{eqnarray}
In the following we will refer to the integers $n$---or more precisely to 
$2\pi n T$---as frequencies.
In this way we obtain the following expansion for the effective action:
\begin{eqnarray}
\hat{\cal S}_{\mathrm{eff}}[\hat{\phi}_n] &=& 
N \{ {\cal S}_{\mathrm{eff}}[\phi] - {\cal S}_{\mathrm{eff}}[\ol\phi]\}
\nonumber \\
&=& {T\over 2}\sum_n\int_{p < \Lambda} 
  {\di^{d} \vp \over (2\pi)^d}
\,P(\vp,n) \hat{\phi}_n(\vp) \hat{\phi}_{-n}(-\vp)
\nonumber\\
&&+\sum_{l \ge 3}{T^{l-1}\over l! N^{l/2-1} }
   \sum_{\{n_i\}}
 \int_{p_i < \Lambda} 
   {\di^{d} \vp_1\over (2\pi)^d} \cdots {\di^{d} \vp_l\over (2\pi)^d}
   (2\pi)^d \delta \Big(\sum_{j=1}^l \vp_{j} \Big)\delta_{\sum_i n_i , 0}
\nonumber\\
&&\qquad \qquad
 \hat{\phi}_{n_1}(\vp_1) \cdots \hat{\phi}_{n_l}(\vp_l)
 V^{(l)}(\vp_{1},n_1;\cdots;\vp_{l},n_l).
\label{action-exp}
\end{eqnarray}
Note the Kronecker $\delta$ on the frequencies that ensures that 
vertices are nonvanishing only for $\sum_i^l n_i = 0$. In particular---this
property will be important in order to obtain the effective action for the
zero mode---if only one frequency is nonvanishing
$V^{(l)}(\vp_{1},n;\vp_{2},0;\cdots;\vp_{l},0) = 0$.

The vertices appearing in expansion (\ref{action-exp}) 
are easily computed. The fermion contribution is obtained by 
considering the one-loop fermionic graphs in theory (\ref{action}). 
If we define the free fermion propagator
\begin{equation}
\Delta_F(\mathbf{p},n;m) \equiv  
{-i\sum_{j=1}^{d}\gamma_j {p}_j-i
\gamma_{d+1} \omega_n +m \over {p}^2+ \omega_n^2 +m^2}
\end{equation}
with $\omega_n \equiv  (2 n + 1) \pi T$, 
the fermion contribution is 
\begin{eqnarray}
&& V^{(l)}_f(\vp_{1},n_1;\cdots;\vp_{l},n_l) = 
\nonumber \\ 
&& \qquad\qquad
  (-1)^l {T\over D} \sum_{a\in \mathbb{Z}}
    \int_{q < \Lambda} {\di^{d} \vq\over (2\pi)^d}\, \hbox{tr}\, 
   \left[ \prod_{i=1}^l \Delta_F \Bigl(\vq + \sum_{j=1}^i \vp_j;
                      a + \sum_{j=1}^i n_j; m + M\Bigr) \right] 
\nonumber \\
   &&  \qquad\qquad
   \vphantom{{\di^{d} \vq\over (2\pi)^d}}
    + \hbox{permutations},
\end{eqnarray}
where the permutations make the vertex completely symmetric 
[there are $(l-1)!$ terms]. Note that the frequencies $\omega_n$ 
never vanish and thus the vertices have a regular expansion 
in powers of $m + M$. Vertices $V_f^{(l)}$ satisfy an important 
symmetry relation. First, note that 
\begin{equation}
\gamma_{d+1} \Delta_F(\vp,n;m) \gamma_{d+1} = 
- \Delta_F(\vp,-n-1;-m).
\end{equation}
It follows
\begin{eqnarray}
&& \sum_{a\in \mathbb{Z}}\hbox{tr}\,
   \left[\prod_{i=1}^l \Delta_F(\vq_i ; a + b_i; m)\right] = 
\sum_{a\in \mathbb{Z}}\hbox{tr}\,
\left[\prod_{i=1}^l 
   \gamma_{d+1} \Delta_F(\vq_i ; a + b_i; m)\gamma_{d+1}\right]=
\\
&& \qquad 
 (-1)^l \sum_{a\in \mathbb{Z}}\hbox{tr}\,
   \left[\prod_{i=1}^l \Delta_F(\vq_i ; - a - b_i - 1; -m)\right] = 
 (-1)^l \sum_{a\in \mathbb{Z}}\hbox{tr}\,
   \left[\prod_{i=1}^l \Delta_F(\vq_i ; a - b_i; -m)\right].
\nonumber 
\end{eqnarray}
In the second step we used $\gamma_{d+1}^2 = 1$, while in the 
last one we redefined $a \to - a + 1$. This relation implies
(we write here explicitly the dependence of the vertices on $m$ and 
$M$)\footnote{If $d$ is odd, one can repeat the same argument 
using $\gamma_{\cal C}$ defined in (\ref{gamma-five}). It shows that 
vertices with $l$ legs are multiplied by $(-1)^l$ if 
one changes the sign of $m + M$ at fixed momenta and {\em frequencies}.}
\begin{equation}
V_f^{(l)}(\vp_1,n_1;\ldots;\vp_l,n_l;m + M) = (-1)^l
V_f^{(l)}(\vp_1,-n_1;\ldots;\vp_l,-n_l;-m - M).
\label{symmetryVf}
\end{equation}
For every $l > 4$, the vertex is due only to the fermion loops, so that 
$V^{(l)} = V^{(l)}_f$. For $l \le  4$ we must also take into account the 
contribution of the bosonic part of the action, so that 
\begin{eqnarray}
 V^{(3)}(\mathbf{p},n_1;\mathbf{q},n_2;\mathbf{r},n_3) 
&=&  \ol\lambda m + 
 V^{(3)}_f(\mathbf{p},n_1;\mathbf{q},n_2;\mathbf{r},n_3),
\nonumber \\
 V^{(4)}(\mathbf{p},n_1;\mathbf{q},n_2;\mathbf{r},n_3;\mathbf{s},n_4) 
&=&  \ol\lambda + 
 V^{(4)}_f(\mathbf{p},n_1;\mathbf{q},n_2;\mathbf{r},n_3;\mathbf{s},n_4).
\end{eqnarray} 
Finally, for the inverse propagator we have 
\begin{eqnarray}
P(\mathbf{p},n) &=& {\mathbf{p}^2\over g^2}+{(2\pi T n)^2
\over g^2}
 +\ol\mu + {\ol\lambda\over 2} m^2 + 
  V_f^{(2)} (-\vp,-n;\vp,n).
\label{prop_gen}
\end{eqnarray}
Vertices $V^{(l)}$ also satisfy the symmetry relation (\ref{symmetryVf}),
while the inverse propagator $P(\vp,n)$ satisfies 
$P(\vp,n;m,M) = P(\vp,-n;-m,-M)$. 
In order to clarify the equation 
of the critical crossover limit we need to explicitly compute 
$P(\vp,n)$,
$V_3(\vp,n)\equiv V^{(3)}(\vp,n;-\vp,-n;\mathbf{0},0) $ and
$V_3(\vp,n)\equiv V^{(4)}(\vp,n;-\vp,-n;\mathbf{0},0;\mathbf{0},0)$ for 
$M,m\to 0$. Using standard formula
$\Tr \gamma_\mu \gamma_\nu =D\, \delta_{\mu,\nu}$, 
$\Tr \gamma_\mu\gamma_\nu\gamma_{\delta} = 0$ and
$\Tr \gamma_\alpha\gamma_\beta\gamma_\gamma\gamma_\delta =
D\, (\delta_{\alpha,\beta}\delta_{\gamma,\delta}-\delta_{\alpha,\gamma}\delta_{\beta,\delta}+\delta_{\alpha,\delta}\delta_{\beta,\gamma})$
and defining
\begin{eqnarray}
\tau_h(\vp,n) &=& 2 v_h(\vp,n)+ v_h(\mathbf{0},0)
\nonumber\\
&&-T\bigg(\vp^2+(2\pi T n)^2\bigg)\sum_{a\in \mathbb{Z}}
\int_{q<\Lambda}
\eta_h(\mathbf{q},a)^2\eta_h(\mathbf{q}+\vp,a+n)
\nonumber\\
v_h(\vp,n) &=& T\sum_{a\in \mathbb{Z}}\int_{q<\Lambda} \eta_h(\mathbf{q},a)
\eta_h(\mathbf{q}+\vp,a+n)
\end{eqnarray}
with  $\eta_h(\mathbf{p},n):=\Big(\mathbf{p}^2+\pi^2T^2(2n+1)^2+h^2
\Big)^{-1}$, we get the final result
\begin{eqnarray}
P(\vp,n) &=& {\vp^2\over g^2}+{(2\pi T n)^2\over g^2}+\ol\mu+{\ol \lambda
\over 2}m^2-T\sum_a \int^{\Lambda}\di^d \mathbf{q} \,\eta_h(\mathbf{q},a)
\nonumber\\
&&+\Big({\vp^2\over 2 }+{(2\pi n)^2\over 2 } \Big)
v_h(\vp,n)+2 h^2 v_h(\mathbf{0},0)
\label{twoc_zero}
\\
V_3(\vp,n) &=& \overline \lambda m+2  (m+M) \tau_{m+M}(\vp,n)
\\
&&-8 (m+M)^3 T \sum_{a\in \mathbb{Z}} \int_{q<\Lambda}\eta_{m+M}(\vq,a)^2\eta_{m+M}(\vq+\vp,a+n)
\label{threec_zero}\\
V_4(\vp,n) &=&\ol \lambda +2   \tau_{m+M} (\vp,n)
+O(m+M)^2
\label{fourc_zero}
\end{eqnarray}
The similarity of (\ref{threec_zero}) with (\ref{fourc_zero})
simplifies the equations of the critical crossover limit with respect to
the general case. This is due to the symmetry of the model
(\ref{symmetryVf}) that is not present in other models investigated
in this work. Indeed in the general case one has to tune the 
model at the critical point recovering the symmetry which is than
dynamically broken. However in this case the critical point remain
fixed at $M=0$ without $1/N_f$ fluctuations, technically this
follows from (\ref{threec_zero}) and (\ref{fourc_zero}).

Finally, note that 
$V^{(4)}(\mathbf{0},0;\mathbf{0},0;\mathbf{0},0;\mathbf{0},0)$ is positive 
at the transition. Indeed, one obtains explicitly 
(note that $\lambda \ge 0$ to ensure the stability of the quartic potential)
\begin{equation}
V^{(4)}(\mathbf{0},0;\mathbf{0},0;\mathbf{0},0;\mathbf{0},0) = 
   \overline{\lambda} + 6 T_c \sum_a \int_{q<\Lambda} 
   {\di^{d}\vq \over (2 \pi)^{d}} 
\,{1\over ({q}^2 + \omega_{c,n}^2)^2} > 0.
\label{V4pos}
\end{equation}
It is easy to verify that $P(\mathbf{0},0)$ vanishes at the 
transition. Indeed, for $m = M = 0$ we have
\begin{eqnarray}
P(\mathbf{0},0) &=&  \ol\mu
 -   T\sum_{n\in\mathbb{Z}}\int_{q<\Lambda} {\di^{d} \mathbf{q}\over (2\pi)^d} 
     {1\over {q}^2 + \omega_n^2}
\nonumber\\
&=&  -g_{01} (T-T_c)+O(T-T_c)^2,
\end{eqnarray}
where we used eqs.~(\ref{Gexp}) and (\ref{eq:Tc}). Thus, the 
mode with $n = 0$ is singular at the critical point.
 It is exactly this singularity that forbids 
a standard $1/N_f$ expansion at $T = T_c$ and gives rise
to the Ising behavior. This type of singular behavior is the main topic 
of this work and has been described
in the first chapter of this works in a general way using
generalized Heisenberg Models. The strategy proposed 
there consists in integrating all the nonsingular modes $\hat{\phi}_n$,
$n\not = 0$, and study the effective theory for the zero mode 
$\hat{\phi}_0$ and is the same we are going to apply
for the model under investigation.

\section{The effective theory of the zero mode}

Following the conclusion of the previous  section and the standard
protocol developed in the first chapter, we  compute now the
effective action for the zero mode integrating the non critical
degrees of freedom.
Integrating  all fields $\hat{\phi}_n$ with $n\neq 0$
we obtain the effective action
\begin{equation}
e^{-\tilde{\cal S}_{\mathrm{eff}}[\hat{\phi}_0]}=
\int \prod_{n\neq 0}\di \hat{\phi}_n \,
e^{-\hat{\cal S}_{\mathrm{eff}}[\hat{\phi}_n]},
\end{equation}
with
\begin{eqnarray}
\tilde{\cal S}_{\mathrm{eff}}&=& \sqrt{N} \tilde H\hat{\phi}_0(\mathbf{0})+
    {T\over 2} \int_{q<\Lambda}\,{\di^{d} \mathbf{p}\over (2\pi)^d} \,
\hat{\phi}_0(\mathbf{p})\, \tilde P(\mathbf{p})\,\hat{\phi}_0(-\mathbf{p})
\\
&&+\sum_{l\ge 3}{T^{l-1}\over l! N^{l/2-1}}
\int_{p_i < \Lambda}  
  {\di^{d}\mathbf{p}_1\over (2\pi)^d} \cdots
  {\di^{d}\mathbf{p}_l\over (2\pi)^d}\, 
  (2\pi)^d \delta\Big(\sum_{i=1}^l\vp_i \Big)\, 
\nonumber \\
&& \qquad \times 
  \tilde V^{(l)}(\mathbf{p}_1,\ldots\mathbf{p}_l)
  \hat{\phi}_0(\mathbf{p}_1) \cdots \hat{\phi}_0(\mathbf{p}_l)
\nonumber
\end{eqnarray}
where $H$, $\tilde{P}$, and $\tilde V^{(l)}$ have an expansion in powers of 
$1/N$. The computation of these quantities is quite simple. 
The contribution of order $1/N^k$ to $\tilde V^{(l)}$ is obtained 
by considering all $k$-loop diagrams contributing to 
the $l$-point connected correlation function of $\hat{\phi}_0$ and 
considering only the nonsingular fields (i.e. propagators with 
$n\not=0$) on the internal lines. 
Frequency conservation $V^{(n)}(\vp,n;\mathbf{0},0;\cdots \mathbf{0},0)\sim \delta_{n,0}$ implies that all tree-level diagrams with more than
one vertex
vanish.\footnote{
For a tree-level graph, the usual topological arguments give the 
relation $\sum_n (n-2) N_n = -2$, where $N_n$ is the number of 
vertices belonging to the graph such that $n$ legs belong to internal
lines. Since $n\ge 1$ if there is more than one vertex, 
the previous equality requires 
$N_1 \ge 2$. But frequency conservation implies that $V_l$ vanishes 
if all frequencies but one vanish. Therefore, each nontrivial 
tree-level diagram vanishes.}
Therefore,  $\tilde{H} = O(N^{-1})$, 
$\tilde{P} (\vp) = P(\vp,0) + O(N^{-1})$,
and $\tilde V^{(l)} = V^{(l)}_{n_i=0} + O(N^{-1})$ 
($V^{(l)}_{n_i=0}$ is the vertex $V^{(l)}$ with all frequencies set to zero).
For the inverse propagator $\tilde{P}$ and for the magnetic field $\tilde{H}$
we shall also need the $1/N$ corrections. We obtain
\begin{eqnarray}
\tilde H&=& {H_1\over N} + O(N^{-2})
\\
\tilde P(\mathbf{p}) &=& P(\mathbf{p},0) + {P^{(1)}
(\mathbf{p})\over N} + O(N^{-2}),
\end{eqnarray}
with
\begin{eqnarray}
H_1&=& {T\over 2}\sum_{n\neq 0}\int_{p<\Lambda}
{\di^{d} \mathbf{p}\over (2\pi)^d} \,
V_3(\mathbf{p},n) P(\mathbf{p},n)^{-1} 
\label{H1expr}
\\
P^{(1)}(\mathbf{0})&=& {T\over 2}\sum_{n\neq 0}\int_{p<\Lambda}
{\di^{d} \mathbf{p}\over (2\pi)^d} \,
\,\Big[
 V_4(\mathbf{p},n) P(\mathbf{p},n)^{-1} -  V_{3}(\mathbf{p},n)^2 
 P(\mathbf{p},n)^{-2}
\Big],
\end{eqnarray}
where $V_l(\mathbf{p},n) \equiv  
V^{(l)}(\mathbf{p},n;-\mathbf{p},-n; \mathbf{0},0;\ldots)$. 
Note that relation (\ref{symmetryVf}) implies an analogous 
symmetry relation for $\tilde{V}^{(l)}$\footnote{
For a diagram $D(m,M)$, with $n_j$ $j-$legs vertices, entering 
in the computation of $V^{(l)}$ we have $D(m,M)=(-1)^{\sum_j n_j j}
D(-m,-M)=(-1)^lD(-m,-M) $. We have used the relation
$\sum_j n_j j=l + 2 n_I$, $n_I$ being the internal line}:
\begin{equation}
\tilde{V}^{(l)}(\vp_1,\ldots,\vp_l;m,M) = 
(-1)^l \tilde{V}^{(l)}(\vp_1,\ldots,\vp_l;-m,-M),
\label{symmetryVtilde}
\end{equation} 
where we have written explicitly the dependence on $m$ and $M$. 
Analogously $\tilde{P}(\vp)$ and $\tilde{H}$ are respectively 
symmetric and antisymmetric under $m,M\to -m,-M$. 

In order to obtain the final effective theory we introduce a new field
$\chi(\vp)$ such that the corresponding zero-momentum three-leg vertex 
vanishes for any value of the parameters. For this purpose we write
\begin{equation}
\alpha \chi(\vp) = T \hat{\phi}_0(\vp) + \sqrt{N} k \delta(\vp),
\end{equation}
where $\alpha$ and $k$ are functions to be determined. If we write 
$a_l \equiv \tilde{V}^{(l)}(\mathbf{0},0;\ldots;\mathbf{0},0)$, $k$ is 
determined by the equation
\begin{equation}
\sum_{l=0} {(-1)^l k(m,M)^l\over l!} a_{l+3}(m,M) = 0,
\end{equation}
where we have written explicitly the dependence on $m$ and $M$.
Now, symmetry (\ref{symmetryVtilde}) implies also
\begin{equation}
\sum_{l=0} {(-1)^l [-k(-m,-M)]^l\over l!} a_{l+3}(m,M) = 0,
\end{equation}
so that $k(m,M) = - k(-m,-M)$. Therefore, $k$ has an expansion of the 
form
\begin{equation}
k = \sum_{ab,a+b\, {\rm odd}} k_{ab} m^a M^b,
\end{equation}
where the coefficients $k_{ab}$ have a regular expansion in powers of $1/N$.
The leading behavior close to the transition is easily computed:
\begin{equation}
k = {a_3\over a_4} + O(m^a M^b, a + b = 3).
\end{equation}
In terms of $\chi$ the effective action can be written as 
\begin{eqnarray}
\tilde{\cal S}_{\mathrm{eff}}&=& N^{1/2} \ol H\chi(\mathbf{0})+
    {1\over 2} \int_{q<\Lambda}\,{\di^{d} \mathbf{p}\over (2\pi)^d} \,
\chi(\mathbf{p})\, \ol P(\mathbf{p})\,\chi(-\mathbf{p})
\\
&&+\sum_{l\ge 3}{1\over l! N^{l/2-1}}
\int_{p_i < \Lambda}  
  {\di^{d}\mathbf{p}_1\over (2\pi)^d} \cdots
  {\di^{d}\mathbf{p}_l\over (2\pi)^d}\, 
  (2\pi)^d \delta\Big(\sum_{i=1}^l\vp_i \Big)\, 
  \ol V^{(l)}(\mathbf{p}_1,\ldots\mathbf{p}_l)
  \chi(\mathbf{p}_1) \cdots \chi(\mathbf{p}_l).
\nonumber
\end{eqnarray}
The quantities $\ol H$, $\ol P$, and $\overline{V}^{(l)}$ have an expansion
in terms of $m$, $M$, and $1/N$. 
Explicitly we have:
\begin{eqnarray}
&& \ol H = \alpha T^{-1}[ \tilde{H} - k \tilde{P}(\mathbf{0}) + 
             {k^2\over2} \tilde{V}_3(\mathbf{0}) -
             {k^3\over6} \tilde{V}_4(\mathbf{0}) + 
             O(m^a M^b,a+b=5)],
\\
&&\ol P(\vp) = \alpha^2 T^{-1}[ \tilde{P}(\vp) - k \tilde{V}_3(\mathbf{p}) +
             {k^2\over2} \tilde{V}_4(\mathbf{p}) + 
             O(m^a M^b,a+b=4)],
\\
&& {\ol V}^{(2l+1)}(\vp_1,\ldots,\vp_{2l+1}) = 
    \alpha^{2l+1} T^{-1} [ {\tilde V}^{(2l+1)}(\vp_1,\ldots,\vp_{2l+1}) - 
     k {\tilde V}^{(2l+2)}(\vp_1,\ldots,\vp_{2l+1},\mathbf{0}) 
\nonumber \\
&& \qquad\qquad + O(m^a M^b,a+b=3)],
\\
&& {\ol V}^{(2l)}(\vp_1,\ldots,\vp_{2l}) = 
    \alpha^{2l} T^{-1} {\tilde V}^{(2l)}(\vp_1,\ldots,\vp_{2l})  + 
             O(m^a M^b,a+b=2).
\end{eqnarray}
Up to now we have not defined the parameter $\alpha$. We will fix it by
requiring
\begin{equation}
   \left. {d\ol P(\vp)\over dp^2} \right|_{p=0} = 1,
\label{alpha}
\end{equation}
for all values of the parameters. The parameter $\alpha$  is a function 
of $m$, $M$, and $1/N$. The symmetry properties of $k$ and of the vertices
imply that $\alpha$ is invariant under $m,M\to -m,-M$. 
As a consequence, under $m,M\to -m,-M$,
the quantities $\ol H$, $\ol P$, and $\overline{V}^{(l)}$ have 
the same symmetry properties as 
$\tilde H $, $\tilde P$, and $\tilde{V}^{(l)}$. 

In the following we shall need the expansions of 
$\ol H$, $\ol P(\mathbf{0})$, and  
${\ol V}^{(3)}(\mathbf{p},-\mathbf{p},\mathbf{0})$ 
close to the critical point. Using (\ref{twoc_zero}), (\ref{threec_zero})
and (\ref{fourc_zero}) we get
\begin{eqnarray}
&& P(\mathbf{0},0)\approx 
   {\ol\lambda \over2} m^2 - (T-T_c) g_{01} - 3 (M + m)^2 g_{10} ,
\nonumber \\
&& V_3(\mathbf{0},0) \approx {\ol\lambda} m - 6 (M+m) g_{10} ,
\nonumber \\[2mm]
&& V_4(\mathbf{0},0) \approx {\ol\lambda} - 6 g_{10},
\label{expan-V2V3V4}
\end{eqnarray}
Notice the relation
\begin{equation}
V_f^{(3)}(\mathbf{p},n;-\mathbf{p},-n;\mathbf{0},0;m) = 
m V_f^{(4)}(\mathbf{p},n;-\mathbf{p},-n;\mathbf{0},0;\mathbf{0},0;m) + 
   O(m^3),
\label{relV3V4}
\end{equation}
(we have explicitly written the mass dependence of the vertices)
that we have just commented in the previous section. In the relations
(\ref{expan-V2V3V4}) we have used $g_{10}=-v_0(\mathbf{0},0)$ that
can easily obtained using the development of the gap equation
near the critical point. 
We expand $\ol H$ and $\ol P(\mathbf{0})$ is powers of $1/N$ as 
\begin{eqnarray}
{T \overline H\over\alpha} = {h}_0 + { {h}_1\over N} + O(N^{-2}), 
\\
{T \overline P(\mathbf{0})\over\alpha^2} = {p}_0 + { p_1\over N} + O(N^{-2}).
\end{eqnarray}
By using expansions (\ref{expan-V2V3V4}) we obtain 
\begin{eqnarray}
{h}_0 &\approx& 
     - {V_3(\mathbf{0},0)\over V_4(\mathbf{0},0)} P(\mathbf{0},0) + 
       {1\over3} {V_3(\mathbf{0},0)^3\over V_4(\mathbf{0},0)^2}
\nonumber \\
    & \approx & - {\ol \mu} M + 
       {g_{01}\ol \lambda\over 6 g_{10} - \ol \lambda} M (T - T_c) - 
       {g_{10}\ol \lambda (6 g_{10} + \ol \lambda)\over 
            (6 g_{10} - \ol \lambda)^2} M^3 + 
        O(m^a M^b,a+b=5),
\nonumber \\
{h}_1 &\approx& 
    H_1 - {V_3(\mathbf{0},0)\over V_4(\mathbf{0},0)} P^{(1)}(\mathbf{0}) 
\nonumber \\
    & \approx & - {\ol \lambda M T \over 2 (6 g_{10} - \ol \lambda)}   
      \sum_{n\neq 0}\int_{p<\Lambda}
      {\di^{d} \mathbf{p}\over (2\pi)^d} \,
      [6 g_{10} + V_f^{(4)}(\mathbf{p},n;-\vp,-n;\mathbf{0},0;\mathbf{0},0)]
      P(\mathbf{p},n)^{-1} 
\nonumber \\
     & &  + O(m^a M^b,a+b=3),
\nonumber \\
{p}_0 &\approx& P(\mathbf{0},0) - 
    {V_3(\mathbf{0},0)^2\over 2 V_4(\mathbf{0},0)} 
\nonumber \\
   &\approx& - g_{01} (T - T_c) + 
      {3 g_{10} \ol \lambda\over 6 g_{10} - \ol \lambda} M^2 + 
       O(m^a M^b,a+b=4),
\nonumber \\
{p}_1 &\approx& P^{(1)} (\mathbf{0},0) \approx e + O(m^a M^b,a+b=2),
\label{expanHP}
\end{eqnarray}
where $e$ is the value of $P^{(1)} (\mathbf{0},0)$ for $M=m=0$.
Note that several terms that are allowed by the symmetry 
$m,M\to -m,-M$ are missing in these expansions. In the case of $h_0$
we used the gap equation to eliminate the term proportional 
to $m^3$. This substitution is responsible for the appearance of the 
term linear in $M$ and cancels the terms proportional 
to $m (T - T_c)$, $m^2 M$, and $m M^2$. In the case of $h_1$ and $p_0$
note that the terms proportional to $m$, and $m^2$, $m M$ cancel out.
Finally, we compute the three-leg vertex. At leading order in $1/N$ 
we obtain 
\begin{eqnarray}
&& {T\over \alpha^3} {\ol V}^{(3)} (\vp,-\vp,\mathbf{0}) = 
            V_3(\vp,0) - {V_3(\mathbf{0},0)\over V_4(\mathbf{0},0)} 
            V_4(\vp,0) 
\nonumber \\
&& \qquad \approx {- \ol \lambda M\over 6 g_{10} - \ol \lambda}
        [6 g_{10} + V_f^{(4)}(\mathbf{p},0;-\vp,0;\mathbf{0},0;\mathbf{0},0)] 
     + O(m^a M^b,a+b=3).
\label{expanV3}
\end{eqnarray}
Note that the term proportional to $m$ is missing as a consequence of 
relation (\ref{relV3V4}).

Having computed the expansion of translated vertices of the theory
with one incoming momenta (\ref{expanHP},\ref{expanV3}) and at zero
external momenta (\ref{expan-V2V3V4}) we are now in a position to 
define the critical crossover limit for the model under investigation.

\section{The critical crossover limit} \label{sec4}

The manipulations presented in the previous section allowed us to 
compute the effective action for the zero mode $\chi(\vp)$. Far from the 
critical point $\ol P(\vp)\not = 0$ for all momenta and thus one can perform
a standard $1/N_f$ expansion. At the critical point instead 
this expansion fails because $\ol P(\mathbf{0}) = 0$. At the critical
point, for $N\to \infty$ the long-distance behavior is controlled by the action
\begin{equation}
\tilde{S}_{\rm eff} \approx 
   \int \di^d {\mathbf x}\, 
   \left[ {1\over2} (\partial \chi)^2 + {u\over 4!} \chi^4 
   \right] + O(N^{-2}),
\label{phi4crit}
\end{equation}
where 
\begin{equation}
u = {1\over N} \overline{V}^{(4)}(\mathbf{0},\mathbf{0},\mathbf{0},\mathbf{0}).
\end{equation}
Here we have used the fact that vertices with an odd number of fields 
vanish at the critical point and the normalisation condition (\ref{alpha}).
Moreover, since the critical mode corresponds to $\vp = \mathbf{0}$,
we have performed an expansion in powers of the momenta, keeping only the 
leading term.
Since for $N\to \infty$, 
$\overline{V}^{(4)}(\mathbf{0},\mathbf{0},\mathbf{0},\mathbf{0}) = 
\alpha^4 T^{-1} {V}^{(4)}(\mathbf{0},\mathbf{0},\mathbf{0},\mathbf{0})$,
inequality (\ref{V4pos}) implies $u > 0$ at the critical point. 
eq.~(\ref{phi4crit}) is the action of the critical $\phi^4$ theory which 
should be studied in the weak-coupling limit $u\to 0$. In this regime
the model shows an interesting scaling behavior---we named it 
{\em critical crossover}---that describes the crossover
 between Mean-Field 
and Ising behavior and has been studied in detail in the
chapter \ref{chapter2} and applied for a specific model
in chapter \ref{chapter3}.
In this chapter we want to recover the results obtained
for  Heisenberg models in chapter \ref{chapter3} 
for Yukawa model (\ref{action}) using the general
$\N=1$ theory presented
in chapter \ref{chapter2}.

\subsection{The general theory} \label{sec4.1}

Following Ref.~\cite{CMP-05} in chapter 
\ref{chapter2} and \ref{chapter3} we extended these 
considerations to the general 
two-dimensional Hamiltonian
\begin{eqnarray}
{\cal S}_{\rm eff}[\varphi] &=& H \varphi(\mathbf{0}) +
      {1\over 2} \int {\di^2 \mathbf{p}\over (2\pi)^2 }\, 
      [K(\mathbf{p}) + r]
      \varphi(\mathbf{p}) \varphi(-\mathbf{p})
\label{ccl} \\
&& +  \sum_{l\ge 3} {u^{l/2-1}\over l!} 
    \int {\di^2 \mathbf{p}_1\over (2\pi)^2 } \ldots 
            {\di^2 \mathbf{p}_l\over (2\pi)^2 }\, 
   (2\pi)^d \delta\left(\sum_i \mathbf{p}_i\right)\, 
                {\cal V}^{(l)}(\mathbf{p}_1,\ldots,\mathbf{p}_l)
                \varphi(\mathbf{p}_1) \ldots
                \varphi(\mathbf{p}_l),
\nonumber 
\end{eqnarray}
where $K(\vp) = p^2 + O(p^4)$, 
${\cal V}^{(3)}(\mathbf{0},\mathbf{0},\mathbf{0}) = 0$, and 
${\cal V}^{(4)}(\mathbf{0},\mathbf{0},\mathbf{0},\mathbf{0}) = 1$. 
The presence of vertices with an odd number of legs requires an 
additional counterterm for the magnetic field. Indeed, we showed that it 
was possible to find functions $r_c(u)$ and $H_c(u)$ such that 
for $t\equiv r - r_c(u)$ (infrared limit), $h\equiv H - H_c(u)$,
$u\to 0$ (weak-coupling limit), at fixed $t/u$, $h/u$ one has 
\begin{equation}
  \chi_n = u^{1-n} f_n(\tilde{h},\tilde{t}),
\end{equation}
where the scaling function $f_n(x,y)$ is the same as that computed in the 
continuum theory. In particular, $\chi_n u^{n-1}$ vanishes in the 
crossover limit if $n$ is odd. The counterterms are regularization-dependent.
In the continuum theory with a cutoff we have 
\begin{eqnarray}
h_c &=& -{\sqrt{u}\over2} \int_{p<\Lambda}
    {\di^{2} \vp \over (2\pi)^2} {{\cal V}_3(\vp)\over K(\vp)},
\label{hc-cutoff}
\\
r_c &=& {u\over 8\pi} \ln {u\over \Lambda^2}  + 
{u\over 2}\int_{p<\Lambda} {\di^{2}\vp \over (2\pi)^2} \,
{{\cal V}_3(\vp)^2\over K(\vp)^2} +  A_0 u,
\label{rc-expr}
\end{eqnarray}
with ${\cal V}_3(\vp) \equiv {\cal V}^{(3)}(\vp,-\vp,\mathbf{0})$ and 
\begin{eqnarray}
A_0&=&-D_2-{3\over 8\pi}+{1\over 8\pi }\log{3 \over 8\pi}-{1\over 2}
\int_{p< \Lambda}{\di^{2}\vp\over (2\pi)^2}\, \left[
   {{\cal V}^{(4)}(\vp,-\vp,\mathbf{0},\mathbf{0})\over K(\vp)}-{1\over p^2}\right].
\nonumber\\
\end{eqnarray}
The nonperturbative constant $D_2$ was estimated in 
Ref.~\cite{PRV-99}: $D_2=-0.0524(2)$.
The relations (\ref{hc-cutoff}) and (\ref{rc-expr})
generalize the lattice expressions found in
chapter \ref{chapter2} [see eqs.\ (\ref{defrc-const})
 (\ref{uc1-1loop})] to a sharp cut off
regularisation. However as will be clear later 
in this case (in which chiral symmetry is not broken
in the starting action)  $h_c$ is irrelevant in the
scaling limit.

\subsection{Scaling behavior} \label{sec4.2}

In this section we wish to use the previous results to compute the 
crossover behavior of model (\ref{action}) in 2+1 dimensions.
Since $u\sim 1/N$ the relevant scaling variables are
\begin{eqnarray}
x_h &=& {N T_c\over \alpha} ({N}^{1/2} \ol H - H_c) \nonumber \\
x_t &=& {N T_c\over \alpha^2} (\ol P({\mathbf 0}) - r_c),
\label{scal}
\end{eqnarray}
where the factors $\alpha/T_c$ and $\alpha^2/T_c$ are introduced for convenience. 
The critical crossover limit is obtained by tuning 
$T$, $M$, and $N$ close to the critical point so that 
$x_h$ and $x_t$ are kept constant. 
The expansions of $\ol H$  and $\ol P({\mathbf 0})$ are reported in
eq.~(\ref{expanHP}). The expansions of $H_c$ and $r_c$ are easily
derived. For $H_c$ we have 
\begin{equation}
H_c = - {1\over 2 \sqrt{N}} 
     \int_{p<\Lambda} {\di^2 \vp\over (2\pi)^2} \,
          {{\ol V}^{(3)}(\vp,-\vp,\mathbf{0}) \over 
      \ol P(\vp)} ,
\end{equation}
where all quantities are computed for $M=m=0$. Using
eq.~(\ref{expanV3}) we have 
\begin{equation}
H_c = {\alpha h_{c0} M\over T_c \sqrt{N}} + O(m^a M^b,a+b=3),
\end{equation}
where $h_{c0}$ is a constant.
Using eq.~(\ref{rc-expr}) we obtain for $r_c$ the expansion
\begin{equation}
r_c = {\alpha^2\over T_c N} (r_0 \ln N + r_1) + O(m^a M^b,a+b=2),
\end{equation}
where 
\begin{eqnarray}
r_0 &=& - {\alpha^2 V_4(\mathbf{0},0)\over 8\pi},
\\
r_1 &=& \alpha^2 V_4(\mathbf{0},0) \left[
    {1\over 8\pi} \ln {3 \alpha^4 V_4(\mathbf{0},0) \over 8\pi T_c\Lambda^2}
    - D_2 - {3\over 8\pi}\right]
\nonumber \\
    && 
    - {1\over2} \int_{p<\Lambda} {\di^2 \vp\over (2\pi)^2} \, 
      \left[ {T_c V_4(\vp,0)\over P(\vp,0)} - 
             {\alpha^2 V_4(\mathbf{0},0) \over p^2} \right].
\end{eqnarray}
Note that the three-leg vertex that appears in eq.~(\ref{rc-expr}) 
does not contribute to this order, since it vanishes for $m=M=0$.
Thus, eqs.~(\ref{scal}) can be written as 
\begin{eqnarray}
x_h &\approx & N^{3/2} \left[ - \ol \mu M + 
     a_0 M (T - T_c) + a_1 M^3 + {a_2 M\over N} + \cdots \right] - 
      h_{c0} M \sqrt{N},
\\
x_t &\approx & N \left[-g_{01} (T-T_c) + a_3 M^2 + {e\over N} + \cdots
      \right] - 
            r_0 \ln N - r_1,
\end{eqnarray}
where $a_0$, $a_1$, $a_2$, $a_3$, and $e$ are coefficients that can be read from 
eq.~(\ref{expanHP}). These expansions show that 
\begin{eqnarray} 
&&\ol \mu M = - {x_h N^{-3/2}},
\label{scalM}
\\
&& T-T_c  = {e - r_0 \ln N - r_1 - x_t\over N g_{01}}.
\end{eqnarray}
The critical point is specified by the condition $x_t = x_h = 0$. 
The symmetry under $m,M\to -m,-M$ guarantees that the critical
point corresponds to $M = 0$. On the other hand, 
$1/N$ fluctuations give rise to a shift of the critical temperature.
If $T_c(N)$ is the finite-$N$ critical temperature, we obtain 
\begin{equation}
T_c(N) \approx T_c + {1\over N} {e - r_0 \ln N - r_1\over g_{01}}.
\end{equation}
Note that, beside the expected $1/N$ correction there is also a 
$\ln N/N$ term that is related to the nontrivial renormalization.
It follows
\begin{equation}
T - T_c(N) = - {x_t\over N g_{01}}.
\label{scalT}
\end{equation}
Note that $g_{01}$ is negative [see eq.~(\ref{defgab})]
and thus we have $x_t > 0$ for $T > T_c(N)$, as expected.
Using the gap equation we can also derive the behavior of $m$ in the 
critical crossover limit.  We obtain 
\begin{eqnarray}
m \equiv {m_0 \over N^{1/2}}, \qquad
\end{eqnarray}
where $m_0$ is a function of $\ln N$ that satisfies the equation
\begin{equation}
{1\over 6} ({\ol\lambda - 6 g_{10}}) m_0^3 + 
(r_0 \ln N  + r_{1} -e + x_t) m_0 + x_h = 0.
\label{eqperm0}
\end{equation}
For $N\to \infty$, $m_0$ has an expansion in inverse powers of $\ln N$,
the leading term being
\begin{equation}
m_0 \approx - {x_h \over r_0} {1\over \ln N} + 
    O(\ln^{-2} N).
\end{equation}
Note that $m_0\to 0$ as $x_h \to 0$.

These results confirm the scaling predictions of Ref.~\cite{KSS-98}. 
For the massless theory with $M = 0$, there are two regimes: 
for $N(T-T_c(N))\ll 1$ one observes Ising behavior, while for 
$N(T-T_c(N))\gg 1$ mean-field behavior occurs. If $M\not=0$ the 
same considerations apply, the relevant variable being 
$M N^{3/2}$. It is important to note the role played in the derivation 
by the symmetry $m,M\to -m,-M$, that is present because the 
regularization preserves chiral invariance. Even though vertices with an
odd number of legs are present, the symmetry makes them irrelevant in the 
crossover limit. Thus, the additional renormalizations computed in 
Ref.~\cite{CMP-05} do not play any role here.

The results reported above can be extended to $d$ dimensions for $d < 4$,
 the relevant scaling variables being eqs.\ 
(\ref{provv-chapter1})
\begin{equation}
x_h \sim M N^{(d + 2)/[2(4-d)]},\qquad\qquad
x_t \sim [T-T_c(N)] N^{2/(4-d)}.\qquad\qquad
\label{scaling-dgen}
\end{equation}
In $d = 3$, on the basis of eq.~(\ref{rcd3}),
we also predict for $T_c(N)$ an expansion of the form
\begin{equation}
T_c(N) \approx T_c + {a\over N} + {b\ln N + c\over N^2},
\end{equation}
where $a$, $b$, and $c$ are constants that can be computed as in the 
two-dimensional case. However notice that, as pointed out in 
sec.\ \ref{chapter2-higher-dimension}, in three dimension
(and in general for $d>2$) one have to translate
the critical field by a constant $k_R$
(\ref{k-ren}) that have to be taken into
account in order to 
cancel some diagram for $\chi_3$ that do not go to
zero in the Critical Crossover Limit eq.\ 
(\ref{kr-esplicito}). However in the symmetric 
case this is not necessary
because the explicit $\mathbb{Z}_2$ symmetry of the
action is protected in the scaling limit. Indeed
the two one loop anomalous contributions 
($\equiv D^{(3)}_1, D^{(3)}_2$) for $\chi_3$ 
[see point c) in \ref{chapter2-higher-dimension}]
are\footnote{In three dimension the integrals
are both Infra Red finite.}
\begin{eqnarray}
D^{(3)}_1 &=& {1\over t^3 N^{3/2}} \int \di^3\vp {\ol V^{(3)}(\vp)^3
\over K(\vp)^3}  \sim {M^3 \over t^3 N^{3/2}}
\\
D^{(3)}_2 &=& {1\over  t^3 N^{3/2}} \int \di^3\vp {\ol V^{(3)}(\vp)^2
\ol V^{(4)}(\vp) \over K(\vp)^2}\sim {M \over t^3 N^{3/2}}.
\end{eqnarray}
In the critical crossover limit in $d=3$ from eq.\
(\ref{scalingrel2}) follows that $\chi_3 \sim N^{9/2}$
and that $t\sim N^{-2}$ so that
\begin{eqnarray}
N^{-9/2} D^{(3)}_1\sim M^3  &\qquad & N^{-9/2} D^{(3)}_2\sim M
\end{eqnarray}
that are obviously  irrelevant in the CCL. We stress that the
reason $D^{(3)}_1$ and $D^{(3)}_2$ are negligible
is the equation (\ref{relV3V4}) that holds for 
all models which are chiral symmetric.
In  general 
 the effect of $k_R$ must be taken
into account for all Hamiltonian which broke
chiral symmetry as for instance in the case
of Wilson fermions. This system is currently under investigation. 

\subsection{Correlation functions} \label{sec4.3}

The results reported in sec.~\ref{sec4.2} allow us to compute the 
scaling behavior of the correlation functions. For instance,
we have
\begin{equation}
\langle \phi(\mathbf{x}_d,x_{d+1}) \rangle 
 = {\ol \phi} - {k \over g} + {\alpha\over g \sqrt{N}} 
   \langle \chi(\mathbf{x}_d) \rangle
\end{equation}
Using eq.~(\ref{scalingrel2}) with $n = 1$ and $d = 2$, we have 
$\langle \chi(\mathbf{x}_d) \rangle = f_1(x_h,x_t)$ in the 
critical crossover limit.
The background term can be neglected in the crossover limit since
\begin{equation}
{\ol \phi} - {k\over g} \approx 
   {1\over g} \left[m - {V_3(\mathbf{0},0)\over V_4(\mathbf{0},0)}\right] 
   \approx 
    {M\over g} {6 g_{10} \over \ol\lambda - 6 g_{10}} \sim N^{-3/2}.
\end{equation}
Thus, we obtain 
\begin{equation}
\langle \phi(\mathbf{x}_d,x_{d+1}) \rangle \approx
    {\alpha\over g \sqrt{N}} f_1(x_h,x_t)
\end{equation}
The factor $1/\sqrt{N}$ is related to the particular normalization 
of $\phi$ used in (\ref{action}) and disappears if we redefine 
$\varphi = \sqrt{N}\phi$ in order to have a canonical kinetic term for 
$\varphi$. The function $f_1(x_h,x_t)$ is the scaling function for the 
magnetization in the Ising model. For instance, for $x_h = 0$ and 
$x_t < 0$ (low-temperature phase), we have 
$f_1(0,x_t) \sim (-x_t)^{\beta_I}$ and 
$f_1(0,x_t) \sim (-x_t)^{\beta_{MF}}$ respectively for 
$|x_t| \ll 1$ and $|x_t| \gg 1$, where $\beta_I = 1/8$ and 
$\beta_{MF} = 1/2$ are the magnetization exponents in the 
Ising and in the Gaussian model. The universality of the crossover 
allows us to compute the scaling functions in any other model in which 
there exists a crossover between the Gaussian and the Ising fixed point.
In particular, we can use the results for systems with medium-range 
interactions \cite{LBB-96,LBB-97} (see also the appendix \ref{App}). 
In Ref.~\cite{LBB-97} (LBB) the authors 
report $\langle |m| R\rangle$ versus $t R^2$ (see their Fig. 9), where 
$m$ is the magnetization, $t$ the reduced temperature, and $R$
the effective interaction range. These results give us
$\langle \phi(\mathbf{x}_d,x_{d+1}) \rangle$ for $x_h = 0$. One only needs
to take into account the different normalizations of the fields, of the 
coupling constant, and of the scaling variable. 
In the crossover limit $N\to\infty$, $T\to T_c(N)$ at
fixed $N(T - T_c(N))$ we have
\begin{eqnarray}
&& g \sqrt{N}\langle \phi(\mathbf{x}_d,x_{d+1}) \rangle  = 
K_{1,\rm LBB} \langle |m| R\rangle_{\rm LBB} \\
&& (t R^2)_{\rm LBB} = K_{\rm LBB} N[T - T_c(N)].
\end{eqnarray}
The nonuniversal constants $K_{\rm LBB}$ and  $K_{1,\rm LBB}$ are 
computed in the app.\ \ref{App}.

It is customary to define an 
effective exponent $\beta_{\rm eff}(T)$ as
\begin{equation}
\beta_{\rm eff}(T) = [T - T_c(N)] {d\over dT} \ln 
   \langle \phi(\mathbf{x}_d,x_{d+1}) \rangle,
\end{equation}
for $M = 0$ and $T < T_c(N)$. In the crossover limit 
$T\to T_c(N)$, $N\to\infty$ at fixed $N[T - T_c(N)]$, 
the exponent $\beta_{\rm eff}(T)$ interpolates 
between the Ising value $\beta_I = 1/8$ and the mean-field $\beta_{MF} = 1/2$.
Again, this effective exponent can be derived from the results of 
Ref.~\cite{LBB-97}.
The curve reported in Fig.~15 of Ref.~\cite{LBB-97}
gives $\beta_{\rm eff}$ in the Yukawa model once $t R^2$ is replaced by 
$K_{\rm LBB} [T - T_c(N)] N$. 

The same considerations apply to the connected zero-momentum $n$-point
function $\chi_n$:
\begin{eqnarray}
\chi_n &=& 
  \int \di^{d+1}\mathbf{x}_2 \ldots  \di^{d+1}\mathbf{x}_n \, 
  \langle \phi(\mathbf{0})\phi(\mathbf{x}_2)\ldots 
          \phi(\mathbf{x}_n) \rangle^{\rm conn} 
\nonumber \\ 
   &=&
  {\alpha^n\over T^{n-1} g^n N^{n/2}} 
  \int \di^{d}\mathbf{x}_2 \ldots  \di^{d}\mathbf{x}_n \, 
  \langle \chi(\mathbf{0})\chi(\mathbf{x}_2)\ldots 
          \chi(\mathbf{x}_n) \rangle^{\rm conn} 
\nonumber \\ 
&=& {\alpha^{4-3n} V_4(\mathbf{0},0)^{1-n} \over g^n} N^{n/2-1} f_n (x_h,x_t),
\end{eqnarray}
For $n = 2$ the crossover function for $x_h = 0$ can be obtained from the 
results of Ref.~\cite{LBB-97}, since 
$g^2 \chi_2 = K_{2,\rm LBB} (\tilde{\chi} R^2)_{\rm LBB}$. The constant
$K_{2,\rm LBB}$ is given in the Appendix.

One can also use field theory to compute the crossover curves and 
thus use the results of Ref.~\cite{PRV-99}.
For instance, in the high-temperature phase, for $M = 0$ we have 
in the crossover limit 
\begin{equation}
g^2 \chi_2 = K_{2,\rm FT} F_\chi (\tilde{t}) ,\qquad\qquad
\tilde{t} = K_{\rm FT} N [T - T_c(N)],
\end{equation}
where $F_{\chi}(\tilde{t})$ is reported in Ref.~\cite{PRV-99} and 
$K_{\rm FT}$, $K_{2,\rm FT}$ are nonuniversal constants computed in the 
appendix \ref{App}.

In the discussion presented above we have focused on the case $d=2$, 
but it is immediate to generalize all these considerations to the 
three-dimensional case. For $d=3$ the universal crossover curves have 
been computed in Refs.~\cite{BB-85,BBMN-87,PRV-99,BB-02} (field theory) and 
in Ref.~\cite{LB-98} (medium-range models). These results apply directly to 
the Yukawa model.

\section{Conclusions} \label{sec5}

In this chapter we have considered the Yukawa model in the limit 
$N_f\to \infty$, focusing on the crossover between mean-field and 
Ising behavior. For this purpose we have determined the action of the 
mode that becomes critical at the transition. In the long-distance limit, 
it becomes equivalent to that of a weakly coupled $\phi^4$ theory. This 
identification allows us to use the results available 
for this model summarized in
chapter \ref{chapter2} (\cite{BB-85} \cite{CMP-05}).
 In particular, we have  identified 
a universal critical crossover occurring for $N_f\to \infty$, 
$M\to 0$, and $T - T_c(N) \to 0$ at fixed $x_t$ and $x_h$, see
eqs.~(\ref{scalM}), (\ref{scalT}), (\ref{scaling-dgen}). 
In field-theoretical terms,
this behavior represents the crossover 
induced by the flow from the 
unstable Gaussian fixed point to the stable Ising fixed point. 
Quantitative results for the Yukawa
 model can be obtained by using 
the field-theoretical results of Refs.~\cite{BB-85}, \cite{PRV-99}, \cite{BB-02},
or the Monte Carlo results available for medium-range models
\cite{LBB-97,LB-98}. The necessary nonuniversal renormalization 
constants 
can be computed in perturbation theory. Results for $d = 2$ are 
reported in the appendix. 

We should stress that our results are not specific of the chosen regularization,
but can be extended to other regularization as well. In particular,
the extension to Kogut-Susskind fermions \cite{Kogut:1974ag}, 
the model considered in Ref.~\cite{KSS-98}, is essentially straightforward and have not been presented here.
The Wilson case is more involved. Indeed, the absence of chiral symmetry 
implies that the symmetry relations satisfied here by the effective 
vertices [see eq.~(\ref{symmetryVf})] are no longer valid. In turn, this 
may imply additional mixing as it happens in the generalized
Heisenberg model \cite{CMP-05}. Also the higher than two dimension
is more complicate for Wilson fermions, due to the necessity
of introducing and tune another parameter $k_R$.

Let us note that all calculations presented here refer 
to the model in infinite spatial volume. However, the crossover 
behavior can also be observed in the finite-size scaling limit. The discussion
in sec.~\ref{sec4.1} can be easily extended to this case too. It is trivial
to verify that the correct scaling variable is $\tilde{L} = L u^{1/(4-d)}$, 
i.e. $\tilde{L} = L N^{-1/(4-d)}$ in the Yukawa model. 
Again, one can use universality
and obtain predictions for the Yukawa model from the results obtained in other 
contexts. In particular, one can use the finite-size scaling results of 
Refs.~\cite{LBB-96,LBB-97,LB-98} that refer to medium-range models at the 
critical point.

Finally, we should mention that one could also
 generalize the model and consider 
fermion fields $\psi_{\alpha f}$ transforming according 
to a representation of 
a group $G$  and a coupling of the form $\ol\psi_f T^a \phi^a \psi_f$, where 
$T^a$ are the generators of the algebra of $G$. 
The discussion is 
essentially unchanged, though in this case one would obtain the 
vector $\phi^4$ theory; several models have been just studied
pointing out the reduction of the critical zone
\cite{SC-05} \cite{S-04} \cite{S-03}. 
Field-theory results relevant for this case are given in 
Refs.~\cite{BB-85}, \cite{PRV-99}, \cite{BB-02}.
More in general we believe that the considerations and the
results obtained in this chapter are not peculiar
of fermion models  but could be recovered for instance 
also for vectorial models at finite temperature.\footnote{
See for instance \cite{S-02}.
} This is not an unexpected result in the light of 
chapter \ref{chapter6} where the large $N$ scheme 
studied in this work is applied to a vectorial
$\phi^6$ interaction. Indeed using standard arguments
one can see as the introduction of a cut off
(the temperature in the $\phi^4$ interaction)
can be resummed in higher order vertices.

We have just stressed  that a careful numerical investigation of  
fermionic models (of the type presented in this chapter) could be 
a very precise test of the idea presented in this work, due to the
fact that a lot of available simulations (also for several $N_f$)
are present \cite{KSS-98} \cite{SC-05} \cite{S-04} \cite{S-03}.
The goals of such a studies would be the determination
of the $1/N$ correction to the critical parameter $T_c(N)$ 
[and $M_c(N)$ in the case
of Wilson fermions] to compare with the simulation data for several $N$.
This step could be require few time and is actually 
under study. On the other hand is not clear to us if the data available
in literature could confirm the claim that
all the crossover considered in this work are universal, i.e.\ if
for instance medium range models can be compared with Yukawa models.
It is our opinion that such a question merit further 
investigations.

%%%%%%%%%%%%%%%%%%%%%%%%%%%%%%%%%%%%%%%%%%%%%%%%%%%%%%%%%%%%
%                                                          %
% CHAPTER 6 - CHAPTER 6 - CHAPTER 6 - CHAPTER 6 -CHAPTER 6 %
%                                                          %
%%%%%%%%%%%%%%%%%%%%%%%%%%%%%%%%%%%%%%%%%%%%%%%%%%%%%%%%%%%%

\chapter{Crossover in tricritical models}\label{chapter6}

In this chapter we want to give another 
example in which our 
technique can be used to elucidate 
some aspects of a tricritical phase diagram 
  in the large $N$ limit. The
Hamiltonian (\ref{hamiltonian-tricritical}) is usually
used in order to investigate systems that exhibit 
a tricritical 
phase transition like for instance metamagnet \cite{F-meta}
or fluid $^3$He-$^4$He mixtures. The large $N$ limit
 has been investigated
long time ago \cite{SF-78}. The standard scenario
predicts (for $H\neq 0$) two lines of continuous 
phase transitions 
(that usually are called
wing lines) that end joining together 
in a tricritical point.
However some questions remain open.
Indeed the ref.\ \cite{SF-78} shows that the 
continuous transitions observed along the wing
lines are Mean Field like while because of 
symmetry argumentation, the authors
 expect a Ising behavior for every $N$. 
In this chapter we want to show as the two
different critical behavior can be reconciled
defining the Critical Crossover Limit
for Hamiltonian (\ref{hamiltonian-tricritical}).
The presentation strictly follows chapter
\ref{chapter1} and \ref{chapter3} and some steps
are missed. In particular we take care 
to define (sec.\ \ref{gap-h3-0}) the $N=\infty$ scaling fields of the theory (with which using results of sec.\ 
\ref{chapter1-sec3}
one can recover the $N=\infty$ phase diagram)
and to study the stability of the weakly
coupled theory (sec.\ \ref{sec-stable}) that is the basic 
point in order to define the Critical Crossover Limit
which implementation for interaction 
(\ref{hamiltonian-tricritical}) is identical
to what done in chapter \ref{chapter3}
for $\N=1$ Heisenberg models.

\section{The model}

In this chapter we will consider  a vectorial 
model with a $\phi^6$
 interaction.
Let us discuss the following lattice regularization
\begin{eqnarray}
{\cal H}[\phi] &=&-
{1\over 2} \sum_{x,\mu} \vec \phi_x \vec \phi_{x+\mu}
-\sum_x \Big(\vec H \vec \phi_x + (\vec H_3\vec \phi_x)
{\vert\vec{\phi}_x\vert^2 \over N}
- {r\over 2}\vert\vec \phi_x\vert^2-{u\over 4!}
{\vert\vec \phi_x\vert ^4\over N}
-{v\over 6!}{\vert \vec \phi_x \vert^6\over N^2}\Big)
\nonumber\\
\label{hamiltonian-tricritical}
\end{eqnarray}
where $\vec H$ and $\vec H_3$ are two vectors with $N$ components  equal 
respectively to $H$ and $H_3$.
Hamiltonian (\ref{hamiltonian-tricritical}) has been
 used to investigate tricritical point. In particular the
common accepted phase diagram exhibits two
lines of continuous phase transition 
(that are usually called wing lines) that
end in a tricritical point. Detecting the tricritical 
point in general is not an easy work and it requires
an accurate fine tuning of $r$, $u$ and $v$ (while
for symmetric reasons at the tricritical point
one has $H=H_3=0$)\footnote{However the study of interaction
(\ref{hamiltonian-tricritical}) in the weak coupling limit
(chap.\ \ref{chapter2} for $\N=3$),  could give
the leading term of the expansion of $r_c(v)$ and
 $u_c(v)$ apart a $\sim v$-factor similar to $A$ defined
in sec.\ \ref{sec6.5}. } (for more details we refer
to \cite{LS-84}).

The large $N$ limit can be studied 
introducing auxiliary
fields as done in chapter \ref{chapter1}.
In this case we need only two auxiliary fields
($\lambda$, $\mu$),
because we relax the unit spin constraint
in order to follow \cite{SF-78}\footnote{
However as we have just noticed during this work, 
in the large $N$ limit unit spin constraint 
is usually irrelevant
\cite{Zinn-Justin} \cite{MZ-03}.}. 
Thus we
can write
\begin{eqnarray}
{\cal Z} &=& \int \prod_x \di \vec \phi_x \, e^{ - \cal{H}[\phi]}
\nonumber\\
&\sim& \int \prod_x \di\vec\phi_x\di \mu_x \di \lambda_x \,
\exp\Bigg[
{1\over 2}\sum_{x,\mu} \vec \phi_x \vec \phi_{x+\mu}
+ \sum_x \Bigg({N\over 2} \lambda_x \Big(\mu_x 
-{\vert\vec \phi_x\vert^2\over N}\Big) 
+\vec H \vec \phi_x + (\vec H_3\vec \phi_x)\mu_x
\nonumber\\
&&-{N r\over 2}\mu_x-{N u \over 4!} {\mu_x}^2 - {N v \over 6!} {\mu_x}^3
\Bigg)\Bigg]
\end{eqnarray}
Integrating the $\phi$ field in the previous expression 
we obtain the effective action $A$ for the auxiliary fields
that is the starting point for every $1/N$
investigation:
\begin{eqnarray}\label{effaction}
{\cal Z}&\sim& \int \prod_x \di \mu_x \di \lambda_x \,
\exp\Bigg[
-{N\over 2}\,\tra \log {\cal O}[\lambda]+{N\over 2}(H+H_3 \mu,
{\cal O}^{-1}[\lambda],H+H_3 \mu)
\nonumber\\
&&+\sum_x\Bigg({N\over 2}\lambda_x\mu_x
-{N r\over 2}\mu_x-{N u \over 4!} {\mu_x}^2 - {N v \over 6!} {\mu_x}^3
\Bigg)\Bigg]
\nonumber\\
&\equiv&\int \prod_x \di \mu_x \di \lambda_x \,e^{-N A[\mu,\lambda]}
\label{cp6-def-A}
\end{eqnarray}
where we have introduced the operator ${\cal O}$
that acts on  scalar fields in the following manner:
\begin{eqnarray}
(\varphi,{\cal O}[\lambda],\varphi)&=&-\sum_{x,\mu}  \varphi_x 
 \varphi_{x+\mu}+\sum_x \lambda_x {\varphi_x}^2
\end{eqnarray}
For constant value of $\lambda_x=\lambda$, ${\cal O}$
can be diagonalized in momenta space so that
\begin{eqnarray}
{\cal O}[\ol\lambda ](\vp) &=& {1\over 2}(
\wi \vp^2 +2 \la -2 d ).
\end{eqnarray}

\section{Gap-Equation. The $H_3=0$ case}\label{gap-h3-0}

\begin{figure}
\begin{center}   
\includegraphics[angle=0,scale=0.5]{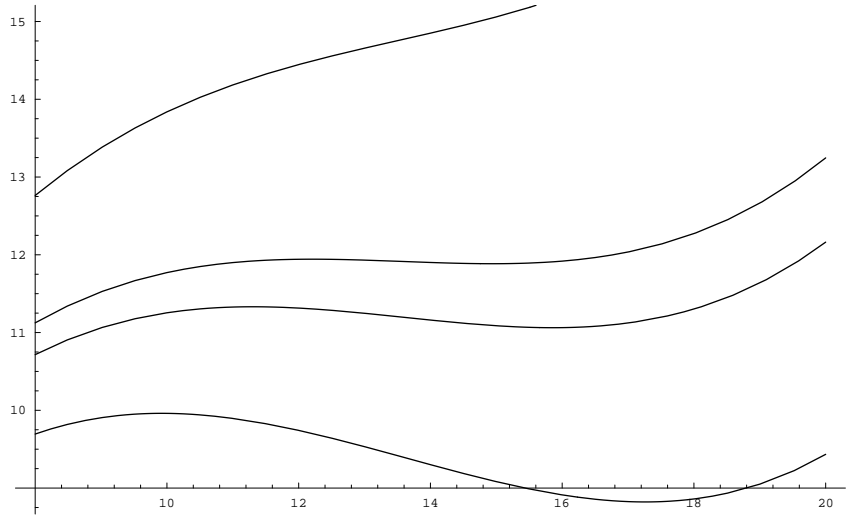}
\caption{$H^2$ versus $\ol\mu$ with  $v=1/4$, $u=-1/2$
 and (from below) $r=5$, $5.05$, $5.07$ and $5.15$}\label{accaq}
\end{center}
\end{figure}

We are interested to understand the crossover from Mean Field criticality
to Ising criticality on the wing lines of the tricritical phase diagram.
In order to do that it  it is expected \cite{SF-78} 
we can put $H_3=0$ without destroy
the qualitative picture.
Indeed imposing the stationarity of the effective
 action (\ref{effaction}) 
we obtain two equations for the fields at the stationary point 
$\ol\mu$ and $\ol\lambda$  
\begin{eqnarray}\label{gapequation}
 \ol\mu &=&2 B_1(\h)+4H^2B_2(\h)
\nonumber\\
\ol\lambda &=& r+{u\over 3!}\ol\mu+{v\over 5!}{\ol\mu}^2
\end{eqnarray}
where we have defined
\begin{eqnarray}
\h &\equiv & 2\ol \lambda -2d.
\label{m-tricritico}
\end{eqnarray}
In eq.\ (\ref{gapequation}) we have taken the notation
of the previous chapters defining
\begin{equation}
B_n(\h)=\int_{\mathbf{p}}{1\over (\wi p^2+\h)^n}
\end{equation}
From the previous equation (\ref{gapequation}) we can obtain $H$ (and
of course $\ol\lambda$) as a function of the $O(N)$ symmetric 
observable $\ol\mu$:\footnote{Indeed using the equations
of motions we get
$\ol\mu=<\vert\vec\phi_x\vert^2/N>$}
\begin{equation}\label{accaqfor}
H^2={\ol\mu-2 B_1(\h)\over 4 B_2(\h)}
\end{equation}
where $\h$  as a function of $\ol\mu$ is obtained
using eq.\ (\ref{m-tricritico}) and the second of
eqs.\ (\ref{gapequation}).
Interesting enough is the fact that 
$H^2$ as a function of $\ol\mu$ in general
cannot be inverted as depicted in fig.\ \ref{accaq}. 
This is the sign of the appearance of a phase transition,
the mechanism being the same of what presented in 
chapter \ref{chapter1} or ref.\ \cite{CP-02} for Heisenberg
models. In this case the magnetic field $H$ plays
the role of the temperature (or spin-spin correlation length)
in Heisenberg models. The key point is that in both model
phase transitions happen at finite temperature ($\beta<\infty$
for the Heisenberg models) and at finite Magnetic
Field ($H\neq 0$ for the model studied in this chapter).
This fact is the reason for which infrared divergences
appear and the standard expansion fails, so that the 
crossover mechanism is observed instead of a one parameter
($N$) family of critical points.

Figure \ref{accaq} gives evidence of the existence of a
 line of critical point (wing line)
near which we have (in the following we reduce 
to the $H>0$ case):
\begin{equation}\label{grap}
H-H_c \sim (\ol\mu-\ol\mu_c)^3  
\end{equation}
At this point the Mean Field considerations of the chapter
\ref{chapter1} and of reference \cite{CP-02} can easily
be applied with analogous results. In order to elucidate this
claim,
if we take as tunable parameters of our theory $r$ and $H$
($u$ and $v$ remaining fixed for instance at the value
reported in fig.\ \ref{accaq}) and expanding the gap-equation
near the critical point $H_c\approx 23$, $r_c \approx 5.07$ and 
$\ol \mu_c\approx 14$ (see fig.\ \ref{accaq})
we get the following expansion
\begin{eqnarray}
H^2-H^2_c &=& \sum_{n,m} a_{nm}(r-r_c)^n (\ol \mu -
\ol\mu_c)^m 
\nonumber\\
a_{0i} &=& 0 \qquad \mathrm{for} \,\, i=0,1,2.
\label{acca-near-cp}
\end{eqnarray}
Equation (\ref{acca-near-cp})
and  the gap equation for $\N=1$ Heisenberg models
(\ref{expansion-gap}) are identical apart
replacing $r$ with $p$ and $\ol\mu$ with $\h$.
In particular proceeding in such a way the $N=\infty$
scaling fields can be obtained using
eq.\ (\ref{scalingfield-uh}) and eq.\
(\ref{chapter1-def-x}). Then following the considerations
of chapter \ref{chapter3} we expect to 
extend the $N=\infty$ analysis including also
$1/N$ fluctuations. In particular this allows us to
explain the crossover between Mean-Field to Ising
behavior using universal crossover functions that have
been investigated in chapter \ref{chapter2} and 
have been used in all the rest of this work.

In the next section \ref{sec-6.3} 
we will obtain the Propagator and the vertices
of the auxiliary  fields $\mu,\lambda$, in particular
we want to show the presence of the zero mode of
the theory. Having done this it will be clear that
 the $1/N$ results obtained for the Heisenberg models
(chap.\ \ref{chapter3})
or for the Yukawa model (chap.\ \ref{chapter5}) 
strictly apply also for Hamiltonian
(\ref{hamiltonian-tricritical}), confirming
the claim that for every $N$ finite on the wing
lines one observes Ising phase transitions that
cross --in a universal way-- on  Mean field phase 
transitions for $N=\infty$.

\section{$1/N$ Expansion and Crossover to Ising Behavior}
\label{sec-6.3} 
Now we are going to show that eq.\ (\ref{grap}) implies the 
existence of a zero mode, so that the picture developed in
chapter \ref{chapter3} can be applied also in this case.
Introducing fluctuations for the auxiliary fields to the
saddle point solutions (\ref{gapequation})
\begin{eqnarray}
\mu_x     &=& \ol\mu + {\wi \mu_x\over \sqrt N}
\nonumber\\
\lambda_x &=& \ol\lambda + {\wi \lambda_x\over \sqrt N}
\end{eqnarray}
the standard $1/N$ expansion [characterized by the
inverse of the propagator $P^{-1}_{AB;\ol\mu}(\mathbf{p})$ and by the effective
vertices $V^{(j)}_{A_1\cdots A_j;\ol\mu}(\mathbf{p_1},\cdots\mathbf{p_j})$]
is easily obtained.
In order to give an outlook on how a zero mode
appears near a gap equation that exhibits criticality like
that presented in fig.\ \ref{accaq} we try
to differentiate the gap-equation (\ref{cp6-def-A})
(rewritten below)
\begin{equation}
{\delta \over \delta \Psi_A(\mathbf{0})} A[\ol\Psi(\ol\mu),H(\ol\mu)]=0
\end{equation}
[$\Psi_1(\mathbf{p}) \equiv \wi \mu (\mathbf{p})$ 
and $\Psi_2(\mathbf{p}) \equiv \wi \lambda(\mathbf{p})$] 
with respect to  $\ol\mu$ obtaining
\begin{equation}
\sum_B {\delta^2 S[\ol\Psi(\ol\mu),H(\ol\mu)]\over \delta \Psi_A(\mathbf{0})
\delta \Psi_B(\mathbf{0})}{\di\ol\Psi_B \over \di \ol\mu }
+{\delta^2 S[\ol\Psi(\ol\mu),H(\ol\mu)]\over \delta \Psi_A(\mathbf{0})
\delta H }{\di H \over \di \ol\mu}=0. 
\label{diff-gap}
\end{equation}
In the previous equation (\ref{diff-gap})
at the critical point the second term cancels [like
$(\ol\mu-\ol\mu_c)^2$] pointing out that $P^{-1}$
is singular at the critical point confirming the
starting claim. In particular (denoting with
$z_A(\vp)$ the zero mode of the theory),
neglecting a normalization factor, from
eq.\ (\ref{diff-gap}) easily follows 
\begin{eqnarray}
\sum_B P^{-1}_{AB}(\mathbf{0}) z_B(\mathbf{0}) &\sim& 
(\ol\mu-\ol\mu_c)^2 z_A(\mathbf{0})
\nonumber\\
z_B(\mathbf{0})  &=& 
{\di\ol\Psi_B \over \di \ol\mu } + O(\ol\mu-\ol\mu_c)^2.
\label{this-is-zero}
\end{eqnarray}
The previous equation is similar to relation
 (\ref{what-is-zero}) found
for Heisenberg models.\footnote{ In this case the
auxiliary fields are two instead of five. However
relation (\ref{this-is-zero}) is general once 
$A$ is taken to run on the number of auxiliary 
fields present in the theory}

To explicitly verify the previous consideration we compute 
$\det P^{-1}$. It is easy to obtain the following
formula
\begin{eqnarray}
P^{-1}_{11}(\mathbf{0})&=&{2 u\over 4!}+{v \ol\mu\over 5!}
={1\over 4} {\di \h \over \di \ol \mu}
\nonumber\\
P^{-1}_{12}(\mathbf{0})&=&-{1\over 2}
\nonumber\\
P^{-1}_{22}(\mathbf{0})&=&-{8 H^2 }B_3(\h)-2B_2(\h)
=\Bigg({\di \h\over \di\ol\mu}\Bigg)^{-1}
\Bigg(1-B_2(\h){\di H^2 \over \di\ol\mu} \Bigg)
\label{inverse-P-metama}
\end{eqnarray}
Using (\ref{accaqfor}) we obtain
\begin{equation}
\det P^{-1}(\mathbf{0})
=- B_2(\h){\di H^2\over \di \ol \mu}
\end{equation}
confirming the presence of the zero mode 
at the critical point (fig.\ \ref{accaq}). Notice
that the sign of $\di H^2/\di\ol\mu$ is positive near the
critical point [i.e.\ $a_{03}>0$ in eq.\
(\ref{acca-near-cp})], this will be relevant to discuss
the stability of the theory.

\section{On the stability of the Weakly-Coupled theory}\label{sec-stable}

In this chapter we want to do some considerations
on the stability of the weakly coupled theory describing
the crossover between Ising to Mean Field behavior. 
This is in principle the unique think that could
destroy the picture we have presented in the previous
sections. We have just noticed in sec.\ 
\ref{sec-hnl-5} how the positivity\footnote{
This is related to the stability of the weakly $\phi^4$ theory
in the sense that if the positivity condition fails
than  one needs to
consider also higher than four legs vertex that have
been discarded in chapter \ref{chapter2} simply 
using scaling arguments.}
 of the zero mode four legs vertex is related to 
the positivity of the zero mode mass near the critical point
 (see eq.\ \ref{id-V4-CP}).

Using equations
(\ref{inverse-P-metama}) it is easy to compute the 
critical eigenvalue $\la_0$ of $P^{-1}$.
If we define
\begin{eqnarray} 
m' &\equiv& {\di \h \over \di \ol\mu}
\end{eqnarray}
we find
\begin{eqnarray}
\la_0 &=& -{B_2(\h)\over 4}
\Bigg( {m'\over 4} +{1\over m'}\Bigg)^{-1}
{\di H^2\over \di \ol\mu}
+O\Bigg({\di H^2\over \di \ol\mu} \Bigg)^2
\end{eqnarray}
so that our stability condition becomes $m'<0$
(we have just noticed that $a_{03}>0$), and
using eq.\ (\ref{m-tricritico}) we get 
the following condition
\begin{eqnarray}
v \ol \mu_c < -10 u
\label{stability}
\end{eqnarray}
Figure \ref{accaq} gives numerical evidence
that the critical point we have considered 
satisfies previous relation (\ref{stability}).
However we expect that the 
same result holds on all the wing lines.

%
% appendici
%

\appendix

\chapter{Properties of $x{\cal G}$  eq.\ (\ref{gap})} \label{app-G-cont}

We want  to proof relations (\ref{property-G}). 
Due to the symmetry ${\cal G}(-x,T)={\cal G}(x,T)$
we will consider here only $x>0$.
Notice that
\begin{eqnarray}
x{\cal G}(x) &:=&{1\over 2}
 \int^{\Lambda}\di \mathbf{p} \, f_{T}(x,\mathbf{p})
\nonumber\\
f_{T}(x,\mathbf{p}) &=&
{\di w_x(\mathbf{p}) \over \di x}
\tanh{w_x(\mathbf{p})\over 2 T } 
\end{eqnarray}
where we have defined:
\begin{eqnarray}
w_x(\mathbf{p}) &=& \sqrt{\mathbf{p}^2+x^2}
\label{app-def-w}
\end{eqnarray}
Then using the previous we have:
\begin{eqnarray}
{\di w_x(\mathbf{p})\over \di  x} &=& {x\over w_x(\mathbf{p})}
\\
{\di^2 w_x(\mathbf{p})\over \di  x^2} &=&{1\over w_x(\mathbf{p})}\Bigg(1-\Big({\di w_x(\mathbf{p})\over \di  x }\Big)^2\Bigg)
\\
{\di^3 w_x(\mathbf{p})\over \di  x^3} &=& -{3\over w_x(\mathbf{p})^2}{\di w_x(\mathbf{p}) \over \di x}\Bigg(1-\Big({\di w_x(\mathbf{p})\over \di  x }\Big)^2\Bigg)
\nonumber\\
&=&-{3\over w_x(\mathbf{p})}{\di w_x(\mathbf{p})\over \di  x}{\di^2 w_x(\mathbf{p})\over \di  x^2}
\label{tredidue}
\end{eqnarray}
We have $x\le w_x(\mathbf{p})$ (\ref{app-def-w}),
 so that both the first and second 
derivative or $w$ are positive while the third derivative is always negative.
Then we have:
\begin{eqnarray}
{\di f_T(x,\mathbf{p})\over \di x} &=&{\di^2 w_x(\mathbf{p})\over \di  x^2}
\tanh{w_x(\mathbf{p})\over 2T}+{1\over 2T}\Big({\di w_x(\mathbf{p})
\over \di  x} \Big)^2 \cosh^{-2}{w_x(\mathbf{p})\over 2T}
\label{intprima}
\end{eqnarray}
which is always positive for the previous considerations.
For the second derivative we find:
\begin{eqnarray}
{\di^2 f_T(x,\mathbf{p})\over \di x^2} &=&{\di^3 w_x(\mathbf{p})
\over \di  x^3} \tanh{w_x(\mathbf{p})\over 2T}+{3\over  2T}{\di w_x(\mathbf{p})
\over \di  x}{\di^2 w_x(\mathbf{p})\over \di  x^2}\cosh^{-2}{w_x(\mathbf{p})\over 2T} 
\nonumber\\
&&-{1\over 2 T^2}\Big( {\di w_x(\mathbf{p}) \over \di  x}\Big)^3 
\sinh{w_x(\mathbf{p})\over 2T}\cosh^{-3} {w_x(\mathbf{p})\over 2T}
\nonumber\\
&=&-{3\over w_x(\vp)}{\di w_x(\mathbf{p})
\over \di  x}{\di^2 w_x(\mathbf{p})\over \di  x^2}
\Bigg(  \tanh {w_x(\mathbf{p})\over 2T} -{w_x(\mathbf{p})\over 2T }
\cosh^{-2}{w_x(\mathbf{p})\over 2T} \Bigg)
\nonumber\\
&&-{1\over 2 T^2}\Big( {\di w_x(\mathbf{p}) \over \di  x}\Big)^3 
\sinh{w_x(\mathbf{p})\over 2T}\cosh^{-3} {w_x(\mathbf{p})\over 2T}
\nonumber\\
&\le & 0
\label{intseconda}
\end{eqnarray}
having used the fact that:
\begin{equation}
\tanh x = {1\over 2}{\sinh 2x\over \cosh^2 x} \ge {x\over \cosh^2 x}
\label{aa}
\end{equation}
Immediately (\ref{intprima}) and (\ref{intseconda}) give us
the relations (\ref{property-G}).
Otherwise similar results show that ${\cal G}(x,T)$ 
\begin{eqnarray}
{\cal G}(x,T) &=&{1\over 2} \int^{\Lambda}\di\mathbf{p}
{1\over w_x(\mathbf{p})}\tanh{w_x(\mathbf{p})\over 2 T}
\end{eqnarray}
is a decreasing function of $x$. Indeed we find
\begin{eqnarray}
{\di{\cal G}(x,T)\over \di x } &=&- {1\over 2} \int^{\Lambda}\di\mathbf{p}
{1\over w_x(\mathbf{p})^2 }{\di w_x(\mathbf{p})\over \di x}
\Bigg(\tanh { w_x(\mathbf{p})\over 2 T}-{w_x(\mathbf{p})\over 2 T}
\cosh^{-2}{w_x(\mathbf{p})\over 2 T}\Bigg)
\label{decreasing}
\end{eqnarray}
and the claim follows using the (\ref{aa}).

%%%%%%%%%%%%%%%%%%%%%%%%%%%%%%%%%%%%%%%%%%%%%%%%
%%%%%%%%%%%%%%%%%%%%%%%%%%%%%%%%%%%%%%%%%%%%%%%%
%%%%%%%%%%%%%%%%%%%%%%%%%%%%%%%%%%%%%%%%%%%%%%%%
%%%%%%%%%%%%%%%%%%%%%%%%%%%%%%%%%%%%%%%%%%%%%%%%
%%%%%%%%%%%%%%%%%%%%%%%%%%%%%%%%%%%%%%%%%%%%%%%%
%%%%%%%%%%%%%%%%%%%%%%%%%%%%%%%%%%%%%%%%%%%%%%%%

\chapter{Relations  among the Yukawa model, medium-range models, 
and field theory} \label{App}

In this appendix we want to relate the weakly coupled
$\varphi^4$ theory, medium-range models, and the Yukawa model 
for $d=2$. This models have been considered in this work
as prototype for large-$\N$ phase transition with infrared
singularities. More interesting the solution proposed 
(i.e.\ to resum the divergences introducing 
$N$-dependent scaling variables with scaling relations given
by a weakly coupled $\varphi^4$ theory) has given us
the possibility to identify the crossover between Ising
behavior to Mean Field behavior in all such models.
In this appendix we want to elucidate how 
one can use numerical available data on medium range model
to obtain numerical informations on Yukawa
models (or vice-versa).

For simplicity, we only consider the case $H=0$, corresponding to 
$M = 0$ in the Yukawa model. The field-theory model has been
discussed in sec.~\ref{chapter2-1}, where it was shown that 
the $n$-point zero momentum connected correlation function $\chi_{{\rm FT},n}$
shows a scaling behavior of the form
\begin{equation}
u^{n-1} \chi_{{\rm FT},n} = f_{{\rm FT},n}(\tilde{t}_{\rm FT})
\qquad\qquad 
\tilde{t}_{\rm FT} \equiv [r - r_c(u)]/u.
\end{equation}
Next, we consider systems with medium-range interactions. Consider a square
lattice, Ising spins $\sigma_x$ at the sites of the lattice, 
and the Hamiltonian
\begin{equation}
{\cal H} = - {1\over2} \sum_{\mathbf{x}\mathbf{y}} 
    J(\mathbf{x}-\mathbf{y}) \sigma_\mathbf{x} \sigma_\mathbf{y}.
\end{equation}
We assume\footnote{One can also consider a much more general class of 
medium-ranged models, see Ref.~\cite{PRV-99}.} that 
$J(\mathbf{x}) = 1$ for $|x| \le  R_m$, $J(\mathbf{x}) = 0$ for $|x| > R_m$. 
The behavior of these
models is very similar to that observed in the Yukawa model, $R_m$ playing the 
role of $N$. For any finite $R_m$, the system belongs to the Ising universality
class, while for $R_m = \infty$ all spins are coupled together and 
one obtains mean-field behavior. In Ref.~\cite{LBB-96} it was shown that 
this model shows a crossover that interpolates between mean-field and Ising
behavior. If one defines an effective interaction range $R$ by 
\begin{equation}
R^2 = {\sum_{\mathbf{x},\mathbf{y}} (\mathbf{x} - \mathbf{y})^2 
        J(\mathbf{x} - \mathbf{y}) \over 
       \sum_{\mathbf{x},\mathbf{y}} 
        J(\mathbf{x} - \mathbf{y}) },
\end{equation}
then for $R,R_m\to \infty$, $t\equiv (T - T_c(R))/T_c(R) \to 0$
at fixed $\tilde{t}_{\rm MR} \equiv  t R^2$ one has 
\begin{equation}
   R^{4-3n} \chi_{{\rm MR},n} = f_{{\rm MR},n}(\tilde{t}_{\rm MR}),
\end{equation}
where $\chi_{{\rm MR},n}$ is the connected zero-momentum $n$-points correlation
function of the fields $\sigma$. In Ref.~\cite{PRV-99} it was shown 
that $f_{{\rm MR},n}(x)$ and $f_{{\rm FT},n}(x)$ are closely related. 
Indeed, we have 
\begin{equation}
    f_{{\rm MR},n}(x) = \mu_{1,\rm MR} \mu_{2,\rm MR}^n 
   f_{{\rm FT},n}(\lambda_{\rm MR} x),
\end{equation}
where $\mu_{i,\rm MR}$ and $\lambda_{\rm MR}$ are model-dependent constants
that reflect the arbitrariness in the definitions of the fields, of the range 
$R$, and of the scaling variable $\tilde{t}$. 
The constants can be computed using the results of Ref.~\cite{PRV-99},
sec.\ 4.2.\footnote{Note that the function 
$f_\chi(\hat{t})$ defined in Ref.~\cite{PRV-99} refers to correlations of the 
fields $\phi$ and not of the original fields $\varphi$. However, 
relation (4.12) of Ref.~\cite{PRV-99} 
shows that in the critical crossover limit 
$\sum_x \langle \varphi_0\varphi_x\rangle \approx 
\sum_x \langle \phi_0\phi_x\rangle$. The same holds for $\chi_4$.
The expression reported here are obtained from those reported in 
Ref.~\cite{PRV-99} by setting $\overline{a}_2 = 1$, $\overline{a}_4 = -2$,
$N=1$, $c_0 = \hat{c}_0 = \tau$, and $\tilde{t}_{\rm MR} = \hat{t} + \hat{c}_0$.}
Explicitly, we have for the two-point and four-point zero-momentum connected 
correlation functions:
\begin{eqnarray}
\chi_{{\rm MR},2} R^{-2} &=& {1\over \tilde{t}_{\rm MR}} + 
    {1\over 4 \pi \tilde{t}^2_{\rm MR}} 
    \left[ \ln \left({4\pi \tilde{t}_{\rm MR}\over3}\right) + 8 \pi D_2 + 3 
    \right] + 
    O(\tilde{t}^{-3}_{\rm MR}),
\nonumber \\
\chi_{{\rm MR},4} R^{-8} &=& -{2\over \tilde{t}^4_{\rm MR}} + 
    O(\tilde{t}^{-5}_{\rm MR}).
\end{eqnarray}
In the field-theory model we have instead
\begin{eqnarray}
u \chi_2 &=& {1\over \tilde{t}} + 
     {1\over 8\pi \tilde{t}^2} \left[ \ln {8 \pi \tilde{t}\over 3} + 
           8 \pi D_2 + 3 \right] + O(\tilde{t}^{-3}).
\\
u^2 \chi_4 &=& -{1\over \tilde{t}^4} + O(\tilde{t}^{-5}).
\end{eqnarray}
Comparing we obtain
\begin{equation}
\mu_{1,\rm MR} = 2, \qquad\qquad
\mu_{\rm MR} = \lambda_{\rm MR} = {1\over2}.
\end{equation}
In sec.~\ref{sec4} we have shown a similar relation for the Yukawa model.
If $x_t \equiv  - g_{01} N [T - T_c(N)]$, we find for the zero-momentum
correlation functions of the field $\phi$ 
\begin{equation}
\tilde{\chi}_n \equiv  g^n \chi_n = 
    N^{n/2-1} f_{Y,n}(x_t),
\end{equation}
and 
\begin{equation}
    f_{Y,n}(x) = \mu_{1,Y} \mu_{2,Y}^n f_{{\rm FT},n}(\lambda_{Y} x).
\end{equation}
In order to compute these constants we compare the one-loop expansions of
the two-point function in field theory and in the Yukawa model.
In the Yukawa model we find
\begin{eqnarray}
\int \di^d \mathbf{x} \langle \chi(\mathbf{0}) \chi(\mathbf{x}) \rangle
   &=& {N T\over \alpha^2 x_t} - \left({N T\over \alpha^2 x_t}\right)^2 r_c 
 \nonumber \\
   && - {1\over 2N} \left({N T\over \alpha^2 x_t}\right)^2
      \int_{p<\Lambda} {\di^2 \vp\over (2\pi)^2} 
      {\overline{V}^{(4)}(\vp,-\vp,\mathbf{0},\mathbf{0}) \over 
       \ol{P}(\vp)} + O(x_t^{-3}).
\end{eqnarray}
Using the explicit expression for $r_c$ we obtain 
\begin{eqnarray}
\tilde{\chi}_2 = {1\over x_t} + 
     {\alpha^2 V_4(\mathbf{0},0) \over 8\pi x_t^2} 
     \left[ \ln {8 \pi x_t\over 3 \alpha^2 V_4(\mathbf{0},0)} + 
            8 \pi D_2 + 3 \right] + O(x_t^{-3}).
\end{eqnarray}
For the four-point function we have instead
\begin{equation}
\tilde{\chi}_4 N^{-1} = - {V_4(\mathbf{0},0)\over x_t^4} 
            + O(x_t^{-5}).
\end{equation}
Comparing we obtain 
\begin{equation}
\mu_{1,Y} = \alpha^4 V_4(\mathbf{0},0), \qquad\qquad
\mu_{2,Y} = {1\over \alpha^3 V_4(\mathbf{0},0)}, \qquad\qquad
\lambda_Y = {1\over \alpha^2 V_4(\mathbf{0},0)}.
\end{equation}
The constants reported in sec.~\ref{sec4.3} are easily derived:
\begin{eqnarray}
&& K_{\rm FT} = - \lambda_Y g_{01}, \qquad \qquad 
   K_{2,\rm FT} = \mu_{1,Y} \mu_{2,Y}^2 , 
\\
&& K_{\rm LBB} = - {\lambda_Y g_{01}\over \lambda_{\rm MR}}, \qquad \qquad 
   K_{n,\rm LBB} = {\mu_{1,Y} \mu_{2,Y}^n \over  
                    \mu_{1,\rm MR} \mu_{2,\rm MR}^n },
\end{eqnarray}
where $n = 1,2$. Note that $g_{01}$ is negative, so that $K_{\rm FT}$ and $K_{\rm LBB}$ are positive as expected.

The generalization to $\N=1$ Heisenberg models is straightforward
and can be obtained repeating the previous considerations
of  the Yukawa model with which shares
the same large $N$ expansion.

%
% bibliography
%

\end{document}

%% file: TXSsymb.tex
\def\frac#1#2{{#1\over#2}}

\def\half{\ifinner {\scriptstyle {1 \over 2}}%
          \else {\textstyle {1 \over 2}}\fi}

\def\simge{%  ``greater than about'' symbol
    \mathrel{\rlap{\raise 0.511ex
        \hbox{$>$}}{\lower 0.511ex \hbox{$\sim$}}}}

\def\simle{%  ``less than about'' symbol
    \mathrel{\rlap{\raise 0.511ex
        \hbox{$<$}}{\lower 0.511ex \hbox{$\sim$}}}}

\def\therefore{% three-dot symbol for ``therefore''
   \setbox0=\hbox{$.\kern.2em.$}\dimen0=\wd0    %
   \mathrel{\rlap{\raise.25ex\hbox to\dimen0{\hfil$\cdotp$\hfil}}%
   \copy0}}

\def\|{\ifmmode\Vert\else \char`\|\fi}

\def\tr{\mathop{\rm tr}\nolimits}       % tr for trace
\def\Tr{\mathop{\rm Tr}\nolimits}       % Tr for functional trace
\def\sc#1{\setbox0=\hbox{$#1$}           % set a box for #1
   \dimen0=\wd0                                 % and get its size
   \setbox1=\hbox{/} \dimen1=\wd1               % get size of /
   \ifdim\dimen0>\dimen1                        % #1 is bigger
      \rlap{\hbox to \dimen0{\hfil/\hfil}}      % so center / in box
      #1                                        % and print #1
   \else                                        % / is bigger
      \rlap{\hbox to \dimen1{\hfil$#1$\hfil}}   % so center #1
      /                                         % and print /
   \fi}                                         %

\def\subrightarrow#1{%                          % #1 under arrow
  \setbox0=\hbox{%                              % set a box
    $\displaystyle\mathop{}%                    % no mathop
    \limits_{#1}$}%                             % just limits
  \dimen0=\wd0%                                 % get width
  \advance \dimen0 by .5em%                     % add a bit
  \mathrel{%                                    % space like =
    \mathop{\hbox to \dimen0{\rightarrowfill}}% % arrow to width
       \limits_{#1}}}                           % text below

\def\vbig#1#2{{\vbigd@men=#2\divide\vbigd@men by 2%
   \hbox{$\left#1\vbox to \vbigd@men{}\right.\n@space$}}}